	\documentclass[10pt,a4paper,oneside,openany]{book}
	\usepackage{graphics,graphicx} 
	\usepackage[utf8]{inputenc} 
	\usepackage{palatino}     
	\usepackage[caption=false]{subfig}
	\usepackage{color}
	\usepackage{pdfpages}
	\usepackage{wrapfig}
	\usepackage{enumerate}
	\usepackage{bm}
	\usepackage[Lenny]{fncychap}
	\usepackage{fancyhdr} 
	\usepackage[all]{xy}
	\usepackage{tensor}
	\usepackage[top=2.0cm,left=2.5cm,right=2.0cm,bottom=2.0cm,a4paper]{geometry}
	\usepackage[version=3]{mhchem}
	\usepackage{epstopdf}
	\usepackage[normalem]{ulem} 

	\usepackage[colorlinks,citecolor=red,linkcolor=blue,urlcolor=blue]{hyperref}

	\usepackage{amsmath} 
	\usepackage{amsthm} 
	\usepackage{amssymb}	
	\usepackage[numbers]{natbib}
	\usepackage[toc,page]{appendix}
	\makeatletter 
	\renewcommand{\v}[1]{\ensuremath{\mathbf{#1}}} 
	 
	\newcommand{\ket}[1]{\left| #1 \right>} 
	\newcommand{\bra}[1]{\left< #1 \right|} 
	\newcommand{\braket}[2]{\left< #1 \vphantom{#2} \right|
	 \left. #2 \vphantom{#1} \right>} 
	\let\baraccent=\= 
	\renewcommand{\=}[1]{\stackrel{#1}{=}} 

	\theoremstyle{definition}
	
	\theoremstyle{remark}

	\usepackage{tikz}
	\usepackage{pgffor}
	\usepackage{pgfplots}
	\usetikzlibrary{calc,trees,positioning,arrows,chains,shapes.geometric,%
		decorations.pathreplacing,decorations.pathmorphing,shapes,%
		matrix,shapes.symbols,plotmarks,decorations.markings,shadows}
	\usetikzlibrary{math}
	\usetikzlibrary{graphs}
	\usetikzlibrary{arrows.meta}
	\usetikzlibrary{decorations}
	\usetikzlibrary{decorations.markings}
	\usetikzlibrary{snakes}
	\usetikzlibrary{patterns}
	\usetikzlibrary{datavisualization.formats.functions}
	\usetikzlibrary{automata,positioning,arrows}
	\usetikzlibrary{external}  
	\definecolor{myblue}{rgb}{0, 0.25, 1}
	\definecolor{myred}{rgb}{1, 0, 0.25002}
	\makeatletter
	\renewcommand{\chaptermark}[1]{\markboth{\textsc{\@chapapp}\ \thechapter:\ #1}{}}
	\makeatother
	
	\renewcommand{\chaptermark}[1]%
	{\markboth{{\chaptername\ \thechapter.\ #1}}{}} 
	
	\renewcommand{\sectionmark}[1]%
	{\markright{{\thesection.\ #1}}}

	\fancypagestyle{plain}
	{%
		\fancyhead{} 
	}
	
	\makeatletter
	\def\cleardoublepage{\clearpage\if@twoside
		\ifodd\c@page
		\else\hbox{}\thispagestyle{empty}\newpage
		\if@twocolumn\hbox{}\newpage\fi\fi\fi}
		
		\newenvironment{alwayssingle}
		{%
		\@restonecolfalse
		\if@twocolumn\@restonecoltrue\onecolumn
		\else\newpage\fi
		}
		{\if@restonecol\twocolumn\else\newpage\fi}
		
\usepackage{epigraph}
\usepackage{float}

\newcommand*{\signatureDhiman}{%
	\par\makebox[5cm][c]{13/08/2020} \hfill \makebox[5cm] {\includegraphics[width=0.24\textwidth]{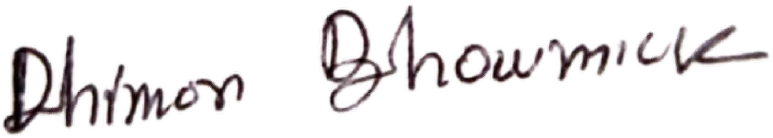}} 
	\par\makebox[5cm] {\dotfill} \hfill \makebox[5cm] {\dotfill} %
	\par\makebox[5cm] {\centering Date}\hfill \makebox[5cm] {\centering Dhiman Bhowmick}%
	}

\newcommand*{\signaturePinaki}{%
	\par\makebox[5cm][c]{13/08/2020} \hfill \makebox[5cm] {\includegraphics[width=0.2\textwidth]{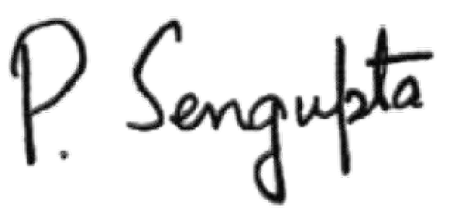}} 
	\par\makebox[5cm] {\dotfill} \hfill \makebox[5cm] {\dotfill} %
	\par\makebox[5cm] {\centering Date}\hfill \makebox[5cm] {\centering Pinaki Sengupta}%
	}

\usepackage{calrsfs}
	\DeclareMathAlphabet{\pazocal}{OMS}{zplm}{m}{n}

\usepackage{esint}

\usepackage{ulem}
\usepackage{todonotes}
\newcommand{\addref}[1]{\todo[color=red!40]{Add reference.}}

\newcommand{\icap}{\hat{i}}
\newcommand{\jcap}{\hat{j}}
\newcommand{\kcap}{\hat{k}}
\newcommand{\uu}[1]{u_{#1,\bold{k}}}
\newcommand{\uuu}[2]{[u_{#1,\bold{k}}]_{#2}}
\newcommand{\dudkx}[1]{\frac{\partial u_{#1,\bold{k}}}{\partial k_x}}
\newcommand{\dudky}[1]{\frac{\partial u_{#1,\bold{k}}}{\partial k_y}}
\newcommand{\gradK}{\vec{\nabla}_\bold{k}}
\newcommand{\ddkx}{\frac{\partial\quad}{\partial k_x}}
\newcommand{\ddky}{\frac{\partial\quad}{\partial k_y}}
\newcommand{\dXdY}[2]{\frac{\partial #1}{\partial #2}}
\newcommand{\ddY}[1]{\frac{\partial }{\partial #1}}
\newcommand{\squeeze}[3]{\left\langle #1\middle| #2\middle| #3\right\rangle}
\newcommand{\bk}{\bold{k}}

\newcommand{\cdag}[1]{\hat{b}^\dagger_{#1,s}}
\newcommand{\cdagU}[1]{\hat{b}^\dagger_{#1,\uparrow}}
\newcommand{\cdagD}[1]{\hat{b}^\dagger_{#1,\downarrow}}
\newcommand{\cndag}[1]{\hat{b}_{#1,s}}
\newcommand{\cndagU}[1]{\hat{b}_{#1,\uparrow}}
\newcommand{\cndagD}[1]{\hat{b}_{#1,\downarrow}}
\newcommand{\CHI}[0]{\hat{\chi}_{ij}}
\newcommand{\CHIU}[0]{\hat{\chi}_{ij,\uparrow}}
\newcommand{\CHID}[0]{\hat{\chi}_{ij,\downarrow}}

\newcommand{\mn}{m\boldsymbol{\alpha}_1+n\boldsymbol{\alpha}_2}
\newcommand{\pq}{p\boldsymbol{\alpha}_1+q\boldsymbol{\alpha}_2}
\newcommand{\squeezeA}[2]{\left\lbrace #1 , #2\right\rbrace}

\newcommand{\capk}[1]{\hat{#1}_{\bold{k}}}
\newcommand{\capmk}[1]{\hat{#1}_{-\bold{k}}}

\newcommand{\capdagk}[1]{\hat{#1}^\dagger_{\bold{k}}}

\newcommand{\expk}[1]{e^{i\bold{k}\cdot\bold{#1}}}
\newcommand{\expmk}[1]{e^{-i\bold{k}\cdot\bold{#1}}}

\newcommand{\doneN}[0]{\delta_1}
\newcommand{\dtwo}[0]{\delta_2}

\newcommand{\squeezeD}[3]{\left\langle #1\middle| #2\middle| #3\right\rangle}

\newcommand{\squeezeN}[2]{\left\lbrace #1 \middle| #2\right\rbrace}

\newcommand{\txdag}[1]{\hat{\tilde{t}}^{x\dagger}_{\bold{r}_i+\boldsymbol{\delta}_{#1}}}
\newcommand{\tx}[1]{\hat{\tilde{t}}^{x}_{\bold{r}_i+\boldsymbol{\delta}_{#1}}}
\newcommand{\tydag}[1]{\hat{\tilde{t}}^{y\dagger}_{\bold{r}_i+\boldsymbol{\delta}_{#1}}}

\newcommand{\tzdag}[1]{\hat{\tilde{t}}^{z\dagger}_{\bold{r}_i+\boldsymbol{\delta}_{#1}}}

\newcommand{\txdagn}{\hat{\tilde{t}}^{x\dagger}_{\bold{r}_i}}
\newcommand{\txn}{\hat{\tilde{t}}^{x}_{\bold{r}_i}}
\newcommand{\tydagn}{\hat{\tilde{t}}^{y\dagger}_{\bold{r}_i}}
\newcommand{\tyn}{\hat{\tilde{t}}^{y}_{\bold{r}_i}}

\newcommand{\tzn}{\hat{\tilde{t}}^{z}_{\bold{r}_i}}
\newcommand{\tmudagn}{\hat{\tilde{t}}^{\mu\dagger}_{\bold{r}_i}}
\newcommand{\tmun}{\hat{\tilde{t}}^{\mu}_{\bold{r}_i}}

\usepackage{listings}
\usepackage{tikz}
\usetikzlibrary{calc}

\newcommand{\braHket}[3]{\left\langle #1 \middle| #2 \middle| #3 \right\rangle}

\usepackage{relsize}

\usepackage{xcolor}

\usepackage[nonumberlist,abbreviations]{glossaries-extra}

\makenoidxglossaries
\newabbreviation{DMI}{DMI}{Dzyaloshinskii-Moriya Interaction}
\newabbreviation{HP}{HP}{Holstein-Primakoff}
\newabbreviation{LSWT}{LSWT}{Linear Spin Wave Theory}
\newglossaryentry{triplon}
{
    name=triplon,
    description={triplons are bosonic qusiparticles representing the quasi-particle correspond to the triplet state of two spin half sites in bond-operator formalism}
}
\newglossaryentry{spinon}
{
    name=spinon,
    description={spinons are spin quasi-particles correspond to the Schwinger boson in Schwinger Boson mean field theory}
}

\newglossaryentry{magnon}
{
    name=magnon,
    description={magnon is a quanta of spin-waves or spin-excitations in a long-range ordered system. In other words magnon is a Holstein-Primakoff boson.}
}

\newabbreviation{sbmft}{SBMFT}{Schwinger Boson mean field theory}
\newabbreviation{bdg}{BdG}{Bogoliubov-de Gennes Hamiltonian}
\newabbreviation{dssf}{DSSF}{Dynamical Spin Structure Factor}
\newabbreviation{ss}{SS}{Shastry Sutherland}

\begin{document}
\frontmatter
\includepdf[pages={1}]{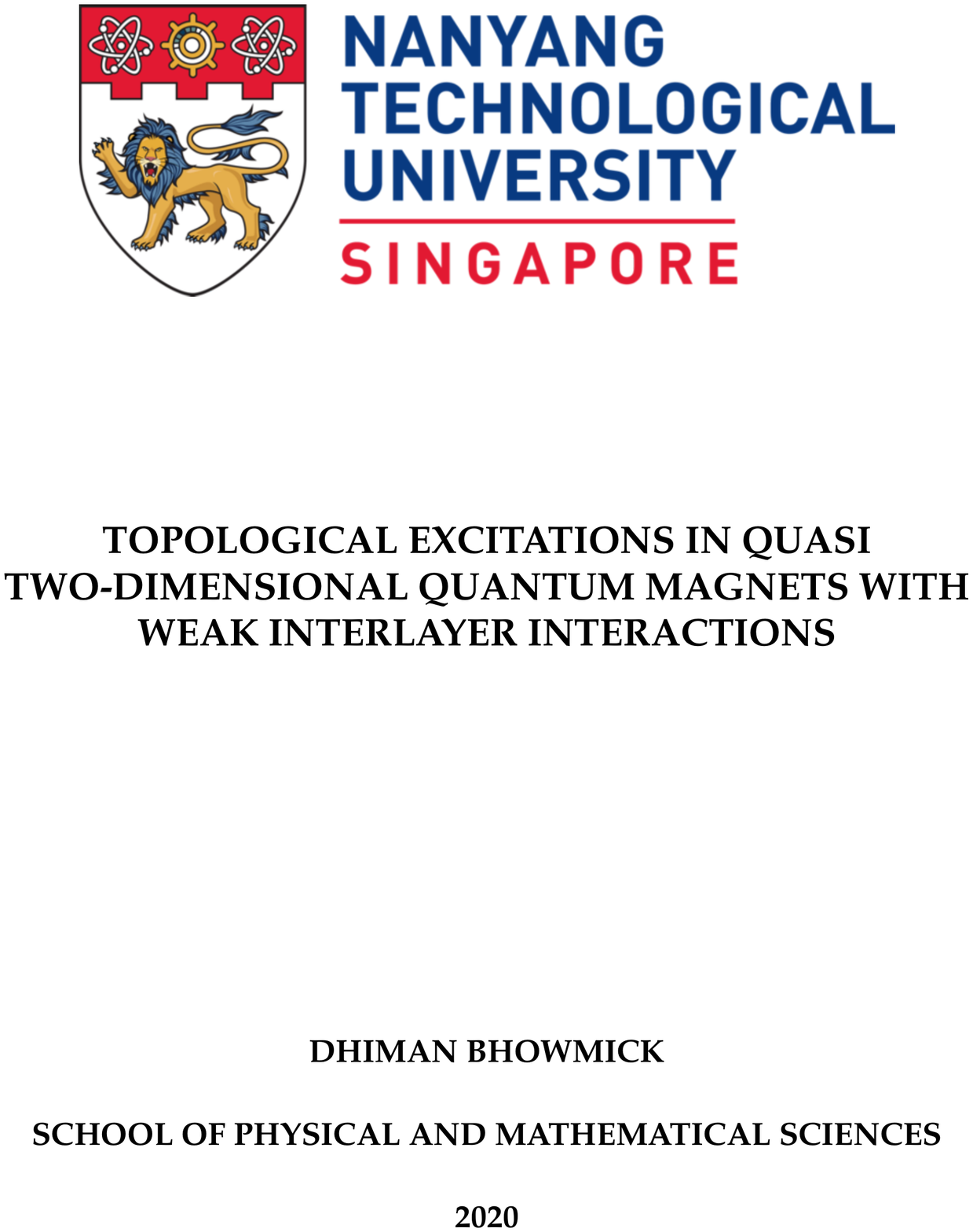}
\includepdf[pages={1}]{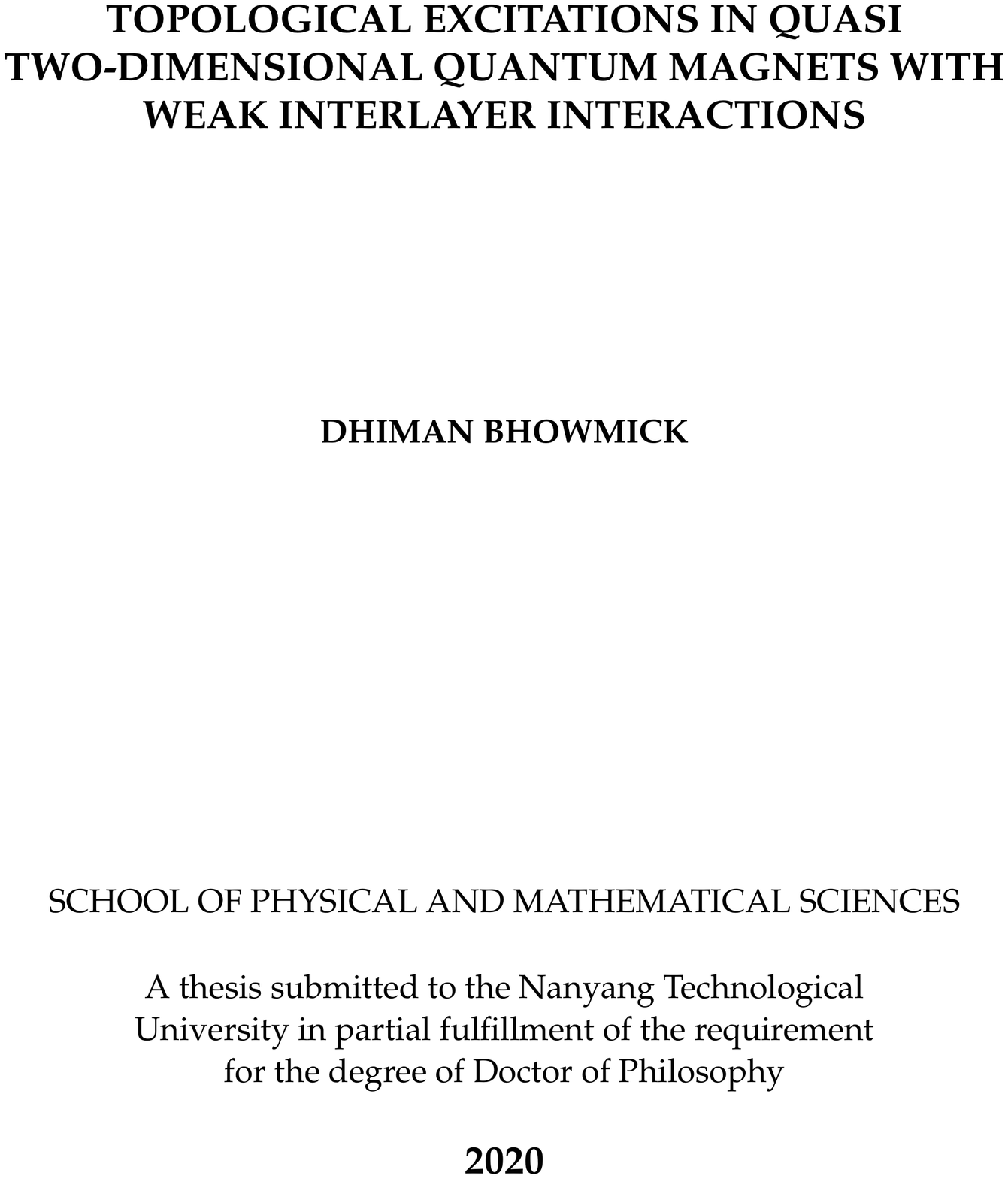}

\chapter*{}
\parbox{\linewidth}{\centering {\Large Statement of Originality}}
I hereby certify that the work embodied in this thesis is the result of original research done by me except where otherwise stated in this thesis. The thesis work has not been submitted for a degree or professional qualification to any other university or institution. I declare that this thesis is written by myself and is free of plagiarism and of sufficient grammatical clarity to be examined. I confirm that the investigations were conducted in accord with the ethics policies and integrity standards of Nanyang Technological University and that the research data are presented honestly and without prejudice.

\vspace{5cm}
\signatureDhiman

\chapter*{}
\parbox{\linewidth}{\centering {\Large Supervisor Declaration Statement}}
I have reviewed the content and presentation style of this thesis and declare it of sufficient grammatical clarity to be examined. To the best of my knowledge, the thesis is free of plagiarism and the research and writing are those of the candidate’s except as acknowledged in the Author Attribution Statement. I confirm that the investigations were conducted in accord with the ethics policies and integrity standards of Nanyang Technological University and that the research data are presented honestly and without prejudice.

\vspace{5cm}
\signaturePinaki

\chapter*{}
\parbox{\linewidth}{\centering {\Large Authorship Attribution Statement}}
This thesis contains material from two papers published in the following peer-reviewed journals. 

Chapter-3 is published as D. Bhowmick and P. Sengupta.  Antichiral edge states in Heisenberg ferromagnet on a honeycomb lattice.  Physical Review B 101, 195133 (2020). DOI: 10.1103/PhysRevB.101.195133.

Chapter-4 is published as D. Bhowmick and P. Sengupta.  Topological magnon bands in the flux state of Shastry-Sutherland lattice model.  Physical Review B 101, 214403 (2020). DOI: 10.1103/PhysRevB.101.214403.

Chapter-5 is on the arXiv repository as D. Bhowmick and P. Sengupta.  Weyl-triplons in \ce{SrCu2(BO3)2}. arXiv identifier: arXiv:2004.11551. (under review)

The contributions of the co-authors in all papers are as follows:
\begin{itemize}
	\item I and Pinaki Sengupta provided the initial project direction.
	\item  All analytical calculations and programming codes were done by me, and I performed the numerical simulations.
	\item The results were analyzed by me with the help of Professor Sengupta.
	\item I and Professor Sengupta prepared the manuscript drafts.
\end{itemize}

\vspace{2cm}
\signatureDhiman

\phantomsection
\addcontentsline{toc}{chapter}{Abstract}
	\chapter*{Abstract}
The study of topological magnetic excitations have attracted widespread attention in the past few years.
	The wide variety of ground state phases realized in different two-dimensional magnets have emerged as versatile platforms for realizing magnetic analogues  of topological phases uncovered in electronic systems over the past two decades.
	Dzyaloshinskii-Moriya(DM) interactions, that are ubiquitous in many quantum magnets, and has been demonstrated to induce a non-trivial topology in the magnetic excitations in many quantum magnets.
	In particular, signatures of DM-interaction induced topological features observed in the dispersion of magnetic excitations in the geometrically frustrated Shastry-Sutherland compound \ce{SrCu2(BO3)2} and honeycomb ferromagnet \ce{CrI3} have provided the motivation behind the bulk of the research presented in this thesis. 
	In the first chapter we introduce the background and motivation of our study.
	We showed that the vastly available materials with a possibility of non-zero symmetry allowed DM-interaction motivate us to study the minimal models of interacting spins related to the materials.

	In the second chapter we have described the methods that is used to study the spin systems.
	To describe the spin excitations in the long-range ordered system and dimerized system, we use Holstein-Primakoff transformation and bond-operator formalism respectively.
	The Schwinger Boson mean field theory is applicable for any generic magnetic ground state and useful to study the system at high temperatures compared with Holstein-Primakoff transformation.
	After transforming the Hamiltonian into a quadratic bosonic Hamiltonian, it is diagonalized by using Bogoliubov-Valatin transformation.
	Different observables like Berry-curvature, Chern-number, Dynamical Spin structure factor etc. are calculated by using the single particle wave-function obtained from Bogoliubov-Valatin transformation.
	A non-zero Berry curvature is a signature of topologically protected edge states in a stripe geometry of a lattice and a systematic way of calculating edge states is also explained in the second chapter.
	Moreover in the second chapter the usefulness of symmetry in obtaining band degeneracies and the allowed spin-spin interaction terms are also described.

	The past theoretical studies show the presence of topological magnons in honeycomb ferromagnetic models due to presence of DM-interaction on a next-nearest neighbour bonds.
	Recently this kind of DM-interaction is detected in the honeycomb ferromagnet \ce{CrI3}.
	This type of DM-interaction induces chiral edge states of magnon in the system and so we named it as chiral DM-interaction.
	In the third chapter we show that breaking of inversion symmetry at the center of honeycomb structure gives rise to a extra DM-interaction which we named as anti-chiral DM-interaction.
	We studied the system by using Schwinger Boson mean field theory and Holstein-Primakoff transformation and show that when the antichiral DM-interaction dominates over the chiral DM-interaction, the direction of velocity of the edge states become in the same direction\,(antichiral edge states).
	Due to conservation of total number of spins in the system a bulk current flows in the opposite direction relative to the antichiral edge current.
	In this study we suggested the way to break the inversion symmetries in the materials \ce{CrSrTe3}, \ce{CrGrTe3}, \ce{AFe2(PO4)2} (A=Ba, Cs, K, La) to achieve antichiral DM-interaction.
	The presence of antichiral edge states can be detected by using inelastic neutron scattering by detecting the band tilting at $K$ and $K'$ points of magnon bands.
	Moreover we showed that spin noise spectroscopy at the edges is useful measurements to detect the presence of antichiral edge modes.
	Magnetic force microscopy is also a promising tool to detect the antichiral magnons at edges.

	The underlying spin-Hamiltonian of the antiferromagnetic systems in the materials like rare-earth tetraborides(\ce{RB4}, R=Er,Tm) and \ce{U2Pd2In} can be described by using extended Shastry-Sutherland models.
	In previous theoretical studies it is shown that there is a non-collinear spin-state known as flux state exists in the Shastry-Sutherland model in presence of out of plane DM-interaction.
	Although the presence of perpendicular component of DM-interactions in the rare-earth materials are not known, it can be induced artificially by using circularly polarized light or using heavy-metal alloys. 
	In the fourth chapter, we studied the magnetic excitations in the flux state on Shastry-Sutherland lattice incorporating realistic in-plane DM-interactions\,(using the DM-interactions present in the low symmetry crystal structure of \ce{SrCu2(BO3)4}) by using Holstein-Primakoff transformation and showed the presence of topological magnon bands with non zero Chern-numbers in the system.
	We found a variety of band topological transitions and showed that each band-topological transition is associated with the logarithmic divergence in the derivative of the thermal Hall conductance. 
	We derived a analytical expression for the temperature dependence of the derivative of thermal Hall conductivity near band topological transition point for a generic spin model.
	This is useful to extrapolate the energy of band touching point during the band topological phase transition by using thermal Hall effect experiment.

	In the fifth chapter the spin systems related to Shatry-Sutherland lattice is described.
	There are three possible phases in Shastry-Sutherland model and those are (i) dimer-phase, (ii) plaquette order phase and (iii) Nèel phase.
	The dimer-phase material \ce{SrCu2(BO3)4} possesses two crystal symmetries; one is low symmetry phase\,(at low temperature) in which case both the in plane and out of plane DM-interactions are allowed; another is high symmetry phase\,(at high temperature) in which only the out of plane DM-interaction is allowed.
	We use bond operator formalism to show the existence of Weyl-triplons in low crystal symmetry structures of Shastry-Sutherland lattice.
	We predict that the low symmetry phase of \ce{SrCu2(BO3)4} at low temperature should contain Weyl-triplons in presence of any finite interlayer perpendicular DM-interactions and at low temperature the Weyl-triplon phenomenon is not altered in presence interlayer in-plane DM-interaction and Heisenberg interaction. 
	There are several band topological transitions happen by changing the external magnetic field and interlayer DM-interaction\,(which might be varied by applying a pressure).
	The topological nature of Weyl triplons are confirmed by the presence of the non-zero Berry-curvature and monopole charge of Weyl-point in the bulk system.
	Again topological non-triviality is further confirmed by showing that at the surface of the material the Weyl-points are connected by Fermi-arc like surface states which is possible to detect by neutron scattering.
	Using Kubo-formula of thermal conductivity it is shown that the thermal Hall conductivity of the triplons has a  quasi-linear dependence as a function of magnetic field in the Weyl-triplon region and this functional feature is absent in topologically trivial or non-trivial gapped triplon bands.
	
\phantomsection
\addcontentsline{toc}{chapter}{Acknowledgement}
\chapter*{Acknowledgements}
\indent\indent First of all, I would like to thank my research supervisor Prof. Pinaki Sengupta for his continuous support and guidance throughout my Ph.D. study on and off the field. 
whenever I went to his office with a problem he always welcomed me and assisted me. 
He has encouraged and inspired me to become an independent researcher and developed my critical thinking. 
I am really grateful to him as a person who introduced me the field of research.
Moreover my scientific writing and presentation skills also improved due to his suggestions.
I am really very thankful for being so flexible and allowing me to choose the work and project which interests me most. 
 I am really lucky to have such fabulous supervisor.

\indent I met many professors and lecturers in my life who inspired me and shaped my thinking through the journey from high school to masters.
I have completed my secondary and higher secondary education from Barisha High School in Kolkata.
I am very glad to be student under the teachers Anup Maity(Math), Brindaban Sardar(English), Arindam Mandal(Physics), Indranil Dutta (Computer Science), Prabir Roy(Biology).
I have achieved my Bachelors degree in physics from Vivekananda College under Calcutta university.
I am also very proud to be a student under former professor Chapal Kumar Chattopadhyay, associate professor Arivind Pan.
I am really motivated by the teaching of former professor Chapal Kumar Chattopadhyay.
According to him, it is easier to forget the equations in physics and the only way to remember physics is to understand the physics in your own way or language.
 I always have a motivation to make myself a influential physics professor like him.
I pursued my masters degree in physics from Indian Institute of Technology(IIT), Kharagpur.
I am very happy to be student under Professors Sugata Pratik Khastgir, Sayan Kar, Debraj Choudhury, Shivkiran B.N. Bhaktha, Ajay Kumar Singh, Somnath Bharadwaj, Maruthi Manoj Brundavanam, Samudra Roy, Samit Kumar Ray, Krishna Kumar and others.
Special thanks to professor Sugata Pratik Khastgir for his advice on choosing coursework in my Master's before pursuing a Ph.D., and thanks for his suggestion on understanding any formalism in physics through examples and thanks for his teaching in group-theory.
Thanks to all other professors and teachers whom I have not mentioned in the acknowledgment, but they are equally important for my learning experience throughout my student life.

\indent I would also like to thank Prof. Justin Song and Prof. Shengyuan Yang for being my thesis advisory committee members. I would like to acknowledge, in no particular order, the current and ex-members of our group as good friends, colleagues and well-wisher, Munir Shahzad, Ong Teng Siang Ernest, Rakesh Kumar, Nyayabanta Swain, Bobby Tan, Ho-Kin Tang. I am thankful to my friends, in no particular order, Shampy Mansha, Md Shafiqur Rahman, Bivas Mondal, Deblin Jana, Jit Sarkar, Mrinmoy Das, Bapi Dutta, Pankaj Majhi, Sourav Mitra, Rakesh Maiti for making my stay at NTU memorable and fruitful. Again I like to thank my friends Bhartendu Satywali and Subhaskar Mandal, sharing their knowledge particularly in the field of my study. In four years of staying in Singapore it was also a great experience making friends with Singaporean, again in no particular order, Ong Teng Siang Ernest, Eugene Tay, Png Kee Seng, Zhao Zhuo, Alex Lim; again thanks to them being part of my journey as a Ph.D. student in Singapore. Moreover I am thankful to Jareena being a special part of my life; taking care and supporting me through this Ph.D. journey. Finally, I would like to express my deepest appreciation to my Family- my mother Minati Bhowmick, my father Dhanu Kumar Bhowmick, my cousin Mampi Guha Majumder and my aunty Jharna Guha Majumder, for their endless support and unconditional love all the way long. I appreciate it a lot and love you all!!!

\phantomsection
\renewcommand{\contentsname}{Table of Contents}
\tableofcontents
\newpage
\phantomsection
\addcontentsline{toc}{chapter}{List of Figures}
\listoffigures
\addcontentsline{toc}{chapter}{List of Tables}
\listoftables
\glsunsetall
\printnoidxglossaries
\glsresetall

\newpage
\mainmatter
	
	\chapter{Introduction}\label{chapter01}
	
	\section{Background and Motivations}\label{sec:chap01-intro}

 Condensed Matter physics is the branch of physics devoted to understanding the emergence of macroscopic properties of matter from (local) microscopic interactions between constituent particles and their arrangement. 
 These can be molecules, atoms, electrons or quasi-particles such as phonons or magnons. 
 On the theoretical front, this involves constructing microscopic Hamiltonians and solving for the emergent properties using exact or (more often) approximate analytica methods, and numerical simulations.
 On the experimental front, this involves measuring definitive physical properties and interpreting them in terms of microscopic modeling.
Although the fundamental building block that make up all materials are atoms, different arrangements of the atoms in different  materials give rise to different properties, this is known as principle of emergence.
Landau\,\cite{Landau} discovered that the key to understand different phases of the materials are related to different symmetries of the system and phase transitions are associated with change in symmetry of that system.
Later Ginzburg and Landau\,\cite{LandauGinzburg} developed a general theory for phase transition based on the idea of order parameter which transforms non-trivially under phase transition or change in the symmetry.
In last three decades a different picture has emerges following the discovery of integer quantum Hall effect which was experimentally detected in two-dimensional electron gas in MOSFET at low temperature and at high magnetic field\,\cite{Klitzing}, just after theoretical prediction of this novel phenomenon in two-dimensional material\,\cite{TheoryOfHall}.
The phase transitions of different plateaus observed in the integer quantum Hall effect can not be described in the framework of Landau's symmetry breaking theory, because in different plateau neither the symmetry of the material changes nor any local order-parameter of the system alters.
Following the work of Laughlin, we know that the integer quantum Hall states arise due to the presence of the edge states present at the boundary of the system.
These edge states emerges due to non-trivial topology of the bulk wave function of the system which is quantified by using Chern or  Thouless–
Kohmoto–Nightingale–den Nijs number\,\cite{Thouless,Kohomoto}.
This topological invariant can only be changed by closing bulk energy gap of the system and that is why the bulk of the integer Hall quantum system become conducting during the transitions between two plateau regions.
It was long thought that topological states are rare and it is only possible under extreme conditions.
However, with the advent of spin-orbit induced topological insulators, it became clear that topological quantum states are more abundant than previously thought.
Haldane’s paradigmatic model\,\cite{Haldane} showed the possibility of topological bands with non-zero Chern-numbers in the band-structure for a tight binding model of spinless particles on a honeycomb lattice with complex next-nearest-neighbor hopping.
Later Kane and Mele\,\cite{Kane} illustrated that the complex hopping term naturally emerges from spin-orbit interactions in the electron system and at the same time it was realized experimentally in the materials HgTe/CdTe semiconductor quantum wells\,\cite{SOIinduced1, SOIinduced2}, in InAs/GaSb heterojunctions sandwiched by AlSb\,\cite{SOIinduced3,SOIinduced4}, in BiSb alloys\,\cite{SOIinduced5}, in Bi2Se3\,\cite{SOIinduced6,SOIinduced7}, and in many other systems\,\cite{SOIinduced8,SOIinduced9}.
Unlike Haldane's model, the spin orbit induced topological electronic band systems are protected by timer-reversal symmetry i.e. in the absence of time-reversal symmetry, it is possible to adiabatically transform the spin-orbit-induced topological insulators into a topologically trivial state without closing the bulk gap.
The topological invariant that correspond to these time-reversal symmetry protected topological state is $\mathbb{Z}_2$-invariant and the boundary of the system manifests gap-less Helical edge modes which consists pair of counter propagating electronic states.
The experimental manifestation of these Helical edge modes is the novel quantum spin Hall effect.
In case of three dimensional materials, topological nodal semimetals are gapless systems which exhibits band topology even when the bulk band gap closes at certain points in the Brillouin zone.
Dirac-semimetals (\ce{Na3Bi}\,\cite{Na3Bi}), Weyl-semimetals (\ce{TaAs}\,\cite{TaAs1,TaAs2}), nodal-line semimetals\,\cite{DiracNodalLine1,DiracNodalLine2,DiracNodalLine3} are the examples of gapless topological semimetals.

Band topological systems such as those mentioned above can be realized within a free electron system without any electron-electron interaction, where spin orbit coupling plays an important role to induce topology.
Interactions also play an important role to generate a avenue of new kinds of topological states, like fractional topological insulators\,\cite{FQH1,FQH2,FQH3,FQH4}, interaction induced topological insulators\,\cite{TopologicalMott1,TopologicalMott2,TopologicalMott3,TopologicalMott4}, quantum spin-liquids\,\cite{QSL1,QSL2,QSL3}, Haldane spin-$1$ chain\,\cite{HaldaneChain1,HaldaneChain2,HaldaneChain3} and so on.

A promising place to look for the interplay between interaction and spin-orbit interaction induced topological states is to study the physics of widely available magnetic ground states in different lattices.
In these systems, the electron's charge degrees of freedom are frozen due to large on-site and neighbouring site electron-electron interactions, but spin degrees of freedom of electrons remain due to the quantum fluctuations present in the system.
These separated spin-degrees of freedoms are represented as bosonic quasi-particles known as \gls{spinon}.
At low temperature specific flavour of \gls{spinon}s Bose-Einstein condensate to form the magnetic ground state of the system and remaining \gls{spinon} flavours constitute magnetic excitations above the magnetic ground state, which are known as spin-waves or \gls{magnon}s.
In magnetic systems the spin-orbit coupling shows up in the form of \gls{DMI} and anisotropic interactions\,\cite{Dzyaloshinsky,Moriya}.
Presence of the \gls{DMI} introduces non-trivial topology in the \gls{magnon} bands.
The charge neutral magnons carry information in the form of magnetic moment and being a charge neutral particle magnons do not respond to an external electric field and consequently, do not exhibit conventional Hall effect. 
However a temperature gradient across the topological magnon material induces transverse thermal magnon Hall conductivity.
Thus the study of toplogical magnons is very important in the emerging field of magnon based spintronic devices.
Transport of magnons does not produce any Joule heating effect in the device because of its' charge neutral nature.
So the \gls{magnon} based spintronic devices are very attractive with regard to low waste energy production and power consumption. 
For their successful engineering, it is necessary to understand the fundamental transport properties of \gls{magnon}s.

All of the particles electrons, phonons and \gls{magnon}s can cause a non-zero thermal Hall effect.
The \gls{magnon} induced Thermal Hall effect was first proposed by Katsura et al.\,\cite{Katsura} in presence of interaction due to spin-chirality (order of $\frac{t^3}{U^2}$, where $t$ is the hopping amplitude and $U$ is the onsite repulsion in Hubbard model) in the Kagome lattice.  
 Later on Onose et al.\,\cite{THE1} first discovered the \gls{DMI} induced \gls{magnon} thermal Hall effect of the insulator \ce{Lu2V2O7}, which is basically a pyrochlore lattice made of ferromagnetically ordered spins on vanadium atoms below the Curie temperature $T_C=70K$.
 Afterwards, several pyrochlore materials \ce{Ho2V2O7}, \ce{In2Mn2O7}\,\cite{THE3} are observed to exhibit the thermal \gls{magnon} Hall effect phenomenon below Curie temperature.
  The two-dimensional projection of a pyrochlore lattice along $(111)$-plane forms the geometrically frustrated kagome lattice.
  However topological \gls{magnon} bands in collinear ground-states in a Kagome lattice have also been discovered in the material Cu(1-3,bdc)\,\cite{KagomeExperiment1,KagomeExperiment2}.
  Generally the Curie-temperatures of most of the magnetic materials are lower than room-temperature and even at higher temperature (less than Curie-temperature) the width of the \gls{magnon} bands broden and possibly a topological phase transition can occur due to band-gap closing.
  So, it was thought previously that topological \gls{magnon} Hall effect is only possible at low temperatures, until recently the discovery of topological \gls{magnon} Hall-effect in Kagome ferromagnet \ce{YMn6Sn6} at room temperature\,\cite{KagomeExperiment3}.
 Kagome lattice with antiferromagnetic nearest neighbour Heisenberg interaction is a frustrated magnetic system and frustration of magnetic lattices is a key ingredient to generate non-coplanar magnetic ground states (e.g. $120^o$ antiferromagnetic order with a out of plane magnetic field).  
 The non-coplanar magnetic structure produces non-zero scalar spin chirality which acts as a source of Berry-phase for \gls{magnon}s\,\cite{TexturedFerromagnet, SpinChirality}.
 Scalar spin chirality induced topological \gls{magnon} bands are theoretically  shown to be present in the $120^o$-antiferromagnetic order of the frustrated kagome\,\cite{FrustratedKagome}, honeycomb\,\cite{FrustratedTriangularHoneycomb} and triangular lattices\,\cite{FrustratedTriangularHoneycomb, triangular}.
 It is also fascinating that several \gls{magnon}-band-topological transitions occur in a non-coplanar magnetically ordered system because change in parameters are responsible for change in the scalar spin chirality resulting in a different Berry-phase of the \gls{magnon}s\,\cite{triangular}.  
 Furthermore Lu et al.\,\cite{SpinChiralityFluctuation} demonstrate that even in absence of scalar spin chirality in the antiferromagnetic systems, the topological Berry-phase of \gls{magnon}s can be generated via quantum fluctuations of scalar spin-chirality. 
 Other studied frustrated lattice systems for topological \gls{magnon} system are frustrated star-lattice\,\cite{StarLattice} and \gls{ss}-lattice\,\cite{ferroIntro} with ferromagnetic ground state.
 
  \begin{figure}[!htb]
	\centering
	{\includegraphics[width=\textwidth]{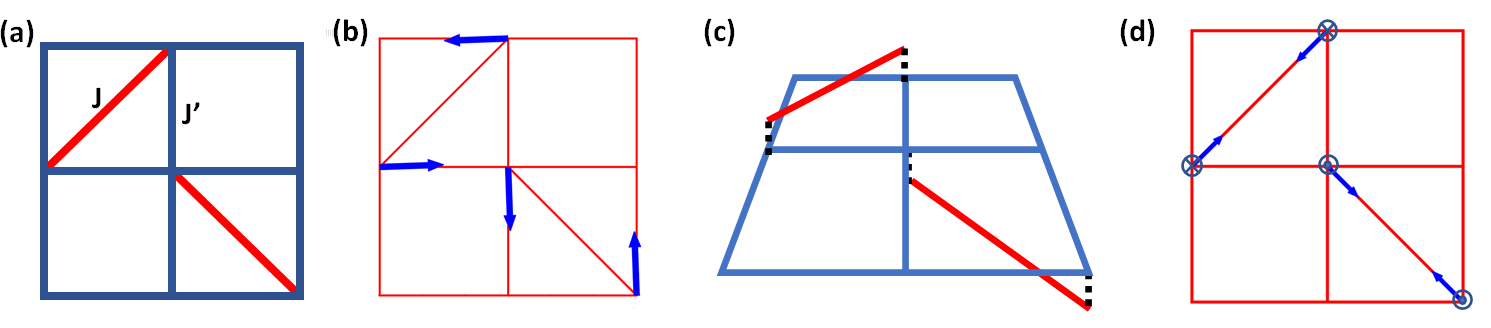}}
	\caption{(a) The Shastry-Sutherland lattice with on-dimer Heisenberg interaction $J$ and inter-dimer Heisenberg interaction $J'$,
	(b) The flux state in presence of out of plane \gls{DMI} on inter-dimer bonds, where the blue arrows denote the direction of the spins on the lattice.
	(c) The \gls{ss}-lattice crystal structure such that the in-plane \gls{DMI}s are allowed,
	(d) The canted flux state in presence of in-plane \gls{DMI}, where the dotted circle and crossed circle represent out of the plane and into the plane components of the spins.}
	\label{IntroFig1}
	\end{figure}

 In chapter Ch.\,\ref{chapter04}, we discuss the toplogical \gls{magnon} band system in a non-coplanar magnetic ground state on the frustrated \gls{ss}-lattice (see Fig.\,\ref{IntroFig1}(a)).
 The rare-earth tetraborides are the material realization for frustrated \gls{ss}-lattices because the antiferromagnetic Heisenberg-interactions on the dimer and axial bonds are nearly equal ($J\approx J'$)\,\cite{RB41,RB42,RB43}.
 Shahzad et al.\,\cite{FluxStateIntro} show that for a classical spin systems in \ce{RB4}, crystal-symmetry allowed perpendicular \gls{DMI} on the axial bond gives rise to coplanar flux state (see Fig.\,\ref{IntroFig1}(b)).
 If the crystal structure is such that the dimer bonds are out of plane as shown in the dimer-bonds are out of plane as shown in the figure Fig.\,\ref{IntroFig1}(c), then in plane DM-interactions are symmetry allowed.
 In presence any in-plane \gls{DMI}, the co-planar flux state become a non-coplanar canted-flux state (see Fig.\,\ref{IntroFig1}(d)) and in presence of magnetic field the \gls{magnon} bands in this state become topologically non-trivial.
 The topology of the \gls{magnon} bands in the system is due to the interplay among \gls{DMI}, scalar-spin chirality and quantum fluctuation of scalar spin chirality.
 This results in a wealth of band topological transitions with varying parameter which gives us the opportunity to study the behaviour of thermal Hall conductivity at band topological transition point.
 It was already known that the derivative thermal Hall conductivity shows a logarithmic divergence at the band topological transition Dirac-point\,\cite{triangular}, but here we found that the logarithmic divergence is quite universal independent of the type of band touching point.
 Finally we derived a simple algebraic equation for Thermal Hall conductivity as a function of temperature, which is applicable for any generic spin model and experimentally useful for determining the band-touching point during band topological transition.

  On the other hand honeycomb lattice is an important lattice that supports Dirac electrons and gained much attention after the discovery of graphene by using mechanical exfoliation\,\cite{AndreGeim}.
  The \gls{magnon} version of Dirac-semimetal can also be realized in the honeycomb ferromagnets\,\cite{DiracMagnon1,DiracMagnon2}.
  Similar to the electronic Dirac-point, the \gls{magnon} Dirac-point in these systems are robust against \gls{magnon}-\gls{magnon} interactions\,\cite{DiracMagnon}.
  The Dirac-\gls{magnon} is possible to gap out by breaking the inversion symmetry resulting in a next-nearest neighbour \gls{DMI} and the low temperature tight binding model of \gls{magnon}s transform into Haldane model\,\cite{CoplanarIntro1} where the complex next nearest neighbour originates from \gls{DMI}.
  The gapped \gls{magnon} bands are topologically non-trivial and thermal \gls{magnon} Hall effect can be realized in this model\,\cite{opendirac2}.
	Experimentally the honeycomb ferromagnet \ce{CoTiO3}\,\cite{CoTiO3} is detected to have Dirac \gls{magnon} point, whereas the honeycomb ferromagnet \ce{CrI3}\,\cite{CrI3} is discovered to be an topological \gls{magnon} material due to presence of \gls{DMI}.	
	The \gls{DMI} induced thermal \gls{magnon} Hall effect is also theoretically shown to be present in the antiferromagnetic Néel\,\cite{CoplanarIntro4}, zigzag and stripe phases\,\cite{StripeZigzag} in honeycomb lattice.	
	It has been shown that a bilayer honeycomb lattice with antiferromagnetic inter-layer Heisenberg interaction realizes both thermal \gls{magnon} Hall effect and \gls{magnon} spin Nernst effect\,\cite{HoneycombBilayer}.
	The spin Nernst effect is also possible in two dimensional honeycomb ferromagnet at higher temperature when the system is theoretically described in terms of \gls{spinon}s.
	At higher temperature The two dimensional honeycomb ferromagnetic system transforms into paramagnet with short range correlations and the \gls{spinon}s with up and down magnetic moment hop around the lattice making the \gls{spinon} tight binding model equivalent to Kane-Mele model in presence of second nearest neighbour \gls{DMI}\,\cite{Kane_Mele_Haldane}.
	Thus honeycomb ferromagnet is a very fascinating system where both Haldane model in terms of \gls{magnon}s and Kane-Mele model in terms of \gls{spinon}s can be realized\,\cite{Kane_Mele_Haldane}.
	Similarly, in paramagnetic regime \gls{spinon} induced spin Nernst effect can also be observed in Néel ordered honeycomb antiferromagnet\,\cite{SBMFT5}.

  Recently Colom\'es et al.\,\cite{Colomes} theoretically study the presence of anti-chiral edge states in the fermionic modified Haldane model.
  In the Haldane model the chiral edge states propagate in the opposite directions at the opposite edges, whereas in modified Haldane model the two edge states propagate in the same direction.
  The realization of these topological edge states in electron system is very unrealistic.
  So, is it possible to realize the antichiral edge states in a realistic spin systems?
  If it is possible how to engineer the material and how to detect it experimentally?

  In chapter Ch.\,\ref{chapter03} we answered the questions.
  We showed that breaking of inversion symmetry at the center of the hexagonal plateau in honeycomb ferromagnet give rise to two different kinds of \gls{DMI}, chiral and anti-chiral \gls{DMI}.
  The chiral \gls{DMI} and anti-chiral \gls{DMI} terms are the source of complex hopping terms in Haldane and modified-Haldane model in the \gls{magnon} or \gls{spinon} picture respectively.
   We showed how to engineering the materials \ce{CrSrTe3}, \ce{CrGeTe3} and \ce{AFe2(PO4)2} (A=Ba, Cs, K, La) to achieve anti-chiral \gls{DMI} in the system.
   Moreover we proposed that inelastic neutron scattering, magnetic force microscopy and spin Hall noise spectroscopy are the promising experimental techniques by which the anti-chiral edge states can be detected.

  In two dimension, the topological band structures are always gapped, whereas in three dimension the underlying topological band-structure have more variations and topological bands might be degenerate forming 0D Weyl-point, 1D nodal-line or 2D nodal-surface.
   These different dimensional degeneracies carry a topological charge and they are topologically protected.
   Moreover the topological nodal systems might also be protected by non-spatial or spatial symmetries.
   Due to bulk-edge correspondence, the topological surface state is also associated with the topological nodal systems.
   The \gls{magnon}ic analogue of the topological nodal systems has recently attracted heightened attention.
 The Weyl \gls{magnon}s are theoretically shown to be present in pyrochlore ferromagnet\,\cite{WeylPyrochlore1}, breathing pyrochlore anti-ferromagnet\,\cite{WeylPyrochlore2} as well as in pyrochlore all-in-all-out ordering\,\cite{WeylPyrochlore3}.
 Pyrochlore ferromagnets are also proposed to be host of nodal line \gls{magnon}s\,\cite{NodalPyrochlore}.
 Weyl \gls{magnon}s also theoretically proposed to be present in the stacked kagome antiferromagnets\,\cite{WeylPyrochlore4}.
   The presence of both Weyl and nodal line \gls{magnon}s are theoretically shown in the pyrochlore iridates\,\cite{WeylNodalPyrochlore1}, 3D honeycomb lattice\,\cite{WeylNodalPyrochlore2} and 3D kitaev magnets (non-symmorphic symmetry protected)\,\cite{WeylNodalPyrochlore3}.
   Moreover the $PT$-symmetry protected magnon nodal-line and loop was proposed to be present in material \ce{Cu3TeO6}\,\cite{Cu3TeO6_Theory}, which was experimentally probed to be exist in this material later\,\cite{Cu3TeO6_Experiment1,Cu3TeO6_Experiment2}.
   Weyl \gls{magnon}s are also detected experimentally in the multiferroic ferrimagnet \ce{Cu2OSeO3}\,\cite{Cu2OSeO3}.

   In the past studies, the Weyl-magnetic excitations were realized for a long-range ordered ferromagnetic or antiferromagnetic ground states.
   There have been theoretical and experimental studies of topological magnetic excitations in one-dimensional\,\cite{Triplon1D_1,Triplon1D_2} and two-dimensional\,\cite{Romhanyi2, triplon2, triplon3, Triplon2D_1,Triplon2D_2,Triplon2D_3} dimerized quantum magnets.
   However the topological Weyl-point in the magnetic excitations for a ground state as dimerized quantum magnets has not been shown to be exist.
   In chapter\,\ref{chapter05} for the first-time, we proposed the presence of Weyl-\gls{triplon}s in stacked dimerized Shastry-Sutherland material \ce{SrCu2(BO3)4}.
   We showed that a interlayer perpendicular-\gls{DMI} transforms the Dirac-nodal line material into \gls{magnon}ic version of Weyl-semimetal. 
   Moreover, neither inter-layer Heisenberg interaction nor interlayer in-plane-\gls{DMI} has any effect in low temperature physics of this material.
   Furthermore we study the canonical model in a extended parameter region by varying the interlayer perpendicular-\gls{DMI} and out-of-plane magnetic field.
   We found a rich topological phase diagram which contains regions with multiple Weyl-points as well gapped topological triplon bands.
   Furthermore, we calculated the thermal Hall conductivity and show the thermal Hall conductivity is quasilinear as a function of magnetic field in presence of Weyl-triplon.

\section{Outline}
The chapters and their contents are arranged in the following manner.
\paragraph{Model, Methods and Physical Observables}
The chapter Ch.\,\ref{chapter02} is organized in a way that first generic spin models of a quantum magnets are discussed; then the methods to study the spins and spin-excitations are illustrated; next the analytic and computational methods to calculate physical observables and their physical significances are discussed in a broader sense; finally the influence of symmetry and breaking of symmetry in the quantum magnet and its' excitations are discussed.
The chapter does not only describe the method of calculations, but also provides the physical significance and relations of physical observables and methods in contexts of physics in a broader sense.
Firstly spin-exchange-Hamiltonians and their physical origins are explained in section Sec.\,\ref{sec2.1}.
Next we discuss the methods of transformations Holstein-Primakoff transformation, bond-operator formalism, \gls{sbmft} which transforms spin Hamiltonian to another bosonic Hamiltonian to study the spin and spin-excitation physics of the system in the section Sec.\,\ref{diagonalization}.
Then, the methods of diagonalization of a quadratic Hamiltonian is discussed in absence and in presence of pair-creation-annihilation terms in the section\,\ref{diagonalization}.
In presence of pair-creation-annihilation terms Bogoliubov-Valatin transformation is explained in section Sec.\,\ref{sec2.4} and in absence of the pair-creation-annihilation terms the procedure of diagonalization become simpler and discussed in section Sec.\,\ref{sec2.3}.
We maintain the similarities of both the sections for diagonalization in terms of physical origin behind the formulations.
Next, the way of calculations of the physical observables Berry-curvature, Chern-number, edge-states, velocity of edge-states, Thermal Hall conductance, spin-Nernst effect and dynamical-structure factor are discussed in section Sec.\,\ref{sec2.5}.
The presence of non-zero Berry curvature or Chern number in the magnetic excitations is the reason behind all the non-trivial phenomena like Hall-effect, spin-Nernst effect, edge state etc.
So in the section Sec.\,\ref{sec2.5.1} we illustrate that in any systems with non-trivial topology is associated with fibre-bundle geometry (in case of condensed matter, k-space is the base space and eigenstates are fibres) of the parameter space and Berry-phase is the holonomy of non-trivial fibre bundle.
Finally in the section Sec.\,\ref{sec2.6}, the importance of symmetry in band degeneracy and determination \gls{DMI} in magnetic system is explained.

\paragraph{Engineering antichiral edgestates in ferromagnetic honeycomb lattices}

  Chapter Ch.\,\ref{chapter03} shows the presence of antichiral edge states in a honeycomb ferromagnet.
 In the introduction section Sec.\,\ref{sec3.1} the past-studies and importance for antichiral edge states have been discussed.
 The model section Sec.\,\ref{sec3.2} introduces the definitions of chiral and anti-chiral \gls{DMI} for a honeycomb ferromagnet along with the nearest neighbour Heisenberg exchange interaction.
 In the next-section Sec.\,\ref{sec3.3} the calculations of the system in spinon-picture as well as in \gls{magnon}-picture are shown.
 In spinon-picture the spin operators are transformed into Schwinger bosons and in comparison with Holstein-Primakoff transformation the Schwinger boson formalism is valid at higher temperature.
 Using spinon picture (in section Sec.\,\ref{sec3.3.1}) we showed that interplay between chiral and anti-chiral \gls{DMI} changes chiral edge magnetic edge states into anti-chiral and vice-versa.
 The edge state phenomenon should exists for a temperature range $k_BT\leq J$.
  In section Sec.\,\ref{sec3.3.2}, we showed that the \gls{magnon} picture correctly overlaps with the spinon picture at low temperature.
 The spin-Nernst effect is not effected by the anti-chiral \gls{DMI} as shown in section\,\ref{sec3.3.1} and so we proposed some experimental techniques to detect antichiral edge states in section Sec.\,\ref{sec3.3.3}.
 We propose magnetic force microscopy as well as inelastic neutron scattering experiments are useful to detect the antichiral edge states.
 Furthermore we show dynamical spin structure factor of the two edges of honeycomb nano-ribbon is a key signature to detect antichiral edge states and it can be measured using spin Hall noise spectroscopy.
 Next we propose the ideal material structures for realization of antichiral edge states in honeycomb ferromagnet in section Sec.\,\ref{sec3.3.4}.
 Then, in section Sec.\,\ref{sec3.4} using more realistic model show that the antichiral edge phenomenon will be still valid in presence of other realistic Hamiltonian terms.
 Finally the conclusion is discussed in section Sec.\,\ref{sec3.5}.

\paragraph{The topological magnon bands in the Flux state in Sashtry-Sutherland lattice}
The chapter Ch.\,\ref{chapter04} illustrates the topological \gls{magnon} bands in the flux state of \gls{ss}-lattice.
The introductory section Sec.\,\ref{sec4.1} describes the background and motivations behind the study of the topological \gls{magnon} bands in the flux state of \gls{ss}-lattice.
In the next section Sec.\,\ref{sec4.2} the model and realization of the model in a realistic material related to rare-earth tetraborides are explained.
  Moreover in that chapter the k-space Hamiltonian in terms Holstein-Primakoff bosons is shown.
  Then in the section Sec.\,\ref{sec4.3} the topological characteristic of the \gls{magnon} bands for the magnetic ground states flux-state and canted-flux state is discussed.
  In the section Sec.\,\ref{sec4.3.1}, the \gls{magnon} bands in the flux state is shown to topologically non-trivial.
  However in-plane \gls{DMI} induced canted-flux state has topologically non-trivial \gls{magnon} bands as shown in section Sec.\,\ref{sec4.3.2}.
  There are many possible topological phase transitions that occur in the parameter space of the Hamiltonian and the derivative of the thermal Hall conductivity is shown to be logarithmically divergent at the point of band-topological transition.
  Moreover we derived a analytical expression for the temperature dependence of thermal Hall conductivity which is applicable to any generic spin model and the expression might be useful in determining the energy of band-touching during band-topological transition.
  At last in the section Sec.\,\ref{sec4.4} the conclusions are discussed.

\paragraph{Weyl-triplons in \ce{SrCu2(BO3)2}}
We propose the presence of Weyl-triplons in the \ce{SrCu2(BO3)2} in the chapter Ch.\,\ref{chapter05}.
The introductory section Sec.\,\ref{sec5.1} explains the background and motivation for the research.
In the result section Sec.\,\ref{sec5.2}, the topological Weyl-points are shown to be present in the material \ce{SrCu2(BO3)2}.
Generally, the low energy magnetic property of  \ce{SrCu2(BO3)2} is well described by using the quasi two-dimensional spin-exchange interactions, although the inter-layer interactions are nonzero and specifically the inter-layer \gls{DMI}s are symmetry allowed.
Taking into account realistic interlayer symmetric and assymetric exchange interactions along with quasi two-dimensional spin-exchange model, we propose a new microscopic model in the section Sec.\,\ref{sec5.2.1}.
The ground state of the model is well described as a product states of dimers and the low-lying magnetic excitations are described in terms of \gls{triplon}s, as illustrated in the section Sec.\,\ref{sec5.2.2}.
In section Sec.\,\ref{sec5.2.3}, we show that the model exhibits different phase regions categorized by the numbers and the position of Weyl-points, in the phase space of inter-layer \gls{DMI} and perpendicular magnetic field.
In the next section Sec.\,\ref{sec5.2.4}, we verified the presence of Fermi-arc like surface states at the surface of the material which in turn proves the Bulk-edge correspondence of the topological system.
Moreover in the section Sec.\,\ref{sec5.2.5}, we theoretically show that the quasilinear dependence of the thermal Hall conductance as a function of magnetic field is a experimental signature of the Weyl-triplons in the system.
Finally the conclusion of the study is elaborated in the section Sec.\,\ref{sec5.3}.

	\chapter{Model, Methods and Physical Observables}\label{chapter02}
	
	\section{Spin-Hamiltonian}\label{sec2.1}
	In the Mott insulating phase of a material the charge degrees of freedom of electron is frozen to a local lattice-site and the ground state of the material is magnetic and described by the electron spin. 
	It is well known that the magnetic ground state (e.g. ferromagnetic, antiferromagnetic etc.) of a material is governed by the spin-spin interactions of magnetic ions. 
	The dominant spin-spin interaction in many magnetic materials is described by the Heisenberg Hamiltonian, which is given by,
	\begin{equation}
		\pazocal{H}=\frac{1}{2} \sum_{i,j,i\neq j} J(i,j)\bold{S}_i\cdot\bold{S}_j,
		\label{2.1}
	\end{equation}	 	
	where $\bold{S}_i$ and $J(i,j)$ are the vector array of spin-operator components at site-$i$ and Heisenberg-exchange coupling constant between sites $i$ and $j$ respectively.
	 The factor-$\frac{1}{2}$ is multiplied to avoid double counting of same bonds between $i$ and $j$ in the summation. 
	There are two kinds of Heisenberg exchange interactions, (i) Direct exchange\,\cite{Heisenberg,Yosida-Book} and (ii) Superexchange interaction\,\cite{Anderson,Yosida-Book} and they are purely quantum mechanical.
	 In both cases the physical origin of the exchange interactions is from Coulomb interaction between the electrons and the Pauili exclusion principle. 
	 Direct exchange interaction is due to nearest neighbour Coulomb repulsion and favours ferromagnetism. 
	 On the other hand Superexchange originates due to onsite Coulomb repulsion and favours antiferrmagnetism.

	From observation of the mathematical structure of Eq.\,\ref{2.1}, we may can generalise the spin Hamiltonian to the following Hamiltonian (in order to include a broader range of interactions)\,\cite{Sandratskii},
	\begin{equation}
		\pazocal{H}=\frac{1}{2} \sum_{i,j,i\neq j} \bold{S}_i\cdot A_{ij}\cdot \bold{S}_j,
	\end{equation}		
	where $A_{ij}$ is a $3\times 3$-matrix and $\bold{S}_i$ is a vector-array of spin-operator components.
	The symmetric and anti-symmetric part of the matrix is given by,
	\begin{equation}
	B_{ij}=\frac{1}{2}(A_{ij}+A_{ji}), \quad C_{ij}=\frac{1}{2}(A_{ij}-A_{ji}),
	\end{equation}
	where the symmetric part $B_{ij}$ represents the Heisenberg exchange coupling and single-ion anisotropies. 
	The single-ion anisotropies denote that the material consist a preferred direction for the spins to be aligned and arise from the crystal field effects in the quantum magnets. 
	If the easy axis of the material is along z-direction, the single ion anisotropy takes the mathematical form of $K\sum_i \bold{S}_i^2$.
	Further details of the term can be found in reference Ref.\,\cite{Yosida-Book}.

	The anti-symmetric part of the matrix-$A_{ij}$ can be rearranged in a mathematical form as,
	\begin{equation}
		\bold{S}_i \cdot C_{ij}\cdot \bold{S}_j=\bold{D}_{ij}\cdot (\bold{S}_i\times\bold{S}_j),
		\label{eq2.4}
	\end{equation}
	where $\bold{D}_{ij}$ is the \gls{DMI}. The \gls{DMI} was first proposed by the Dzyaloshinsky on the basis of symmetry of a crystal structure\,\cite{Dzyaloshinsky}.
	 Moriya showed that spin-orbit coupling in the Superexchange interaction proposed by Anderson describes the microscopic origin of the \gls{DMI}\,\cite{Moriya}.
	 There is a symmetric anisotropy tensor in Moriya's derivation, but it can be neglected for any physical system
	 \footnote{ As an exception, the Kitaev-honeycomb magnet $\alpha$-\ce{RuCl3} is proposed to have a symmetric-anisotropic tensor with order of magnitude similar to Heisnberg-exchange interaction and Kitaev interaction. Moriya's derivation is based on single band Hubbard model for electrons which results in antiferromagnetic Heisenberg interaction along with symmetric and anti-symmetric anisotrpic spin exchange interaction. Thus the order of magnitude provided in Moriya's work may not be suitable for all magnetic materials.
	 }
	 because the order of magnitude of the term is $(\frac{\Delta g}{g})^2 J$, where $g$ is the gyromagnetic ratio, $\Delta g$ is the deviation of it from the free electron value and $J$ is the Heisenberg-exchange coupling.
	 On the other hand the order of magnitude of the \gls{DMI} is\,\cite{Moriya},
	 \begin{equation}
		D\sim{~}\frac{\Delta g}{g}J.
	\end{equation}	  
So, the order of magnitude of \gls{DMI} is generally small compared with the Heisenberg-interaction for a physical system and thus in general the \gls{DMI} often plays a sub-dominant role in determining the ground state of the system.
 However, the \gls{DMI} plays an important role to make magnetic excitations topologically non-trivial which will be further described in the models and results of chapters Chap.\,\ref{chapter03} and Chap.\,\ref{chapter04}.
 
 The Heisenberg-exchange interaction is independent of the symmetry of crystal structure and present in any magnetic material.
  But the nature \gls{DMI} depends on the symmetry of the crystal and sometime it vanishes on the bond which has a inversion center at the middle. The symmetry dependence of the \gls{DMI} is further elaborated in the section Sec.\,\ref{sec2.6}.

	\section{Spins and spin-wave quasiparticle}
	In an insulating quantum magnet due to quantum fluctuation the spin degrees of freedom can traverse through the system without any movement of the charge degrees of freedom of an electron. 
	This is very important in spintronic applications, because Joule heating effect is absent in case of transport of only spins and so dissipation is reduced during transportation. 
	In this section, we describe different formalisms to describe the spins and spin-excitations as bosonic quasi-particles and obtain a tight binding model of these quasi-particles in the mean field limit to investigate the properties of the spin systems in quantum magnets.
	These quasi-particles carry a finite quantized magnetic moment and thus the transportation of the quasi-particles is equivalent to the spin-transportation.
	
	\subsection{Holstein-Primakoff transformation}
	\label{sec2.2.1}
We use the standard \gls{HP}-transformation to investigate the low lying magnetic excitations above a long range magnetically ordered ground state.
The \gls{HP} transformations map these quantized excitations into a system of (interacting) quasiparticles, viz., \gls{magnon}s that obey Bose-Einstein statistics.
	The well-known \gls{HP}-method transforms the spin-operators to a single species of boson operators, as follows\,\cite{Holstein-Primakoff,Yosida-Book},
	\begin{align}
		S_i^+ &=\sqrt{2S} \left(1-\frac{a_i^\dagger a_i}{2S}\right)^\frac{1}{2} a_i \nonumber\\
		S_i^- &=\sqrt{2S} \left(1-\frac{a_i^\dagger a_i}{2S}\right)^\frac{1}{2} a_i^\dagger \nonumber\\
		S^z_i &=S-a_i^\dagger a_i,
		\label{eq2.6}
	\end{align}
	where $a_i^\dagger$ and $a_i$ are the creation and annihilation operators of the \gls{HP}-bosons respectively and $S_i^+=S^x_i+iS^y_i$, $S_i^-=S^x_i-iS^y_i$.
	The \gls{HP}-bosons are known as \gls{magnon}s and magnons are low energy spin wave excitations above a magnetically ordered ground state and constitute slowly varying spin configurations. 
	The number of \gls{magnon}s follows a non-holonomic constraint $0\leq a_i^\dagger a_i\leq 2S$.
	A \gls{magnon} carries a magnetic moment of value $g\mu_B$, where g is the Lande g-factor, $S$ is the spin at each site and $\mu_B$ is Bohr magneton (which follows simply from the third equation of Eq.\,\ref{eq2.6}, because presence of one \gls{magnon} changes the $S_z$ quntum number by one).
	Thus \gls{magnon} is equivalent to a spin one particle.
	The presence of square root in the expression in Eq.\,\ref{eq2.6} makes the use of \gls{HP}-method trickier to be applied and we need to make assumptions for concrete calculations.
	At low temperature, it is natural to assume that number of \gls{magnon}s is small so that $a_i^\dagger a_i\ll 2S$. With this assumption the terms under square root is expanded in a series expansion and the series expansion is known as $\frac{1}{S}$-expansion or spin-wave expansion.
	The first two terms of $\frac{1}{S}$-expansion is given as,
	\begin{align}
		S_i^+ &\approx\sqrt{2S} a_i-\frac{1}{2\sqrt{2S}} n_i a_i+\cdots \nonumber\\
		S_i^- &\approx\sqrt{2S} a_i^\dagger-\frac{1}{2\sqrt{2S}} n_i a_i^\dagger+\cdots \nonumber\\
		S_z &= S-a_i^\dagger a_i,
		\label{eq2.7}
	\end{align}	 
	 where $n_i=a_i^\dagger a_i$ is number of \gls{magnon}s at lattice site-i.
	 Collecting the first terms of the expansion,
	 \begin{equation}
	 S_i^+ \approx\sqrt{2S} a_i,\quad S_i^- \approx\sqrt{2S} a_i^\dagger,\quad S_z = S-a_i^\dagger a_i,
	 \end{equation}
	 this expressions are called \gls{LSWT}.

	 \begin{figure}[!htb]
	\centering
	{\includegraphics[width=0.49\textwidth]{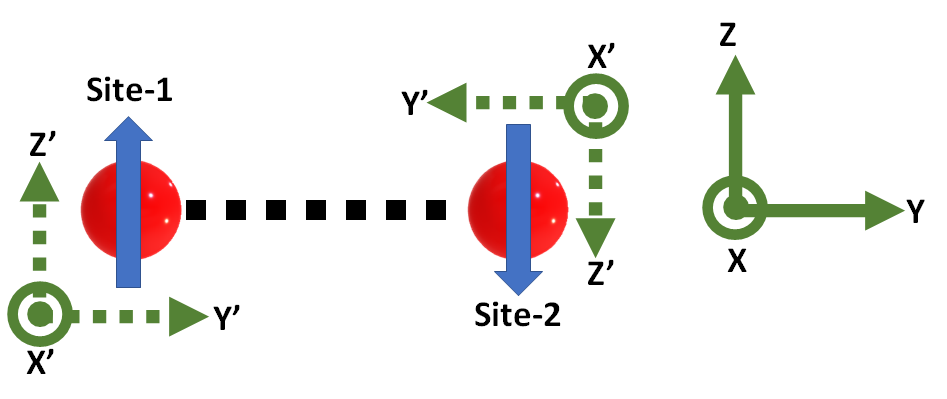}}
	\caption{The illustration of co-ordinates transformation at each spin sites before application of \gls{HP}-transformation.}
	\label{Holstein_Primakoff}
	\end{figure}
For spin-wave expansion we have approximated that the number of \gls{magnon}s is small i.e. $a_i^\dagger a_i\ll 2S$, which is true at low temperature.	
	 But to make this assumption to be true there is another underlying assumption that is local z-axis at each lattice site is aligned along the local spin-axis.
	 In other words we can perform the \gls{HP}-transformation in any arbitrary co-ordinate system, but the derived \gls{HP}-bosons in this transformation do not necessarily  represent the magnetic excitations. 
	 One needs a co-ordinate transformation to map the HP bosons to localised quasiparticles.
	The co-ordinate transformation before \gls{HP}-transformation is illustarted in the figure Fig.\,\ref{Holstein_Primakoff}. 
	There are two spins at site-1 and site-2.
	The unprimed and primed co-ordinates denote the global and local co-ordinates respectively.
	The spin at site-1 is aligned along the direction of the z-axis of global co-ordinate and so the local co-ordinate is same as the global co-ordinate.
	On the other hand, the spin at site-2 align in the opposite direction of the z-axis of the global co-ordinate and so the global co-ordinates are rotated $180^o$ about the x-axis to obtain the local co-ordinates of site-2.
	The transformation relation among the spin components in the global and local co-ordinates are given by,
	\begin{align}
	S^x_1&=S^{\prime x}_1, \quad S^y_1=S^{\prime y}_1, \quad S^z_1=S^{\prime z}_1, \nonumber\\
	S^x_2&=S^{\prime x}_2, \quad S^y_2=-S^{\prime y}_2, \quad S^z_2=-S^{\prime z}_2.
	\end{align}
	Accordingly the \gls{HP}-transformation of each site is given by,
	\begin{align}
	S^+_1&=\sqrt{2S}a_1, \quad S^-_1=\sqrt{2S}a_1^\dagger, \quad S^z_1=S-a_1^\dagger a_1,\nonumber \\
	S^+_2&=-\sqrt{2S}a_2^\dagger, \quad S^-_2=-\sqrt{2S}a_2, \quad S^z_2=-S+a_2^\dagger a_2.
	\end{align}
	
In case of \gls{HP}-transformation the knowledge of the directions of local spin moments are necessary. 
	Thus the ground state of the spin-system should have a classical counterpart.
	In brief in the \gls{HP}-method the spins are treated as classical spins and quantum fluctuations are perturbatively added through the spin-wave expansion.
	 The following flowchart describes the formalism of \gls{LSWT},
	\begin{itemize}
		\item[1.] Determine the classical ground state of the spin system.
		\item[2.] Align the z-direction of the local co-ordinate along the direction of the classical spins at each sites.
		\item[3.] Perform the \gls{HP}-transformation on the spin operators in local co-ordinate system.
	\end{itemize}

	\subsection{Bond-operator formalism\label{sec2.2.2}}

	\begin{figure}[!htb]
	\centering
	{\includegraphics[width=\textwidth]{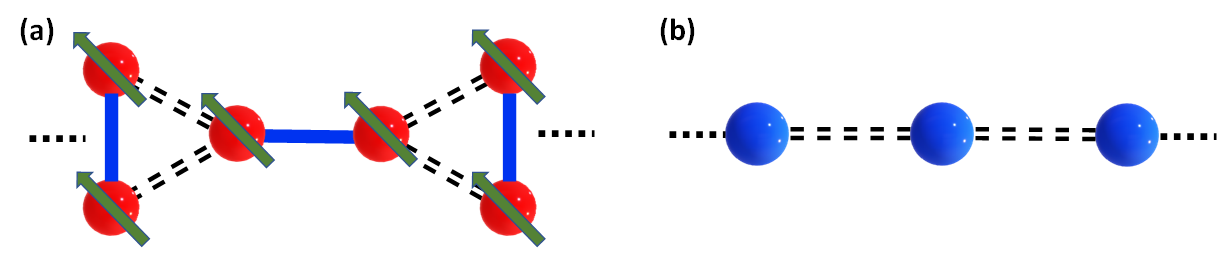}}
	\caption{(a) Schematic of orthogonal dimer chain. The blue thich bonds are dimer bonds. The double dotted bonds denote weak ferromgnetic or antiferromagnetic coupling. The single dotted lines represent that the lattice is infinite. (b) The transformation of Orthogonal dimer chain to a one dimensional lattice, where the blue-thick dimer bonds transforms into one single site.}
	\label{OrthogonalDimer}
	\end{figure}

	The bond operator formalism was first introduced by Sachdev and Bhatt to describe the phase transitions between dimerized and magnetically ordered phases in antiferromagnets\,\cite{SachdevBhatt}.
	 A dimerized state means a product state of singlet states  on a particular bonds and a singlet (Eq.\,\ref{eq2.11}) is formed due to antiferromagnetic interaction between spin-$\frac{1}{2}$ sites of those bonds. 
	 As an example of dimerized state of an orthogonal dimer chain is shown in Fig.\,\ref{OrthogonalDimer}(a)\,\cite{OrthogonalDimer}.
	 The thick-blue bonds in the figure represent that the antiferromagnetic Heisenberg interactions on those bonds are stronger compared with the ferromagnetic or antiferromagnetic interactions on other bonds represented by dotted double lines.
	  Thus singlet states exist on each bold bonds, making the system as dimerized state.
	  The wave-function of the singlet state is given by,
	  \begin{equation}
	  	\left|s\right\rangle=\frac{1}{\sqrt{2}}\left( \left|\uparrow\downarrow\right\rangle-\left|\downarrow\uparrow\right\rangle\right).
	  	\label{eq2.11}
	  \end{equation}
	  There are three degenerate states which are higher in energy compared to singlet, which are known as triplets. 
	  The triplets are degenerate in energy unless any other interactions (e.g. \gls{DMI}) on the dimer bond or external magnetic field present.
	   The wavefunction of the triplet state is given as,
	  \begin{equation}
	  	\left|t_x\right\rangle=\frac{i}{\sqrt{2}}\left(\left|\uparrow\uparrow\right\rangle-\left|\downarrow\downarrow\right\rangle\right),\quad
	  	\left|t_y\right\rangle=\frac{1}{\sqrt{2}}\left(\left|\uparrow\uparrow\right\rangle+\left|\downarrow\downarrow\right\rangle\right),\quad
	  	\left|t_z\right\rangle=-\frac{i}{\sqrt{2}}\left( \left|\uparrow\downarrow\right\rangle+\left|\downarrow\uparrow\right\rangle\right),
	  	\label{eq2.12}
	  \end{equation}
	  where due to degeneracy the triplet wavefunction can be represented in any linear combination of the defined wavefunctions in Eq.\,\ref{eq2.12} and the wavefunctions must be orthogonal with respect to each other.
	   The triplet wavefunctions in eq.\,\ref{eq2.12} are chosen to be time-reversal invariant\,\cite{Romhanyi1}.

	    The spin operators in the bond operator formalism are given by,
	    \begin{align}
	    	S^\gamma_{i,j}= &\left(\left\langle s_i\right| S^\gamma_{i,j} \left|s\right\rangle\right)\left|s_i\right\rangle\left\langle s\right|+
	    	\sum_\alpha \left(\left\langle s_i\right| S^\gamma_{i,j} \left|t_{\alpha,i}\right\rangle\right)\left|s_i\right\rangle\left\langle t_{\alpha,i}\right|\nonumber\\
	    	&+\sum_\alpha \left(\left\langle t_{\alpha,i}\right| S^\gamma_{i,j} \left|s_i\right\rangle\right)\left|t_{\alpha,i}\right\rangle\left\langle s_i\right|+
	    	\sum_{\alpha,\beta} \left(\left\langle t_{\alpha,i}\right| S^\gamma_{i,j} \left|t_{\beta,i}\right\rangle\right)\left|t_{\alpha,i}\right\rangle\left\langle t_{\beta,i}\right|,
	    \end{align}
	    where $\alpha$, $\beta$ and $\gamma$ indices represent $x$, $y$ or $z$. 
	    Again $S^\gamma_{i,j}$ is the $\gamma$-th component of the spin operator at $j$-th site of the $i$-th dimer-bond.
	    $\left|s_i\right\rangle$ and $\left|t_{\alpha,i}\right\rangle$ are singlet and triplet states at $i$-th bond.
	     After evaluating the matrix elements (e.g. $\left(\left\langle s\right| S^x_{i,j} \left|s\right\rangle\right)$, $\left(\left\langle s\right| S^x_{i,j} \left|t_\alpha\right\rangle\right)$ etc.) the explicit form of the spin operators in terms of bond operators are given by,
	    \begin{align}
	    	S^\alpha_{i,1}&=\frac{i}{2}(t_{\alpha,i}^\dagger s_i-s_i^\dagger t_{\alpha,i})-\frac{i}{2} \epsilon_{\alpha,\beta,\gamma} t_{\beta,i}^\dagger t_{\beta,i} \nonumber\\
	    	S^\alpha_{i,2}&=-\frac{i}{2}(t_{\alpha,i}^\dagger s_i-s_i^\dagger t_{\alpha,i})-\frac{i}{2} \epsilon_{\alpha,\beta,\gamma} t_{\beta,i}^\dagger t_{\beta,i},
	    	\label{eq2.14}
	    \end{align}
	    where $s_i^\dagger$ and $t_{\alpha,i}^\dagger$ are the singlet and triplet creation operator at $i$-th bond and creates singlet and \gls{triplon}s from vacuum state $\left|0\right\rangle$,
	\begin{align}
		s^\dagger\left|0\right\rangle =\left|s\right\rangle,\quad
		t_x^\dagger\left|0\right\rangle =\left|t_x\right\rangle,\quad
		t_y^\dagger\left|0\right\rangle =\left|t_y\right\rangle,\quad
		t_z^\dagger\left|0\right\rangle =\left|t_z\right\rangle,
	\end{align}		        
	    in literature the triplet states are known as \gls{triplon} quasi-particle in bond operator formalism.

	    The singlets are essentially spin-$0$ particle.
	    On the other hand, the \gls{triplon}s are equivalent to spin-1 particle which means it carries a magnetic moment $g\mu_B$, where g is Lande g-factor and $\mu_B$ is the Bohr magneton.
	    The bond-operator formalism transforms the dimer-bonds into an effective lattice site, transforming the lattice to a new lattice.
	    This is illustrated in the figure Fig.\,\ref{OrthogonalDimer}(b).
	    In the figure the orthogonal dimer lattice transforms into a effective one dimensional lattice.
	     The singlets and triplons reside on each site of this effective lattice.
	    Furthermore the number of \gls{triplon}s and siglet qusiparticles on each effective lattice site follows a hard-core boson constraint,
	    \begin{equation}
	    	s_i^\dagger s_i +\sum_\alpha t_{\alpha,i}^\dagger t_{\alpha,i}=1,
	    \end{equation}
	    because there is only one possible triplet or singlet possible on each dimer bond.
	    For further simplification at low temperature, we assume that the number of \gls{triplon} is low and the ground state is vacuum state made of singlets and so,
	    \begin{equation}
	    	\left\langle s^\dagger_i\right\rangle=\left\langle s_i\right\rangle=s,
	    \end{equation}
	where $s$ is mean-field parameter and for more simplification we assume $s\approx 1$.

	The dimension of Hilbert-space of N spin-half sites is $2^N$, because only two states are possible at each spin-half sites which are  $S_z=\pm \frac{1}{2}$. 
	After bond-operator transformation each strong antiferromagnetically coupled bonds turns into one site with 4 possible states (one singlet and three triplons). 
	So the dimension of Hilbert space in bond operator formalism is equal to $4^{\frac{N}{2}}=2^N$. 
	Thus the bond operator formalism transforms the spin Hamiltonian to a new Hamiltonian which represents the same spin system without any approximation. This is further discussed in Ref.\,\cite{SachdevBhatt} in a Group theoretical manner and here it is also discussed briefly.  
	A spin operator of non-relativistic spin-half particle is given by a two dimensional representation of $SU(2)$ group (in which the Pauli matrices are the basis of Lie algebra of $SU(2)$). 
	Thus the tensor product of two spin half operators is representation of $SU(2) \times SU(2)$ group.
	There is a well known ismorphism between the groups $SO(4)$ and $SU(2)\times SU(2)$ (i.e. $SU(2)\times SU(2)\cong SO(4)$).
	The translational and rotational generators of $SO(4)$ group are $\bold{S}_1+\bold{S}_2$ and $\bold{S}_1-\bold{S}_2$, where $\bold{S}_1$ and $\bold{S}_2$ are spin operators at the two sites of a dimer.
	From equations in\,\ref{eq2.14}, it is noticeable that the bond operators $s^\dagger$ and $t_\alpha^\dagger$ form the representations of generators of $SO(4)$ group.
	Thus bond operators canonically transforms the spin Hamiltonian into a equivalent Hamiltonian which is easier to treat in mean-field level for the dimerized phase a system.

	There are many other formalisms which are based on the similar idea of bond operator formalism.
	For example, plaquette-operator formalism is used to analyse the plaquette phase of Shastry-Sutherland model which is an intermediate phase between the dimer-phase and Neel phase\,\cite{Zhifeng}.
	Furthermore, similar formalism is also applied to study the trimarized phase of Kagome lattice\,\cite{Brijesh}.
	Moreover, sometimes the bond operator formalism needs to be modified, if any other kinds of interaction like \gls{DMI} is present on the dimer bonds.
	In this case the singlet and triplet wavefunctions deviates from their actual form and as a consequence the equation Eq.\,\ref{eq2.14} needs to be modified.
	This modified bond operator formalism is applied due to presence of \gls{DMI} on the dimer bonds for the material \ce{SrCu2(BO3)2}\,\cite{Romhanyi1, Romhanyi2} and this modified formalism is also adapted for the study of Weyl-\gls{triplon}s in \ce{SrCu2(BO3)2} in chapter Ch.\,\ref{chapter05}.

 	\subsection{Schwinger-Boson mean field theory}\label{SBMFT}
Similar to \gls{HP}-method, the \gls{sbmft} is well known in the study of quantum spin systems\,\cite{SchwingerBososnXXZModel2,SBMFT_Pathology,SBMFT_intro1,SBMFT_intro2,SBMFT_intro3}.
While the magnon picture is valid at low temperatures where 
the system is ordered, it fails at higher temperatures comparable to the exchange strength $J$.
In this regime, the \underline{\it Schwinger boson} representation
provides an alternative framework to study the topological features of the spin excitations.  
	In contrast with the \gls{HP}-method, the spin operators are defined in terms of two species of boson operators as,
	\begin{equation}
		S^+_i=b_{i\uparrow}^\dagger b_{i_\downarrow},\quad
		S^-_i=b_{i\downarrow}^\dagger b_{i\uparrow},\quad
		S^z_i=\frac{1}{2}\left(b_{i\uparrow}^\dagger b_{i\uparrow}-b_{i\downarrow}^\dagger b_{i\downarrow}\right),
		\label{eq2.18}
	\end{equation}
	where $b_{i\uparrow}^\dagger$ and $b_{i\downarrow}^\dagger$ are creation operators of the two species of Schwinger bosons at i-th site of the lattice.
	The Schwinger bosons are also known as \gls{spinon}s, which carries a magnetic moment equivalent $\frac{1}{2}g\mu_B$, making them spin-half bosons.
	In a more compact form the spin operators in terms of Schwinger boson can be written as,
	\begin{equation}
		\bold{S}_i=\frac{1}{2} \sum_{\alpha,\beta} b_{i\alpha}^\dagger \bold{\sigma}_{\alpha,\beta} b_{i\beta}.
	\end{equation}
	Furthermore the number of Schwinger boson is constrained by the following holonomic constraint,
	\begin{equation}
		b_{i\uparrow}^\dagger b_{i_\uparrow}+b_{i\downarrow}^\dagger b_{i_\downarrow}=2S,
		\label{eq2.20}
	\end{equation}
	So, the constraint on the number of bosons in \gls{sbmft} is holonomic making it easier to handle, whereas in \gls{HP}-method the constraint was non-holonomic.
	A holonomic constraint can be taken into account in the Hamiltonian by using Larange's undetermined multiplier, which is further discussed later.

	In comparison with \gls{HP}-transformation (Eq.\,\ref{eq2.6}), the Schwinger boson transformation does not involve the square root.
	 The \gls{HP}-method is applicable only for broken symmetry phases of a magnetic system, because in those phases the ground state of the system can be represented classically. 
	 However, \gls{sbmft} is applicable for short range ordered as well as for long range ordered system, the magnetic phase does not need to be broken symmetry phase.
	 The long range magnetic order  in a magnetic system is described as the condensation of the Schwinger bosons.

	  Generally a spin Hamiltonian is quadratic in terms of spin operators.
	   In this situation, the Hamiltonian contains only quartic terms in terms of Schwinger boson operators.
	    The main tricky part of this formalism is to decouple the quartic terms into quadratic terms of bosonic operators to treat the Hamiltonian in a mean-field level.
	    Here, three types of Hamiltonians are discussed for the decoupling from the quartic to quadratic forms.
	    The Hamiltonians which are discussed here are,
	    \begin{itemize}
	    	\item[(1)] Isotropic Heisenberg model,
	    	\item[(2)] $XXZ$ Heisenberg model,
	    	\item[(3)] Isotropic ferromagnetic Heisenberg model with \gls{DMI}-interaction along $z$-direction.
	    \end{itemize}

	    A isotropic Heisenberg Hamiltonian on a lattice is given as,
	\begin{equation}
		H_1=J\sum_{\left\langle i,j\right\rangle} \bold{S}_i \cdot\bold{S}_j,
		\label{eq2.21}
	\end{equation}
	where $J$ can take values positive or negative values giving rise to antiferromagntic and ferromagnetic Heisenberg exchange interaction.
	Moreover $\left\langle\cdots\right\rangle$ denote the nearest neighbour bonds.
	    There are two possible choices of quadratic operators, which preserves the $SU(2)$-symmetry of a isotropic Heisenberg Hamiltonian, which are,
	    \begin{equation}
	    	A_{ij}^\dagger=\frac{1}{2} \left(b_{i\uparrow}^\dagger b_{j\downarrow}^\dagger-b_{i\downarrow}^\dagger b_{j\uparrow}^\dagger\right), \quad
	       	B_{ij}^\dagger=\frac{1}{2} \left(b_{i\uparrow}^\dagger b_{j\uparrow}+b_{i\downarrow}^\dagger b_{j\downarrow}\right),
	    	\label{eq2.22}
		\end{equation}	     
	where $A_{ij}$ and $B_{ij}$ are bond operators because the operators contain operators of two sites of a bond. To avoid the ambiguity among the bond operators of bond operator formalism and Schwinger boson formalism, in this text we call it Schwinger boson bond operator. 
	In the appendix.\,\ref{appendixA}, the $SU(2)$-invariance of the Schwinger boson bond operators is shown.
	Moreover, the two bond operators in eq.\,\ref{eq2.22} are related to two complementary phases of a magnetic system.
	In the appendix.\,\ref{appendixA}, it has been shown that,
	\begin{equation}
		:B_{ij}^\dagger B_{ij}:=\frac{1}{4}(\bold{S}_i+\bold{S}_j)^2-\frac{S}{2},\quad
		:A_{ij}^\dagger A_{ij}:=\frac{1}{4}(\bold{S}_i-\bold{S}_j)^2-\frac{S}{2},
		\label{eq2.23}
	\end{equation}
	where, ``$:\quad:$" denotes the normal ordering.	
	Thus the bond operator $B_{ij}^\dagger$ is related to ferromagnetism and the bond operator $A_{ij}^\dagger$ is related to antiferromagnetism.
	Using the equations Eq.\,\ref{eq2.22}, the Heisenberg Hamiltonian Eq.\,\ref{eq2.21} in terms of Schwinger boson bond operators can be represented in the following forms,
	\begin{align}
	H_1&=J\sum_{\left\langle i,j\right\rangle} \left[2:B_{ij}^\dagger B_{ij}:-S^2\right] \label{2.24a}\tag{2.24a} \\
	&=J\sum_{\left\langle i,j\right\rangle} \left[S^2-2:A_{ij}^\dagger A_{ij}:\right]\label{2.24b}\tag{2.24b} \\
	&=J\sum_{\left\langle i,j\right\rangle} \left[:B_{ij}^\dagger B_{ij}: -:A_{ij}^\dagger A_{ij}:\right]\label{2.24c}\tag{2.24c}.
	\end{align}
	The different forms of the Hamiltonian are applicable to different magnetic systems and the different forms can be achieved by using equations Eq.\,\ref{eq2.23}, which is shown in appendix.\,\ref{appendixA}.
	 Different forms of the Hamiltonian are related to each other through the relation, $:B_{ij}^\dagger B_{ij}:+:A_{ij}^\dagger A_{ij}:=S^2$, which is also derived from the equations Eq.\,\ref{eq2.23}.
	The equations Eq.\,\ref{2.24a} and Eq.\,\ref{2.24b} are useful particularly for ferromagnetic and antiferromagnetic ground states respectively.
	 Wheareas, the equation Eq.\,\ref{2.24c} is compatible for any magnetic ground state.

	A more general case is $XXZ$-Heisenberg model, which is described by,
	\begin{equation}
		H_2=J \text{sgn}(\Delta)\sum_{\left\langle i,j\right\rangle} \left[S_i^xS_j^x+ S_i^yS_j^y+|\Delta| S_i^zS_j^z\right],
	\end{equation}
	where we assume $J>0$ and so $\Delta<0$ denotes ferromagnetic phase, whereas $\Delta>0$ represents antiferromagnetic phase.
	In the reference Ref.\,\cite{SchwingerBososnXXZModel2}, it has been shown that the $XXZ$-Heisenberg model in terms of bond operators can be represented as,
	\begin{equation}
		H_2=-J\sum_{\left\langle i,j\right\rangle} \left[(1+\Delta):A_{ij}^\dagger A_{ij}:+(1-\Delta) :B_{ij}^\dagger B_{ij}:-S^2\right].
	\end{equation}		
	It is noticeable that in $XXZ$-model the $SU(2)$-symmetry is not present, but it is still possible to represent the spin Hamiltonian using the $SU(2)$-symmetric bond operators. 
	It is because in the limiting case $\Delta=\pm 1$, the Heisenberg Hamiltonian retains it's $SU(2)$-symmetry and in the limiting cases it merges exactly with the limiting forms described by the equations Eq.\,\ref{2.24a} and Eq.\,\ref{2.24b}.

	Finally, the \gls{DMI}-terms which preserves the $U(1)$-symmetry are discussed here.
	The Hamiltonian with both the ferromagnetic Heisenberg interaction and \gls{DMI} along the $z$-direction is given as,
	\begin{equation}
		H_3=-J\sum_{\left\langle i,j\right\rangle} \bold{S}_i\cdot\bold{S}_j + \bold{D}\cdot\sum_{\left\langle i,j\right\rangle} (\bold{S}_i\times \bold{S}_j),
		\label{eq2.26}
	\end{equation}
	where both interactions are on the nearest-neighbour bonds and $J>0$.
	In presence of \gls{DMI} along the $z$-direction the $SU(2)$-symmetry of the Hamiltonian is broken, but the $U(1)$-symmetry of Hamiltonian (which denotes $S_z$-quantum number conservation) still preserved.
	In such case we define the following bond-operators,
	\begin{equation}
		\chi_{ij,\uparrow}=b_{i\uparrow}^\dagger b_{j\uparrow},\quad
		\chi_{ij,\downarrow}=b_{i\downarrow}^\dagger b_{j\downarrow},	
	\end{equation}
	where the bond operators $\chi_{ij,s}$ preserves the $U(1)$-symmetry.
	In terms of these bond operators the Hamiltonian Eq.\,\ref{eq2.26} transforms as,
	\begin{align}
		H_3=&-\frac{J}{2}\sum_{\left\langle i,j\right\rangle} \left[:\chi_{ij\uparrow}^\dagger\chi_{ij,\downarrow}:+
		:\chi_{ij,\downarrow}^\dagger\chi_{ij,\uparrow}:+
		:\chi_{ij,\uparrow}^\dagger\chi_{ij,\uparrow}:+
		:\chi_{ij,\downarrow}^\dagger\chi_{ij,\downarrow}:\right]\nonumber \\
		&-\frac{D}{2}\sum_{\left\langle i,j\right\rangle}i\nu_{ij}
		\left[:\chi_{ij\uparrow}^\dagger\chi_{ij\downarrow}:-
		:\chi_{ij\downarrow}^\dagger\chi_{ij\uparrow}:\right]+
		zNJS^2,
	\end{align}
	where $\nu_{ij}=\pm$ depending on the directionality of the bond connected which is associated with the \gls{DMI}. 
	This directionality of the bonds will be further clear in the model sections of the chapters Ch.\,\ref{chapter03} and Ch.\,\ref{chapter04}. Again $z$ and $N$ are the co-ordination number and number of lattice sites respectively.
	In absence of \gls{DMI}, the Hamiltonian Eq.\,\ref{eq2.26} should automatically transform into Eq\,\ref{2.24a} which is the case of ferromagnetic limit.
	  The ferromagnetic bond operators are connected with the bond operators $\chi_{ij,s}$ in the following way,
	\begin{equation}
		B_{ij}=\frac{1}{2}\left( \chi_{ij,\uparrow}+\chi_{ij,\downarrow}\right),
	\end{equation}
	thus the limiting ferromagnetic form of the Hamiltonian Eq.\,\ref{eq2.26} is automatically achieved.

	There are more complicated Hamiltonians including more general types of \gls{DMI}. 
	In this text the methods of transformation from the spin Hamiltonian to the Schwinger boson bond operator transformation is not further discussed for other cases.
	Interested readers are referred to the references Ref.\,\cite{SBMFT1,SBMFT2,SBMFT3,SBMFT4,SBMFT5}.

	 The constraint in the equation Eq.\,\ref{eq2.20} can also be taken into account by adding it into Hamiltonian using the Lagrange's undetermined multiplier in the following way,
	 \begin{align}
	 	H_{\text{const.}}&=\sum_i \lambda_i \left(b_{i\uparrow}^\dagger b_{i\uparrow}+b_{i\downarrow}^\dagger b_{i\downarrow}-2S\right) \nonumber\\
	 	&=\lambda \sum_i \left(b_{i\uparrow}^\dagger b_{i\uparrow}+b_{i\downarrow}^\dagger b_{i\downarrow}\right)-2SN\lambda,
	 \end{align}
	where $\lambda$ is the Lagrange's undetermined multiplier and $N$ is the number of atoms in the lattice.
	It is noticeable that, generally the multiplier is assumed static and constant at each site i.e. $\lambda_i=\lambda$ and so the system retains it's translational invariance.
	The $H_{const.}$ is added with the Hamiltonian and $\lambda$ is a free parameter which is varied along with other mean-field parameters associated with the Schwinger boson bond operators to minimize the free energy of the system.

	The \gls{sbmft} is discussed here for $SU(2)$-spins. 
	There are approaches which treats $SU(2)$ spin system as either $SU(\pazocal{N})$ or $Sp(\pazocal{N})$ spin system with a large $\pazocal{N}$-limit. 
	A $SU(\pazocal{N})$ or $Sp(\pazocal{N})$ system consists of $\pazocal{N}$-flavours of Schwinger boson.
	 In the $\pazocal{N}\rightarrow \infty$ limit, the mean-field parameters and constraints are globally static throughout the system.
	 But for a large but finite $\pazocal{N}$, there are quantum-fluctuations in the mean-field parameters and constraints.
	 There is one or several parameter (mean-field parameter asscosiated with Schwinger boson bond operators and Lagrange's undetermined multiplier) sets in which case the mean-field parameters become static throughout the system.
	 These sets of parameters define the saddle points of the system.
	 The quantum fluctuations are added to the system through the $\frac{1}{\pazocal{N}}$-expansion about this saddle point, which is also known as large-$\pazocal{N}$ or saddle point expansion.
	 The reason of using a saddle point expansion or $\frac{1}{\pazocal{N}}$-expansion, instead of using a perturbative expansion as in case of $\frac{1}{S}$-expansion in \gls{HP}-method is that each terms in $\frac{1}{N}$-expansion respect the symmetry of the spin system.
	 	 Further details of the large-$\pazocal{N}$-expansion or $\frac{1}{\pazocal{N}}$-expansion or saddle-point expansion can be found in references Ref.\,\cite{Assa-Book,FrustratedMagnetismBook,Assa-Review}.

	\section{Diagonalization of quadratic Hamiltonian\label{diagonalization}}
	\subsection{Tight binding Hamiltonian and it's diagonalization}
	\label{sec2.3}
	\begin{figure}[!htb]
	\centering
	{\includegraphics[width=\textwidth]{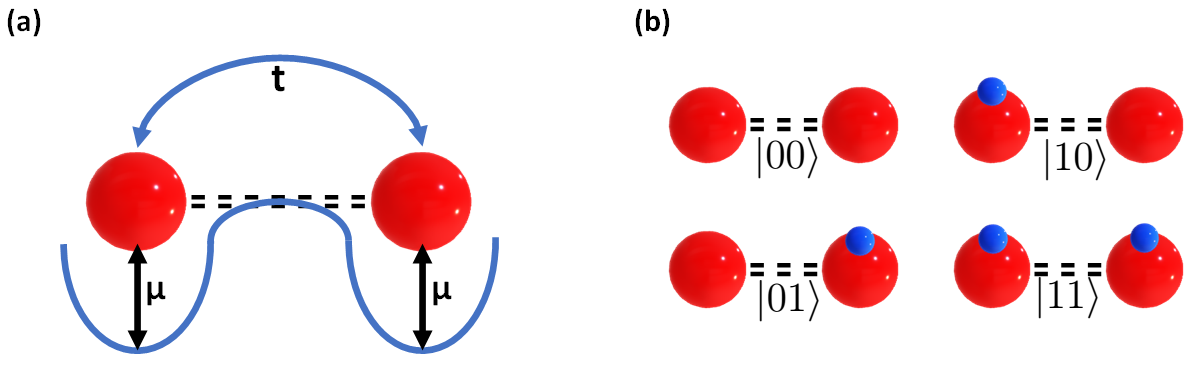}}
	\caption{(a) The schematic of a two-site system with tight binding Hamiltonian. (b) The Hilbert space of two site system.}
	\label{Hopping}
	\end{figure}
	
	In this section the way to diagonalize a tight binding Hamiltonian for non-interacting particles is described using simple examples and after that the generic formalisms are described.
	First of all, a simple tight binding Hamiltonian as described in figure Fig.\,\ref{Hopping}(a), is described in the following way,
	\begin{equation}
		H_1=t(c_1^\dagger c_2 +c_2^\dagger c_1)+\mu (c_1^\dagger c_1 +c_2^\dagger c_2 ),
		\label{eq2.31}
	\end{equation}
	where the first term is represents hopping of particles between two sites with a hopping amplitude $t$, as shown in figure Fig.\,\ref{Hopping}(a).
	The second term in the Hamiltonian represents on-site potential $\mu$ at each site.
	The particle might be either fermion or boson.
	This kind of Hamiltonian is used to describe the motion of electrons in band-insulator, semiconductor or a semimetal.  
	This origin of this kind of Hamiltonian for electrons in graphene is well described in the section Sec.4 of the reference Ref\,\cite{GQD_Book}.
	The Hilbert space of the system is of dimension $4$ and the convenient choice of basis is shown in the figure Fig.\,\ref{Hopping}(b). 
	The basises in the Dirac braket notation are $\ket{00}$, $\ket{01}$, $\ket{10}$, $\ket{11}$, where the first and second number in the braket denote the occupation number of the first and the second site respectively.

	Now, we re-arrange the Hamiltonian Eq.\,\ref{eq2.31} in the following way,
	\begin{equation}
		H_1=\begin{pmatrix}
		c_1^\dagger & c_2^\dagger
		\end{pmatrix}
		\underbrace{
		\begin{pmatrix}
			\mu & t\\
			t & \mu
		\end{pmatrix}
		}_{M'}
		\begin{pmatrix}
			c_1 \\ c_2
		\end{pmatrix}.
	\label{eq2.36}		
	\end{equation}
 A single particle Hamiltonian can be represented very simply in the form of Eq.\,\ref{eq2.36} because the single particle states are related with the vacuum state (zero-particle state) in the following way,
	\begin{equation}
		\ket{10}=c_1^\dagger\ket{00},\quad \ket{01}=c_2^\dagger\ket{00}.
	\end{equation}	
	All other eigenstates are just product states of the single particle eigenstates, because of the absence of interaction among particles in the Hamiltonian $H_1$.
	 So, these single particle states provide all the physics of the system.

	 Thus for any tight binding Hamiltonian of the following form,
	 \begin{equation}
	 	H=\sum_{i,j,\alpha,\beta} t_{(i,j)}^{(\alpha,\beta)}c_{i\alpha}^\dagger c_{j\beta},
	 \end{equation}
	  a matrix similar to $M'$ (in Eq.\,\ref{eq2.36}) is achieved by taking the row matrix of creation operators and column matrix of  annihilation operators at the left and right side respectively, as shown in Eq.\,\ref{eq2.36} for a simpler Hamiltonian $H_1$.
	  Here $i,j$ are site indices and $\alpha,\beta$ represents different species of particles.
	
	\subsubsection{Diagonalization in k-space}\label{sec2.3.1}
	 For simplicity a single particle system is assumed to be periodic and so the system retains it's translational symmetry. 
	Due to the periodicity the creation and annihilation operators can be Fourier transformed as follows,
	\begin{equation}
		c^\dagger_{\bold{k}\alpha}=\frac{1}{\sqrt{N}} \sum_{\bold{R}_i} c_{i,\alpha}^\dagger \exp(i\bold{k}\cdot \bold{R}_i),\quad
		c^\dagger_{i\alpha}=\frac{1}{\sqrt{N}} \sum_{k} c_{\bold{k},\alpha}^\dagger \exp(-i\bold{k}\cdot \bold{R}_i),
	\end{equation}
	where $\bold{k}$ is the wave-vector and take specific discrete values depending on the periodicity.
	$\bold{R}_i$ denotes the position of i-th site of the lattice and $N$ is the total number of lattice sites.

	Let's assume that a Hamiltonian in k-space is represented in the following way,
	\begin{align}
		H=\sum_{\bold{k}} \Psi_\bold{k}^\dagger h(\bold{k}) \Psi_\bold{k},
		\label{eq2.40}
	\end{align}
	where, $\Psi_{\bold{k}}^\dagger=\left(c_{\bold{k},1}^\dagger c_{\bold{k},2}^\dagger,\cdots ,c_{\bold{k},M}^\dagger \right)$ and $M$ is the number of species or degrees of freedom a particle.
	$h(\bold{k})$ is $M\times M$ matrix.
	According to Bloch's theorem $\bold{k}$ is conserved quantity for a non-interacting system with translational invariance and the Hamiltonian can be diagonalized in the following form,
	\begin{equation}
		H=\sum_{\bold{k},n} E_n(\bold{k}) \eta_{n\bold{k}}^\dagger \eta_{n\bold{k}},
	\end{equation}
	where the energies $E_n(\bold{k})$ are given by the eigenvalues of the matrix $h(\bold{k})$ in equation Eq.\,\ref{eq2.40} (see Appendix.\,\ref{appendixB}).
	 $\eta_{n\bold{k}}^\dagger$ is the creation operator for the single particle state correspond to the energy $E_n(\bold{k})$.
	The operators $\eta_{n\bold{k}}^\dagger$ can be written as linear superposition of the operators $c_{\bold{k},\alpha}$ (in other words the single particle eigenstates of the system are linear superposition of the conventional single particle basis in),
	\begin{equation}
		\eta_{n\bold{k}}^\dagger=\sum_\alpha u_{\alpha,n}(\bold{k}) c_{\bold{k},\alpha}^\dagger,
	\end{equation}
where the coefficients $u_{\alpha,n}(\bold{k})$ is given by $\alpha$-th element of the $n$-th eigenvector (which corresponds to the eigenvalue $E_n(\bold{k})$) of $h(\bold{k})$ (see Appendix.\,\ref{appendixB}).

	\subsection{Bogoliubov Hamiltonian and Bogoliubov-Valatin transformation}\label{sec2.4}
	In this section, a diagonalization formalism of a much more general Hamiltonian compared with the tight-binding Hamiltonian in previous section is treated.
	Before going into the procedure, we revisit the historical reasons behind the nomenclature of the diagonalization method and the Hamiltonian.
		The formalism to diagonalize the bosonic quadratic Hamiltonian which appears in case of super-fluidity was first introduced by  Bogoliubov in 1947\,\cite{Bogoliubov}.
		Later, this formalism was used to diagonalize the Hamiltonian for fermions in superconductors by Bogoliubov\,\cite{Bogoliubov2,Bogoliubov3} as well as Valatin\,\cite{Valatin,Valatin2}.
		So the procedure to diagonalize the Hamiltonian is known as Bogoliubov-Valatin transformation.
		This procedure of diagonalization is applied in many other fields of physics\,\cite{OtherBogoliubov, OtherBogoliubov2}.
		A superconducting state is a many body state with Hilbert space of dimension $2^N$, where $N$ is the number lattice site.
		Using mean field approximations the Hamiltonian is transformed into quadratic Hamiltonian with a Hilbert space of dimension $2N$ and this mean-field quadratic Hamiltonian for superconductors is known as Bogoliubov-de Gennes(BdG) Hamiltonian\,\cite{DeGennes_Book}.

\begin{figure}[!htb]
	\centering
	{\includegraphics[width=0.8\textwidth]{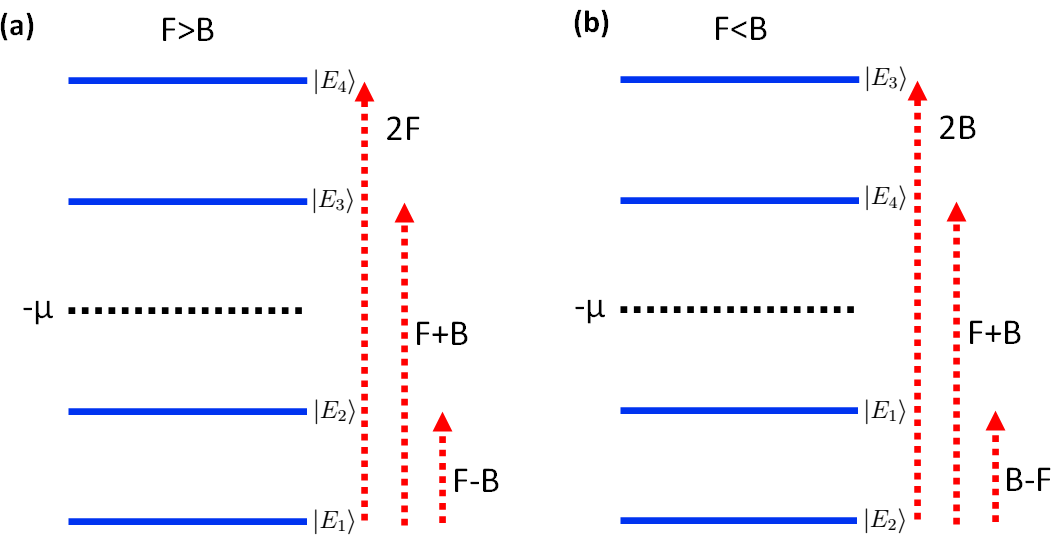}}
	\caption{(a) The schematic of energy-levels when $F>B$. (b) The schematic of energy levels when $F<B$.}
	\label{EnergySchematic}
\end{figure}

		The Bogoliubov Hamiltonian is same as the tight binding Hamiltonian with extra pair-creation and pair annihilation operators as discussed later.
		More technical details of Bogoliubov transformation is discussed  in the reference Ref.\,\cite{BogoliubovValatin} and in the appendix of the reference Ref.\,\cite{Romhanyi1} and also in the method section of reference Ref.\,\cite{BogoliubovMethod}.
		In this section, a simple example is first discussed to gain more insight into the physics and then generic formalism is derived.
		A simple Bogoliubov Hamiltonian of a single site superconductor is given by,
		\begin{equation}
			H_1=
-\mu(c_{\uparrow}^\dagger c_{\uparrow}+c_{\downarrow}^\dagger c_{\downarrow})
+B(c_{\uparrow}^\dagger c_{\uparrow}-c_{\downarrow}^\dagger c_{\downarrow})
+\Delta c_{\uparrow}^\dagger c_{\downarrow}^\dagger
+\Delta^* c_{\downarrow} c_{\uparrow},
\label{eq2.44}
		\end{equation}
		where $c_{\uparrow}^\dagger$ and $c_{\downarrow}^\dagger$ are creation operators of the up-spin and down-spin electrons respectively.
		The first term in Hamiltonian denotes the onsite potential.
		The second term represents Zeeman-coupling between electron-spin and magnetic field.
		The third and fourth term are the Cooper-pair creation and annihilation term in the system.
		The Cooper-pairs of the electrons are possible due to phonon-electron coupling at low temperature and further details of the Hamiltonian can be found in any standerd text-book\,\cite{Tinkham}.
		The Hilbert space of the system is spanned by four basis states,$\ket{0}$, $\ket{\uparrow}$, $\ket{\downarrow}$, $\ket{\uparrow\downarrow}$.
		Thus the Hamiltonian of the system is given by the following matrix-form,
		\begin{equation}
			H_1=\begin{pmatrix}
			\ket{0} & \ket{\uparrow\downarrow} & \ket{\downarrow} &\ket{\uparrow}
			\end{pmatrix}
			\underbrace{
			\begin{pmatrix}
			0 & \Delta^* & 0 & 0\\
			\Delta & -2\mu & 0 & 0\\
			0 & 0 & -\mu-B & 0\\
			0 & 0& 0 &-\mu+B
			\end{pmatrix}
			}_M
			\begin{pmatrix}
			\bra{0} \\ \bra{\uparrow\downarrow} \\ \bra{\downarrow} \\ \bra{\uparrow}
			\end{pmatrix}.
			\label{eq2.45}
		\end{equation}
		 The parity of the system is conserved which means the states $\ket{\uparrow\downarrow}$ and $\ket{0}$ can not mix with the states $\ket{\uparrow}$ and $\ket{\downarrow}$.
		The parity conservation is always true for any general Bogoliubov Hamiltonian.
		Diagonalizing the matrix $M$, we get the following eigenvalues and corresponding eigenvectors of the Hamiltonian as,
		\begin{equation}
			\begin{array}{*2c}
			\multicolumn{1}{c|}{E_1=-\mu-F} & \quad \ket{E_1}=-\Delta^*\ket{0}+(\mu+F) \ket{\uparrow\downarrow} \\
			\multicolumn{1}{c|}{E_2=-\mu-B} & \quad \ket{E_2}=\ket{\uparrow}\\
			\multicolumn{1}{c|}{E_3=-\mu+B} &\quad \ket{E_3}=\ket{\downarrow}\\
			\multicolumn{1}{c|}{E_4=-\mu+F} & \quad \ket{E_4}=-\Delta^*\ket{0}+(\mu-F) \ket{\uparrow\downarrow}
			\end{array},
			\label{eq2.46}
		\end{equation}
		where $F=\sqrt{|\Delta|^2+\mu^2}$.
		The energy levels are schematically represented in the figure Fig.\,\ref{EnergySchematic}.
		The Hamiltonian in Eq.\,\ref{eq2.45} is a many-body Hamiltonian. 
		However, the quadratic nature of the Hamiltonian assures that the system can be described using single-particle states due to absence of interaction between the up- and down-spins on the same site. (which is also true for the Hamiltonian in section Sec.\,\ref{sec2.3}).

		The Hamiltonian in Eq.\,\ref{eq2.44} can be rearranged as follows,
		\begin{equation}
			H_1=\frac{1}{2}
			\begin{pmatrix}
				c_\uparrow & c_\downarrow & c_\uparrow^\dagger & c_\downarrow^\dagger
			\end{pmatrix}
			\underbrace{
			\begin{pmatrix}
					\mu-B & 0 & 0 & -\Delta^*\\
					0 & \mu+B & \Delta^* & 0\\
					0 & \Delta & -\mu+B & 0\\
					-\Delta & 0 & 0 & -\mu-B
			\end{pmatrix}}_{M'}
			\begin{pmatrix}
				c_\uparrow^\dagger \\ c_\downarrow^\dagger \\ c_\uparrow \\ c_\downarrow
			\end{pmatrix}.
			\label{eq2.47}
		\end{equation}
		This way of representing a Hamiltonian is called Bogoliubov representation and the matrix $M'$ is called coefficient matrix.
		After diagonalizing the coefficient matrices $M^\prime_1$, we get the eigenvalues and eigenvectors as,
		\begin{equation}
			\begin{array}{*2c}
			\multicolumn{1}{c|}{\Delta E_1=-B-F,} & \quad d_1^\dagger= (F-\mu) c_\uparrow^\dagger +\Delta c_\downarrow ,  \\
			\multicolumn{1}{c|}{\Delta E_2=F-B,} & \quad d_2^\dagger=-(F+\mu) c_\uparrow^\dagger +\Delta c_\downarrow ,  \\
			\multicolumn{1}{c|}{\Delta E_3=B+F,} &\quad d_3^\dagger=(F+\mu) c_\downarrow^\dagger +\Delta c_\uparrow\\
			\multicolumn{1}{c|}{\Delta E_4=B-F,} & \quad d_4^\dagger=(-F+\mu)c_\downarrow^\dagger +\Delta c_\uparrow
			\end{array},
		\end{equation}
It is noticeable that the energy difference in figure Fig.\,\ref{EnergySchematic} look similar to the eigenvalues.
So the transformation gives the relative energies.

For example for the case $F>B$, $\Delta E_2$ and $\Delta E_3$ are the positive energy difference.
 The corresponding single particle operators are $d_2^\dagger$ and $d_3^\dagger$ respectively.
 So the energies and many-body states are given for the case $F>B$,
 \begin{equation}
 	\begin{array}{*2c}
 	\multicolumn{1}{c|}{E_1=E_{GS}} & \quad \ket{E_1}=\ket{GS} \\
 	\multicolumn{1}{c|}{E_2=E_{GS}+F-B} & \quad \ket{E_2}=d_2^\dagger\ket{GS}\\
	\multicolumn{1}{c|}{E_3=E_{GS}+F+B} &\quad \ket{E_3}=d_3^\dagger\ket{GS}\\
	\multicolumn{1}{c|}{E_4=E_{GS}+2F} &\quad \ket{E_4}=d_2^\dagger d_3^\dagger\ket{GS}
 	\end{array}.
 	\label{eq2.50}
 \end{equation}
 If we use $E_{GS}=-\mu-F$ and $\ket{GS}=-\Delta^* \ket{0}+(\mu+F)\ket{\uparrow\downarrow}$ then the  eigenstates and eigenvalues of Eq.\,\ref{eq2.50} matches exactly with the eigenstates and eigenvalues as in the figure Fig.\,\ref{EnergySchematic}(a).
The procedure can also be verified for the case $F<B$ as in figure Fig.\,\ref{EnergySchematic}(b). For bosonic particles the Bogoliubov transformation is different due to commutation relations.

		A most generic form of quadratic Hamiltonian independent of particle statistics is given by,
		\begin{equation}
			H=\sum_{i,j} \left(\alpha_{ij} c_i^\dagger c_j
			+\frac{1}{2} \gamma_{ij} c_i^\dagger c_j^\dagger
			+\frac{1}{2} \gamma_{ji}^* c_i c_j\right)
		\end{equation}
		The coefficient matrix for the fermions and bosons are respectively,
		\begin{equation}
			M^\prime_{f}=\begin{pmatrix}
			-\alpha^T & \gamma\\
			\gamma^\dagger & \alpha
			\end{pmatrix}
			\quad
			M^\prime_{b}=\begin{pmatrix}
			\alpha^T & \gamma\\
			\gamma^\dagger & \alpha
			\end{pmatrix}.
		\end{equation}
		Thus the dynamical matrices for fermions and bosons are $M^\prime_f$ and $\sigma_z M^\prime_b$ respectively where,
		\begin{equation}
			\sigma_z=\begin{pmatrix}
			\mathbb{I} & 0\\
			0 & -\mathbb{I}
			\end{pmatrix}.
			\label{sigmaz}
		\end{equation}
		If the coefficient matrix is $2N\times 2N$, then $\mathbb{I}$ is $N\times N$ identity matrix.
		The Bogoliubov-Valatin transformation is given by diagonalizing the dynamic matrices.
		
		\subsubsection{Bogoliubov-Valatin transformation in k-space}
		As discussed in the section Sec.\,\ref{sec2.3.1}, due to translational invariance of the non-interacting periodic system the creation and annihilation operators can be Fourier transformed. 
		A general quadratic Hamiltonian in k-space is given by,
		\begin{equation}
			H=\frac{1}{2}\sum_{ij} \sum_{\bold{k}} \left[
			\alpha_{ij}(\bold{k})a_{\bold{k}i} a_{\bold{k}j}^\dagger
			 \pm\alpha_{ij}(-\bold{k}) a_{-\bold{k}j}^\dagger a_{-\bold{k}i}
			 +\gamma_{ij}(\bold{k}) a_{-\bold{k}i}^\dagger a_{\bold{k}j}^\dagger
			 +\gamma^*_{ij}(\bold{k}) a_{\bold{k}j} a_{-\bold{k}i}
			  \right].
			  \label{eq2.57}
		\end{equation}
		The positive and negative signs in the equation is for the bosons and fermions respectively.
		The Bogoliubov representation of the Hamiltonian is given as,
		\begin{equation}
			H=\frac{1}{2}
			\begin{pmatrix}
				a_{\bold{k}} & a_{-\bold{k}}^\dagger 
			\end{pmatrix}
			\underbrace{
			\begin{pmatrix}
				\alpha(\bold{k}) & \gamma^\dagger(\bold{k})\\
				\gamma(\bold{k}) & \pm\alpha^T(-\bold{k})
			\end{pmatrix}
			}_{M'(\bold{k})}
			\begin{pmatrix}
				a_{\bold{k}}^\dagger \\ a_{-\bold{k}}
			\end{pmatrix},
			\label{eq2.58}
		\end{equation}
		where $M'(\bold{k})$ is coefficient matrix and
		\begin{equation}
			\gamma(\bold{k})=\pm \gamma(-\bold{k})^T,\quad
			\alpha(\bold{k})=\alpha(\bold{k})^\dagger.
		\end{equation}
		The condition $\gamma(\bold{k})=\pm \gamma(-\bold{k})^T$ is for simplicity to define the Bogolibov-Valatin transformation and the condition $\alpha(\bold{k})=\alpha(\bold{k})^\dagger$ is due to hermiticity of the Hamiltonian.
		Even if the first condition is not valid then it is easy to construct a new matrix following $\gamma^\prime(\bold{k})=\frac{1}{2}\left(\gamma(\bold{k})+\gamma(-\bold{k})^T\right)$ and then construct the coefficient matrix using $\gamma^\prime(\bold{k})$ in place of $\gamma(\bold{k})$.

		The justification of choosing the Bogoliubov representation as in Eq.\,\ref{eq2.58} can be understood in the following way. 
		The single particle operator after diagonalization is linear combination of the $a_{\bold{k},i}^\dagger$ operators as follows,
		\begin{equation}
			\eta_{n\bold{k}}=\sum_{\bold{k}',i} \left[
			u_{n,i}(\bold{k}') a_{\bold{k}',i}^\dagger
			+v_{n,i}(\bold{k}') a_{-\bold{k}',i}
			+w_{n,i}(\bold{k}') a_{-\bold{k}',i}^\dagger
			+z_{n,i}(\bold{k}') a_{\bold{k}',i}			
			\right].
			\label{eq2.59}
		\end{equation}
		The operations $a_{\bold{k},i}^\dagger$ and $a_{-\bold{k},i}$ are equivalent, in the sense that creation of particle at momentum $\bold{k}$ is equivalent to the annihilation of particle at momentum $-\bold{k}$.
		Thus according to Bloch's theorem this two operators can be linearly superimposed to create the Bloch's state.
		But the third and fourth contributions in Eq.\,\ref{eq2.59} should be zero and the summation over $\bold{k}'$ is also reduce to a single term at $\bold{k}$. So the operator after diagonalization should be of the following form,
		\begin{align}
			\eta_{n\bold{k}}&=\sum_{i} \left[
			u_{n,i}(\bold{k}) a_{\bold{k},i}^\dagger
			+v_{n,i}(\bold{k}) a_{-\bold{k},i}		
			\right].\nonumber\\
             &=\begin{pmatrix}
             	u_n(\bold{k}) & v_n(\bold{k})
				\end{pmatrix}             	
				\begin{pmatrix}
					a_{\bold{k}}^\dagger \\ a_{-\bold{k}}
				\end{pmatrix}						
			\label{eq2.60}
		\end{align}
The choice of Bogoliubov representation in Eq.\,\ref{eq2.58} can be understood from observation of equation Eq.\,\ref{eq2.60}.
The Hamiltonian in diagonalized form is,
\begin{equation}
	H=\sum_{n,\bold{k}}E_n(\bold{k})\eta^\dagger_{\bold{k}n}\eta_{\bold{k}n}
	\label{eq2.61}
\end{equation}
The energies in Eq.\,\ref{eq2.61} and the matrix $\begin{pmatrix}           u_n(\bold{k}) & v_n(\bold{k})\end{pmatrix}$ in Eq.\,\ref{eq2.60} are the eigenvalues and eigenvectors of a dynamic matrix.
In Appendix.\,\ref{appendixB}, it is shown that dynamic matrix of fermions is same as coefficient matrix $M'(\bold{k})$, but the dynamic matrix of the bosons is $\sigma_z M'(\bold{k})$ where, $\sigma_z$ is the $N$-dimensional Pauli matrix defined in the equation Eq.\,\ref{sigmaz}.

 \subsubsection{Bogoliubov-Valatin transformation of Bosons as para-unitary transformation}
 In this section a more technical details of Bogoliubov-Valatin transformation specifically shown for the bosons as para-unitary transformation.
 In presence of pair-creation and annihilation operator in the Hamiltonian, the Hamiltonian can be represented as,
\begin{equation}
\pazocal{H}=\frac{1}{2} \sum_{\alpha, \beta, \bold{k}} \Psi^\dagger_{\alpha,\bold{k}} H_{\alpha,\beta}(\bold{k}) \Psi_{\alpha,\bold{k}},
\end{equation}
where, $\Psi_{\alpha,\bold{k}}=(\hat{a}_{\alpha\bold{k}},\hat{a}^\dagger_{\alpha ,-\bold{k}})^T$ is the Nambu-spinor. Here $\hat{a}_{\alpha\bold{k}}$ is the set of all annihilation operator.

Any Matrix is diagonalized by using a similarity transformation as,
\begin{equation}
T^\dagger(\bold{k}) \pazocal{H}(\bold{k}) T(\bold{k})=\epsilon(\bold{k}),
\label{eqI}
\end{equation}
In case of Bogoliubov-Valatin transformation for the Bosons the matrix $T(\bold{k})$ is a para-unitary matrix which follows the relations,
\begin{equation}
T^\dagger(\bold{k}) \sigma_3 T(\bold{k})=T(\bold{k}) \sigma_3 T^\dagger(\bold{k})=\sigma_3,
\label{eqIII}
\end{equation}
After para-unitary transformation the Eigen-values are given as,
\begin{equation}
\epsilon(\bold{k})=\text{diag}\begin{pmatrix}
 E_1(\bold{k}), & 
 E_2(\bold{k}), & 
 ... & 
 E_{N}(\bold{k}), & 
 E_1(-\bold{k}),  &
 E_2(-\bold{k}),  &
 ...&
 E_{N}(-\bold{k})
\end{pmatrix}
\end{equation}
where $N$ is the number of sub-lattices multiplied by the number of types of bosons at each site and $\text{diag}()$ denotes a diagonal matrix.
Using the equation Eq.\,\ref{eqIII}, the Eq.\,\ref{eqI} is transformed as,
\begin{equation}
\sigma_3 \pazocal{H}(\bold{k}) T(\bold{k})=T(\bold{k})\sigma_3\epsilon(\bold{k})
\label{eqII}
\end{equation}
		It is noticeable that according to the the above equation the dynamic matrix of $\pazocal{H}(\bold{k})$ is the matrix $\sigma_3\pazocal{H}(\bold{k})$ which is discussed in the previous section.
		
		The equation Eq.\,\ref{eqII} takes the following form (see Appendix.\,\ref{appendixB}),
\begin{equation}
(\sigma_3 H_\bold{k})\ket{\uu{n}}=(\sigma_3\epsilon_\bold{k}) \ket{\uu{n}}
\label{Geq1},
\end{equation}
where $\ket{u_{n,\bold{k}}}$ is the $n$-th column of the paraunitary matrix $T(\bold{k})$ i.e. it represents the eigen-vectors after diagonalization. 
The transpose-conjugate of the equation gives,
\begin{equation}
\bra{\uu{n}}\pazocal{H}_\bold{k}\sigma_3=\bra{\uu{n}}(\sigma_3 \epsilon(\bold{k})),
\label{Geq2}
\end{equation}
where $\bra{u_{n,\bold{k}}}$ is the $n$-th row of the paraunitary matrix $T^\dagger(\bold{k})$ i.e. it represents the eigen-vectors in reciprocal Hilbert-space.
	According to equation\,\ref{eqIII}, the eigenvectors follow the orthonormality relation,
	\begin{equation}
	\bra{\uu{n}}\sigma_3\ket{\uu{m}}=\delta_{nm} (\sigma_3)_{nn}.
	\label{eq2.71V2}
	\end{equation}	
	  The completeness relations of eigenvectors are (see Appendix.\,\ref{appendixB})	,
	\begin{align}
	\sum_n (\sigma_3)_{nn} \sigma_3 \ket{\uu{n}}\bra{\uu{n}}&=\mathbb{I}
	\nonumber\\
	\sum_n (\sigma_3)_{nn} \ket{\uu{n}}\bra{\uu{n}}\sigma_3 &=\mathbb{I}
	\end{align} 
	
	\section{Physical Observable}
	\label{sec2.5}
	\subsection{Berry Phase, Berry-curvature and Chern-number}
	\label{sec2.5.1}
	In this section the concepts of Berry phase, Berry-curvature and Chern-number are discussed for different contexts in physics to provide a in depth physical insight of these quantities.
	First the concepts are discussed for the case of closed 2D-manifolds embedded in 3D-space (e.g. sphere, toroid, etc.). 
	Secondly I discuss the origin of the phenomenon in the context of quantum mechanics.
	In the end of the section, it is shown that the physical quantities are related to the non-trivial fibre-bundle structure of parameter space and so phenomenon related to Berry-phase can arise even in classical system.

\begin{figure}[H]
	\centering
	{\includegraphics[width=0.8\textwidth]{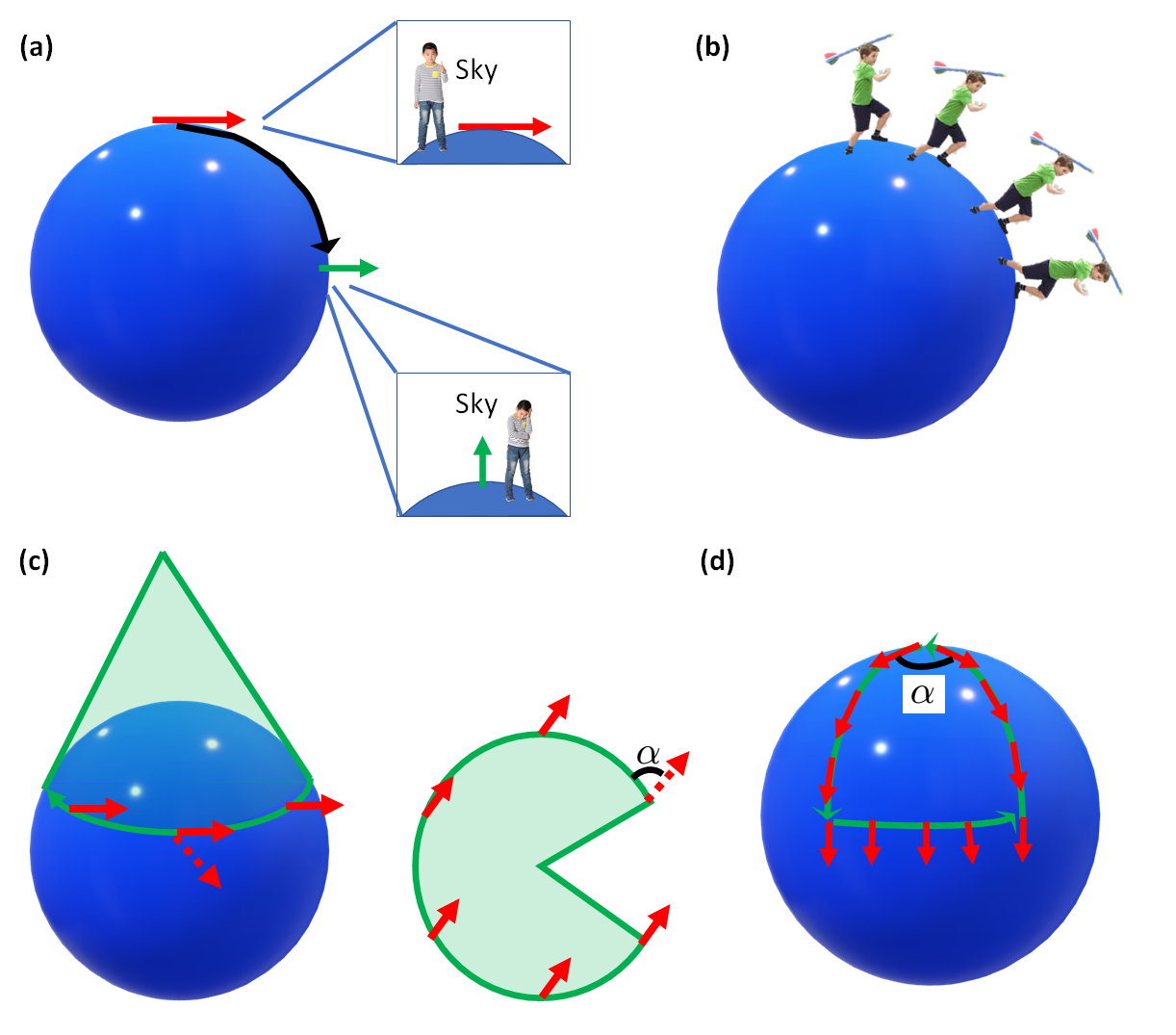}}
	\caption{(a) Schematic of transport of vector without any constraint on a spherical surface. The vectors after transportation (green arrow) is exactly same with the initial vector (red-arrow). In the magnified figures it is shown that the vector is parallel to the ground at North-pole but the vector point towards the sky after transportation. So for a flatlander on earth the two vectors are different. (b) The illustration of a parallel transportation, where the child carrying an arrow and parallelly transporting the arrow from the North-pole to the equator. (c) The demonstration of deficit angle after parallel transport along a latitude of the sphere. The red arrow is parallel transported along the latitude and the final vector after transport shown by dotted arrow. The figure shows a cone that touches the latitude of the sphere. The right hand figure shows that the cone is transformed into a 2D-surface to simply demonstrate the parallel transport. (d) Another example of parallel transport to demonstrate the deficit angle for the transport along a closed curve on a sphere is equal to the solid angle made by the closed curve at the center of the sphere.}
	\label{sphere}
	\end{figure}

	Firstly, we discuss the concepts in case of closed 2D-manifolds (Riemannian manifold) embedded inside a 3D-space.
	The words in italics inside the parenthesis are the corresponding physical quantities in quantum mechanics and that will be discussed later.
	A sphere is the simplest closed 2D-manifold embedded inside 3D-space as shown in figure\,\ref{sphere}(a).
	Let us assume that the sphere represent the earth.
	If a vector from north-pole to equator is transported using the way described in figure Fig.\,\ref{sphere}(a), then for the local resident of the sphere or earth the two vectors are not same, because at the north-pole the vector is parallel to the earth surface and at the equator the vector is pointing towards the sky (Fig.\,\ref{sphere}(a)).
	So, a different process of transport is defined which is known as parallel transport (\textit{adiabatic process}).
	The parallel transport (\textit{adiabatic process}) is defined in such a way that the covariant derivative (see Ref.\,\cite{ParallelTransportYouTube} for details) of the vector during the transport is zero (Fig.\,\ref{sphere}(b)).
	The intuitive definition of covariant derivative is that the rate of change of the vector along a certain direction on the manifold subtracted less the component of the rate of change of the vector along the normal direction of the surface.
	The logic of the subtracting the normal component is to avoid the situation as in Fig.\,\ref{sphere}(a), where the vector at north-pole is parallel to the surface, but at equator the vector is perpendicular to the surface.
	In figure Fig.\,\ref{sphere}(c), a vector is parallel transported along a closed loop on the surface of the earth.
	
	 \noindent The fascinating outcome of this parallel transport is that the final vector at end of the parallel transport is not the same as the initial vector, although locally the transportation is such that the nearby vectors are parallel to each other.
	 The initial and final vector after parallel transport differs by an angle $\alpha$ known as deficit angle (\textit{Berry phase}).
	 The deficit angle (\textit{Berry phase}) is intuitively understood by using a cone on the sphere as in figure Fig.\,\ref{sphere}(c).
	 The cone touches the latitude of the sphere and describes the local tangent plane of the sphere near to the latitude.
	 It is easy to understand the origin of the deficit angle by parallel transport by projecting the cone to a 2D-plane as in the right figure in Fig.\,\ref{sphere}(c).
	  According to Gauss-Bonnet theorem the deficit angle is given by,\,\cite{ParallelTransport},
	 \begin{equation}
	 	\alpha=\int\int_{\sigma} K dS,
	 	\label{eq2.64}
	\end{equation}
	where $K$ is the Gaussian-curvature (\textit{Berry-curvature}) and the integration is over the surface closed by the loop along which the parallel transport (\textit{adiabatic process}) is done.
	For the case of sphere the Gaussian curvature at each point is equal to $\frac{1}{R^2}$, where $R$ is the radius of the sphere.
	Thus from equation Eq.\,\ref{eq2.64}, the deficit angle (\textit{Berry phase}) is just the solid angle at the center of the sphere produced by the loop for the parallel transportation (\textit{adiabatic process}) (see figure Fig.\,\ref{sphere}(c)).
	Furthermore, according to Gauss-Bonnet theorem if the integration in equation Eq.\,\ref{eq2.64} is done over a whole 2D-manifold (e.g. sphere or toroid) then it provides a integer number which defines the topology of the 2D manifold,
	\begin{equation}
		n=\frac{1}{2\pi}\oiint_\sigma K dS.
	\end{equation}	
The integer $n$ is known as Euler-characteristic (\textit{Chern-number}) of the 2D manifold.
The Euler-characteristic $n=2$ for a sphere, because the total solid angle traversed by a whole sphere is $4\pi$.
 Moreover, the Euler-characteristic (\textit{Chern-number}) is related with the number of genus or hole $g$ (e.g. sphere has no hole so $g=0$, but toroid has one hole so $g=1$) present in the 2D-manifold by the following relationship, $n=2(1-g)$.

 The adiabatic process, wave-function, Berry phase, Berry curvature and Chern-number in quantum mechanics are equivalent to the Parallel-transport, vector(which is parallel transported), deficit angle, Gaussian curvature and Euler-characteristic in 2D-manifold embedded in 3D-space respectively.
 The definition of Chern-number which is equivalent to the Euler-characteristic only exists in the closed parameter space, for example a sphere is represented as a closed parameter space of polar angle ($\theta$) and azimuthal angle ($\phi$).
 Similarly the Brillouin-zone is a toroid in terms of crystal momentum due to periodicity and so Chern-number can be defined in the k-space, which is further discussed in the section Sec.\,\ref{sec2.5.2}.

 An adiabatic process (\textit{parallel transport}) in quantum mechanics is a type of process to evolve the Hamiltonian in a way that the eigenstates in the system do not change locally.
 The process of evolution is adiabatic (\textit{parallel transport}) if the following condition is fulfilled (see Appendix\,\ref{appendixC}) ,
 \begin{equation}
 	\frac{\bra{\psi_m(t)}\dot{H}(t)\ket{\psi_n(t)}}{|E_m-E_n|}\rightarrow 0,
 \end{equation}
 where $m\neq n$,  $\dot{H}(t)=\frac{dH(t)}{dt}$ and $\ket{\psi_m}$ is the $m$-th eigenstate of the system.
 The left hand side of the expression is the characteristic time of transition between the $m$-th and $n$-th state. 
 The condition need to be fulfilled for all the states $\ket{\psi_m}$ and $\ket{\psi_n}$.
  The condition require an energy-gap among all the eigen-states as well as requires no enrgy level crossing (or mixing) during the evoluion of the Hamiltonian.
 So we assume all the eigenstates of the system are gapped.
 If the eigenstates are not gapped then a net non-abelian Berry-phase can be calculated for the degenerate states, which is not discussed here.

 If initially the quantum system is in the state $\ket{\Psi}$ such that,
 \begin{equation}
 	\ket{\Psi}=\sum_n c_n \ket{\psi_n},
\end{equation}  
then the evolved state of the system under adiabatic process is (see Appendix\,\ref{appendixC}),
\begin{equation}
	\ket{\Psi(\bold{R}(t))}=\sum_n c_n(0) e^{i\gamma_n(\bold{R}(t))} e^{\theta_n(t)}\ket{\psi_n(\bold{R}(t))},
\end{equation}
  where $\bold{R}(t)$ represents a set of parameters of the Hamiltonian and the system is adiabatically evolved by varying the parameters of the Hamiltonian.
  The phases $\theta_n(t)$ and $\gamma_n(\bold{R})$ are known as the dynamical phase and geometric phase respectively,
  	\begin{align}
  		\theta_n(t)=-\frac{i}{\hbar}\int^t_0 E_n(\tau) d\tau,
  		\quad
  		\gamma_n(\bold{R})&=i\int_{\bold{R}(0)}^{\bold{R}(t)} \bra{\psi_n(\bold{R})}\nabla_\bold{R}\ket{\psi_n(\bold{R})}\cdot d\bold{R} \nonumber\\
		&= \int_{\bold{R}(0)}^{\bold{R}(t)} \pazocal{A}(\bold{R})\cdot d\bold{R},  		
		\label{eq2.69}
	\end{align} 
	where $\pazocal{A}(\bold{R})=\bra{\psi_n(\bold{R})}\nabla_\bold{R}\ket{\psi_n(\bold{R})}$ is the Berry-connection.  
It is interesting to note that the dynamic phase changes with time, but the geometric phase only depends on the path of the adiabatic process in the parameter space $\bold{R}$ and so theoretically the geometric-phase does need to be associated with a dynamical process with change in real time. That is why the name is geometric phase.
The geometric phase under adiabatic approximation is known as the Berry-phase (\textit{deficit angle}). 
In general Berry phase as well as Berry connections are not physical observable because the quantity is not Gauge invariant.
The Berry-connection and Berry-phase after gauge transformation are given as, 
\begin{align}
	\pazocal{A}^\prime(\bold{R})&=\pazocal{A}(\bold{R})-\nabla_\bold{R} \alpha(\bold{R})\nonumber\\
	\gamma^\prime_n(\bold{R(t)})&=\gamma_n(\bold{R(t)})-\alpha(\bold{R}(t))+\alpha(\bold{R}(0)),
	\label{eq2.70}
\end{align} 
where $\alpha(\bold{R}(t))$ is the $U(1)$-gauge transformation introduced as $\ket{\psi^\prime_n(\bold{R})}=e^{i\alpha(\bold{R})}\ket{\psi_n(\bold{R})}$.
But it is obvious that for a closed path $\alpha(\bold{R}(t))=\alpha(\bold{R}(0))$ (because $\bold{R}(t)=\bold{R}(0)$ for a closed path) and so the Berry-phase is gauge invariant for a closed path (see equation Eq.\,\ref{eq2.70}).
Thus Berry-phase is a physical quantity for a cyclic or closed adiabatic process.
  For a closed loop in three-dimensional parameter space the expression for geometric phase in equation Eq.\,\ref{eq2.69} is re-expressed using Stoke's theorem as,
  \begin{equation}
  	\gamma_n(\bold{R})=\oiint_{S_{\bold{R}}} (\nabla \times\pazocal{A}(\bold{R}))\cdot dS_{\bold{R}},
  	\label{eq2.71}
  \end{equation}
where $S_{\bold{R}}$ is the surface closed by the loop in 3D-parameter space.
The quantity $\boldsymbol{\Omega}(\bold{R})=\nabla \times\pazocal{A}(\bold{R})$ is known as Berry-curvature (\textit{Gaussian curvature}).
The Berry curvature (\textit{Gaussian curvature}) is gauge invariant quantity as can be seen by taking curl ($\nabla \times$) at the both sides of the first equation in Eq.\,\ref{eq2.70}.
For a 2D-parameter space the expression of Berry curvature simply become,
\begin{equation}
\Omega_z(\bold{R})=\frac{\partial \pazocal{A}_y}{\partial x}-\frac{\partial \pazocal{A}_x}{\partial y}.
\label{eq2.72}
\end{equation}
Furthermore there is no definition of Berry-curvature in one-dimensional parameter space.
But Berry-phase in one dimension still exists if the wave-function is complex\,\cite{BerryPhaseYouTube}.
However for the one-dimensional material the Berry-phase exists which is known as Zak phase which is no-zero for a complex wave-function for a one-dimensional tight-binding model with real hopping parameters and this is to have more than one sub-lattice in the material.    
So the most simplest example is one dimensional bipartite lattice where Zak phase exists\,\cite{ZakPhase}.

  \begin{figure}[H]
	\centering
	{\includegraphics[width=0.8\textwidth]{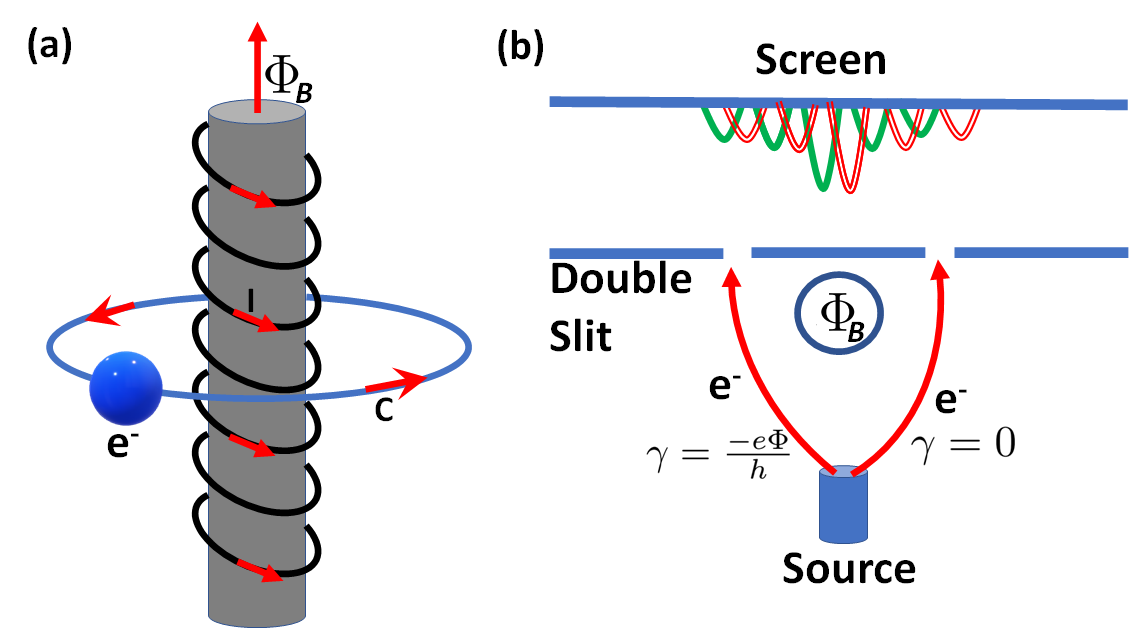}}
	\caption{(a) Schematic of Aharnov-Bohm effect. The electron gains a phase $\frac{e\Phi}{\hbar}$ by traversing along the path-$C$ which revolves around a infinite solenoid with a magnetic flux $\Phi_B$. (b) The schematic of electron double slit experiment with a solenoid in the middle of the double-slits and the flux through the solenoid is in the direction outward and perpendicular to the plane of paper. The green interference fringes are in absence of magnetic flux through the solenoid and the red interference fringes are in presence of magnetic flux $\Phi_B$ through the solenoid. Because of Aharnov-Bohm effect two electron beams has different phases and so the red interference fringes are shifted right with respect to the green interference fringes.}
	\label{solenoid}
	\end{figure}
  
  To further illustrate the concepts of Berry-phase, here the Aharonov–Bohm effect is discussed.
  A more experimental and theoretical description of this effect can be foun in reference Ref.\,\cite{AharnovBohm}.
  The Aharnov-Bohm effect is a quantum mechanical phenomenon in which a charged particle gains a phase while travelling around a infinite solenoid, as shown in the figure Fig.\,\ref{solenoid}(a) for an electron.
  The interesting feature of the effect is that even if outside a infinite solenoid there is no magnetic field, but the charged particle gains a phase after traversing a closed loop around the solenoid.
The physical observable phase-difference of the charged particle is solely due to the presence of physical non-observable (because not a gauge invariant) magnetic vector potential outside the infinite solenoid.  
  After solving the Schrödinger's equation of a particle in a zero magnetic field $\bold{B}$ but with a non-zero magnetic potential $\pazocal{A}$ (where $\nabla \times\pazocal{A}=\bold{B}=0$) the wave-function is given by (see Appendix\,\ref{appendixC}),
  \begin{equation}
  	\ket{\Psi_B(\bold{r},t)}=e^{ig(\bold{r})}\ket{\Psi_0(\bold{r},t)},
  \end{equation}
   where $\ket{\Psi_0(\bold{r},t)}$ is the wave function in absence of magnetic vector potential and the phase,
	\begin{equation}    
    g(\bold{r})=\frac{e}{\hbar}\int_0^\bold{r} \pazocal{A}(\bold{r}')\cdot d\bold{r}'.
    \label{eq2.73}
    \end{equation} 
    It is well known that magnetic-vector potential is not a gauge-invariant quantity and it can also be shown that the phase $g(\bold{r})$ is also not a gauge invariant quantity similar to the Berry-phase.  
    Comparing the equations Eq.\,\ref{eq2.69} and Eq.\,\ref{eq2.73}, we find that the magnetic vector potential $\pazocal{A}(\bold{r})$ and the phase $g(\bold{r})$ are equivalent to the Berry-connection and the Berry-phase.  
    For a closed path $C$ the Berry-phase,
   \begin{align}
   		\gamma=g(\bold{R})-g(\bold{0})&=\frac{e}{\hbar}\oint_C \pazocal{A}(\bold{r}')\cdot d\bold{r}'\quad\left[\text{$\bold{0}$ and $\bold{R}$ are same points}\right]
   		\nonumber\\
   		&=\frac{e}{\hbar}\oiint_{S} (\nabla\times \pazocal{A})\cdot d\bold{S} \quad\left[\text{Using Stoke's theorem. $S$ is the surface of loop $C$}\right]
   		\nonumber\\
   		&=\frac{e}{\hbar} \oiint_{S} \bold{B}\cdot d\bold{S}
   		\nonumber\\
   		&=\frac{e\Phi_B}{\hbar}.
   \end{align}
   Thus the total Berry-phase for a closed loop is proportional to the flux of the magnetic field through the infinite solenoid which is a physical observable.
   This effect is experimentally observed in a double slit experiment for electrons as shown in figure Fig.\,\ref{solenoid}(b).
   In between the double slit there is an infinite solenoid.
   When the magnetic flux through the solenoid is finite the interference fringes are shifted with respect to the interference fringes in absence of magnetic flux.

 In case of magnetic systems, \gls{magnon}s can also acquire Berry-phase in presence of non-trivial electric field through Aharonov-Casher effect.
If a circularly polarized light is implemented on a magnetic material a Berry-phase is acquired by the magnon which is given by\,\cite{Aharnov_Casher},
\begin{equation}
	\gamma_{AC}=\frac{g\mu_B}{\hbar c^2}
	\int^{\bold{r}_\beta}_{\bold{r}_\alpha} \bold{A}(t)
	\cdot d\bold{l},
\end{equation}
where $\bold{A}(t)=A_0\left(\tau \sin(\omega t), -\cos(\omega t), 0\right)$ is a vector potential with $A_0=\frac{E}{\omega}$ and the electric field is given by $\bold{E}(t)=-\frac{\partial \bold{A}(t)}{\partial t}=E_0(\tau \cos{\omega t}, \sin(\omega t), 0)$.
 It can be shown that the circularly polarized light driven magnetic system with zero \gls{DMI}, can be mapped into a effective time-independent spin-system with a non-zero \gls{DMI}\,\cite{Floquet}.

\begin{figure}[H]
	\centering
	{\includegraphics[width=\textwidth]{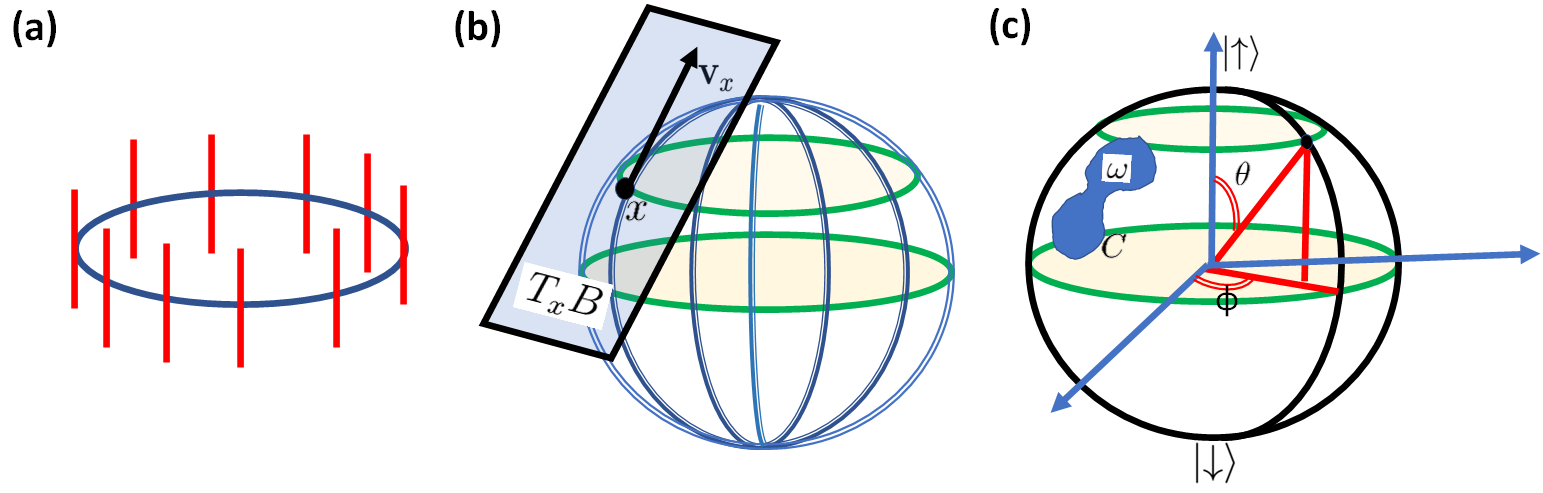}}
	\caption{(a) Illustration of cylinder as fibre-bundle structure. The circle (blue-line) is base space and the fibre is a set of real numbers $\mathbb{R}$ which is denoted as red-line.
			(b) The illustration of tangent-bundle of a sphere. The base space is a sphere. The tangent plane at point $x$ is $T_xB$. $T_xB$ is the fibre at point x on base. The tangent vectors $\bold{v}_x$ are the components of the fibre $T_xB$.
			(c) The schematic of Bloch sphere. Each point on the sphere represents the linear superposition of the states $\ket{\uparrow}$ and $\ket{\downarrow}$ as $\ket{\Psi}=e^{i\Phi}\left[\cos(\frac{\theta}{2})\ket{\uparrow}+e^{i\phi} \sin(\frac{\theta}{2})\ket{\downarrow}\right]$. The north and south-pole are the pure states $\ket{\uparrow}$ and $\ket{\downarrow}$ respectively. Bloch-sphere is the base-space $\mathbb{S}^2$ of the fibre-bundle and the wave function with different global phase $\Phi$ correspond to different points on fibre $\mathbb{S}^1$ and according to Hopf-fibration it constructs a non-trivial $\mathbb{S}^3$ fibre bundle. Along a closed loop $C$ the global-phase changes by $-\frac{1}{2}\omega$, where $\omega$ is solid angle made by the surface closed by $C$.}
	\label{fibrebundle}
	\end{figure}

The phenomena of parallel transport on a 2D manifold and adiabatic process in Hilbert space is very similar.
The similarity arises because both of the phenomena can be described using a mathematical structure known as fibre-bundle.
The details of the fibre-bundle and its connection with the geometric phase are elaborately described in the following references Ref.\,\cite{FibreBundleBook,FibreBundleThesis,FibreBundleLectureNote}
A fibre bundle is locally a product space, but globally it may or may not be a product space (For example sphere is locally flat ($\mathbb{R}\times\mathbb{R}$) but globally different).
A fibre-bundle is non-trivial when globally it can not be described as a product space and the Berry phase can be interpreted as holonomy of a non-trivial fibre bundle.
The fibre bundle contains the following elements,
\begin{itemize}
	\item Manifolds: A total space $E$, A base space $B$ and a fibre $F$.
	\item A structure Lie group $G$ which acts freely on $F$ and homeomorphically transforms $F$ into itself. 
	The elements of $G$ give a way to traverse from one fibre to another fibre or one point to another point on the same fibre. 
	\item A bundle projection which is a surjective map $\pi:E\rightarrow B$.
	The inverse image $F_x=\pi^{-1}(x)$ is homeomorphic to $F$ and $F_x$ is fibre at $x$.
\end{itemize}    
Again fibre-bundle follows the condition of local trivialization,
\begin{itemize}
	\item If $\{U_\alpha \}$ is open-covering of base $B$ then there is a mapping $\phi_\alpha: \pi^{-1}(U_\alpha)\rightarrow U_\alpha \times F$.
\end{itemize}
A cylinder as fibre bundle is shown in the figure Fig.\,\ref{fibrebundle}(a).
 The base-space of cylinder is $\mathbb{S}^1$ (circle) and the fibre is a set of real numbers $\mathbb{R}$.
 So the cylinder is a fibre bundle $\mathbb{S}^1\times\mathbb{R}$.
 The local trivialization of cylinder is $\mathbb{S}^1\times\mathbb{R}$ which is valid globally, so cylinder is a trivial fibre-bundle.

The deficit-angle between initial and final tangent vectors on a sphere after parallel transport as in figure Fig.\,\ref{sphere}(c) and Fig.\,\ref{sphere}(d) is a result of non-trivial tangent-bundle of the sphere.
A tangent plan at point $x$ of a sphere schematically shown in figure Fig.\,\ref{fibrebundle}(b).
The sphere is the base space $B$ and the tangent-space at point $x$ is fibre $T_xB$.
The tangent vectors $\bold{v}_x$ are the elements of fibre $T_xB$.
The union of all tangent planes $T_xB$ gives the tangent-bundle of the sphere.
The local trivialization of the tangent-bundle of sphere is $\mathbb{R}\times \mathbb{R}$ which is equivalent to say that the locally the sphere is a flat-surface.
However the tangent bundle of the sphere globally can not be represented as product space $\mathbb{R}\times\mathbb{R}$, so tangent bundle of sphere is a non-trivial fibre-bundle.
That is why, in figure Fig.\,\ref{sphere}(c) and Fig.\,\ref{sphere}(d) the final tangent vector and initial tangent vector are different after the parallel transport around a closed loop.  
In terms of fibre-bundle the initial and final points on the fibre after parallel transport are different.
Another important concept in fibre-bundle is bundle-connection which gives the notion of parallel transportation on the manifold.
In case of Riemannian manifold (like sphere, toroid are 2D-Riemannian manifold), the bundle connection is given by Levi-Civita connection (\textit{Berry-connection}).
Using the tangent-bundle the Euler-characteristic can be defined for a any even-dimensional or  $2k$-diemnsional ($k\in \mathbb{Z}^+$) manifold $X$ and given by,
\begin{equation}
	\chi(X)=\left(\frac{-1}{2\pi}\right)^k \int_X Pf(F_{\nabla}),
\end{equation}
where $Pf$ is Pfaffian 2k-form and $F_\nabla$ is curvature 2-form for any Levi-Civita connection $\nabla$ on its' tangent-bundle.

In quantum-mechanics the Hilbert-space is a fibre bundle structure with parameter space as base space.
All the eigenstates as well as eigenstates with relative $U(1)$ phase-difference are the elements of the same fibre at a particular point in parameter or bas-space.
This is exactly like the tangent bundle where the tangent vectors (\textit{eiegnstate with relative $U(1)$-phases}) are elements of the tangent-space $T_xB$ which is the fibre at point $x$ of base $B$.
The Bloch-sphere is a good example to illustrate the context of the fibre-bundles in quantum mechanics maintaining the comparison with parallel transport on a sphere.
Bloch-sphere is a unit-sphere which represents parameter space of the parameters $\theta$ and $\phi$ as shown in figure Fig.\,\ref{fibrebundle}(c).
At each point of the sphere a fibre is attached and each point of the fibre is an eigenstate,
\begin{equation}
	\ket{\Psi}=e^{i\Phi}\left[\cos(\frac{\theta}{2})\ket{\uparrow}+e^{i\phi} \sin(\frac{\theta}{2})\ket{\downarrow}\right],
	\label{eq2.77}
\end{equation}
where $\ket{\uparrow}$ and $\ket{\downarrow}$ are the spin up and spin down states of a spin-half.
The north and south-pole of Bloch sphere represents the up-spin and down-spin states respectively.
The phase $\Phi$ is a global phase and for each value of $\Phi$ the eigenstate is a point on the fibre at a particular point $\left(\theta,\phi\right)$ on the base.
In general the equation Eq.\,\ref{eq2.77} represents the eigenstate for any two level quantum system which is known as qubit.
The Berry-connection is the bundle-connection in case of quantum mechanics which provides the way for parallel transport (\textit{adiabatic process}) along a path in fibre-bundle.
Using equation Eq.\,\ref{eq2.69} and Eq.\,\ref{eq2.77} we get the following Berry-connection for Bloch sphere,
\begin{equation}
	\pazocal{A}_\theta=0,\quad
	\pazocal{A}_\phi=-\frac{1}{2}(1-\cos(\theta)).
\end{equation}
Because Berry-connection is not a gauge invariant quantity the Berry curvature is calculated using the equation Eq.\,\ref{eq2.72} and the Berry-curvature is given by,
 \begin{equation}
 	\Omega_z(\theta,\phi) =
 	\frac{\partial \pazocal{A}_\phi}{\partial \theta}
	-\frac{\partial \pazocal{A}_\theta}{\partial \phi}
	=-\frac{1}{2}\sin(\theta).
	\label{eq2.79}
\end{equation}  
Due to non-zero Berry-curvature, if a closed path is traversed on the Bloch-sphere a Berry-phase is expected, as shown in figure Fig.\,\ref{fibrebundle}(c).   
 For spin-half system a closed path is traversed on the Bloch-sphere by rotating a uniform magnetic field in a closed manner.
Using equation Eq.\,\ref{eq2.71} and Eq.\,\ref{eq2.79} we get the Berry-phase,
\begin{align}
	\Delta\Phi &=-\frac{1}{2}\oiint_{S} \sin(\theta) d\theta d\phi 
	\quad\left[\text{S is surface of closed loop C}\right]
	\nonumber\\
	&= -\frac{1}{2} \omega,
	\label{eq2.80}
\end{align}
where $\omega$ is the solid angle made by the surface  $S$ at the center of the sphere.
The solid-angle on the Bloch sphere is equivalent to the solid angle made by an classical-spin around a loop due to rotation of a uniform magnetic field.
It is further noted that the deficit angle for parallel transport on a sphere is equal to solid angle due to the surface of the loop, see figure Fig.\,\ref{sphere}(d).
However a factor-$\frac{1}{2}$ in equation Eq.\,\ref{eq2.80} is due to spin-half particle.
 In terms of fibre-bundle, after parallel transport the end point (\textit{eigenvector with final global phase}) on the fibre is different compared with initial point (\textit{eigenvector with initial global phase}).
This is because the fibre-bundle of Bloch sphere (\textit{base-space}) and eigen-vectors ($fibre$) is non-trivial and which is equivalent to $\mathbb{S}^3$-sphere.
A $\mathbb{S}^3$ sphere is made of base-space $\mathbb{S}^2$ (\textit{Bloch-sphere}) and fibre $\mathbb{S}^1$ (\textit{eigenvectors along with their different $U(1)$ global phase counterparts}) which is given by Hopf-fibration.
According to Hopf-fibration local trivialization of $\mathbb{S}^3$ is $\mathbb{S}^2\times \mathbb{S}^1$ but globally it is not a product space.
That is the reason of non-zero Berry phase (\textit{deficit angle}) in the adiabatic process (\textit{parallel transport}) governed by the Berry-connection (\textit{bundle-connection}).

In case of non-interacting electrons or \gls{magnon}s in two dimensional system, the reciprocal space is a fibre bundle space.
The base-space is made of parameters $(k_x, k_y)$, where $k_x$ and $k_y$ are $x$ and $y$-component of crystal-momentum.
The eigen-states in the reciprocal space forms the fibre.
If the fibre-bundle in the reciprocal space is non-trivial then non-trivial topological effect is expected to be observed in these systems.
Generally topological invariant like Chern-number denotes the nature of fibre-bundle in reciprocal space.

The physical phenomena related to Berry phase is not limited to the quantum mechanics.
The phenomenon can also be observed in classical systems like photonics, acoustics and mechanics\,\cite{Classical1,Classical2,Classical3,Classical4,Classical5}.
The reason that the phenomena also appear in classical mechanics is that the parameter space in all this system are base space and the the normal-modes along with their global phase counterpart are the points on the fibre.  
If the fibre-bundles constructed in this classical systems are non-trivial then a Berry-phase is expected for closed adiabatic process.

	\subsection{Numerical calculation of Berry-curvature and Chern-number in k-space}\label{sec2.5.2}
	According to Bloch's theorem the eigenstates of a particle in a periodic crystal system is given by,
	\begin{equation}
		\psi_{n,\bold{k}}=e^{i\bold{k}\cdot\bold{r}} u_{n,\bold{k}}(\bold{r}).
	\end{equation}
	The form of the eigenstate is same as the plane waves but modulated by a function which has a periodicity same as the lattice.
	$\bold{k}$ is the reciprocal lattice vector.
	 $n$ is the band-index, which denotes the $n$-th eigenstate at a particular crystal momentum (reduced-zone scheme).
	From the discussion of the section\,\ref{sec2.5.1} the expression of Berry-curvature should be of the form $\pazocal{A}^\prime_{n,\bold{k}}=i\bra{\psi_{n,\bold{k}}}\nabla_{\bold{k}}\ket{\psi_{n,\bold{k}}}$.
	However the Berry connection in reciprocal space in a condensed matter system is given by,
	\begin{equation}
		\pazocal{A}_{n,\bold{k}}=i\bra{u_{n,\bold{k}}}\nabla_{\bold{k}}\ket{u_{n,\bold{k}}}.
		\label{eq2.82}
	\end{equation}
	The position operator in reciprocal-space and the semi-classical equation of motion of the position for a particle at $n$-th band  at crystal-momentum $\bold{k}$ with a band dispersion $\epsilon_n(\bold{k})$ are respectively (see Appendix.\,\ref{appendixC}),
	\begin{equation}
		\hat{r}=i\nabla_{\bold{k}}+\pazocal{A}_{n,\bold{k}},\quad
		\frac{d\bold{r}}{dt}=\nabla_{\bold{k}} \epsilon_n(\bold{k})+
		\frac{d\bold{k}}{dt}\times \left(\nabla_{\bold{k}}\times \pazocal{A}_{n,\bold{k}}\right),
		\label{eq2.83}
	\end{equation}		
	which is similar to the momentum operator in presence of magnetic vector potential $\bold{A}$ and the equation of motion of a charged particle in free space in presence of electric and magnetic field respectively,
	\begin{equation}
		\hat{p}=-i\nabla_{\bold{r}}-\frac{q}{c}\bold{A}(\bold{r}),\quad
		\frac{d\bold{r}}{dt}=-q\nabla_{\bold{r}}\phi+\frac{q}{c}\frac{d\bold{r}}{dt}\times \left(\nabla_{\bold{r}}\times \bold{A}(\bold{r})\right).
		\label{eq2.84}
	\end{equation}
The similarity of magnetic vector-potential and Berry-connection is discussed in the section Sec.\,\ref{sec2.5.1} for the case of Aharnov-Bohm effect. 	
	Comparing the equations Eq.\,\ref{eq2.83} and Eq.\,\ref{eq2.84}, the Berry-connection in equation Eq.\,\ref{eq2.82} is the correct definition in reciprocal space.

	The Berry-connection is not a gauge invariant quantity but Berry-curvature is a gauge invariant physical quantity. 
	The form of Berry-curvature for 2D-materials is given by,
	\begin{equation}
		\boldsymbol{\Omega}_n(\bold{k})=\nabla_{\bold{k}}\times \pazocal{A}_{n,\bold{k}}=i\left[\braket{\frac{\partial u_{n\bold{k}}}{\partial k_x}}{\frac{\partial u_{n\bold{k}}}{\partial k_y}}-\braket{\frac{\partial u_{n\bold{k}}}{\partial k_y}}{\frac{\partial u_{n\bold{k}}}{\partial k_x}}\right].
		\label{eq2.85}
	\end{equation}
	For this thesis most of the calculations of Berry-curvature is done using equation Eq.\,\ref{eq2.85}.
	If the Hamiltonian in $\bold{k}$-space can be diagonalized analytically and the analytical expressions of eigen-values and eigen-vectors are available then the equation Eq.\,\ref{eq2.85} gives the analytical expression of the Berry-curvature.
	But Eq.\,\ref{eq2.85} can not be implemented directly to calculate the Berry-curvature for a numerical calculation.
	In a numerical calculation the reciprocal-space Hamiltonian is diagonalized independently for each $\bold{k}$-point independently and so the gauge degrees of freedom of the eigenstates or eigenvectors for different $\bold{k}$-points are arbitrary and not related with each other.
	So the numerical derivative of eigenvectors can not be calculated correctly.
The equation Eq.\,\ref{eq2.85} can be transformed into the following equation (see Appendix.\,\ref{appendixC}),
\begin{equation}
	\boxed{\Omega_n(\bold{k})=i\mathlarger{\mathlarger{\sum}}_{n\neq m} \frac{\braHket{u_{m,\bold{k}}}{\frac{\partial H}{\partial k_x}}{u_{n,\bold{k}}}\braHket{u_{n,\bold{k}}}{\frac{\partial H}{\partial k_y}}{u_{m,\bold{k}}}-(k_x\leftrightarrow k_y)}{\left(E_m(\bold{k})-E_n(\bold{k})\right)^2}.}
\end{equation}	
The above equation is independent of the choice of the gauge as the terms like $\ket{u_{n,\bold{k}}}$ and $\bra{u_{n,\bold{k}}}$ are present, any choice of phase factor or $U(1)$-gauge cancels out.

For Bogoliubov Valatin transformation the Berry-curvature for 2D-materials is given by\,\cite{BogoliubovBerryCurvature},
\begin{align}
	\Omega_{n,\bold{k}}&=i\epsilon_{\mu\nu} \left[\sigma_3\frac{\partial T_{\bold{k}}^\dagger}{\partial k_{\mu}}\sigma_3 \frac{\partial T_{\bold{k}}}{\partial k_\mu}\right]_{nn}\nonumber\\
	&=i\left[(\sigma_3)_{nn} \frac{\partial (T_{\bold{k}}^\dagger)_{n\alpha}}{\partial k_x} (\sigma_3)_{\alpha\alpha} \frac{\partial (T_{\bold{k}})_{\alpha n}}{\partial k_y}-(k_x\leftrightarrow k_y)\right],
	\label{eq2.87}
\end{align}
where $T_{\bold{k}}$ is the para-unitary matrix for the transformation and each column of the matrix represents the  eigenvectors of both positive and negative energy eigenvalues after the transformation as discussed in section Sec.\,\ref{sec2.4}.
 $\sigma_3$ is z-component of the Paulli matrix as defined in equation Eq.\,\ref{sigmaz}.
 Eq.\,\ref{eq2.87} is not suitable for the numerical calculation for the same reason of unsuitability of the equation Eq.\,\ref{eq2.85} discussed before.
The gauge-independent form the equation Eq.\,\ref{eq2.87} is given by (see Appendix.\,\ref{appendixC}),
\begin{equation}
	\boxed{\Omega_{n,\bold{k}}=\sum_{m\neq n} i (\sigma_3)_{nn} (\sigma_3)_{mm} \left[\frac{\braHket{u_{m,\bold{k}}}{\frac{\partial H}{\partial k_x}}{u_{n,\bold{k}}}\braHket{u_{n,\bold{k}}}{\frac{\partial H}{\partial k_y}}{u_{m,\bold{k}}}-(k_x\leftrightarrow k_y)}{\left(E_m(\bold{k})-E_n(\bold{k})\right)^2}\right].}
	\label{eq2.88}
\end{equation}
The sum is over all the eigenstates with positive and negative eigenvalues of dynamic matrix of bosonic Bogoliubov-Hamiltonian.

Furthermore the first Brillouin-zone has a toroidal geometry which is similar to the closed 2D-manifold discussed in the section Sec.\,\ref{sec2.5.1}.
For a closed parameter space a topological invariant like Euler-characteristic can be defined.
The topological invariant defined for each band in the Brillouin-zone is known as Chern number which is given as,
\begin{equation}
	\boxed{C_n=\frac{1}{2\pi}\oiint_{BZ} \Omega_{n,\bold{k}} dk_x\, dk_y}.
\end{equation}
The Chern-number is always is a positive or negative integer for a band.
However when the gap between two band is finite but very small, in general the Berry-curvature is mostly concentrated around the point of direct-gap between the bands (see equation Eq.\,\ref{eq2.88}).
So in this situation, to calculate a integer-valued Chern-number a very fine grid of momentum space is necessary in the numerical calculation which increases the computational complexity.
This problem of calculating Chern-number can be avoided by using the method described in the reference\,\cite{Hatsugai}. 
The Chern number is given by,
\begin{equation}
	\boxed{\tilde{C}_n=\frac{1}{2\pi i}\sum_l \tilde{F}^n_{12}(k_l),}
	\label{eq2.90}
\end{equation}
	where,
	\begin{align}
		\tilde{F}^n_{12}(k_l)&=\ln[U^n_1(k_l)U^n_2(k_l+\hat{1})U^n_1(k_l+\hat{2})^{-1}U^n_2(k_l)^{-1}] \nonumber\\
		U^n_{\hat{\mu}}(k_l)&=\frac{\braket{u_{n,k_l}}{u_{n,k_l+\hat{\mu}}}}{|\braket{u_{n,k_l}}{u_{n,k_l+\hat{\mu}}}|}.
	\end{align}
Here $U^n_{\hat{\mu}}$ is $U(1)$-link variable for $n$-th band.
$k_l$ is the discretized Brillouin-zone and $k_l+\mu$ is the $k$-point next to the point $k_l$ in the direction $\hat{\mu}$, where $\hat{\mu}=\hat{1},\,\text{or}\,\hat{2}$ denoting $k_x$ or $k_y$ direction for a square-lattice Brillouin-zone (This can be generalized to Brillouin-zone of any lattice).
In this method, the critical number of points required in Brillouin zone to calculate the Chern number is,
\begin{equation}
N^c_B\approx \pazocal{O}\left(\sqrt{\frac{2C_n}{q_1q_2}}\right),
\end{equation} 
where $q_1$ and $q_2$ is the periodicity of the lattice in $\hat{1}$ and $\hat{2}$ direction respectively.

If two bands are degenerate, it is not possible to define the Chern number for the bands separately.
However the Chern number of the multiple bands can be calculated simultaneously, it is known as non-abelian Chern number.
The same equation Eq.\,\ref{eq2.90} is used to calculate the non-abelian Chern number with the following definition of $U(1)$-link variable,
\begin{equation}
	U_{\mu}^n(k_l)=\frac{det\left(\psi^\dagger(k_l)\psi(k_l+1)\right)}{|det\left(\psi^\dagger(k_l)\psi(k_l+1)\right)|},
\end{equation}
 where $\psi(k_l)$ is a row matrix which contains $u_{n,k_l}$ for M-degenerate bands and given by $\psi(k_l)=\left(u_{n1,k_l},u_{n2,k_l},\cdots,u_{nM,k_l}\right)$.

	\subsection{Edge states on a strip geometry}
	\label{sec2.5.3}
	\begin{figure}[H]
	\centering
	\includegraphics[width=\textwidth]{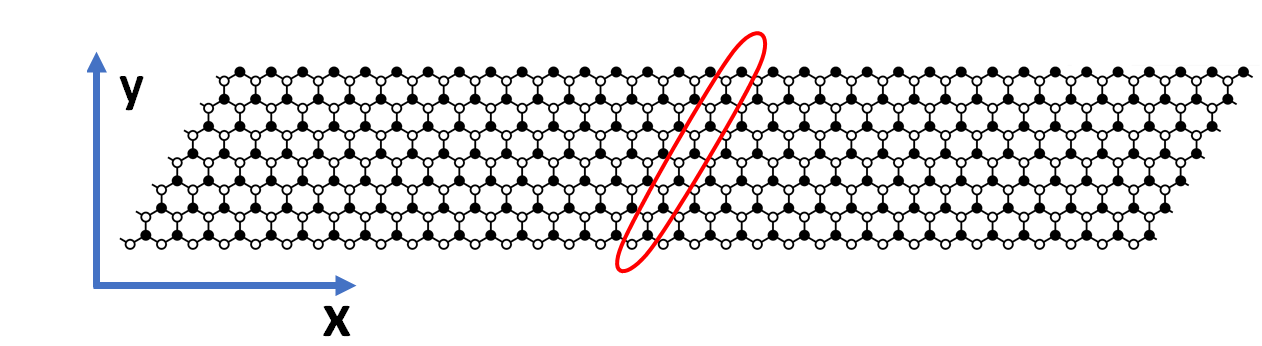}
	\caption{The honeycomb lattice structure of size ($20\times 14$) with periodic boundary condition along x-axis and open boundary condition along y-axis. The encircled sites are the basis of unit cell and mentioned as a stripe in the text.}
	\label{StripeGeometry}
\end{figure}
For a non-trivial topology of bulk band structure, edge-states are present in the band structure for a strip-geometry of a system (Fig.\,\ref{StripeGeometry}), which consists of periodic (open) boundary condition along the length (width) of the (two dimensional) lattice.	
	The number of surface or edge-states are related to the bulk topological invariants (e.g. Chern number) and it is known as Bulk edge correspondence.
	Although the formalism for calculating the band structure for a strip geometry is similar to the section Sec.\,\ref{sec2.3} and Sec.\,\ref{sec2.4}, a more organized approach for making a matrix-form for a Hamiltonian is described in this section\,\cite{GQD_Book}.
	The Hamiltonian in tight binding form can be written as,
\begin{equation}
    \pazocal{H}=\sum_{ij}t_{ij} \ket{i}\bra{j},
\end{equation}
where, i and j are the sites of lattice and $t_{ij}$ is the hopping amplitude. More explicitly the Hamiltonian can be written as a matrix form as\,\cite{GQD_Book},
\begin{equation}
    \pazocal{H}=\sum_{m,s} \left[\Psi_{m}^\dagger U\Psi_{m}+\Psi_{m}^\dagger T\Psi_{m+1}+\Psi_{m+1}^\dagger T^\dagger\Psi_{m}\right],
    \label{MatrixForm}
\end{equation}
where, $\Psi_{m}^\dagger=(\hat{b}_{1,m}^\dagger,\hat{b}_{2,m}^\dagger,...,\hat{b}_{N,m}^\dagger)$ and $N_s$ is the number of sites along the stripe(the sites inside the red circle of Fig.\,\ref{StripeGeometry} makes one stripe) and $m$ denotes the stripe index.
 $U$ and $T$ are $N_s\times N_s$ matrices.
  Matrix-$U$ contains all the on-site and intra-stripe hopping elements(inside the red circle of Fig.\,\ref{StripeGeometry}) and matrix-$T$ contains all the inter-stripe hopping elements.  
Imposing periodic boundary condition along the x-direction as shown in the figure Fig.\,\ref{StripeGeometry}, one can Fourier transform the Hamiltonian with a 1D Bloch-wave vector, given by $\Psi_m=(1/\sqrt{M})\sum_{k=0}^{M-1} \Psi_k e^{-i2\pi km/M}$, where $M$ is the number of stripes along the x-direction in Fig.\,\ref{StripeGeometry}. After Fourier transform the Hamiltonian is given by,
\begin{align}
H &=\sum_k \Psi^\dagger_k \left[ U + \left(Te^{i\frac{2k\pi}{M}}+\text{H.c.}\right)\right]\Psi_k\nonumber\\
&=\sum_k \Psi^\dagger_k H_k \Psi_k,
\end{align}
The Hamiltonian is diagonalized as,
\begin{equation}
\epsilon_k=P^\dagger H_k P,
\end{equation}
where, $P$ is an unitary matrix and the corresponding eigenvector is given by,
\begin{equation}
\Psi_k^d=P^\dagger\Psi_k.
\label{eigenvector}
\end{equation}
 Diagonalizing the momentum space Hamiltonian, we obtain the bands for the strip geometry.
 Presence of edge states (eigenstates confined to the edges of the system) in the band-structure signifies a non-trivial topology of the bulk system.

In two dimensional band-topological system there are mainly two types of edge-state.
One is known as helical edge-state and another is known as chiral edge state.
In case of chiral edge-modes, there is one edge state at each edges and the velocity at the two edges are equal and opposite.
On the other hand, Helical edge-states are two copies of chiral edge-modes with opposite velocity and in general each chiral edge modes are associated with opposite spin degrees of freedom.

The velocity operator is expressed in terms of the k-space eigenstates as\,\cite{GDMahan,StackExchange},
\begin{equation}
    \hat{v}=\sum_{ij}v_{ij} \ket{i}\bra{j},
    \label{VelocityOperator}
\end{equation}
where the coefficients $v_{ij}$ is given by,
\begin{equation}
    v_{ij}=\bra{i}\hat{v}\ket{j}=-\frac{i}{\hbar}\bra{i}[\hat{r},\pazocal{H}]\ket{j}=-\frac{i}{\hbar}(\boldsymbol{r}_i-\boldsymbol{r}_j)t_{ij}.
\end{equation}
Using the velocity operator the velocity distribution across the width of the strip can be determined.
The contribution to $x$-component of the velocity at co-ordinate $y$ from the eigenstate at $k_x$-point of $n$-th band is expressed as,
\begin{equation}
    v_n(k_x,y)=\rho_{n,k_x} P_n(k_x,y) v_n(k_x),
\end{equation}
where $v_n(k_x)$ is the velocity eigenvalue, $\rho_{n,k_x}$ is the Bose-Einstein distribution, $P_n(k_x,y)$ is the probability of the eigenstate at $y$-coordinate along the strip.
Thus the total velocity in $x$-direction of particles at $y$-coordinate is given by,
\begin{equation}
    v(y)=\sum_{n,k_x} v_n(k_x,y)
\end{equation}
A similar formalism can be developed for a Bogoliubov Hamiltonian, where
\begin{equation*}
\Psi_{m}^\dagger=(\hat{b}_{1,m}^\dagger,\hat{b}_{2,m}^\dagger,...,\hat{b}_{N,m}^\dagger,\hat{b}_{1,m},\hat{b}_{2,m},...,\hat{b}_{N,m})    
\end{equation*}
 and a factor $\frac{1}{2}$ need to be multiplied in the equation Eq\,\ref{MatrixForm} to avoid double counting of the terms in the summation.

	\subsection{\label{sec2.4.4}Thermal Hall Conductance and Spin Nernst effect}
	\begin{figure}[H]
	\centering
	\includegraphics[width=0.8\textwidth]{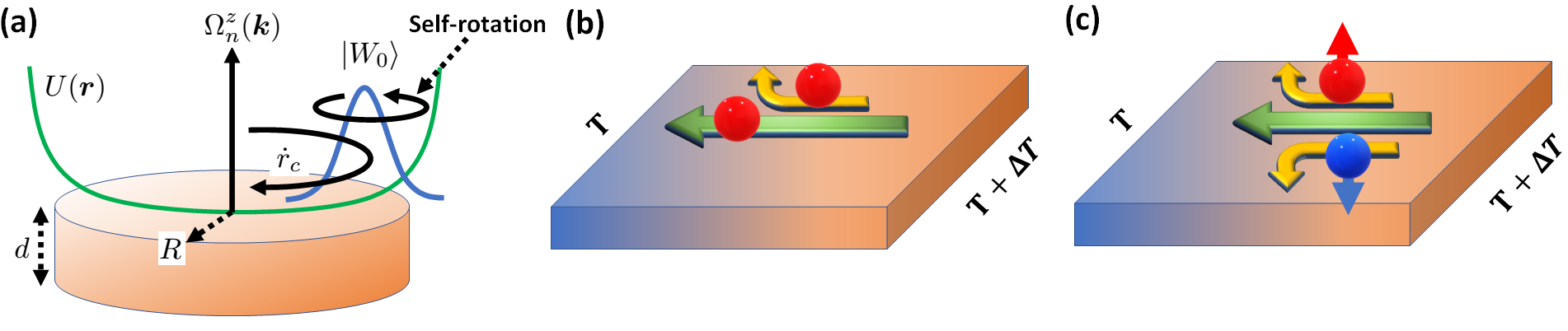}
	\caption{(a) Schematic picture of particle orbital magnetization. The particle orbital magnetization has two contributions, one is from self-rotation and another is from the flow of particle along the edge in presence of non-zero Berry-curvature.
	(b) The thermal Hall effect for one species of particle (red sphere) due to temperature gradient ($\nabla T$). The green-arrow shows the longitudinal current and the yeollow arrow shows the transverse Hall current.
			(c) The spin-Nernst effect for a temperature gradient $\nabla T$. The spin-up (red sphere) and spin-down (blue sphere) moments flow in the opposite directions (shown by yellow arrows, resulting in a net flow of magnetic moment in the transverse direction.
	}
	\label{THEV9}
\end{figure}	
	Being charge neutral particles, \gls{magnon}s are not effected by external electric field and so conventional electric field driven Hall-effect can not be observed.
	But a thermal gradient along a topological magnon system would drive a transverse magnon current known as thermal \gls{magnon} Hall effect (see Fig.\,\ref{THEV9}(b)).
	The reason of transverse current can be explained by the presence of chiral-edge states.
	 The occupation number of magnons in the edge-state at edge at high temperature increases compared with the edge-state at edge at low temperature.
	 In turn a transverse current flows in the direction of the edge state with higher occupation number.
	 Here we recall the derivation of the thermal Hall conductance following the references Ref.\,\cite{SemiClassical,SemiClassical2,SemiClassical3,SemiClassical4}. 
	 Assuming perturbations in the system is small such that inter-band transition can be neglected and semi-classical dynamics of a wave-packet formed by the wave-functions of $n$-th band can be studied independent of the other bands and the wave-function of the wave-packet is given by,
	 \begin{equation}
	     \ket{W_0}=\int d^3\bk w(\bk,t) \ket{\psi_n(\bk)},
	     \label{wavepacket}
	 \end{equation}
	 where Bloch function $\ket{\psi_n(\bk)}=e^{i\bk\cdot\bold{r}}\ket{u_n(\bk)}$. The complex function $w(\bk,t)$ is an envelope function such that the peak of the wave-packet in the momentum and real-space are at $\bk_c$ and $\bold{r}_c$ respectively. Thus,
	 \begin{align}
	     \bk_c&=\int |w(\bk,t)|^2 \bk d^3\bk \nonumber\\
	     \bold{r}_c&=\squeezeD{W_0}{\bold{r}}{W_0}=-\boldsymbol{\nabla}_{\bk_c} \text{arg}[ w]+\bold{A}_n(\bk_c),
	     \label{peak}
	 \end{align}
	 where Berry-connection,$\bold{A}_n(\bk)=i\squeezeD{u_n}{\boldsymbol{\nabla}_{\bk}}{u_n}$. 
	 The equation of motion of the wave-packet is given by (see Appendix.\,\ref{appendixD} for the derivation and the subscript $c$ from $\bk_c$ and $\bold{r}_c$ is dropped),
	 \begin{align}
	     \dot{r} &=\frac{1}{\hbar}\boldsymbol{\nabla}_\bk \epsilon_n(\bk)+\frac{1}{\hbar} \boldsymbol{\nabla} U(\bold{r})\times \boldsymbol{\Omega}_n^z(\bk) \nonumber \\
	     \dot{k} &=-\frac{1}{\hbar}\boldsymbol{\nabla}U(\bold{r})
	 \end{align}
	 The schematic of the equation of motion is depicted in the figure.\,\ref{THEV9}(a).
	 If we assume the potential throughout the system is uniform, the Berry-curvature contribution to the velocity $\dot{r}$ exists only near the edges due to potential gradient and so in presence of non-zero Berry-curvature the wave-packet gains a rotational motion around the edge.
	 The thermal orbital magnetization per unit volume due to the rotational motion along the edge is given as (see Appendix.\,\ref{appendixD}, here chemical potential $\mu=0$ for magnons),
	 \begin{equation}
	     M_Q^\text{edge}=\frac{1}{V\hbar}
	     \sum_{n,\bk}
	     \Omega_n^z(\bk)
	     \int_{\epsilon_n(\bk)}^\infty
	     \epsilon \rho(\epsilon) d\epsilon,
	 \end{equation}
	 where $V$ is the volume of the system and $\rho(\epsilon)=1/(\exp(\epsilon/T)-1)$ is the Bose-Einstein distribution.
	 Moreover the wave-packet self-rotates around itself as shown in the figure Fig.\,\ref{THEV9}(a). As a consequence the total thermal current of \gls{magnon} has two contributions, one from the motion of center of mass of wave-packet and another from rotational motion of the wave packet, which is expressed as,
	 \begin{equation}
	     J_Q^{\text{total}}=
	     \frac{1}{V}\sum_{n,\bk} \rho_{n,\bk} \epsilon_n(\bk) \dot{r}(n,\bk) +\boldsymbol{\nabla}\times M_Q^\text{self}.
	 \end{equation}
	 The part $\boldsymbol{\nabla}\times M_Q$ (where $M_Q=M_Q^\text{edge}+M_Q^\text{self}$) of the total current $J_Q^{\text{total}}$ is divergence free and this divergence-less part does not contribute to the current measured by conventional transport experiment\,\cite{SemiClassical5}. Therefore, the transport current is given by,
	 \begin{align}
	     J_Q^{\text{tr}}&=J_Q^{\text{total}}-\boldsymbol{\nabla}\times M_Q\nonumber\\
	     &=\frac{1}{V}\sum_{n,\bk}\rho_{n,\bk} \epsilon_n(\bk) \dot{r}(n,\bk) -\boldsymbol{\nabla}\times M_Q^\text{edge}.
	 \end{align}
	 If a linear temperature gradient is applied, the transverse current is achieved due to the second term in the expression as it can be re-expressed as $\boldsymbol{\nabla}T\times \frac{\partial M_Q^{\text{edge}}}{\partial T}$. Therefore, the thermal Hall conductance is given by,
	 \begin{align}
	     \kappa_H&=\frac{\partial M_Q^\text{edge}}{\partial T} \label{something}\\
	     &=\frac{1}{V\hbar T}\sum_{n,\bk} 
	     \left[\int_{\epsilon_n(\bk)}^\infty 
	     \epsilon^2 \frac{\partial \rho(\epsilon)}{\partial \epsilon} d\epsilon\right]\Omega_n^z(\bk) \nonumber\\
	     &=\frac{1}{V\hbar T}\sum_{n,\bk} 
	     \left[-k_B^2 T^2 \int_0^{\rho_{n,\bk}} (\log(1+\rho^{-1}))^2 d\rho\right] \Omega_n^z(\bk) \nonumber\\
	     &=\frac{1}{V\hbar T}\sum_{n,\bk} 
	     \left[-k_B^2 T^2 c_2(\rho_{n,\bk})\right] \Omega_n^z(\bk) \nonumber\\
	     &=-\frac{k_B^2 T}{V\hbar }\sum_{n,\bk} 
	     c_2(\rho_{n,\bk}) \Omega_n^z(\bk)
	 \end{align}

	The scaled thermal Hall conductivity for a two-dimensional material is given by\,\cite{THE1,THE2} (where the summation $\frac{1}{d}\sum_{k_z}$ is omitted and $\frac{1}{A}\sum_{k_x,ky}=\frac{1}{(2\pi)^2}\int d^2\bk$, where $A$ is area.),
\begin{equation}
	\kappa'_{xy}=\frac{\kappa_{xy}\hbar}{k_B}=-\frac{T'}{(2\pi)^2}\sum_n 	\int_{BZ} c_2(\rho_{n,\bk}) \Omega^n_{xy}(\bk) d^2\bk,
	\label{Eq:THE}
\end{equation}
where $\kappa_{xy}$ is the thermal Hall conductivity for a two-dimensional material, $T'$ is the scaled temperature, $T'=k_B T$, and $\rho_{n,\bk}=1/(\exp(\epsilon_n(\bk)/T')-1)$ is the Bose-Einstein distribution function with $\epsilon_n(\bk)$ as the energy of the $n$-th \gls{magnon} band at $\bk$-point in Brillouin zone, $\Omega^n_{xy}(\bk)$ is the Berry-curvature of the n-th band at the $\bk$-point and $c_2(x)=(1+x)\left(\log \frac{1+x}{x}\right)^2-(\log(x))^2-2\text{Li}_2(-x)$ where $\text{Li}_2(x)$ is a polylogarithmic function.

	The thermal Hall effect in a magnetic system is equivalent to the quantum Hall effect in an electronic system with one species of electron (either spin-up or spin-down).
	In a similar manner, spin Nernst effect in a spin system is similar to the quantum spin Hall effect in a electron system (see Fig.\,\ref{THEV9}(c)).
	As described before that the thermal Hall effect is a consequence of chiral edge states.
	In a similar manner spin-Nernst effect is result of the Helical edge states. 
	The Nernst conductivity for a two-dimensional material is calculated using the  
expression\,\cite{Nernst},
\begin{equation}
\mathbin{\alpha^s_{xy}}{=\frac{k_BT}{2A}\sum_{\bold{k},s,n}s c_1[\rho^n_s(\bold{k})]\Omega^n_s(\bold{k})},
\label{Eq:Nernst}
\end{equation}
where $A$ is the area or volume of the system. Again, $\rho^n_s(\bold{k})$ and $\Omega^n_s(\bold{k})$ are the Bose-Einstein distribution and Berry-curvature of $n$-th band respectively. $s=\pm 1$ represents a spin $\pm\frac{1}{2}$ quasi-particle. The function $c_1(x)=(1+x)\ln(1+x)-x\ln x$.
 The Nernst conductivity can also be derived using semi-classical formalism in a similar manner and it can be shown that Nernst conductivity is proportional to the temperature derivative of particle orbital magnetization instead of thermal orbital magnetization as in equation Eq.\,\ref{something}\,\cite{SemiClassical}.

	\subsection{Dynamical Spin Structure Factor}
	\label{sec2.5.5}
	\begin{figure}[H]
	\centering
	\includegraphics[width=0.6\textwidth]{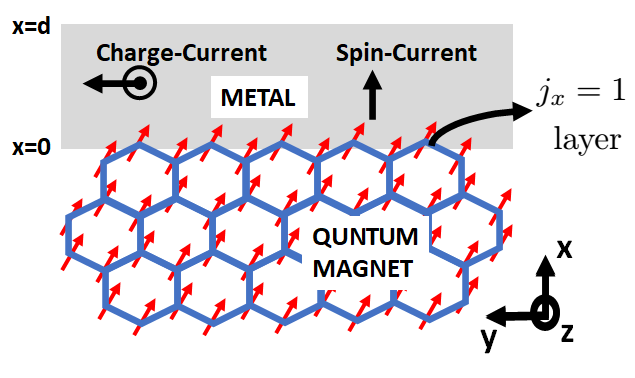}
	\caption{The schematic of experimental arrangement of spin Hall noise spectroscopy.
	 The edges of the spin system is attached with a metal.
	 The quantum fluctuation at the edges generate a fluctuating pure spin current along x-axis.
	 Due to inverse spin Hall effect a transverse fluctuating charge current flows  along $y-$ or $z-$direction.
	 The double derivative of the noise spectrum of the charge current is proportional to the dynamical spin structure factor at the edge of the spin system as discussed in main text.
	 The dynamical spin structure factor contains the information related to the edge state dispersion.
	 	}
	\label{SHNS}
\end{figure}
	The non-trivial topology of the bulk of a system denotes presence of states at the edges or surfaces of that system.
	The reference Ref.\,\cite{DSSF} shows that the edge states can be detected via spin Hall noise spectroscopy.
	The experimental arrangement of spin Hall noise spectroscopy is shown in the figure Fig.\,\ref{SHNS}.
	The quantum fluctuations of spins at the edges generate a fluctuating pure spin current in the metal and as a result of inverse spin Hall effect a fluctuating transverse charge current is generated.
	If the x-axis is perpendicular to the common edge of the metal and spin system, the double derivative of noise spectrum $\pazocal{I}^{z(y)}_c(\Omega)$ of charge current in the z (or y) direction in the metal is proportional to the \gls{dssf} $\pazocal{S}^{z(y)}(\Omega)$ in the following way,
	\begin{equation}
		\frac{d^2 \pazocal{I}^{z(y)}_c(\Omega)}{d\Omega^2}\propto \pazocal{S}^{y(z)}(\Omega),
		\label{newone}
	\end{equation}
	where,
	\begin{equation}
	\pazocal{S}^{\alpha}(\Omega)=i\sum_{j\in edge} \left[\pazocal{S}^{\alpha\alpha}_{jj}(\Omega)+\pazocal{S}^{xx}_{jj}(\Omega)\right],
	\quad
	\pazocal{S}^{\alpha\alpha}_{jk}(\Omega)=-i\int^{\infty}_{-\infty} dt e^{-i\Omega t}\left\langle S_j^\alpha(t)S_k^\alpha(0)\right\rangle_0.
	\label{Eq:DSSF}
	\end{equation}
	The sign $\left\langle\cdots\right\rangle_0$ denotes equilibrium thermal average and $S_j^{\alpha}$ is the $\alpha$-th component of spin-operator at site-$j$.

	In this section, we briefly outline the derivation of Eq.\,\ref{newone} following the reference Ref.\,\cite{DSSF}.
	The quantum spin-fluctuation at edge gives rise to a spin current in metal in the direction perpendicular to the magnet-metal interface\,($x$-direction in Fig.\ref{SHNS}) and at $x=0$ the current is given by,
	\begin{equation}
	    \bold{I}^{\boldsymbol{\sigma}}_s(t)=I^{\boldsymbol{\sigma}}_s(t)\hat{x},
	\end{equation}
	where $\boldsymbol{\sigma}$ denotes the direction of spin-polarization. 
	The magnitude of the spin-current decays towards the other boundary of the metal (boundary at $x=d$ in Fig.\,\ref{SHNS}) and considering vanishing spin-current at $x=d$, the spin current-density profile throughout the metal is given by,
	\begin{equation}
	    \bold{j}^{\boldsymbol{\sigma}}_s(\bold{x},t)=\bold{I}^{\boldsymbol{\sigma}}_s
	    \frac{\sinh[(d-x)/\lambda]}{A_i \sinh(d/\lambda)},
	    \label{add2.116}
	\end{equation}
	where $A_i$ is the interfacial area between metal and magnet, $\lambda$ is the spin-diffusion length of isotropic metal. 
	The fluctuating spin-current induces a fluctuating charge current in the metal due to inverse spin Hall effect and the direction of charge current is perpendicular to both direction of spin-current and spin-polarization.
	The charge current at time $t$ is obtained by integrating the spin current-density in Eq.\,\ref{add2.116} over the volume of metal and multiplying appropriate factor related to spin-Hall effect and written as,
	\begin{equation}
	    \bold{I}_c(t)=\Theta\frac{2e}{\hbar} 
	    \frac{\lambda A_m}{d A_i}\tanh\left(\frac{d}{2\lambda}\right)(\hat{x}\times\boldsymbol{\sigma})
	    \bold{I}_s^{\boldsymbol{\sigma}},
	\end{equation}
	where $A_m$ is the area perpendicular to the charge current and $\Theta$ is the spin-Hall angle.
	The noise spectrum for spin-current and charge current respectively,
    \begin{align}
        \pazocal{I}_s^{z(y)}(\Omega)&=\int dt
        \left\langle \hat{I}_s^{z(y)}(0) \hat{I}_s^{z(y)}(t)\right\rangle \exp(-i\Omega t),
        \nonumber\\
        \pazocal{I}_c^{y(z)}(\Omega)&=\int dt
        \left\langle \hat{I}_c^{y(z)}(0) \hat{I}_c^{y(z)}(t)\right\rangle \exp(-i\Omega t) 
        =\left[\Theta\frac{2e}{\hbar} 
	    \frac{\lambda A_m}{d A_i}\tanh\left(\frac{d}{2\lambda}\right)\right]^2
	    \pazocal{I}_s^{z(y)}(\Omega),
	    \label{add2.118}
    \end{align}
    where $\left\langle\right\rangle$ denotes equilibrium thermal average.
    The interaction between metal and magnet is considered as isotropic Heisenberg interaction,
    \begin{equation}
    \hat{H}_c=-\eta a^3\sum_{j}\delta_{j_x,1}\hat{\bold{s}}(x=0,\bold{R}_j)\cdot\hat{\bold{S}}_j,
    \,\,\text{where,}\,\,
    \hat{\bold{s}}(\bold{R})=\frac{1}{2}\hat{\psi}_s^\dagger(\bold{R}) \boldsymbol{\tau}_{ss'} \hat{\psi}_s(\bold{R}).
    \label{add2.119}
    \end{equation}
    $\hat{\bold{s}}(\bold{R})$ and $\hat{\bold{S}}_j$ are the local spin-density operator in metal and spin-operator in magnet respectively.
    $\hat{\psi}_s(\bold{R})$ is the electronic field operator in metal, $\boldsymbol{\tau}_{ss'}$ is the vector representing Pauli-matrices, $\eta$ is the exchange constant and $a$ is the lattice scaling in the metal.
     Considering the spin-quantization axis in metal (axis for $\hat{s}^z$) aligned along the spin-polarization axis $\boldsymbol{\sigma}$ in quantum magnet, the $y$-component of spin current operator entering into the metal is calculated using Hamiltonian Eq.\,\ref{add2.119}  (see Appendix.\,\ref{appendixE}),
    \begin{equation}
        \hat{I}^y_s=\frac{i\eta a^3}{2} \hat{T}_y +\text{h.c.},
        \,\,\text{where,}\,\,
        \hat{T}_y=\sum_j \delta_{j_x,1} \hat{\psi}^\dagger_{\uparrow}(\bold{R}_j) \hat{\psi}_\downarrow(\bold{R}_j)
        (\hat{S}_j^z-i\hat{S}_j^x)
        \label{add2.120}
    \end{equation}
	
	Using equations Eq.\,\ref{add2.120} and Eq.\,\ref{add2.118} the noise spectrum in spin current is given by\,(see Appendix.\,\ref{appendixE}),
	\begin{equation}
	 \pazocal{I}_s^y(\Omega)\approx 2i 
	 \left(\frac{\eta a^3 m k_F}{2\pi^2\hbar}\right)^2
	 \int d\nu \sum_j \delta_{j_x,1} \left[\pazocal{S}_{jj}^{xx}(\nu)+\pazocal{S}_{jj}^{zz}(\nu)\right]
	 \frac{\Omega-\nu}{1-\exp{-\beta\hbar(\Omega-\nu)}},
	\end{equation}
	where,
	$\pazocal{S}_{jj}^{xx}(\nu)$ is the \gls{dssf} as defined in Eq.\,\ref{Eq:DSSF} and $k_F$ is the Fermi-vector for the metal and $m$ is the mass of electron.
	The double-derivative of spin-noise spectrum is given by,
	\begin{align}
	    \frac{d^2 \pazocal{I}_s^y(\Omega)}{d\Omega^2}
	    &=2i\left(\frac{\eta a^3 m k_F}{2\pi^2\hbar}\right)^2
	    \int d\nu \sum_j \delta_{j_x,1} \left[\pazocal{S}_{jj}^{xx}(\nu)+\pazocal{S}_{jj}^{zz}(\nu)\right]
	 \frac{d^2}{d\Omega^2}\frac{\Omega-\nu}{1-\exp\left[-\beta\hbar(\Omega-\nu)\right]}
	 \nonumber\\
	 &\approx 2i\left(\frac{\eta a^3 m k_F}{2\pi^2\hbar}\right)^2
	    \sum_j \delta_{j_x,1} \left[\pazocal{S}_{jj}^{xx}(\Omega)+\pazocal{S}_{jj}^{zz}(\Omega)\right],
	\end{align}
	where it is assumed that the double derivative at the right hand side is a Dirac-delta function because at low temperature  it is highly peaked at $\nu=\Omega$.
	According to equations Eq.\,\ref{add2.118} the double-derivative of the noise spectrum of charge current is proportional to the double derivative of noise spectrum of spin-current and so the proportionality between double-derivative of charge noise spectrum and \gls{dssf} as in equation Eq.\ref{newone} is proved.

	The non-zero \gls{dssf} at the edges not only proves the presence of edge states, but also it gives the information of the dispersion of the edge states.
	If the edge state is dispersionless then \gls{dssf} has a single Dirac delta peak at the energy equal to the energy of edge state (considering the edge has only one edge state).
	Otherwise the \gls{dssf} has a distribution in the frequency space with a finite width.
	This is due to many Dirac delta peaks at each (quasi-continuous) energy levels of the edge state.
	The width of the distribution denotes that the energy levels of the edge state is within that energy range.

	\section{Symmetry}
	\label{sec2.6}
\begin{figure}[H]
	\centering
	\includegraphics[width=\textwidth]{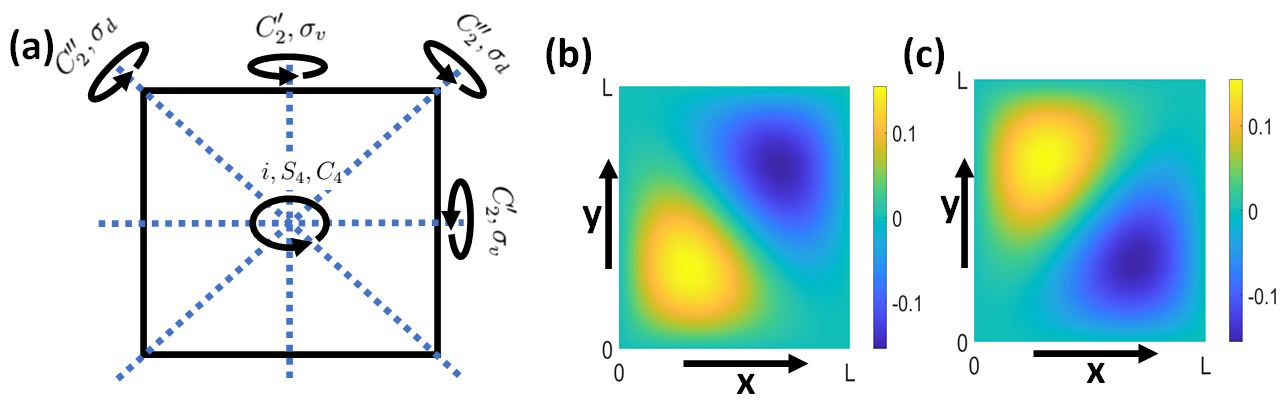}
	\caption{(a) Symmetries of square box and symmetry operations constitute $D_{4h}$-point group.
			The wavefunctions (a) $\frac{1}{N}(\psi_{12}+\psi_{21})$ and (b) $\frac{1}{N}(\psi_{12}-\psi_{21})$ as a function of space. 
			$N$ is the normalization factor.
			}
	\label{ParticleInBox}
\end{figure}		
		
	The symmetry of a system is paramount in understanding its mathematical structure.
	In the classical systems symmetry of the system defines conserved quantities (Noether's theorem) and in quantum mechanics it translates into the conserved quantum numbers.
	For example one of the most important result in condensed matter physics is the Bloch's theorem which states that in a periodic crystal the crystal momentum is conserved and eigenstates of the system is associated with a specific crystal momentum.
	The conservation of crystal momentum is a consequence of translational symmetry of the periodic system.

	In quantum mechanics the energies of a system are discrete in general. 
	Degeneracies may arise due to certain symmetry protection or be accidental in nature (which, in turn may or may not reflect some hidden symmetry).
	The symmetry protected degenerate eigenstates of the quantum system is related to the two or higher dimensional irreducible representation of the symmetry-group of the system.
	For example a square box can be classified by point group symmetry $D_{4h}$ (see Fig.\,\ref{ParticleInBox}(a)), where a point group is a group of symmetry operations which do not alter the position of particular point.
	Moreover the energies and wavefunctions of particle in a square box is given by the following expression $E_{n_x,n_y}=\frac{\hbar^2\pi^2}{2mL^2}(n_x^2+n_y^2)$ and $\psi_{n_1,n_2}=\frac{2}{L}\sin(\frac{n\pi x}{L})\sin(\frac{n\pi y}{L})$ respectively.
	 So the energy levels for unequal $n_x$ and $n_y$ are two-fold degenerate.
	This degeneracy is due to two dimensional irreducible representation of the point group $D_{4h}$.
	As an example the linear superposition of wave-functions for the states $n_x=1$, $n_y=2$ and $n_x=2$, $n_y=1$ are shown in the figure Fig.\,\ref{ParticleInBox}(b).
	The wave-functions has the same symmetry as the square box.
	The traces of the matrices correspond to the symmetry operators assuming the wavefunctions as the basis is equal to the characters of $E_u$ representation of the $D_{4h}$ point group as shown below in the table,
	
	\begin{center}
	\begin{tabular}{|c|c|c|c|c|c|c|c|c|c|c|}
\hline
$D_{4h}$ & $E$ & $2C_4(z)$ & $C_2$ & $2C_2^\prime$ & $2C_2^{\prime\prime}$ & $i$ & $2S_4$ & $\sigma_h$ & $2\sigma_v$ & $2\sigma_d$ 
\\
\hline
$E_u$ & +2 & 0 & -2 & 0 & 0 & -2 & 0 & +2 & 0 & 0
\\
\hline
\end{tabular}	
\end{center}
So the degeneracy of the states is protected by the $D_{4h}$ symmetry group of the system.
Additionally there is an accidental degeneracy between the states $n_x=1$, $n_y=7$
and $n_x=5$, $n_y=5$.

In crystal there are several point groups associated with different points on the lattice.
All these point groups along with the translational symmetries form a group known as space group.
There are 32 possible point groups and 230 possible space groups.
Moreover it is possible to include time-reversal symmetry operation for a magnetic system, which flips the spins on the sites.
Inclusion of antiunitary time-reversal operation extends the 32 point groups into 122 magnetic point group(32 point group, 32 grey group, 58 color point group) and 230 space group into 1651 magnetic space group (230 space group, 230 grey group, 1191 color space group).
The degenerate states at the high symmetry points are due to the two or higher dimensional irreducible representations(corepresentation) of the little-group(little-magnetic group), where little-group(little magnetic group) is the point-group(magnetic-point group) valid for the high symmetric point in the Brillouin zone\,\cite{Hamermesh,Corepresentation1,Corepresentation2,Corepresentation3}.
Sometimes the presence of anti-unitary symmetry in the system gives rise to the Kramers degeneracy if the square of the operator is negative one at certain points of the Brillouin-zone. 
Sometimes the energies at the Brillouin zone boundary are degenerate, this is due to presence of nonsymmorphic space group.
The Herring's method is useful to show the nonsymmorphic symmetry protection of band degeneracies\,\cite{Herring1,Herring2}.

\begin{figure}[H]
	\centering
	\includegraphics[width=0.8\textwidth]{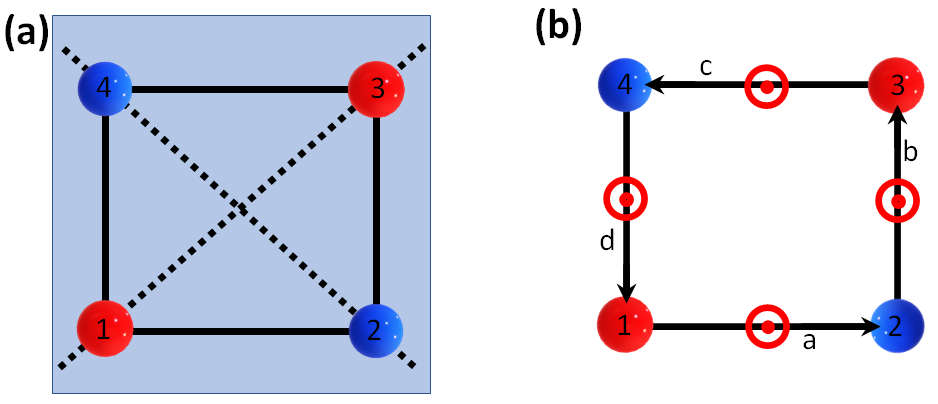}
	\caption{(a) The schematic of square shaped molecule with spins sitting on two different atoms denoted by blue and red sphere.
	The inversion symmetry on each nearest neighbour bond is broken.
	The horizontal mirror plane $\sigma_h$ is denoted by the blue plane and the dotted line represents the mirror plane $\sigma_d$.
			(b) The \gls{DMI} on each bonds due to the broken inversion symmetry and presence of other symmetries. 
			The red dotted circle denotes the direction of DM-interaction.
	}
	\label{Square}
\end{figure}
	
	Breaking of symmetry plays an important role from fundamental physics to applied physics. 
	For example presence of Higgs boson is a consequence of spontaneous breaking of electroweak symmetry and the fundamental particles gain mass due to interaction with the Higgs boson field.
	In a similar manner in a magnetic system breaking of inversion symmetry in the lattice introduces \gls{DMI} (see Eq.\,\ref{eq2.4}).
	As an example the \gls{DMI} on the bonds of the square shaped molecule as in the figure Fig.\,\ref{Square}(a) is determined here by considering symmetries of the molecule.
	Firstly the bonds do not posses any inversion symmetry due to two different atoms denoted by blue and red sphere.
	On the bond-a the \gls{DMI} is given by,
	\begin{align*}
		&\bold{D}\cdot(\bold{S}_1\times\bold{S}_2)
		\\
		&=\begin{vmatrix}
		D_{a,x} & D_{a,y} & D_{a,z}\\
		S_{1,x} & S_{1,y} & S_{1,z}\\
		S_{2,x} & S_{2,y} & S_{2,z}
		\end{vmatrix}
		=\sigma_h\begin{vmatrix}
		D_{a,x} & D_{a,y} & D_{a,z}\\
		S_{1,x} & S_{1,y} & S_{1,z}\\
		S_{2,x} & S_{2,y} & S_{2,z}
		\end{vmatrix}\sigma_h^{-1}
		=\begin{vmatrix}
		D_{a,x} & D_{a,y} & D_{a,z}\\
		-S_{1,x} & -S_{1,y} & S_{1,z}\\
		-S_{2,x} & -S_{2,y} & S_{2,z}
		\end{vmatrix}
		=\begin{vmatrix}
		-D_{a,x} & -D_{a,y} & D_{a,z}\\
		S_{1,x} & S_{1,y} & S_{1,z}\\
		S_{2,x} & S_{2,y} & S_{2,z}
		\end{vmatrix},
	\end{align*}
	where, $\sigma_h$ is the horizontal-mirror symmetry operator.
	\begin{equation}
	\therefore D_{a,x}=0, D_{a,y}=0.
	\end{equation}
	Similarly the $x$ and $y$ components of \gls{DMI} at all bonds are zero.
   Further presence of the mirror symmetry $\sigma_{d1}$ at the diagonal of the square structure transforms the \gls{DMI} on the bond-a to bond-b in the following way,
    \begin{align*}
		\begin{vmatrix}
    		0 & 0 & D_{b,z}\\
    		S_{2,x} & S_{2,y} & S_{2,z}\\
    		S_{3,x} & S_{3,y} & S_{3,z}
    	\end{vmatrix}=    	
    	\sigma_{d1}\begin{vmatrix}
    		0 & 0 & D_{a,z}\\
    		S_{1,x} & S_{1,y} & S_{1,z}\\
    		S_{2,x} & S_{2,y} & S_{2,z}
    	\end{vmatrix}\sigma_{d1}^{-1} 
    	=\begin{vmatrix}
    		0 & 0 & D_{a,z}\\
    		S_{3,y} & S_{3,x} & S_{3,z}\\
    		S_{2,y} & S_{2,x} & S_{2,z}
    	\end{vmatrix}
    	=\begin{vmatrix}
    		0 & 0 & D_{a,z}\\
    		S_{2,x} & S_{2,y} & S_{2,z}\\
    		S_{3,x} & S_{3,y} & S_{3,z}
    	\end{vmatrix}
    \end{align*}
	\begin{equation}
		\therefore D_{bz}=D_{az}
	\end{equation}
	Similarly using other symmetry operators the \gls{DMI} at other bonds,
	\begin{equation}
		D_{b,z}=D_{c,z}=D_{d,z}=D_{a,z}
	\end{equation}
	The figure Fig.\,\ref{Square}(b) shows the resultant \gls{DMI} due to the symmetry of the system.

	
		
	\chapter[Engineering antichiral edgestates in ferromagnetic honeycomb lattices]{Engineering antichiral edgestates in ferromagnetic honeycomb lattices}\label{chapter03}
	
	\epigraph{The results in this chapter is published as D. Bhowmick and P. Sengupta.  Antichiral edge states in Heisenberg ferromagnet on a honeycomb lattice.  Physical Review B 101, 195133 (2020). DOI: 10.1103/PhysRevB.101.195133}
	
\section{\label{sec3.1}Introduction}
	Haldane's paradigmatic model\,\cite{Haldane} of tight binding on a honeycomb lattice with complex next-nearest neighbor hopping -- constitutes the foundation of many of the electronic topological phases.
It has a natural realization in (quasi-) 2D insulating ferromagnets such as \ce{CrI3}\,\cite{CrI3} and \ce{AFe2(PO4)2} (A=Ba,Cs,K,La) \cite{IronDirac} in terms of magnetic excitations called \gls{magnon}s. 
In many of these materials, the dominant Heisenberg exchange is supplemented by a next-nearest neighbour anti-symmetric \gls{DMI}. 
Furthermore these systems can be described by two species 
of quasi-particles known as \gls{spinon}s with up and down spins.
 The spin model translates into the Kane Mele model for \gls{spinon}s -- has been proposed to describe the dynamics of the system for a wide 
range of temperatures\,\cite{Kane_Mele_Haldane}. 
The \gls{spinon} bands acquire a non-trivial topology due to Berry phase arising from the \gls{DMI}. 
This results in a spin Nernst effect (SNE) where a thermal gradient drives a transverse spin current, a \gls{spinon} version of the spin Hall effect\,\cite{Kane_Mele_Haldane,SNE1,SNE2,SBMFT5}. 
In a finite sample, the up(or down) \gls{spinon} species generate two counter propagating spin currents along the edges that are protected by chiral symmetry of the Hamiltonian -- analogous to two copies of the thermal Hall effect (THE) of \gls{magnon}s that has been observed in many insulating magnets\,\cite{THE1,THE2,THE3,THE4}.

Recently, there has been growing interest in engineering systems with co-propagating edge currents\,\cite{Colomes,AntiChiral1,AntiChiral2}, through an ingenious, yet physically unrealistic, modification of the Haldane model.
The conservation of net current is satisfied by counter propagating bulk current.
So in contrast to conventional topological insulators the bulk is not insulating anymore.
In this work, we demonstrate that anti-chiral states arise naturally in \gls{spinon}s on a honeycomb magnet comprised of two different magnetic ions, with unequal \gls{DMI} for the two sublattices.
In the absence of \gls{DMI}, the \gls{spinon} dispersion consists of two doubly degenerate bands with linear band crossings at {\bf K} and {\bf K$^\prime$}\,\cite{DiracMagnon}.
 The degeneracy between the two \gls{spinon} branches are lifted for a finite {\it symmetric} \gls{DMI}\,\cite{opendirac2,opendirac3,opendirac4}. 
 For {\it asymmetric} \gls{DMI}, The two bands for each \gls{spinon} species are shifted in opposite directions relative to each other at the {\bf K} and {\bf K$^\prime$} points in the Brillouin zone.
As a consequence gapless edge modes in up-\gls{spinon} or down-\gls{spinon} sector achieve similar dispersion at both edges, giving rise to co-propagating edge states for each \gls{spinon}-sectors.
The spin current due to up-\gls{spinon} and down-\gls{spinon} channels at a particular edge are in opposite directions.
Thus the two edges contribute to the net flow of spin-momentum in the same direction (see Fig.\,\ref{band}(b)).
So, this yields effective anti-chiral edge states for the \gls{spinon}s in addition to normal chiral ones.
We present a detailed characterization of the nature of the edge and bulk \gls{spinon} states and suggest suitable experimental signatures to detect these novel topological states.

\section{\label{sec3.2}Model}

\begin{figure}[H]
	\centering
	\includegraphics[width=0.5\textwidth]{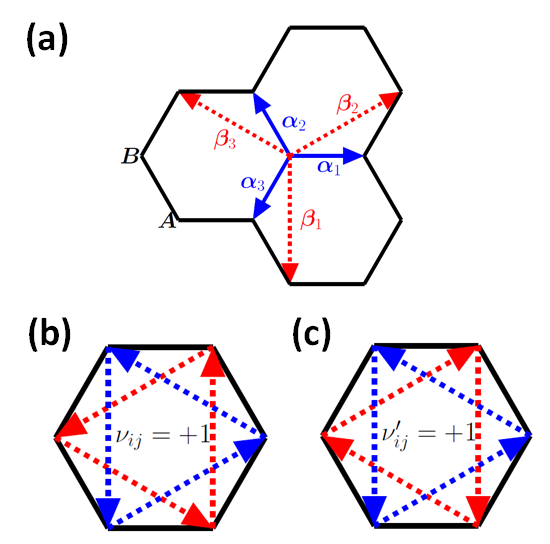}
	\caption{(a) The honeycomb lattice structure. (b) The directions along which $\nu_{ij}=+1$ has been shown, otherwise $\nu_{ij}=-1$. (c) The directions along which $\nu_{ij}'=+1$ has been shown, otherwise $\nu_{ij}'=-1$.}
	\label{lattice3}
\end{figure}

We consider a Heisenberg ferromagnet on the honeycomb lattice with
unequal \gls{DMI} ($D_A$ and $D_B$) on the two sub-lattices. 
Introducing the symmetric and anti-symmetric combinations of $D_A$ and $D_B$ as,
$D~=~\frac{1}{2}\left( D_A+D_B \right)$, $D'~=~\frac{1}{2} \left( D_A-D_B \right)$ -- 
termed chiral and anti-chiral \gls{DMI} respectively for reasons that will become
clear later -- the Hamiltonian is given by,

\begin{align}
\pazocal{H}=& -J\sum_{\left\langle i,j \right\rangle} \bold{S}_i\cdot\bold{S}_j+D_A\sum_{\left\langle\left\langle i,j \right\rangle\right\rangle_A} \nu_{ij} \hat{z}\cdot \left(\bold{S}_i\times \bold{S}_j\right)
 +D_B\sum_{\left\langle\left\langle i,j \right\rangle\right\rangle_B} \nu_{ij}' \hat{z}\cdot \left(\bold{S}_i\times \bold{S}_j\right)-B\sum_i S_i^z,
 \nonumber\\
&=-J\sum_{\left\langle i,j \right\rangle} \bold{S}_i\cdot\bold{S}_j+D\sum_{\left\langle\left\langle i,j \right\rangle\right\rangle} \nu_{ij} \hat{z}\cdot \left(\bold{S}_i\times \bold{S}_j\right)
 +D'\sum_{\left\langle\left\langle i,j \right\rangle\right\rangle} \nu_{ij}' \hat{z}\cdot \left(\bold{S}_i\times \bold{S}_j\right)-B\sum_i S_i^z,
\label{eq:hamil}
\end{align}
where, $J>0$ is the nearest neighbor Heisenberg interaction and $\nu_{ij}=+1$ when $i$ and $j$ are along the cyclic arrows shown in Fig.\,\ref{lattice3}(b). Again, $\left\langle...\right\rangle$ and $\left\langle\left\langle...\right\rangle\right\rangle$ denote the nearest-neighbour and next-nearest neighbour bonds respectively.
Moreover, $\left\langle\left\langle...\right\rangle\right\rangle_A$ ($\left\langle\left\langle...\right\rangle\right\rangle_B$) represents the next-nearest neighbour bonds with among sublattice-A (sublattice-B).
  Finally, $\nu_{ij}'=+\nu_{ij}$ for sublattice-A and $\nu_{ij}'=-\nu_{ij}$ for sublattice-B (Fig.\,\ref{lattice3}(c)). The zero-temperature ground state of the Hamiltonian (Eq.\,\ref{eq:hamil}) is ferromagnetic for
$J > -{3\sqrt{3} \over 2}\sum_p \left | D + pD'\right |,\, p=\pm 1$ in absence of magnetic field. The magnetic field $B$ is introduced in a Zeeman coupling term to stabilize the ferromagnetic ground state at finite temperature. The energy scale is set
by choosing $J=1$ --  all other parameters in the Hamiltonian are in units of $J$. 

\section{\label{sec3.3} Results}
\subsection{Spinon-Picture\label{sec3.3.1}}
The Hamiltonian (\,\ref{eq:hamil}) has a ferromagnetic ground state. This can 
be described as the condensation of one of the spin species. At low
temperatures ($T \ll J$), the only dynamics comes from \gls{magnon} excitations that
carry the opposite spin. However, at finite temperatures ($T\sim J$), both
spin species contribute to the dynamics of the system. We start the discussion of
our results with the general consideration of both spins contributing to the 
dynamics and later demonstrate that this naturally reduces to the \gls{magnon} 
picture at low temperatures. We apply the \gls{sbmft} to study the topological
character of the magnetic system at finite temperatures.
A detailed description of \gls{sbmft} is given in the section Sec.\,\ref{SBMFT} in "Model and Numerical Methods" chapter.
The \gls{spinon} representation consists of the mapping the spin 
operators into \gls{spinon}s as,
$
S_i^+ =b_{i,\uparrow}^\dagger b_{i,\downarrow}$,
$S_i^- =b_{i,\downarrow}^\dagger b_{i,\uparrow}$,
$S_i^z =\frac{1}{2}\left(b_{i,\uparrow}^\dagger b_{i,\uparrow}-b_{i,\downarrow}^\dagger b_{i,\downarrow}\right),
$
where $b_{i,s}$ and $b_{i,s}^\dagger$ are the annihilation and creation operators 
of spin-1/2 up$(s=+1)$ or down$(s=-1)$ \gls{spinon}s respectively. The constraint 
$\sum_{s} b_{i,s}^\dagger b_{i,s}=2S$, $\forall i$ on the bosonic operators 
ensures the fulfillment of the spin-S algebra.

 Defining the bond-operators, $\hat{\chi}_{ij,s}=\cdag{i}\cndag{j}$ and $\CHI =\left(\CHIU +\CHID\right)/2$, we can re-write the Hamiltonian in terms of Schwinger bosons bond operators as,
\begin{align}
    \pazocal{H}_{sp}=&-2J\sum_{\left\langle i,j\right\rangle} :\CHI^\dagger\CHI:
    -\frac{D}{2}\sum_{\left\langle\left\langle i,j\right\rangle\right\rangle} i\nu_{ij} \left(:\CHIU^\dagger\CHID:-:\CHID^\dagger\CHIU:\right)\nonumber \\
    &-\frac{D'}{2}\sum_{\left\langle\left\langle i,j\right\rangle\right\rangle} i\nu'_{ij} \left(:\CHIU^\dagger\CHID:-:\CHID^\dagger\CHIU:\right)
    -\frac{B}{2} \sum_{i} \left(\cdagU{i}\cndagU{i}-\cdagD{i}\cndagD{i}\right) \nonumber\\
    &+\sum_{is} \lambda_i \cdag{i}\cndag{i}-4SN\lambda+3NJS^2
\end{align}
The bond operators is chosen such that the total number of \gls{spinon} is conserved in the mean field Hamiltoian which is equivalent to $S_z$-conservation in terms of spin\,\cite{SBMFT3}.
Furthermore new bond operators are defined so that corresponding mean field parameters are real:  $\hat{A}_{ij,s}=\frac{1}{2}\left(\hat{\chi}_{ij,s}+\hat{\chi}_{ji,s}\right)$, $\hat{B}_{ij,s}=\frac{\nu_{ij}}{2i}\left(\hat{\chi}_{ij,s}-\hat{\chi}_{ji,s}\right)$, $\hat{B'}_{ij,s}=\frac{\nu'_{ij}}{2i}\left(\hat{\chi}_{ij,s}-\hat{\chi}_{ji,s}\right)$.
$\lambda_i$ is the Lagrange undetermined multiplier introduced to implement the local constraint.
We define mean field parameters $\eta=\left\langle \CHI^\dagger \right\rangle=\left\langle \CHI \right\rangle$, $\zeta_s=\left\langle\hat{A}_{ij,s}^\dagger\right\rangle=\left\langle\hat{A}_{ij,s}\right\rangle$, $\xi_s=\left\langle\hat{B}_{ij,s}^\dagger\right\rangle=\left\langle\hat{B}_{ij,s}\right\rangle$,
$\xi'_s=\left\langle\hat{B'}_{ij,s}^\dagger\right\rangle=\left\langle\hat{B'}_{ij,s}\right\rangle$.

 Thus applying Schwinger Boson transformation along with the constraint, and
 using a mean field approximation to reduce the 4-body operators to bilinear forms,
 the spin model Eq.\,\ref{eq:hamil} is mapped to the the mean field hamiltonian,

\begin{align}
\pazocal{H}=&-\eta J\sum_{\left\langle i,j \right\rangle,s} \left[ \cdag{i} \cndag{j}+\mathtt{H.c.} \right] +\sum_{i,s} \left(\lambda-\frac{sB}{2}\right) \cdag{i}\cndag{i}\nonumber\\
&+\frac{D}{2}\sum_{\left\langle\left\langle ij \right\rangle\right\rangle, s}\left[ \left(i \nu_{ij} s \zeta_{-s} + s\xi_{-s} \right)  \cdag{i}\cndag{j}+\mathtt{H.c.}\right]
+\frac{D'}{2}\sum_{\left\langle\left\langle ij \right\rangle\right\rangle, s}\left[ \left( i \nu'_{ij} s \zeta_{-s} + s\xi'_{-s} \right) \cdag{i}\cndag{j}+\mathtt{H.c.}\right]
\label{eq:spinonH}
\end{align}
where the mean field parameters are defined as, $\eta=\left\langle \hat{\chi}_{ij} \right\rangle$ evaluated 
on the nearest neighbour-bonds,  and $\zeta_s=\frac{1}{2}\left\langle \hat{\chi}_{ij,s}+\hat{\chi}_{ji,s} \right\rangle$,  
$\xi_s=\frac{\nu_{ij}}{2i}\left\langle \hat{\chi}_{ij,s}-\hat{\chi}_{ji,s} \right\rangle$ 
and $\xi'_s=\frac{\nu'_{ij}}{2i}\left\langle \hat{\chi}_{ij,s}-\hat{\chi}_{ji,s} \right\rangle$, 
evaluated on next nearest neighbour bonds. The terms associated with the parmeters $\eta, \nu_{ij}\zeta_{-s} $ of \gls{spinon} Hamiltonian
Eq\,\ref{eq:spinonH} constitute the Kane-Mele model\,\cite{Kane_Mele_Haldane}. 
The term with parameter $\nu'_{ij}\zeta_{-s}$ corresponds to the anti-chiral hopping term introduced in Ref.\,\cite{Colomes}. The terms with the parameters $\xi_s$ and $\xi'_s$ have no effect on the energy or 
the topological character of the bands, as the parameters are found to be much
smaller compared to other mean field parameters.    
The constraint $\lambda_i=\lambda$ is considered uniform throughout the lattice to retain the translational symmetry of the lattice. 
 Fourier transformation of the mean field Hamiltonian  in 
momentum-space yields,
\begin{equation}
\pazocal{H}_{sp}^{mf}=\sum_{\boldsymbol{k}\in \mathtt{B.Z.},s} \Psi^\dagger_{\boldsymbol{k},s} \left[g_s(\boldsymbol{k})I+\boldsymbol{h}_s(\boldsymbol{k})\cdot \boldsymbol{\sigma}\right] \Psi_{\boldsymbol{k},s}+E_0,
\label{eq7}
\end{equation} 
where, $\Psi_{\boldsymbol{k},s}^\dagger=\left(\hat{a}_{\boldsymbol{k},s}^\dagger,\hat{b}_{\boldsymbol{k},s}^\dagger\right)$. $\hat{a}_{\boldsymbol{k},s}^\dagger$ and $\hat{b}_{\boldsymbol{k},s}^\dagger$ are the creation operators for \gls{spinon}s on sublattice-A and sublattice-B~(see Fig.\,\ref{lattice3}(a)), respectively. $\boldsymbol{\sigma}_{\alpha}$($\alpha=x,y,z$) represents the Pauli matrices. The other terms are given by,
\begin{align}
g_s(\boldsymbol{k})=&-\frac{sB}{2}+\lambda+sD\xi_{-s} \gamma_c^\beta
+sD'\left(\xi'_{-s}  \gamma_c^\beta-\zeta_{-s}\gamma_s^\beta\right),\nonumber \\
h_s(\boldsymbol{k})=&\begin{pmatrix}
-J\eta\gamma_c^\alpha, &
J\eta\gamma_s^\alpha, &
-D s \zeta_{-s}^c\gamma_s^\beta
\end{pmatrix},\nonumber \\
E_0=& 6N_uJ\eta^2-6DN_u\sum_s s\zeta_s \xi_{-s}
-6D' N_u\sum_s s\zeta_s \xi'_{-s}-4SN_u\lambda+3N_uJS^2,
\end{align}
where, $\gamma_c^\beta=\sum_{j} \cos(\boldsymbol{k}\cdot\boldsymbol{\beta}_j),\gamma_s^\beta=\sum_{j}\sin(\boldsymbol{k}\cdot\boldsymbol{\beta}_j),\gamma_c^\alpha=\sum_{j}\cos(\boldsymbol{k}\cdot\boldsymbol{\alpha}_j),\gamma_s^\alpha=\sum_j \sin(\boldsymbol{k}\cdot\boldsymbol{\alpha}_j)$ and the vectors $\boldsymbol{\beta}_j$ and $\boldsymbol{\alpha}_j$ are shown in figure Fig.\,\ref{lattice3}(a). $N_u$ is the number of unit cells in lattice. $E_0$ is the energy of the ground state and the energies of \gls{spinon}s are considered with respect to the ground state energy. The mean field parameters are obtained by solving a set of self-consistent equations, derived by minimizing the Helmholtz free energy at a particular temperature.

After diagonalizing the k-space Hamiltonian we get,
\begin{equation}
    \pazocal{H}_{sp}^{mf}=E_0+\sum_{\boldsymbol{k},s,\tau}E_s^\tau(\boldsymbol{k})\cdag{\boldsymbol{k},\tau}\cndag{\boldsymbol{k},\tau},
\end{equation}
where, the relative energies,
\begin{equation}
E_s^\tau(\boldsymbol{k})=g_s(\boldsymbol{k})+\tau \left|h_s(\boldsymbol{k})\right|,
\end{equation}
refer to the upper ($\tau=+1$) and the lower ($\tau=-1$) band for each \gls{spinon} sectors $s=\pm 1$.

\begin{figure}[H]
	\centering
	\includegraphics[width=0.6\textwidth]{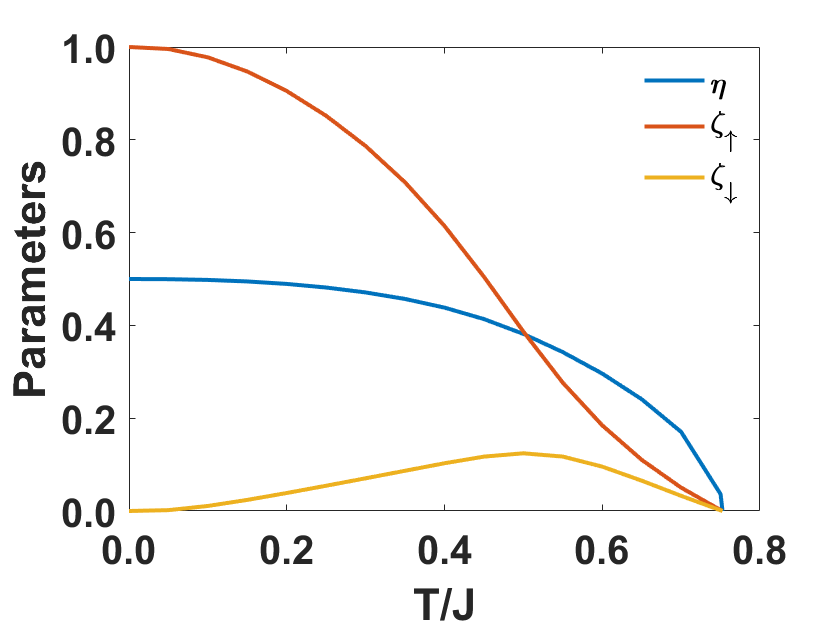}
	\caption{Plot of mean-field parameters, for $J=1.0, D=0.1, D'=0.05, B=0.1$}
	\label{parameter}
\end{figure}

From this we get the internal energy and the entropy of the non-interacting bosonic system as,
\begin{align}
    U&=E_0+\sum_{\boldsymbol{k},s,\tau} \rho_s^\tau(\boldsymbol{k})E_s^\tau(\boldsymbol{k}), \nonumber\\
    S&=k_B\sum_{\boldsymbol{k},s,\tau} [\left(1+\rho_s^\tau(\boldsymbol{k})\right) \ln\left(1+\rho_s^\tau(\boldsymbol{k})\right)-\rho_s^\tau(\boldsymbol{k})\ln \rho_s^\tau(\boldsymbol{k})],
\end{align}
where, $\rho_s^\tau(\boldsymbol{k})=\left[\exp\left(E^\tau_s(\boldsymbol{k})\right)-1\right]^{-1}$ 
is the Bose-Einstein distribution of spin-s \gls{spinon}s in the $\tau$-band. The Helmohltz-free-energy is given by,
\begin{align}
    G&=U-TS
    =E_0-{k_BT}\sum_{\boldsymbol{k},s,\tau}\ln\left(\frac{1}{1-\exp\left(\frac{-E_s^\tau(\boldsymbol{k})}{k_BT}\right)}\right)
\end{align}
After minimizing the Helmholtz free energy with respect to the mean field parameters, we get six self consistent equations, given by,
\begin{align}
2S=&\frac{1}{2N_u}\sum_{\boldsymbol{k},\tau,s} \rho_{s}^\tau(\boldsymbol{k}), \nonumber\\
1=&-\frac{J}{12N_u}\sum_{\boldsymbol{k},s,\tau} \tau \frac{\rho_s^\tau(\boldsymbol{k})}{\left| h_s \right|} \left|\sum_j e^{i\boldsymbol{k}\cdot\boldsymbol{\alpha}_j}\right|^2\nonumber \\
D\xi_s+D'\xi'_s = & \frac{1}{6N_u} \sum_{\boldsymbol{k},\tau} \left(D'- \tau s \frac{D^2 \zeta_{-s} \gamma}{h_s}\right)\rho_s^\tau(\boldsymbol{k}) \gamma_s^\beta,\nonumber\\
\zeta_s=&\frac{1}{6N_u}\sum_{\boldsymbol{k},\tau} \rho^\tau_s(\boldsymbol{k}) \sum_j \cos(\boldsymbol{k}\cdot\boldsymbol{\beta}_j),
\end{align}
The mean field parameters are obtained by 
solving these six self-consistent equations. It is notable that the mean field parameter $\eta$ can be chosen as real, absorbing the complex phase factor into operator $\cndag{i}$. All other mean field parameters are defined in a way that those are real valued. Using the parameters, we plot the band structure and evaluate corresponding topological information. For a fixed set of $J,D,D',B$, the mean field parameters are solved and plotted against temperature $T$ in Fig\,\ref{parameter}. The parameters 
$\eta$ and $\zeta_s$ represent short range correlations identifying magnetic
ordering and serve as order parameters for the transition from paramagnet with short-range correlations to completely uncorrelated paramagnet.  

 At low temperatures,
finite, non-zero values of $\eta$ and $\zeta_\uparrow$ denote finite ferromagnetic short-range correlations. A positive $B$ determines that the spins are all aligned along the +ve
z-direction at $T=0$. In other words, the system is populated with up-\gls{spinon}s.
As the temperature increases, thermally excited down-\gls{spinon}s are generated, resulting
in a finite, non-zero $\zeta_\downarrow$.
Finally, at high temperatures, a vanishing of all the mean field parameters denotes a transition from paramagnetic phase with finite short range correlations to a totally uncorrelated paramagnetic phase.
 The paramagnetic phase transition with all zero correlations to be expected to be an outcome of large-N expansion. 
 It has been shown for Heisenberg model that taking into account of the quantum fluctuations in the mean field parameter removes the phase transition\,\cite{mean_field_artifact}.

\begin{figure}[H]
	\centering
	\includegraphics[width=0.8\textwidth]{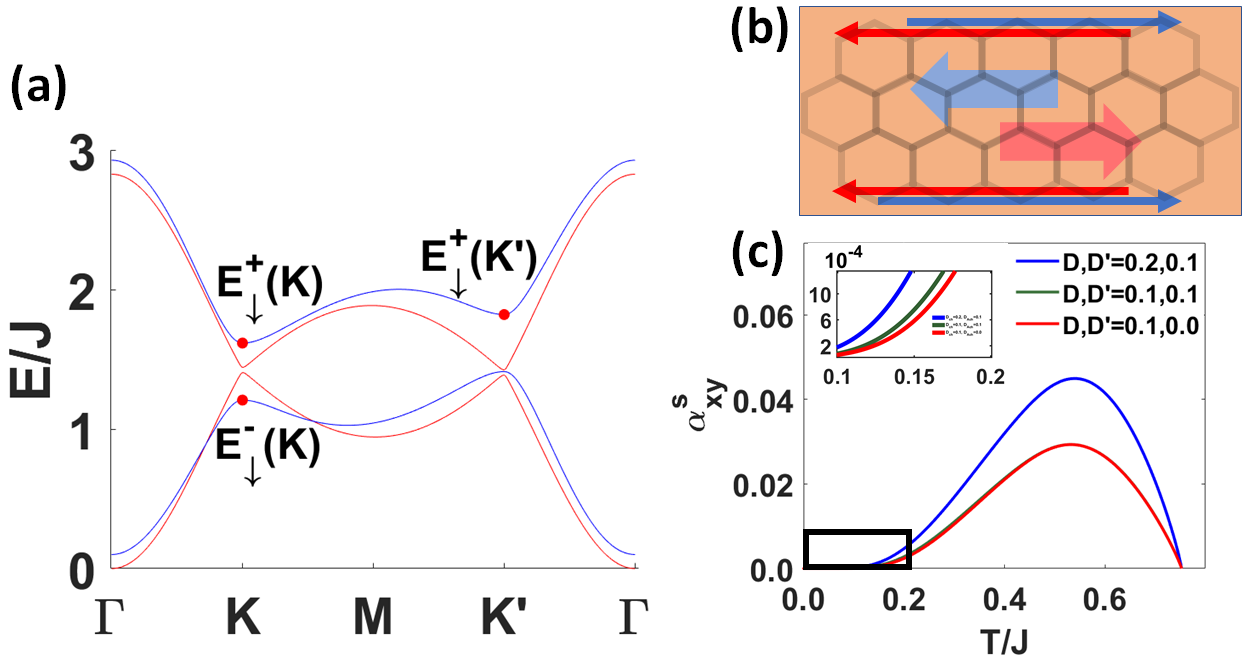}
	\caption{(a) Band along symmetry lines 
	$\Gamma K$, $KM$,$MK'$,$K'\Gamma$ for $J=1.0, B=0.1, D_{ch}=0.1,D_{Ach}=0.05, T=0.25$. The blue band is for down-\gls{spinon} band and the red band is for up-\gls{spinon} band. (b) The schematic of the antichiral contribution($D=0, D'\neq 0$) of the current from edge-states(thin-arrows) and bulk-states(thick-arrows). The color red and blue are the contributions from the up-\gls{spinon} and down-\gls{spinon} respectively. The direction of the thin arrows for each \gls{spinon} sectors are opposite and so there is a net spin-current along the edges. Moreover, for each \gls{spinon} sectors the direction of the thick arrows and thin arrows are opposite, which in turn reduces the total spin-current flow along the edges. So, a net spin-current flow along the two edges are the evidence of anti-chiral state. (c) Nernst-conductivity v.s. temperature plot for different \gls{DMI}s. The inset shows magnified figure of the rectangular portion of the figure.}
	\label{band}
\end{figure}

Band structure for \gls{spinon}s at a temperature T=0.25J is shown in  Fig.\,\ref{band}(a).
In the absence of \gls{DMI}, the two bands
cross linearly at the Dirac points {\bf K} and {\bf K$^\prime$}\,\cite{DiracMagnon}. A finite \gls{DMI} opens up a gap with magnitude 
$\Delta_{s}=3\sqrt{3}\left|D\zeta_{-s}\right|$ in each \gls{spinon} sector at {\bf K} 
and {\bf K$^\prime$}\,\cite{opendirac2,opendirac3,opendirac4,Kane_Mele_Haldane}. 
For {\it anisotropic} system ($D_A\neq D_B$) considered here, the gap opening
is not symmetric and leads to a {\it tilting} of the \gls{spinon} bands near the Dirac 
momenta. The band tilting for each 
band in each \gls{spinon} sector, defined as the energy difference between 
two Dirac-points in the same band, is given by, $T_{s}^\tau=3\sqrt{3}\left|D'\zeta_{-s}\right|$. While the anti-chiral 
\gls{DMI} drives the tilting of the bands, it has no effect on 
the magnitude of the band gap. 
Crucially, the tilting is opposite for the two species of \gls{spinon}s. 
For the parameters chosen in Fig.\,\ref{band}(a), the gap and tilting for 
the up-\gls{spinon} bands are smaller than those for the 
down \gls{spinon} bands. This is because in the presence of positive magnetic 
field $B=0.1$ considered here, there are
fewer down \gls{spinon}s and consequently, $\zeta_{\downarrow}<\zeta_{\uparrow}$.

\subsubsection{Nernst effect}

The bands in each \gls{spinon} sector carry non-zero Berry curvature, $\Omega^\tau_s(\bold{k})$ given by.
 \begin{align}
  \Omega^\tau_s(\bold{k})=i\sum_{\tau' \neq \tau}&\frac{\squeeze{\uu{\tau}}{\dXdY{\pazocal{H}}{k_x}}{\uu{\tau'}}\squeeze{\uu{\tau'}}{\dXdY{\pazocal{H}}{k_y}}{\uu{\tau}}}{(E^\tau_s(\bold{k})-E^{\tau'}_s(\bold{k}))^2}
  -(k_x\leftrightarrow k_y),
\end{align}   
where $\tau=+1(-1)$ represents upper(lower) band and $s=+1(-1)$ denotes the up-\gls{spinon}(down-\gls{spinon}) sector. $\uu{\tau}$ ($E^\tau_s(\bold{k})$) is the eigen-vector (energy) of $\tau$-th band at reciprocal space point-$\bold{k}$ for \gls{spinon} setor-$s$. Detecting Berry curvature of \gls{spinon} bands is more challenging than that for
electrons. The absence of Pauli exclusion principle for the (bosonic) 
quasiparticles as well as their charge neutral character make the standard
approaches used for electronic systems inapplicable. Spin Nernst effect has been proposed as a 
physical phenomenon to identify Berry curvature of \gls{spinon} bands 
when there are comparable numbers of up 
and down \gls{spinon}s. Here we explore whether it can detect the existence of
anti-chiral \gls{DMI}. The Nernst conductivity has been calculated using the equation Eq.\,\ref{Eq:Nernst} in the "Model and Numerical Methods" chapter.
 The results are plotted in Fig.\,\ref{band}(c) for some representative values of $D$ and $D'$. Increase in $D$ increases the band gap as well as the Berry curvature away from the Dirac-points. 
As a result, the Nernst conductivity is substantially affected by $D$ (Fig.\,\ref{band}(c)). Conversely, since the Berry curvature is independent of 
$D'$, the anti-chiral \gls{DMI} has very little effect in the Nernst conductivity. The effect of $D'$ on Nernst conductivity can be observed at low temperature due to tilting of the band structure(inset of Fig.\,\ref{band}(c)). But at higher temperature, the $D'$ has no influence in Nernst conductivity, because the contributions from higher bands overshadows the effects of band tilting.  So, the presence of $D'$ in the system is very hard to detect using Nernst conductivity. Instead, we suggest an alternative way to detect the presence of antichiral \gls{DMI}.

\subsubsection{Spin current}
The gapped bands are topologically non-trivial with Chern numbers $C_{\uparrow}^-=+1,C_{\uparrow}^+=-1, C_{\downarrow}^-=-1, C_{\downarrow}^+=+1$\,\cite{Hatsugai}. 
Due to bulk-edge correspondence, we expect to observe edge states in a 
finite

\begin{figure}[H]
	\centering
	\includegraphics[width=\textwidth,height=0.76\textwidth]{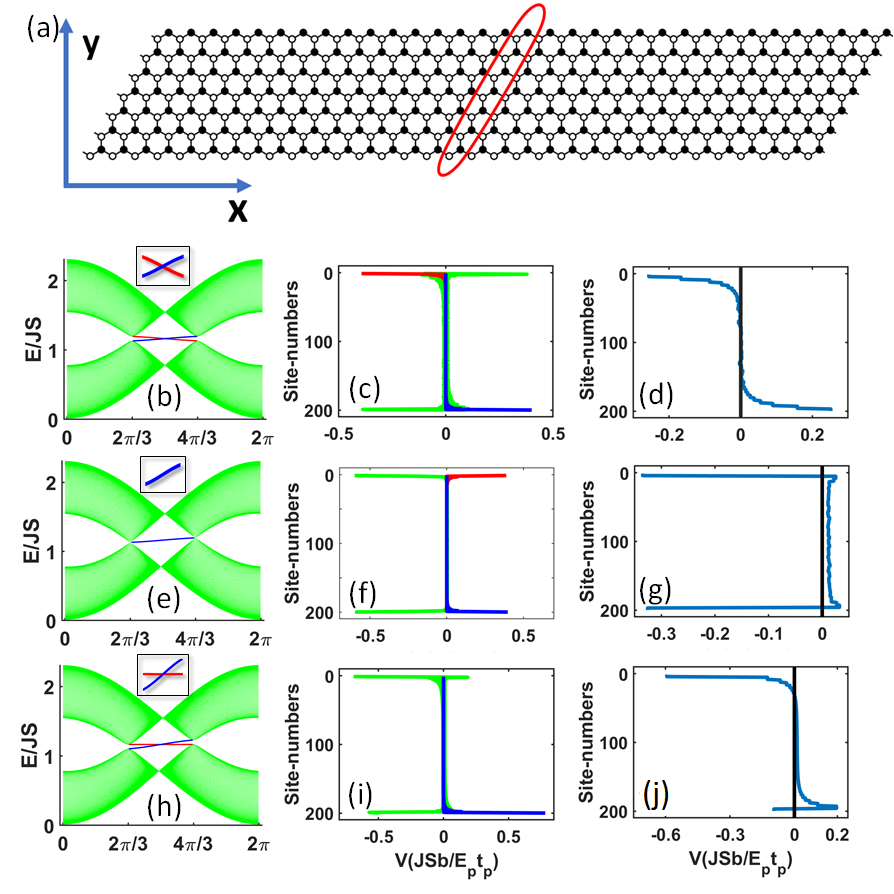}
	\caption{(a) A honeycomb ribbon. The encircled sites are the basis of unit cell. The figure sets $\left\lbrace(b)-(d)\right\rbrace$,$\left\lbrace(e)-(g)\right\rbrace$,$\left\lbrace(h)-(j)\right\rbrace$ represents result for down-\gls{spinon} from $200\times 500$ (width$\times$length) stripe with \gls{DMI}s $\left\lbrace D=0.1,D'=0.001\right\rbrace$,$\left\lbrace D=0.001,D'=0.1\right\rbrace$,
	$\left\lbrace D=0.1,D'=0.1\right\rbrace$ respectively. The other parameters are $J=1.0,  B=0.1, T=0.5$ for all the plots. The figure sets $\left\lbrace(b),(e),(h)\right\rbrace$ shows the band structure and inset of the figures shows the magnified dispersion of the edge states. The figure sets $\left\lbrace(c),(f),(i)\right\rbrace$ gives the spatial current distribution along width of the stripe, where $b$, $E_p$ and $t_p$ are the lattice constant, Planck-energy and Planck-time respectively. The figure sets $\left\lbrace(d),(g),(j)\right\rbrace$ shows the average of spatial current distribution over four sites respectively. In the figure sets $\left\lbrace(b),(c),(e),(f),(h),(i)\right\rbrace$ the green, red and blue plots corresponds to bulk-state, upper edge edge-state and lower-edge edge-state respectively. A non-zero net spin current along the edges in figures (g) and (j) are definitive signatures of anti-chiral edge states. The results for up-\gls{spinon} are qualitatively same, with the dispersion being opposite for the bulk and edge states relative to the down \gls{spinon}s. Hence the up-\gls{spinon}s further contribute to the net spin current along the edges in the presence of anti-chiral edge states.}
	\label{edge_state}
\end{figure}

\noindent system. In the isotropic limit ($D'=0$, i.e., $D_A=D_B$),
the edge states are topologically protected by a chiral symmetry. The \gls{spinon} currents
along the two edges are equal and opposite for the up and down 
\gls{spinon}s. This results in a net flow of spins along the two edges in 
opposite directions -- any scattering to the bulk states is prevented
by symmetry constraints. For the asymmetric system considered here,
$D'$ induces an anti-chiral edge current of \gls{spinon}s where each
species of \gls{spinon} flows in the same direction along the two edges.
This is balanced by counterflow current of \gls{spinon}s in the opposite direction
carried by the bulk modes. The anti-chiral \gls{DMI} breaks the chiral symmetry 
protecting the edge states and enables scattering between edge and bulk states. This edge-to-bulk scattering produces the bulk current that balances the
anti-chiral edge current.
In the following we discuss how the bulk and edge state dispersion changes
due to interplay between the chiral and anti-chiral \gls{DMI}.

Fig.~\,\ref{edge_state} shows the \gls{spinon} bands for a honeycomb nano-ribbon 
with dimension $200\times 500$ lattice sites with zigzag edges, together
with the spin current profile
along the width of the ribbon. 
The procedure of numerical calculations for band structures and spin current profile is described in Sec.\,\ref{sec2.5.3} in chapter "Model and Numerical Methods".
Three different sets of $(D, D')$
are chosen to illustrate the evolution of band dispersion and spin
currents with changing \gls{DMI}. For clarity of presentation, only one
species of \gls{spinon}s is shown. Along with the total spin current, the 
contributions from the bulk and two edge modes are calculated separately to 
identify the effects of $D'$ on each component. The \gls{spinon}
bands and the individual spin currents are color coded for easy 
identification. Green represents the bulk bands and their
contribution to the spin current at each position along the width of the 
ribbon; red (blue) denotes the localized \gls{spinon} mode and the associated
spin current at the top (bottom) edge. A negative (positive) value 
of the spin current denotes \gls{spinon} transport to the left (right) along
the length of the ribbon.

For $D>D'$, the tilting of the bands is
small and the dispersion of edge states at 
upper and lower edges are opposite, as shown in the  Fig.\,\ref{edge_state}(b).
The edge states are predominantly chiral in nature, 
and the spin current at the two edges are opposite in direction (Fig.\,\ref{edge_state}(c), though
not equal in magnitude since $D'\neq 0$ breaks chiral symmetry).
For large $D'$  
($D'~\gg~D$), the tilting of the bands at the Dirac points 
is much greater and yields identical dispersion for the two edge states 
(Fig.~\,\ref{edge_state}(e)). This results in {\it anti}-chiral edge states
where the spin current is in the same direction along both edges of the
ribbon (Fig.~\,\ref{edge_state}(f)).
Finally, when $D~\approx~D'$, one of the edge
states (the top edge in the present case) acquires a dispersionless 
character (Fig.~\,\ref{edge_state}(h)). In other words, the edge state 
at the top is localized with no \gls{spinon} transport while the bottom edge 
has a finite dispersion with a finite edge current 
(Fig.~\,\ref{edge_state}(i)).
Because of $U(1)$-symmetry of each \gls{spinon} sector, there is a counter-propagating bulk current to compensate the imbalance between edge states.
The bulk current is not
uniform across the width of the ribbon. Instead, it is primarily confined 
to a small region near the edges.
At each edge, the bulk current opposes the edge current, with its 
magnitude decreasing rapidly away from the edges. 
 To summarize, a non-zero net spin current along the two edges is a definitive 
signature of the existence of anti-chiral edge states.
In principle, the anti-chiral edge states persist over the entire temperature range for which the short range correlations exist for the system ($k_B T \lessapprox J$).
However, at very high temperature the short range correlations $\eta$ and $\zeta_s$ are very small (see Fig.\,\ref{parameter}),  which in turn makes the antichiral edge phenomenon experimentally undetectable at this limits. So we suggest the temperature range  $k_B T\leq 0.6J$ is ideal for experimental detection of Anti-chiral edge modes.

\subsection{\label{sec3.3.2}Magnon Picture}
At low temperature one of the \gls{spinon} species (say, the up-\gls{spinon}) condense at the lowest energy and form the ferromagnetic ground state. The down-\gls{spinon}s are now excitations above the ground state. In contrast  to the high temperature para-magnetic regime, at low temperatures only spin-excitations above the ground state  contribute to the dynamics of the system. Thus the system at low temperature can be described more simply using linearized \gls{HP} bosons, which is given as,
\begin{equation}
    S_i^+=\sqrt{2S}\hat{a}_i,\quad S_i^-=\sqrt{2S}\hat{a}_i^\dagger,\quad S_i^z=S-\hat{a}_i^\dagger\hat{a}_i,
    \label{eq3.13}
\end{equation}
where, $\hat{a}_i$ is the annihilation operator of \gls{HP} Boson. The \gls{HP}-transformation is discussed in detail in the section Sec.\,\ref{sec2.2.1} in the chapter "Model and Numerical Method". The real space Hamiltonian in terms of \gls{HP} Boson is given by,
\begin{align}
    \pazocal{H}=&-JS\sum_{\left\langle i,j\right\rangle} \left( \hat{a}_i^\dagger \hat{a}_j+\hat{a}_j^\dagger\hat{a}_i\right)-iDS\sum_{\left\langle\left\langle i,j\right\rangle\right\rangle} \nu_{ij}\left(\hat{a}_i^\dagger\hat{a}_j-\hat{a}_j^\dagger\hat{a}_i\right)\nonumber\\
    &-iD'S\sum_{\left\langle\left\langle i,j\right\rangle\right\rangle} \nu'_{ij}\left(\hat{a}_i^\dagger\hat{a}_j-\hat{a}_j^\dagger\hat{a}_i\right)+(B-3JS)\sum_i \hat{a}_i^\dagger\hat{a}_i
\end{align}

\begin{figure}[H]
	\centering
	\includegraphics[width=0.7\textwidth]{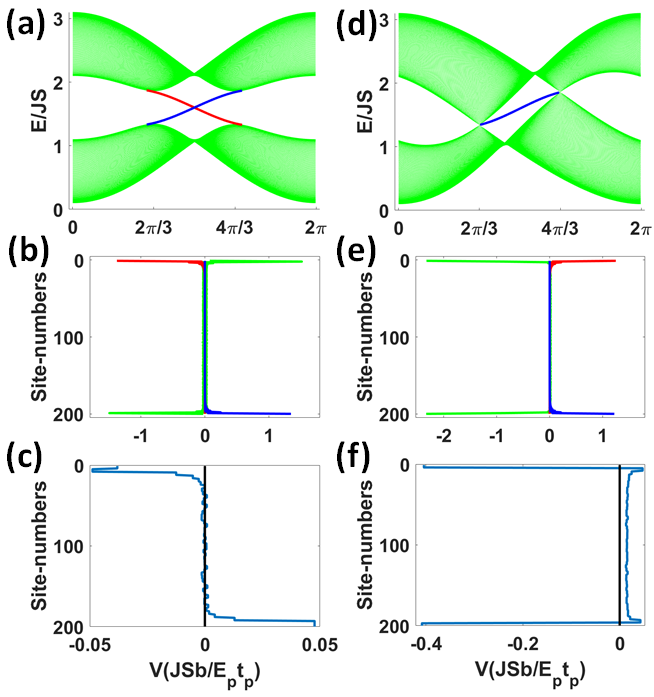}
	\caption{The figure sets $\left\lbrace(a)-(c)\right\rbrace$, $\left\lbrace(d)-(f)\right\rbrace$ represents result for down-\gls{spinon} from $200\times 500$ (width$\times$length) stripe with \gls{DMI}s $\left\lbrace D=0.1,D'=0.001\right\rbrace$,$\left\lbrace D=0.001,D'=0.1\right\rbrace$
	respectively. The other parameters are $J=1.0,  B=0.1, T=0.5$ for all the plots. The figure sets $\left\lbrace(a),(d)\right\rbrace$ shows the \gls{magnon} band structure of the stripe geometry. The figure sets $\left\lbrace(b),(e)\right\rbrace$ gives the spatial current distribution along width of the stripe, where $b$, $E_p$ and $t_p$ are the lattice constant, Planck-energy and Planck-time respectively. The figure sets $\left\lbrace(c),(f)\right\rbrace$ shows the average of spatial current distribution over four sites respectively. In all the figures the green, red and blue plots correspond to bulk-state, upper edge edge-state and lower-edge edge-state respectively. A non-zero net \gls{magnon} current along the edges in figures (f) denotes the existence of anti-chiral edge states.}
	\label{magnonpicture}
\end{figure}

In Fourier space the Hamiltonian is given by,
\begin{equation}
\pazocal{H}_{sp}^{mf}=\sum_{\boldsymbol{k}\in \mathtt{B.Z.}} \Psi^\dagger_{\boldsymbol{k}} \left[g(\boldsymbol{k})I+\boldsymbol{h}(\boldsymbol{k})\cdot \boldsymbol{\sigma}\right] \Psi_{\boldsymbol{k}},
\label{eq11}
\end{equation} 
where, $\Psi_{\boldsymbol{k}}^\dagger=\left(\hat{a}_{\boldsymbol{k}}^\dagger,\hat{b}_{\boldsymbol{k}}^\dagger\right)$. $\hat{a}_{\boldsymbol{k}}^\dagger$ and $\hat{b}_{\boldsymbol{k}}^\dagger$ are the creation operators for \gls{HP} Boson on sublattice-A and sublattice-B~(see Fig.1(a) of main text), respectively. $\boldsymbol{\sigma}_{\alpha}$($\alpha=x,y,z$) represents the Pauli matrices. The other terms are given by,
\begin{align}
g(\boldsymbol{k})=&B-3JS+2D'S\gamma_s^\beta,\quad
h_s(\boldsymbol{k})=\begin{pmatrix}
-JS\gamma_c^\alpha, &
JS\gamma_s^\alpha, &
2DS\gamma_s^\beta
\end{pmatrix},
\end{align}
where, $\gamma_c^\beta=\sum_{j} \cos(\boldsymbol{k}\cdot\boldsymbol{\beta}_j),\gamma_s^\beta=\sum_{j}\sin(\boldsymbol{k}\cdot\boldsymbol{\beta}_j),\gamma_c^\alpha=\sum_{j}\cos(\boldsymbol{k}\cdot\boldsymbol{\alpha}_j),\gamma_s^\alpha=\sum_j \sin(\boldsymbol{k}\cdot\boldsymbol{\alpha}_j)$ and the vectors $\boldsymbol{\beta}_j$ and $\boldsymbol{\alpha}_j$ are shown in figure Fig.1(a).
The \gls{magnon} Hamiltonian is identical to the down-\gls{spinon} Hamiltonian at low temperatures, as the mean field parameters of the \gls{spinon} Hamiltonian reduces to the following values, $\eta=S$, $\zeta_\uparrow=2S$, $\zeta_\downarrow=\xi_s=\xi'_s=0$(see Fig.\,\ref{parameter}) and the Lagrange's undetermined multiplier becomes $\lambda=3JS+B/2$. In other words, the \gls{spinon} Hamiltonian naturally converges to the
\gls{magnon} Hamiltonian at low $T$. The figure Fig.\,\ref{magnonpicture} shows the results obtained using \gls{magnon} picture. The results agree well with those for the down \gls{spinon} sector in the \gls{spinon} picture.

\subsection{\label{sec3.3.3}Experimental detection}
How does one detect anti-chiral spin currents experimentally? 
We suggest  a suite of experimental probes that, taken together, can provide a "smoking-gun" signature of the existence of anti-chiral edge states. First, magnetic force microscopy(MFM) offers a promising experimental technique to measure the \gls{spinon} current across the nano-ribbon and hence can detect the presence of anti-chiral edge states. A non-zero value of net spin current along the two edges provides a direct evidence of anti-chiral edge states. Current MFM techniques can probe the local spin current in a finite sample to a resolution of a few nm. Since the topological character of the \gls{spinon} bands for the different ranges of anisotropic \gls{DMI} is reflected in distinct current profile across the ribbon, we believe MFM provides a promising experimental technique to identify anti-chiral edge states in real quasi-2D materials.
Second, inelastic neutron scattering spectra can indirectly detect the presence of anti-chiral edge modes, by probing the \gls{magnon} band structure. If the bands are tilted or the energy at $K$ and $K'$-point are unequal, it will suggest the presence of anti-chiral edge modes.


\begin{figure}[H]
	\centering
	\includegraphics[width=0.7\textwidth]{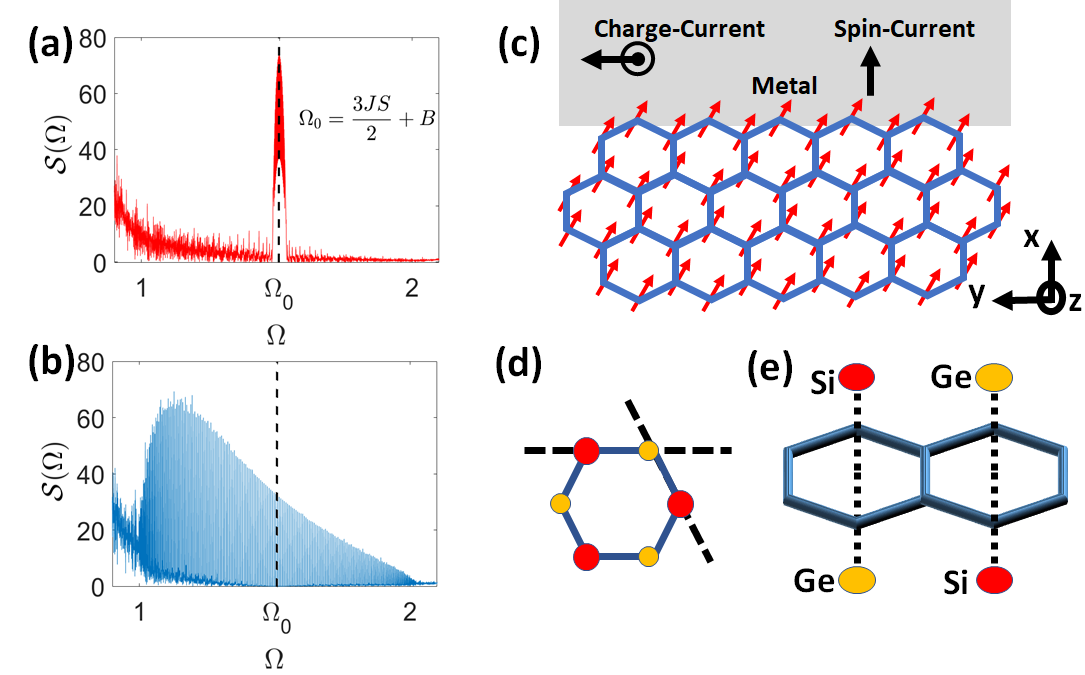}
	\caption{(a)-(b) Dynamical spin structure factor for two edges at temperature $T=0.4J$, with parameters $D=0.1, D'=0.09, B=0.01$. Red and blue colors denote upper and lower edges. (c) The experimental setup for spin Hall noise spectroscopy. (d) A ferromagnetic material with two different sub-latices with a mirror symmetric plane along the dashed lines. (e) A proposed material based on real materials \ce{CrGeTe3} and \ce{CrSiTe3} to realize antichiral edge states.}
	\label{ExperimentalRealization}
\end{figure}

Finally, we show that the probing the low energy excitation spectrum offers a 
powerful probe to detect the asymmetry of the two edges in the limit of $D=D'$, when one of the edge state acquire a dispersionless character compared to the other edge. This low energy spectrum is readily obtained from calculating the \gls{dssf}. 
The \gls{dssf} at the edge of a finite system is defined in the equation Eq.\,\ref{Eq:DSSF} and the schematic of experimental set-up to measure \gls{dssf} via spin-Hall-noise-spectroscopy is shown in the the figure Fig.\,\ref{ExperimentalRealization}(c) which is further explained in the section Sec.\,\ref{sec2.5.5} in "Model and Numerical Methods" chapter.
This experimental technique offers a promising route for detecting anti-chiral edge states. The \gls{dssf}\,($\pazocal{S}(\Omega)$), shown
in the figure Fig.\,\ref{ExperimentalRealization}(a)-(b) for different edges, can be interpreted as the number of edge \gls{magnon}s present in a given energy level, and is proportional to the product of the density of states and Bose-Einstein distribution for the corresponding energy level. The signature of the edge states is
reflected in the features of the \gls{dssf}  near $\Omega=\frac{3J}{2}+B$, which is
the energy of the edge states in the absence of any \gls{DMI}. Our results show that
$\pazocal{S}(\Omega)$ is dramatically different for the two edges for strong anti-chiral edge states, viz., when $\left|D\right|\approx \left|D'\right|$(equivalently $D_A\gg D_B$ or $D_B\gg D_A$).

For simplicity, to calculate the dynamical spin structure factor, we have used the \gls{HP} transformation as in Eq.\,\ref{eq3.13}. For the case of ferromagnet with up spin at each site, the \gls{HP} boson represents the down-\gls{spinon}s in \gls{spinon} picture at low temperature. The \gls{dssf} is calculated at the edges of the stripe geometry as in figure Fig.\,\ref{edge_state}(a). 
For the site-1 of m-th stripe,
\begin{align}
\hat{a}_{1,m}(t)=&\frac{1}{\sqrt{M}}\sum_{k=0}^{M-1} \hat{a}_{1,k}(t) e^{-i\frac{2\pi k m}{M}}
=\frac{1}{\sqrt{M}}\sum_{k=0}^{M-1}\sum_n P_{1,n}  \hat{a}_{n,k}^d(t) e^{-i\frac{2\pi k m}{M}},
\label{bosonic}
\end{align}
where, $P$ is the unitary matrix for diagonalization of Hamiltonian and $\hat{a}^d_{n,k}$ is the bosonic operator after diagonalization and $M$ is the periodicity of the stripe along x-direction. The Hamiltonian can be written as,
\begin{equation}
H_k=\sum_n\epsilon_n(k) \hat{a}_{n,k}^{d\dagger}\hat{a}_{n,k}^{d},
\end{equation}
Using the above relation and Heisenberg's equation of motion it can be proved that,
\begin{align}
\hat{a}_{n,k}^{d\dagger}(t)&=\hat{a}_{n,k}^{d\dagger}(0)e^{i\epsilon_n(k)t},\quad
\hat{a}_{n,k}^{d}(t)=\hat{a}_{n,k}^{d}(0)e^{-i\epsilon_n(k)t}
\label{Heisenberg}
\end{align}

 The dynamical spin structure factor is given by,
\begin{align}
\pazocal{S}(\Omega)&=\sum_{m\in \text{edge}} \left[\pazocal{S}^{xx}_{mm}(\Omega)+\pazocal{S}^{zz}_{mm}(\Omega)\right]\nonumber\\
&=\sum_{m\in \text{edge}} \left[\int^\infty_{-\infty} dt e^{-i\Omega t}\left(-i\left\langle \hat{S}^x_{1,m}(t) \hat{S}^x_{1,m}(0)\right\rangle\right)\right.
\left.+\int^\infty_{-\infty} dt e^{-i\Omega t}\left(-i\left\langle \hat{S}^z_{1,m}(t) \hat{S}^z_{1,m}(0)\right\rangle\right)\right]\nonumber\\
=&-i \sum_m \int^{\infty}_{-\infty} dt e^{-i\Omega t} \left[ \frac{S}{2}\left\langle \hat{a}_{1,m}(t)\hat{a}_{1,m}(0) \right.\right. 
\left.\left.+ \hat{a}_{1,m}(t)\hat{a}_{1,m}^\dagger(0)+ \hat{a}_{1,m}^\dagger(t)\hat{a}_{1,m}(0) + \hat{a}_{1,m}^\dagger(t)\hat{a}_{1,m}(0)\right\rangle\right.\nonumber\\
&\qquad  \left. +\left\langle S^2-S\hat{a}_{1,m}^\dagger (0)\hat{a}_{1,m}(0)-S\hat{a}^\dagger_{1,m}(t)\hat{a}_{1,m}(t)\right\rangle\right] 
\quad
\text{[Neglecting the higer order terms]}.
\label{chi}
\end{align}

Using Eq.\,\ref{bosonic} and Eq.\,\ref{Heisenberg}, we get,
\begin{align}
\left\langle \hat{a}_{1,m}^{\dagger}(t) \hat{a}_{1,m}(0)\right\rangle &=\left\langle \frac{1}{\sqrt{M}}\sum_{k=0}^{M-1}\sum_n P^{*}_{1,n} \hat{a}_{n,k}^{d\dagger}(t) e^{i\frac{2\pi k m}{M}} \right.
\left.\frac{1}{\sqrt{M}}\sum_{k'=0}^{M-1}\sum_{n'} P_{1,n'} \hat{a}_{n',k'}^{d}(0) e^{i\frac{2\pi k' m}{M}} \right\rangle \nonumber \\
&=\frac{1}{M} \sum_{k=0}^{M-1} \sum_n \sum_{k'=0}^{M-1} \sum_{n'} P_{1,n}^{*} P_{1,n'} e^{i\epsilon_n(k)t}
 \left\langle \hat{a}_{n,k}^{d\dagger}(0)\hat{a}_{n',k'}^{d}(0)\right\rangle\nonumber\\
&=\frac{1}{M} \sum_{k=0}^{M-1} \sum_n |P_{1,n}|^2 e^{i\epsilon_n(k)t} n(\epsilon_n(k)),
\label{equation}
\end{align}
where, $n(\epsilon_n(k))$ is the Bose-Einstein distribution for energy $\epsilon_n(k)$.

Similarly,
\begin{align}
\left\langle \hat{a}_{1,m}(t) \hat{a}_{1,m}^{\dagger}(0)\right\rangle &= \frac{1}{M} \sum_{k=0}^{M-1} \sum_n |P_{1,n}|^2 e^{-i\epsilon_n(k)t} n(\epsilon_n(k)),\quad
\left\langle \hat{a}_{1,m}^{\dagger}(t) \hat{a}_{1,m}^{\dagger}(0)\right\rangle =\left\langle \hat{a}_{1,m}(t) \hat{a}_{1,m}(0)\right\rangle=0\nonumber\\
\left\langle \hat{a}_{1,m}^{\dagger}(t) \hat{a}_{1,m}(t)\right\rangle &=\frac{1}{M}\sum_{k=0}^{M-1} \sum_n |P_{1,n}|^2 n(\epsilon_n(k)),\quad
\left\langle \hat{a}_{1,m}^{\dagger}(0) \hat{a}_{1,m}(0)\right\rangle =\frac{1}{M} \sum_{k=0}^{M-1} \sum_n |P_{1,n}|^2 n(\epsilon_n(k))
\label{equations}
\end{align}

Using Eq.\,\ref{equation} and Eq.\,\ref{equations} in Eq.\,\ref{chi}, we get,
\begin{align}
\pazocal{S}(\Omega)
&=-i\pi (M-1) \left[ \frac{S}{2} \sum_{k=0}^{M-1} \sum_n |P_{1,n}(k)|^2 n(\epsilon_n(k)) \delta(\Omega+\epsilon_n(k))\right.\nonumber\\
&\qquad\qquad\qquad+\frac{S}{2} \sum_{k=0}^{M-1} \sum_n |P_{1,n}(k)|^2 n(\epsilon_n(k)) \delta(\Omega-\epsilon_n(k))\nonumber\\
&\qquad\qquad\left.+\left\lbrace MS^2-2S \sum_{k=0}^{M-1} \sum_n |P_{1,n}(k)|^2 n(\epsilon_n(k))\right\rbrace \delta(\Omega)\right]
\end{align}

\subsection{\label{sec3.3.4}Material realization}
Symmetry plays an important role in determining the \gls{DMI} of a spin system as discussed in the section Sec.\,\ref{sec2.6}.
The presence of anti-chiral \gls{DMI} requires two in-equivalent sub-lattices in the 2D-honeycomb lattice, as shown in Fig.\,\ref{ExperimentalRealization}(d). 
The presence of two different types of atoms will result in asymmetric \gls{DMI},
leading to a broken inversion symmetry and non-zero $D'$. 
Mirror symmetry along the dotted lines prevents any non-zero perpendicular \gls{DMI} on nearest-neighbour bonds, whereas in-plane mirror symmetry suppresses any in-plane \gls{DMI}. While we  are not aware of any such material at present, the recent
discovery of ferromagnetic order in 2D limit of several 
\ce{Cr}-based compounds including \ce{CrI3}~\,\cite{CrI3}, \ce{CrBr3}~\,\cite{CrBr3,CrBr32,CrSiTe3}, \ce{CrSrTe3}\,\cite{CrSiTe3} and \ce{CrGeTe3}\,\cite{CrGeTe3} as well as the \ce{Fe}-based family of compounds
\ce{AFe_2(PO_4)_2} (A=Ba,Cs,K,La)\,\cite{IronDirac}
offer great promise. These quasi-2D
materials consist of weakly Van Der Waals-coupled honeycomb ferromagets. Presence of chiral \gls{DMI} in some members of this family~\,\cite{CrI3}has been established using inelastic neutron scattering spectroscopy. In materials like \ce{CrSrTe3} and \ce{CrGeTe3}, presence of inversion center at the center of honeycomb cell makes the two sub-lattices equivalent. The inversion symmetry can be removed by replacing every other \ce{Ge} atom by an \ce{Si} atom as depicted in Fig.\,\ref{ExperimentalRealization}(e). In a similar vein, replacement of \ce{P} atom by another Group V element in \ce{AFe_2(PO_4)_2 } (A=Ba,Cs,K,La)\,\cite{IronDirac} will break the inversion symmetry of the lattice. The breaking of inversion symmetry may, in principle, give rise to additional interactions in these materials, e.g.,
nearest neighbor \gls{DMI}. 
However, we have verified that inclusion of additional interactions, including nearest neighbor \gls{DMI} as well as 2nd and 3rd nearest neighbor Heisenberg interactions only modifies the linear dispersion of the edge states, and
does not suppress the appearance of anti-chiral edge states as shown in the next section.

\section{\label{sec3.4}Modulation of edge state dispersion in presence of other interactions in a honeycomb ferromagnet}
\label{STABILITY}
\begin{figure}[H]
	\centering
	\includegraphics[width=0.3\textwidth]{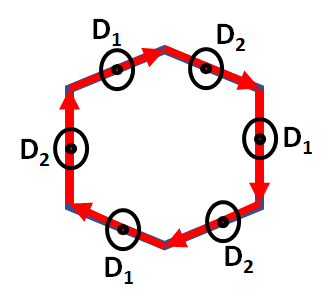}
	\caption{DM-interactions on nearest neighbour bonds.}
	\label{DMinteraction}
\end{figure}
The spin model discussed so far is very ideal. For example, the long range Heisenberg interactions are neglected. Moreover, the criteria to get the anti-chiral DM-term $D'$ is to break the inversion symmetry at the center of each honeycomb plaquette. The breaking of such symmetry might arise the nearest neighbour out of plane DM-interactions and also in plane DM-interactions. At low temperature, the in plane interactions can be neglected, which gives rise to three \gls{magnon} interactions in terms of \gls{HP} Bosons. But, at high temperature, where the spin-system can be treated in terms of \gls{spinon}s, the in plane DM-interaction gives rise to the mixing of two \gls{spinon} sectors, in presence of which the quasi-particles are mixed eigenstates of spin-up and spin-down sectors. Neglecting, any presence of in plane DM-intration at low temperature, we can re-write a more general Hamiltonian of the material as,
\begin{align}
\pazocal{H}&=J_1 \sum_{\left\langle i,j \right\rangle} \boldsymbol{S}_i \cdot \boldsymbol{S}_j+
 J_2 \sum_{\left\langle\left\langle i,j \right\rangle\right\rangle} \boldsymbol{S}_i \cdot \boldsymbol{S}_j+
 J_3 \sum_{\left\langle\left\langle\left\langle i,j \right\rangle\right\rangle\right\rangle} \boldsymbol{S}_i \cdot \boldsymbol{S}_j
 +D_1\sum_{\left\langle ij\right\rangle _A} \nu_{ij}\hat{z}\cdot(\boldsymbol{S}_i \times \boldsymbol{S}_j)\nonumber\\
& + D_2\sum_{\left\langle ij\right\rangle_B} \nu_{ij}\hat{z}\cdot(\boldsymbol{S}_i \times \boldsymbol{S}_j)
+D\sum_{\left\langle\left\langle ij\right\rangle\right\rangle_B} \nu_{ij}\hat{z}\cdot(\boldsymbol{S}_i \times \boldsymbol{S}_j)+D'\sum_{\left\langle\left\langle ij\right\rangle\right\rangle_B} \nu'_{ij}\hat{z}\cdot(\boldsymbol{S}_i \times \boldsymbol{S}_j)
 +B\sum_i \boldsymbol{S}_i,
 \label{Ham}
\end{align}
where, the DM-interactions $D_1$ and $D_2$ are defined on the nearest neighbour bonds $\left\langle ij\right\rangle _A$ and $\left\langle ij\right\rangle _B$ as shown in Fig.\,\ref{DMinteraction}. Furthermore, single ion anisotropy like terms acts as chemical potential for the spin-excitation, which is taken into account into the magnetic field. To consider more realsitic situation,

\begin{figure}[H]
	\centering
	\includegraphics[width=0.8\textwidth]{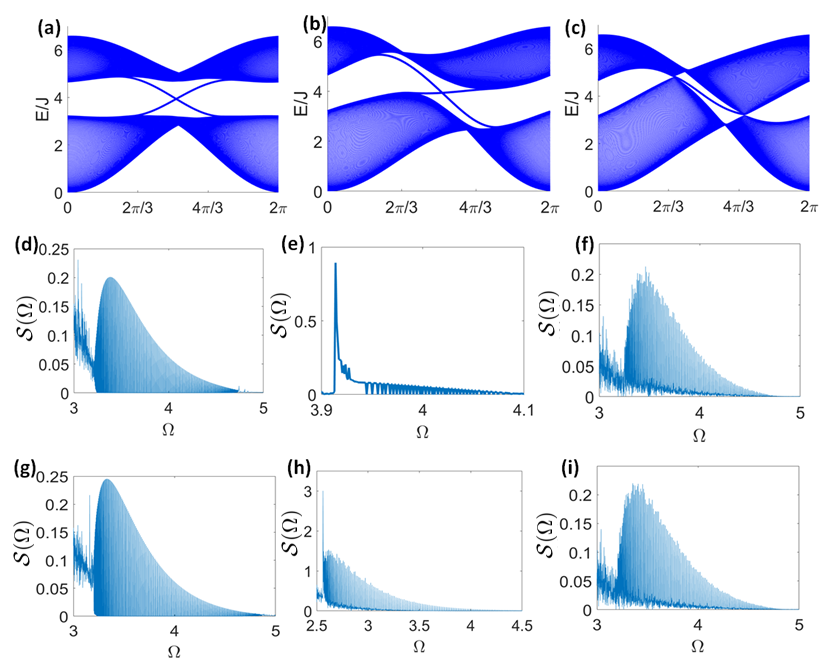}
	\caption{Band structure of stripe geometry for parameters (a) $D=0.31 meV$, $D'=0.01 meV$, (b) $D=0.31 meV$, $D'=0.28 meV$, (c) $D=0.01 meV$, $D'=0.31 meV$. The dynamical spin structure factor for upper edge at $T=0.4$ for parameters, (d) $D=0.31 meV$, $D'=0.01 meV$, (e) $D=0.31 meV$, $D'=0.28 meV$, (f) $D=0.01 meV$, $D'=0.38 meV$. The dynamical spin structure factor for lower edge at $T=0.4$ for parameters, (g) $D=0.31 meV$, $D'=0.01 meV$, (h) $D=0.31 meV$, $D'=0.28 meV$, (i) $D=0.01 meV$, $D'=0.38 meV$.}
	\label{Modified}
\end{figure}

\noindent we have fixed the Hesenberg interactions present in the material \ce{CrI3}\,\cite{CrI3}, $J_1=2.09$ meV, $J_2=0.16$ meV, $J_3=0.18$ meV. The nearest neighbour DM-terms and magentic field are fixed as $D_1=0.1$ meV, $D_2=0.15$ meV, $B=g\mu_B B_z=0.01$ meV. We transformed the spin Hamiltonian into \gls{magnon} Hamiltonian using \gls{HP} transformation defined in Eq.\,\ref{eq11}. Then, plotted the band structure and dynamical spin structure factor in Fig\,\ref{Modified}. It is noticeable, the qualitative behaviour of the edge states is mostly depend on the DM-interactions $D$ and $D'$. The presence of other interaction terms in the Hamiltonian just distorts the linear dispersion of the edge states.

\section{\label{sec3.5}Conclusion}
In conclusion, we have studied a Heisenberg ferromagnet with additional
next nearest neighbor \gls{DMI}s on a honeycomb lattice with broken
sublattice symmetry. The unequal \gls{DMI} between atoms on different 
sublattices, together with the broken chiral symmetry results in the emergence 
of anti-chiral edge states, in addition to the normal chiral modes. This is 
manifested in unique spin current distribution across the width of a finite system
with ribbon geometry. Interestingly, a uncompensated anti-chiral edge current exists, when antichiral DM-interaction $D'$ is larger than the chiral DM-interaction $D$ at a temperature compared to the ferromagnetic Heisenberg interaction $J$(i.e. $k_B T\approx J$). We have shown that the anti-chiral
edges states result in a number of observable physical signatures, including a non-zero net spin current along the edges (that is only compensated by counter-flowing bulk spin current) and strong anisotropy in the dynamic spin structure factor at the opposite edges. These unique features serve as smoking-gun signatures for the existence of anti-chiral edge states. We propose experimental probes to detect the presence of
anti-chiral edge states via these features as well as a potential material where such states may be
realized experimentally. In this work, the stability of the ferromagnetism in the proposed material is not studied. So in future, further studies need to be done to identify the presence of ferromagnetism and anti-chiral DM-interaction in the proposed materials.

\chapter[The topological magnon bands in the Flux state in Shastry-Sutherland lattice]{The topological magnon bands in the Flux state in Shastry-Sutherland lattice}\label{chapter04}

	\epigraph{The results in this chapter is published as D. Bhowmick and P. Sengupta.  Topological magnon bands in the flux state of Shastry-Sutherland lattice model.  Physical Review B 101, 214403 (2020). DOI: 10.1103/PhysRevB.101.214403.}

\section{\label{sec4.1}Introduction}
Quantum magnets have served as a versatile test bed for realizing novel bosonic
phases, including bosonic topological phases. The topological character of the
magnetic phase is manifested through the behavior of \gls{magnon}s.
Magnons are charge neutral quasi-particle which represent the magnetic excitations of a condensed matter system. Analogous to electrons in standard topological insulators, \gls{magnon}s in 
magnetic insulators exhibit thermal Hall effect\,\cite{THEIntro}, spin-Nernst 
effect\,\cite{CoplanarIntro3,CoplanarIntro2,SNEIntro3}, and \gls{magnon}-driven spin Seebeck 
effect\,\cite{SpinSeebackIntro}. The interest in these systems is driven by both
fundamental reasons and potential for technological applications. The recent use
of Skyrmions in spintronics for efficient magnetic storage and read/write 
devices with minimal Joule heating effect\,\cite{Skyrmion1,Skyrmion2} underscores
the potential practical applications of topologically no-trivial magnetic states. 
The wide range of quantum magnets with 
varying interactions and lattice structures as well as the ability to control
the number of quantized excitations with an external magnetic field make them 
ideal for exploring novel magnetic phases. Geometrically frustrated quantum magnets 
are particulalry promising in realizing and controlling topologically non-trivial
spin textures\,\cite{frustration1,NonCoplanar1, FluxStateIntro,frustration2,frustration3,frustration4}. The interplay between competing interactions, geometric frustration
and external magnetic field result in a wide variety of magnetic phases that are
not commonly observed in their non-frustrated counterparts. In most cases, topological 
excitations in quantum magnets are driven by \gls{DMI}\,\cite{CoplanarIntro1,CoplanarIntro2, 
CoplanarIntro3,CoplanarIntro4}, although topological \gls{magnon} bands can 
exist without \gls{DMI} as well~\,\cite{NonCoplanar1,triangular} due to non-coplaner chiral spin-texture. Most strikingly, the change 
in spin-texture by changing the parameters in the non-coplanar spin systems, gives rise 
to variety of topological \gls{magnon} bands in the same system\,\cite{triangular,NonCoplanar3}.

The \gls{ss} model is a paradigmatic model for the study of frustrated
magnetism. Since the degree of frustration can be tuned by varying the ratio of the 
diagonal and axial bonds, the model exhibits a wide range of novel magnetic phases\,\cite{phase1,phase2,phase3}.
The existence of a number of materials with underlying \gls{ss} geometry of the magnetic ions offers the prospect
of observing theoretically predicted phases and phenomena in real materials\,\cite{MaterialRealization,RB41,RB42,RB43}. Since 
\gls{DMI} is ubiquitous in all of these materials, it is natural to supplement the 
canonical \gls{ss} model with \gls{DMI}. This results in an even richer variety of magnetic
orderings including collinear, coplanar and non-coplanar spin configurations, several of 
which host topological \gls{magnon}s\,\cite{FluxStateIntro,Shahzad_2017}. Previous studies of topological \gls{magnon}s for the 
\gls{ss} lattice were restricted to the dimer~\,\cite{Romhanyi2} and the ferro-magnetic 
phases~\,\cite{ferroIntro}. In recent past there has been a growing interest in
studying topological \gls{magnon}s in non-collinear spin configurations in frustrated
lattices. Here we present the results of our investigation of topological \gls{magnon}s
in the recently proposed flux state, that is stabilised by \gls{DMI} perpendicular to the lattice plane  in the \gls{ss}
lattice\,\cite{FluxStateIntro}.In presence of in plane \gls{DMI}, 
this evolves to the canted flux state -- the resulting \gls{magnon} bands in an external longitudinal magnetic field, carry non-zero
Chern numbers that determine the topological character of the \gls{magnon} bands. Varying
the different components of the \gls{DMI} result in a sequence of band topological 
transitions where the Chern number for the \gls{magnon} bands change over a wide range of
possible values. 


The transitions in the band topology of the system can be detected from the first derivative of the thermal Hall conductivity which exhibits a logarithmic divergence at the  transition\,\cite{triangular,divergence}. The peak height of the logarithmic divergence increases with temperature following an  algebraic relation. The interpolation of the first derivative of the thermal Hall conductance as a function of temperature yields information on the nature of band touching (gap closing) at the band topological transition.

\section{\label{sec4.2}Model Hamiltonian and Method}
\begin{figure}[H]
	\centering
		\includegraphics[width=0.7\textwidth]{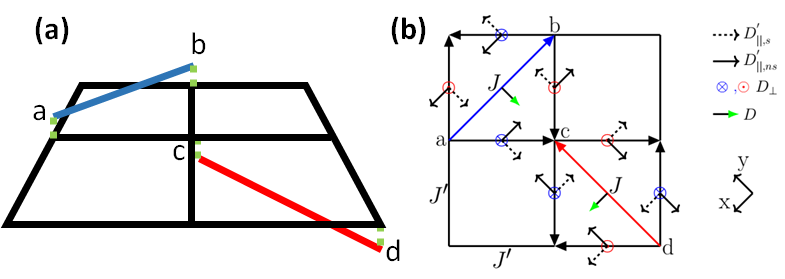}
	\caption{(a) The unitcell of the Shastry-Sutherland lattice with non-coplaner positions of the basis-sites a,b,c,d. (b) The Heisenberg and crystal symmetry allowed Dzyaloshinskii–Moriya interactions on Shastry-Sutherland lattice.}
	\label{lattice2}
\end{figure}

We study a generic microscopic Hamiltonian describing the magnetic properties of
the family of \gls{ss} compounds including the rare earth tetraborides 
and \ce{SrCu2(Bo3)2}. 
We have carefully restricted our choice of Hamiltonian parameters to realistic 
ranges. We use nearly equal Heisenberg exchange interactions on the diagonal and axial 
bonds($J\approx J'$) as observed in 
the rare earth tetraborides
(\ce{RB4}, R=Er,Tm)\,\cite{RB41,RB42,RB43}. The choice is motivated by the
large magnitude of effective localized moments in these compounds which render
the \gls{HP} approach applicable.
The nature of DM-interaction chosen for the 
study(Fig.\,\ref{lattice2}(b)) is based on the symmetry of the \gls{ss} lattice. We have
used a realistic range of the magnitudes of different DM vectors to stabilize
the flux state as the magnetic ground state. The possibility of forming Van der Waals heterostructures with heavy metals offers
the ability to induce and tune DM interactions over a extended range. 
The bonds along the two
inequivalent diagonals of the unit cell are shifted out of the plane in opposite
directions as shown in Fig.\,\ref{lattice2}(a) as observed in the canonical \gls{ss}-compound, \ce{SrCu2(BO3)2} in its low temperature phase\,\cite{Romhanyi1, structure2, Romhanyi2}. 
 The model Hamiltonian is given by,
\begin{align}
\pazocal{H}=&J\sum_{\left\langle\left\langle i,j \right\rangle\right\rangle} \bold{S}_i\cdot\bold{S}_j + J' \sum_{\left\langle i,j \right\rangle} \bold{S}_i\cdot\bold{S}_j 
  + \bold{D}\cdot\sum_{\left\langle\left\langle i,j \right\rangle\right\rangle} \left(\bold{S}_i\times\bold{S}_j\right)+ \bold{D'}\cdot\sum_{\left\langle i,j \right\rangle} \left(\bold{S}_i\times\bold{S}_j\right) 
  -B\sum_i\bold{S}_i^z,
  \label{eq1Ch4}
\end{align}
where $J$ and $J'$ ($>0$)  are anti-ferromagnetic Heisenberg spin-exchange  on the 
axial and diagonal bonds of the \gls{ss} lattice respectively (henceforth referred to as \gls{ss} 
bonds).   $\bold{D}$ and $\bold{D'}$ are DM-vectors on the \gls{ss} bonds as shown in Fig.
\,\ref{lattice2}(b). As discussed above, the nature of DM interactions are chosen consistent with the symmetry
constraints of the \gls{ss} lattice~Fig.\,\ref{lattice2}(a)\,\citep{Romhanyi2,triplon2,triplon3}.

We start with a classical ground state and investigate 
quantized low energy excitations, \gls{magnon}s, focusing on identifying any 
topological character of the \gls{magnon} bands. The classical ground state of the 
Hamiltonian Eq.\,\ref{eq1Ch4} is derived by replacing the local spin moments by classical vectors of 
unit magnitude. The state of each spin is specified by the polar co-ordinates,
\begin{equation}
\bold{S}=S(\sin(\theta)\cos(\phi), \sin(\theta)\sin(\phi),\cos(\theta)).
\label{eq2}
\end{equation} 
The ground state spin configuration is obtained by minimizing the energy of the 
Hamiltonian w.r.t the angles $\theta$ and $\phi$. For the ground state phases of interest, 
viz., the flux and the canted flux states, the magnetic unit cell is of the same size as the unit cell of the \gls{ss} lattice and consists of four sites 
as shown in Fig.\,\ref{lattice2}. 

The classical phases are further discussed in Sec.\,\ref{sec4.3}. Since we are primarily interested
in the flux (Fig.\,\ref{FluxState}(a)) and the (in-plane \gls{DMI}-induced) canted flux states (Fig.\,\ref{3Q-order}(a)), we start by 
identifying the parameter ranges where these are realized. The ground state phases of the
Hamiltonian Eq.(\,\ref{eq1Ch4}) for classical spins has been investigated in Ref.\,\cite{FluxStateIntro}
-- with vanishing in-plane component of \gls{DMI}, the flux state (Fig.\,\ref{FluxState}(a)) is 
stabilized above a critical value of the normal component of the \gls{DMI} along the axial bonds, $D_\perp$(Fig.\,\ref{lattice2}(b)) (the \gls{DMI} on the diagonal bonds are constrained by the symmetry
of the lattice to lie on the plane of the lattice). The continuous $U(1)$ symmetry of the Hamiltonian is spontaneously broken in the flux state and Fig.\,\ref{FluxState}(a) shows one of the degenerate ground states. Interestingly, the flux state state is also realized on 
the square lattice, but for a much stronger \gls{DMI}. The geometric frustration of the \gls{ss} lattice
facilitates the appearance of the flux state at a more moderate (and realistic) strength
of DM interaction. Moreover, DM-interactions can also be induced by incident circularly polarized optical wave\,\cite{Floquet}, further enhancing the
possibility of stabilizing the flux state. However, the symmetry of the \gls{ss} lattice also allows additional in-plane components of \gls{DMI}, denoted in this work by $D$ on the diagonal bonds and $D_{||,s}$, $D_{||,ns}$ on the axial bonds (see Fig.\,\ref{lattice2}(b)). Any non-zero in-plane \gls{DMI} tilts the spins out of plane keeping the in-plane spin component of nearest-neighbour sites perpendicular to each other. The spins on the two distinct diagonal bonds cant in opposite
direction -- for one of the diagonals, the spins cant out of the plane, whereas for the other diagonal, they cant into the plane of the lattice. The in-plane components are aligned along the diagonal bonds, as depicted in Fig.\,\ref{3Q-order}(a). This ground state spin configuration is referred to
as the canted-flux state. In the presence of the in-plane components of \gls{DMI}, the U(1)
symmetry of the Hamiltonian is explicitly broken and there is no spontaneous breaking of U(1) symmetry in the in-plane \gls{DMI} driven canted flux state.
The flux state and canted flux states have not been observed in any real material yet, but recent advances in enhancing \gls{DMI} by interfacing with heavy-metal thin-films holds the promise to realize such systems in artificial hetero-structures.

To study low energy excitations (\gls{magnon}s) above the magnetic ground state, we have used the linearized 
\gls{HP} transformation\,\cite{CoplanarIntro1,CoplanarIntro4}. The \gls{HP} transformation is a  versatile
and extensively used approach to study low energy \gls{magnon}s\,\cite{new,new2,CoplanarIntro1,ferroIntro,NonCoplanar1,CoplanarIntro4,CoplanarIntro3,SBMFT5,StripeZigzag}.
In this work, we have extended the 
\gls{HP} approach to study \gls{magnon} excitations above complex magnetic
orders with longer periodicity. Here we present a brief discussion of the method. First, the local co-ordinate axis at each site of the lattice is rotated such that the $S_z$ axis is aligned along the local spin direction. For low temperature excitations the linearized \gls{HP} transformation is given by,
$
\hat{S}'^+_{i,a}=\sqrt{2S} \hat{a}_i,\;
\hat{S}'^-_{i,a}=\sqrt{2S} \hat{a}^\dagger_i ,\;
\hat{S}^z_{i,a}=S-\hat{a}^\dagger_i\hat{a}_i,
$
where, we consider $\hbar=1$ and $\hat{a}_i^\dagger (\hat{a}_i)$ represent creation (annihilation)
operators for quantized excitations above the magnetic ground state at site $i$. These obey bosonic
commutation relations,
$
[\hat{a}_i,\hat{a}_j^\dagger]=\delta_{i,j}
$
and 
$ 
[\hat{a}_i,\hat{a}_j] = 0 = [\hat{a}_i^\dagger,\hat{a}_j^\dagger] 
$. Since the unit cell consists of 4 sites, there are four species of bosons corresponding
to each inequivalent lattice site. 
The quadratic \gls{magnon} Hamiltonian is given by,
\begin{align}
\pazocal{H}=S&\sum_{\bk}\left[Q_\bk(\doneN)\capdagk{a}\capk{c}+Q_\bk(-\dtwo)^*\capdagk{a}\capk{d}\right.
+Q_\bk(\dtwo)^*\capdagk{b}\capk{c}+Q_\bk(-\doneN)\capdagk{b}\capk{d}
\nonumber
\\
&-\frac{1}{2}(J+\epsilon_{A})\expmk{x}\capdagk{a}\capk{b}
\left.-\frac{1}{2}(J-\epsilon_{B})\expk{y}\capdagk{c}\capk{d})\right]+\mathtt{h.c.}
\nonumber
\\
+S&\sum_{\bk}\left[P_\bk(\doneN)^*\capk{a}\capmk{c}+P_\bk(-\dtwo)\capk{a}\capmk{d}\right.
+P_\bk(\dtwo)\capk{b}\capmk{c}+P_\bk(-\doneN)^*\capk{b}\capmk{d}
\nonumber
\\
&+\frac{1}{2}\left(J-\epsilon_{A}\right)\expk{x}\capk{a}\capmk{b}
\left.+\frac{1}{2}\left(J+\epsilon_{B}\right)\expmk{y}\capk{c}\capmk{d}\right]+\mathtt{h.c.}
\nonumber
\\
+S&\sum_{\bk}\left[\left(2E-\epsilon_{A}+\frac{B}{S}\cos(\theta_1)\right)\capdagk{a}\capk{a}\right.
+\left(2E-\epsilon_{A}+\frac{B}{S}\cos(\theta_1)\right)\capdagk{b}\capk{b}
\nonumber
\\
&+\left(2E+\epsilon_{B}+\frac{B}{S}\cos(\theta_2)\right)\capdagk{c}\capk{c}
\left.+\left(2E+\epsilon_{B}+\frac{B}{S}\cos(\theta_2)\right)\capdagk{d}\capk{d}\right],
\label{eq4.3}
\end{align}
where,
\begin{align}
E&=-2 J' C^X_\theta+2 D_\perp S^X_\theta -D_{||,ns}(\zeta_{\theta_1}-\zeta_{\theta_2})-D_{||,s}(\zeta_{\theta_2}-\zeta_{\theta_1})
\nonumber\\
\epsilon_{A}&=J\cos(2 \theta_1)+D\sin(2 \theta_1)
,\,
\epsilon_{B}=-J\cos(2\theta_2)+D\sin(2\theta_2)
\nonumber\\
Q_\bk(\delta)&=\lambda_{J'}^+(\bk,\delta)-\lambda_{D_\perp}^+(\bk,\delta)
-D_{||,ns}(\xi_{\theta_1}\expk{\delta}-\xi_{\theta_2}\expmk{\delta})
+D_{||,s}(\xi_{\theta_1}\expmk{\delta}-\xi_{\theta_2}\expk{\delta})
\nonumber\\
P_\bk(\delta)&=\lambda_{J'}^-(\bk,\delta)-\lambda_{D_\perp}^-(\bk,\delta)
-D_{||,ns}(\xi_{\theta_1}^*\expk{\delta}-\xi_{\theta_2}\expmk{\delta})
+D_{||,s}(\xi_{\theta_1}^*\expmk{\delta}-\xi_{\theta_2}\expk{\delta})
\nonumber\\
\lambda_{J'}^{\pm}(\bk,\delta)&=J'(i C^{\pm}_\theta+S^X_\theta)\cos(\bk\cdot\delta)
,\,
\lambda_{D_\perp}^{\pm}=D_\perp (C^X_\theta \pm 1) \cos(\bk\cdot\delta)
\nonumber \\
\xi_{\theta_i}&=\frac{i}{2}\sin(\theta_i)+\frac{1}{2}\cos(\theta_i)\sin(\theta_{\bar i})
,\,
\zeta_{\theta_i}=\sin(\theta_i)\cos(\theta_{\bar i}),\;\; (i,\bar{i}) = (1,2)\; \text{or}\; (2,1)
\nonumber \\
C^X_\theta &=\cos(\theta_1)\cos(\theta_2)
,\,
C^{\pm}_\theta =\cos(\theta_1)\pm\cos(\theta_2)
,\,
S^X_\theta =\sin(\theta_1)\sin(\theta_2)
\label{eq9}
\end{align}
where $\theta_1$ and $\theta_2$ are the (different) canting angles made by the spins into and out of the plane of the lattice (see Fig.\,\ref{3Q-order}(a)).
The Hamiltonian equation Eq.\,\ref{eq4.3} is diagonalized by using Bogoliubov-Valatin transformation as described in section Sec.\,\ref{sec2.4}.
In the next section, the detailed classical ground state and corresponding \gls{magnon} bands and their topological properties are discussed.

\section{\label{sec4.3} Results and Discussion}
\subsection{\label{sec4.3.1} Flux State}

\begin{figure}[H]
	\centering
		\includegraphics[width=0.7\textwidth]{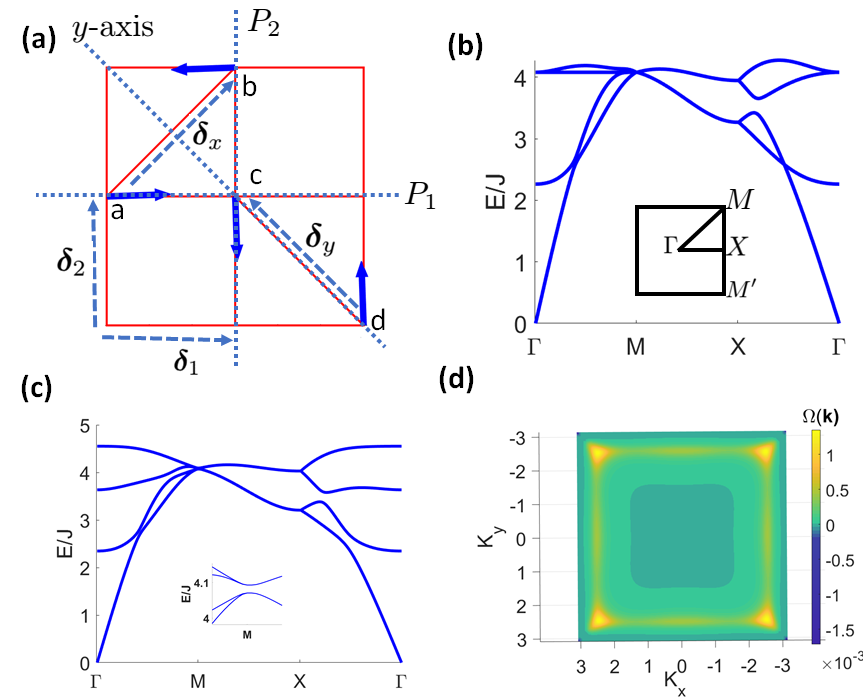}
	\caption{(a) The spin-configuration in flux state on the Shastry-Sutherland lattice. (b) The \gls{magnon} band structure for flux state in absence of magnetic field where $J=1.0, J'=1.1, D_\perp=0.8, B_z=0.0$. The inset of the figure shows the Brillouin zone and high symmetry lines. (c) The \gls{magnon} band structure in presence of magnetic field where $J=1.0, J'=1.1, D_\perp=0.8, B_z=0.5$. Inset of the figure shows the magnified \gls{magnon} band structure to show the lifting of four-fold degeneracy at M-point. (d) The non-abelian Berry curvature of lowest and second-lowest bands for flux state where $J=1.0, J'=1.1, D_\perp=0.8, B_z=0.5$. For convenience the negative Berry-curvature which is concentrated at M-point is divided by 100, to increase the visibility of the Berry-curvature distribution throughout the Brillouin zone.}
	\label{FluxState}
\end{figure}

The flux state is stabilized as the ground state when the component of the \gls{DMI}
normal to the plane of the lattice on the axial bonds, $D_\perp$, exceeds a critical magnitude (e.g. when $J\approx J'$, flux state is stable for $D_\perp\gtrapprox 0.6J$) and the in-plane components vanish for all \gls{DMI} (the \gls{DMI} on the diagonal bonds are constrained by symmetry requirements to be strictly in-plane).
The flux state is comprised of  the nearest-neighbour spins aligned perpendicular to each other and parallel
to the plane of the lattice (Fig.\,\ref{FluxState}(a)), which is energetically favored by the 
perpendicular DM-component $D_\perp$. The state is characterized by the spontaneous breaking of continuous U(1) spin-rotation symmetry about the z-axis. In the bosonic (\gls{magnon}) language, the ground state (flux state) is the vacuum and the \gls{HP} bosons represent quantized low energy excitations above this ground state. The \gls{magnon} bands at zero magnetic field are shown in 
the Fig.\,\ref{FluxState}(b). At the $\Gamma$ point the lowest band becomes gapless 
revealing the presence of Goldstone-mode associated with U(1) symmetry breaking. The bands along line-$\overline{\text{MX}}$ are 

\noindent twofold degenerate -- these can be understood in terms of Kramer's degeneracy\,\cite{StripeZigzag,Symmetry}. The operator $\hat{m}_2=\left\lbrace \tilde{M}_2 \tau|\boldsymbol{\delta}_2\right\rbrace$ commutes with the Hamiltonian where $\tilde{M}_2$ is the reflection operator along the $P_2$ axis
and $\boldsymbol{\delta}_2$ is the translation by have lattice parameter along the
same axis, as shown in Fig.\,\ref{FluxState}(a); $\tau$ is the time-reversal operator. But, $\hat{m}_2$ is not the symmetry operator for the classical ground state shown in the Fig.\,\ref{FluxState}(a). Instead,  symmetry operator for the ground state is given by $\hat{m}'_2=\left\lbrace \tilde{M}_2 e^{i\pi\hat{S}^y} \tau|\boldsymbol{\delta}_2\right\rbrace$, which contains an additional rotation of spin by $\pi$ about the $y-$axis. On the line-$\overline{\text{MX}}$ in the Brillouin zone, $(\hat{m}'_2)^2=e^{ik_x}=-1$. Hence $\hat{m}'_2$ is anti-unitary operator with a squared value of $-1$, which in turn, results in the Kramer's degeneracy. $\hat{m}'_2$ maps one Kramer's degenerate wavefunction along the  $\overline{\text{MX}}$-line to the other Kramer's degenerate wavefunction along the $\overline{\text{M}^\prime\text{X}}$-line. Further, the symmetry operation $\hat{c}_{2,1}=\left\lbrace \tilde{C}_{2,1} e^{i\pi\hat{S}^y} |\boldsymbol{\delta}_1\right\rbrace$, maps the Kramer's degenerate state from $\overline{\text{M}^\prime\text{X}}$ to $\overline{\text{MX}}$, where $\tilde{C}_{2,1}$ is the two-fold rotation about the $P_1$ axis and $\boldsymbol{\delta}_1$ as shown in the Fig.\,\ref{FluxState}(a). Thus, the band degeneracy along $\overline{\text{MX}}$-line is protected by symmetries $\hat{m}_2$ and $\hat{c}_{2,1}$. Additionally, there is four-fold degeneracy at the M-point due to the presence of the symmetry $\hat{m}_y=\left\lbrace \tilde{M}_y\tau|\bold{0}\right\rbrace$, where $\tilde{M}_y$ is the reflection operator about the $y-$axis as shown in Fig.\,\ref{FluxState}(a). The symmetry $\hat{m}_y$ 
maps one pair of Kramer's degenerate state to the other pair of Kramer's degenerate state at M-point. The Berry-curvature is not well defined for bands, because the bands are degenerate.

Application of magnetic field results in a ground state with a finite out of plane spin component, which breaks the $\hat{m}_y$-symmetry. The band structure in the presence of an external longitudinal magnetic field is shown in Fig.\,\ref{FluxState}(c). The four-fold degeneracy reduces to two fold degeneracy at the M-point, due to the breaking of the $\hat{m}_y$-symmetry, as shown in the inset of Fig.\,\ref{FluxState}(c). But the band sticking along $\overline{\text{MX}}$-line and Goldstone modes at $\Gamma$-point are preserved. 
The degeneracy of the bands prevents the calculation of the standard Berry curvature 
and Chern numbers of the bands, although there is no symmetry constraints to make 
Berry curvature zero. One can, however, calculate the non-abelian Berry-curvature and non-abelian Chern number of the degenerate bands following the procedure outlined in Ref.\,\cite{Hatsugai}(see section Sec.\,\ref{sec2.5.2} for details). The results for the non-abelian Berry curvature of the lowest and second-lowest band is shown in Fig.\,\ref{FluxState}(d). The negative part of Berry-curvature is highly concentrated at the M-point and the positive berry curvature is distributed in the remaining Brillouin zone. The inversion symmetry $\hat{i}$ of the system with an inversion center at point-c in Fig.\,\ref{FluxState}(a) is also reflected in the Berry-curvature. 
Our results show that the negative and positive contribution of the non-abelian Berry curvature cancels, yielding a vanishing non-abelian Chern number for the lower (or the upper) pair of bands. 

\subsection{\label{sec4.3.2} Canted flux state}
\subsubsection{\textbf{Topological \gls{magnon} bands and topological phase diagram}}
\begin{figure}[H]
	\centering
		\includegraphics[width=0.8\textwidth]{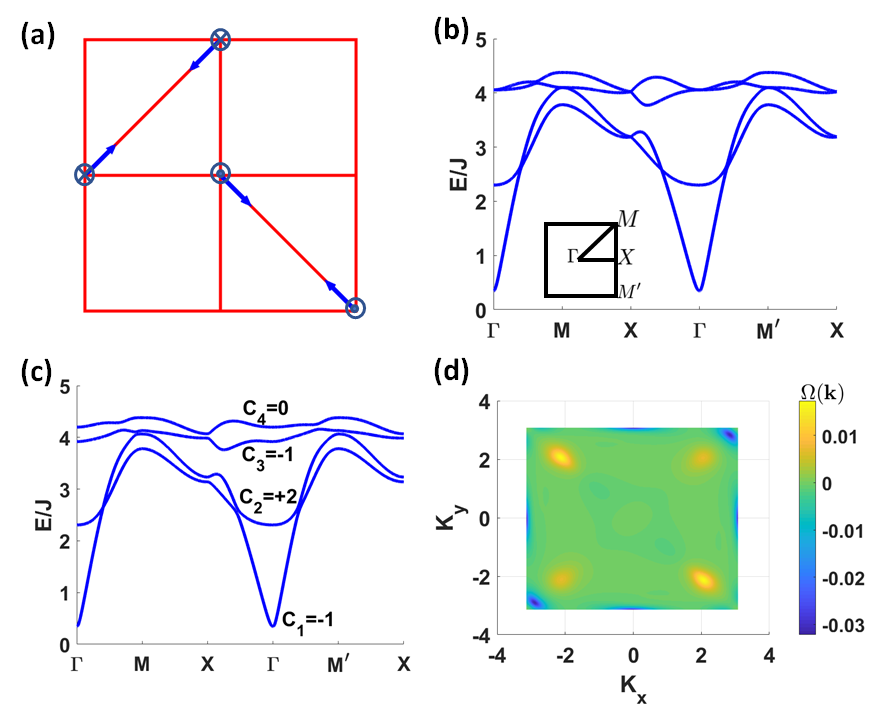}
	\caption{(a) The spin configuration in canted flux state. Circle with dot and the circle with cross represent spin component out of plane upward and downward directions respectively. (b) The \gls{magnon} band structure for canted flux state where $J=1.0, J'=1.1, D_\perp=0.8, D=0.2, D_{||,s}=0.05, D_{||,ns}=0.1, B_z=0.0$. The inset of the figure shows the Brillouin zone and high symmetry lines and high symmetry points. (c) The \gls{magnon} band structure for canted flux state where $J=1.0, J'=1.1, D_\perp=0.8, D=0.2, D_{||,s}=0.05, D_{||,ns}=0.1, B_z=0.3$. The Chern numbers of the bands are also shown in the Figure. (d) The Berry curvature of third \gls{magnon} Band for canted flux state where  $J=1.0, J'=1.1, D_\perp=0.8, D=0.2, D_{||,s}=0.05, D_{||,ns}=0.1, B_z=0.3$.}
	\label{3Q-order}
\end{figure}

When in-plane components of the DM-interactions (both along the axial as well as the diagonal bonds) are included, a canted-flux state, as shown in the Fig.\,\ref{3Q-order}(a), is
realized as the ground state, even in the absence of a magnetic field. The in-plane components of spins are directed along the diagonal bonds; additionally, the spins acquire an out-of-plane component. There is two-fold degeneracy in the ground state configuration -- one of them is shown in Fig.\,\ref{3Q-order}(a) with the spins pointing inwards along the diagonal bonds; the other degenerate state is obtained by flipping the spins so that they point outward along the diagonal bonds and the out-of-plane spin components are also flipped. The spins are canted away of the plane of lattice at an angle given by ,
\begin{equation}
\theta=\frac{\pi}{2}+\frac{1}{2}\tan^{-1}\left(\frac{4D+8(D_{||,s}-D_{||,ns})}{4J+8D_\perp-8J'}\right),
\end{equation}
where the spins along one diagonal are canted out of the plane while those along the 
other diagonal are canted into the plane. 

 The \gls{magnon} bands for the zero-field canted-flux state is shown in the Fig.\,\ref{3Q-order}(b), where the bands are observed to be degenerate at $\Gamma$, M and X-point in the Brillouin zone. These degeneracies are protected by symmetries as discussed here.
The symmetry protection of the band sticking at X-point in the Brillouin zone in the canted flux state without magnetic field is explained using Herring’s method\,\cite{Herring1,Herring2}.
The symmetry operators which keep the X-point in the Brillouin zone invariant or change it by a reciprocal lattice vector are,
\begin{align*}
\left\lbrace M_2\middle| \boldsymbol{\delta}_2\right\rbrace,\;\left\lbrace M_1\middle|\boldsymbol{\delta}_1 \right\rbrace,
\left\lbrace C_{2z}\middle|\boldsymbol{\delta}_{y}\right\rbrace,
\end{align*}  
  where,
  \begin{align*}
 M_2=\tilde{M}_2 e^{i\pi\hat{S}_z} e^{i\pi\hat{S}_1},\; M_1= \tilde{M}_1 e^{i\pi\hat{S}_z} e^{i\pi\hat{S}_2},
   \;C_{2z}=\tilde{C}_{2z} e^{i\pi\hat{S}_z}e^{i\pi\hat{S}_1}e^{i\pi\hat{S}_z}e^{i\pi\hat{S}_2}.
\end{align*}   
The symmetry operators $\tilde{M}_2$, $\tilde{M}_1$ and $\tilde{C}_{2z}$ are mirror reflection  along axis-$P_1$, reflection along axis-$P_2$ and twofold rotation around z-axis at the sublattice-c respectively(Fig.\,\ref{FluxState}(a)).
 The set of translational operators makes the invariant subgroup,
\begin{equation}
\pazocal{T}_\bold{k}=\left\lbrace E\middle| m\boldsymbol{\alpha}_1+n\boldsymbol{\alpha}_2\right\rbrace,\quad m\in \text{even},\; n\in \text{integer},
\end{equation}
where the translational operator follows the constraint $\exp(i\bold{k}\cdot\bold{t})=1$ at X-point($\bold{t}=\left(\frac{\pi}{a},0\right)$).
Then the factor group $\pazocal{G}_\bold{k}/\pazocal{T}_\bold{k}$ can be obtained by deriving the coset of invariant subgroup $\pazocal{T}_\bold{k}$, where $\pazocal{G}_\bold{k}$ is the set of non-translational symmetry operators which keeps the X-point invariant or change it by reciprocal lattice vector. The factor group $\pazocal{G}_\bold{k}/\pazocal{T}_\bold{k}$ is given by, 
\begin{align*}
&e=\squeezeN{E}{\mn}, \quad e'=\squeezeN{E}{\pq} \\
&m_2=\squeezeN{M_2}{\mn+\boldsymbol{\delta}_1}, \quad m'_2=\squeezeN{M_2}{\pq+\boldsymbol{\delta}_1}\\
&m_1=\squeezeN{M_1}{\mn+\boldsymbol{\delta}_2}, \quad m'_1=\squeezeN{M_1}{\pq+\boldsymbol{\delta}_2}\\
&c_2=\squeezeN{C_{2z}}{\mn+\boldsymbol{\delta}_x}, \quad c'_2=\squeezeN{C'_{2z}}{\pq+\boldsymbol{\delta}_x},
\end{align*} 
where $m\in \text{even}$, $p\in$ odd and $n,q\in$ integer.
Next, we have derived the character table using package named "GAP"\,\cite{GAP4}, where the code has been written below, with redefined symbols $e'\rightarrow e,m_2\rightarrow c$
\begin{lstlisting}[frame=single]
gap>f:=FreeGroup("e","c","g");;
gap> AssignGeneratorVariables(f);;
#I Assigned the global variables [e,s,g]
gap>  r:=ParseRelators([e,c,g],
>"e^2=c^2=1,g^2=e,es=se,ge=eg,cg=gce");;
gap> g:=f/r;;
gap> Size(g);
8
gap> LoadPackage("ctbllib");;
gap> tbl:=CharacterTable(g);;
gap>Display(tbl);;
**Displays The Character Table**
class:=ConjugacyClasses(tbl);
**Displays The Order of Conjugacy classes
 in the table head**
\end{lstlisting}
The derived character table is given by,
\begin{table}[H]
\centering
\caption{Character Table of unitary subgroup}
 \begin{tabular}{||c c c c c c||} 
 \hline
 & $e$ & $e'$ & $\squeezeA{m_2}{m'_2}$ & $\squeezeA{m_1}{m'_1}$ & $\squeezeA{c_2}{c'_2}$ \\ [0.5ex] 
 \hline\hline
$\Gamma_1$ & 1 &  1 &  1 &  1 &  1 \\ 
$\Gamma_2$ & 1 &  1 & -1 & -1 &  1 \\ 
$\Gamma_3$ & 1 &  1 &  1 & -1 & -1 \\ 
$\Gamma_4$ & 1 &  1 & -1 &  1 & -1 \\ 
$\Gamma_5$ & 2 & -2 & 0 & 0 & 0 \\ 
 \hline
\end{tabular}
\label{TableI}
\end{table}
The only valid representation is $\Gamma_5$, since the translational operator $e'=\squeezeN{E}{\pq}$ should follow the relation,
\begin{align*}
\exp(i\bold{k}\cdot\bold{t})=-\pazocal{I},
\end{align*}
where, $\pazocal{I}$ is the identity matrix with a dimension equals to the dimension of representation.
Because  the only valid representation is $\Gamma_5$ which is two dimensional, the bands are doubly degenerate at the X-point.
In the similar manner, it can be shown that the degeneracy at the $\Gamma$ and M-points are also symmetry protected.

In the presence of an external longitudinal magnetic field, the canting angles of the 
two pairs of spins are no longer identical. The energy of the canted flux state in a
magnetic field is given by,
\begin{align}
\frac{E_{cl}}{NS^2}=&J(cos(2\theta_1)+\cos(2\theta_2))+J'\cos(\theta_1)\cos(\theta_2)
+D(\sin(2\theta_1)-\sin(2\theta_2))\nonumber\\
&-8D_\perp\sin(\theta_1)\sin(\theta_2)
+4(D_{||,s}-D_{||,ns})\sin(\theta_2-\theta_1)
-\frac{2B}{S}(\cos(\theta_1)+\cos(\theta_2)),
\label{eq6}
\end{align}
where $\theta_1$ and $\theta_2$ are the (different) canting angles made by the spins into
and out of the plane of the lattice. The energy is minimized for $\theta_1\neq \theta_2$
-- the application of magnetic field not only lifts the non-symmorphic symmetries 
but also renders the magnetic symmetry group of the ground state trivial. As a consequence, all the four bands are gapped out, as shown in Fig.\,\ref{3Q-order}(c). The \gls{magnon} Hamiltonian in presence of all DM-interactions and magnetic field can be found in the equation Eq.\,\ref{eq4.3}.

\begin{figure}[H]
	\centering
		\includegraphics[width=0.8\textwidth]{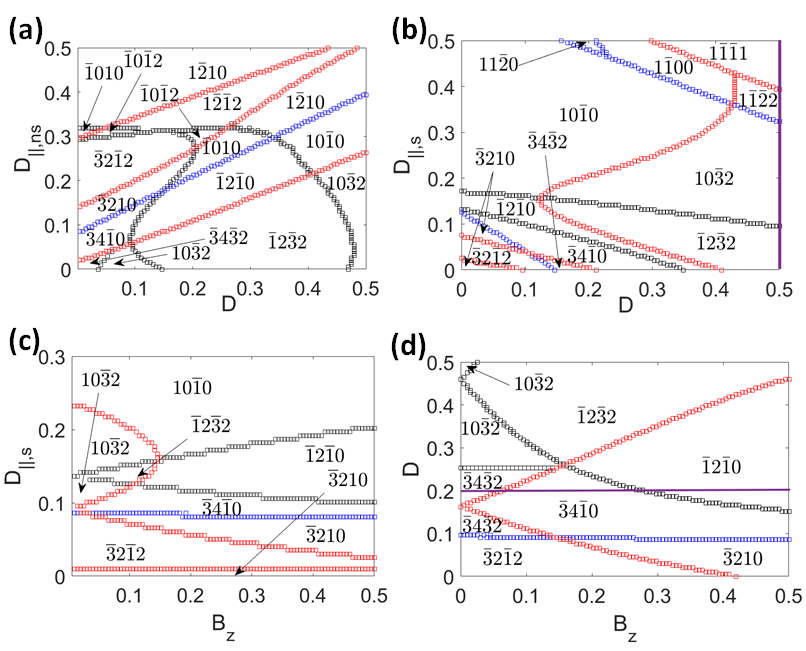}
	\caption{Topological transitions of \gls{magnon} bands in the parameter space (a) $D$ and $D_{||,ns}$, with $D_{||,s}=0.1, B_z=0.2$; (b) $D_{||,s}$ and $D$, with $D_{||,ns}=0.1, B_z=0.2$; (c)  $D_{||,s}$ and $B_z$, with $D_{||,ns}=0.1, D=0.05$; and (d) $D$ and $B_z$, with $D_{||,s}=0.05, D_{||,ns}=0.1$. For all the plots the remaining parameters are $J=1, J'=1.1, D_{\perp}=0.8$. The four digits from left to right denotes Chern-number from lower band to upper band (numbers with bar above them denote negative Chern-numbers). The band topological transitions are denoted by the lines red, blue and black boxed-lines. The red, blue and black denotes the band gap closing between upper, middle and lower pairs of bands respectively. The thermal conductivity is plotted in Fig.\,\ref{THE} along the purple lines in (c) and (d). The purple lines in (c) corresponds to $D=0.5$. The line in (d) corresponds to $D=0.2$ .  }
	\label{Phases}
\end{figure}

The lifting of the symmetry-protected degeneracies in the energy bands allows us to
calculate the Berry curvature and Chern number associated with each band separately
in the usual manner(see section Sec.\,\ref{sec2.5.2}). 
The results reveal that the energy bands acquire topological character for the
parameter set chosen in Fig.\,\ref{3Q-order}. A representative Berry-curvature 
distribution for the third band is shown in Fig.\,\ref{3Q-order}(d). The Berry
curvature is concentrated near the M and M' points of the Brillouin zone. The 
existence of the two-fold rotational symmetry at the center of the diagonal bond 
is reflected in the Berry-curvature. Unlike the flux state, the Berry curvature across 
the entire Brillouin zone do not cancel and this results in a non-zero Chern number
of the four bands(Fig.\,\ref{3Q-order}(c))(see section Sec.\,\ref{sec2.5.2} for details).

The interplay between competing Heisenberg and DM interactions, together with
geometric frustration and external magnetic field results in
topologically ordered energy eigenstates. As the relative strengths of the 
different competing interactions are varied, the energy levels shift and the bands
cross / touch in pairs at different points in the Brillouin zone. The accompanying 
phase transitions are topological in nature as they are characterized by the change 
in Chern number of the pair of bands involved.  
By identifying the state of the system with the band topology\,\cite{NonCoplanar1}, we find a wide variety 
of topological \gls{magnon} bands in different parameter regimes, shown in Fig.\,\ref{Phases}. The four numbers in the figure represent the four Chern numbers from lower to upper \gls{magnon} bands. The bar above the number denotes the negative Chern number. The color of the boundaries identify the pair of bands involved in the transition. Most strikingly, tuning the strength of the different components of the \gls{DMI} 
and applied magnetic field over a small range result in multitude of topologically distinct
set of single \gls{magnon} bands, even though the ground state remains the unaltered (canted flux state). While
this is driven by the non-coplanarity of the ground state spin configuration, the exact
mechanism of the change in topology of the \gls{magnon} bands, or their robustness against interaction effects is not clear\,\cite{NonCoplanar1}.


Transition in band topology occurs due to gap reopening after closing at the the high symmetry points $\Gamma$, X, M and points along line $\overline{\Gamma\text{M}}$ and $\overline{\Gamma\text{M}^\prime}$. Except for the $\Gamma$-point, all other k-points in the Brillouin zone can be mapped into another k-point using two-fold rotational symmetry. Thus, the Chern number of the bands changes by $\pm 1$, if band touching happens at the $\Gamma$-point. It is noticeable that this kind of transition happen in the upper right region of Fig.\,\ref{Phases}(b).  Otherwise, the the Chern numbers changes by $\pm 2$, because accidental band touching take places at two points in the Brillouin zone due to two-fold rotational symmetry of the system. Most of the topological transition is associated with change in Chern number $\pm 2$ in Fig.\,\ref{Phases}.

\subsubsection{\textbf{Thermal Hall conductance and its derivative}}

\begin{figure}[H]
	\centering
		\includegraphics[width=0.8\textwidth]{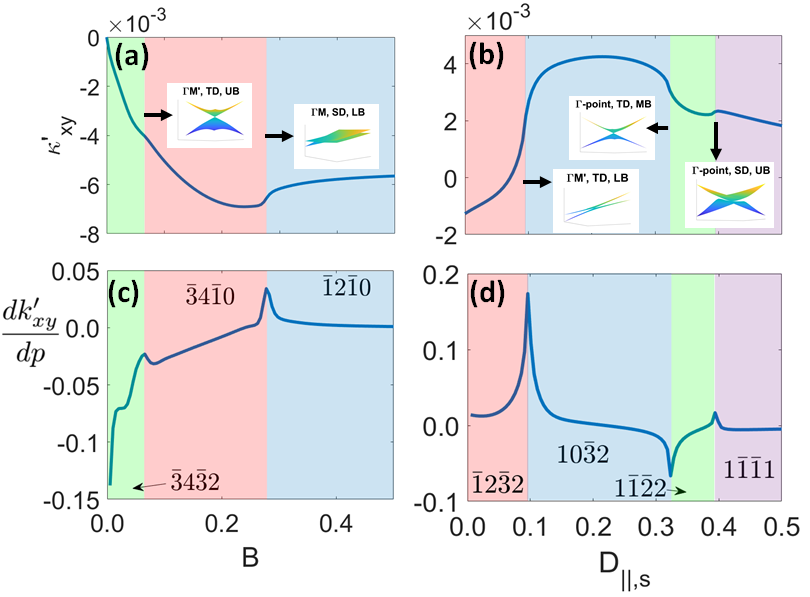}
	\caption{(a)-(b) The scaled thermal conductivity as a function of Magnetic field $B$ and DM-term $D_{||,s}$ along the purple lines in Fig.\,\ref{Phases}(d) and Fig.\,\ref{Phases}(b) respectively. Inset of the figure shows the band touching points at the boundaries of the band topological region. The first, second and third letters denote the band touching point, type of band touching, and the energy wise ordering of bands.For example, "$\Gamma$M', TD, UB", band touches at $\Gamma$M' line , the type of band touching is tilted Dirac-type and pair of upper bands touching respectively. "SD" means the Semi-Dirac point and "LB" means pair of lower bands touching and "MB" means pair of middle bands touching etc. The magnetic field is a scaled quantity and connected with experimental magnetic field as $h_z=\frac{BJ}{g\mu_B}$, where $g$ is Lande-g factor and $\mu_B$ is the Bohr magneton. (c)-(d) The derivative of the thermal Hall conductivity w.r.t. magnetic field and DM-term $D_{||,s}$  as a function of Magnetic field $B$ and DM-term $D_{||,s}$ respectively. The numbers in the  figures are Chern numbers with convention given in Fig.\,\ref{Phases}.}
	\label{THE}
\end{figure}

The non-trivial topology of \gls{magnon} bands give rise to thermal Hall effect in the magnetic system.
The expression of scaled thermal Hall conductivity is given by\,\cite{THE5,THE6},
\begin{equation}
\kappa'_{xy}=\frac{\kappa_{xy}\hbar}{k_B}=\frac{T'}{(2\pi)^2}\sum_n \int_{BZ} c_2(\rho_{n,\bk}) \Omega^n_{xy}(\bk) d^2k,
\end{equation}
where $\kappa_{xy}$ is the thermal Hall conductivity, $T'$ is the scaled temperature, $T'=k_B T$, and $\rho_{n,\bk}=1/(\exp(\epsilon_n(\bk)/T')-1)$ is the Bose-Einstein distribution function with $\epsilon_n(\bk)$ as the energy of the $n$-th \gls{magnon} band at $\bk$-point in Brillouin zone, $\Omega^n_{xy}(\bk)$ is the Berry-curvature of the n-th band at the $\bk$-point and $c_2(x)=(1+x)\left(\log \frac{1+x}{x}\right)^2-(\log(x))^2-2\text{Li}_2(-x)$ where $\text{Li}_2(x)$ is a polylogarithmic function. The scaled temperature $T'$, \gls{magnon} energies $\epsilon_n(\bk)$ as well as scaled-thermal Hall conductivity are normalized in unit of $JS$. The calculation of Berry-curvature $\Omega^n_{xy}(\bk)$ is further discussed in section Sec.\,\ref{sec2.5.2}. In Fig.\,\ref{THE}(a) and\,\ref{THE}(b), the results for the scaled thermal Hall conductivity is plotted as a function of Magnetic field $B_z$ and $D_{||,s}$ respectively, along the purple lines in Fig.\,\ref{Phases}(b) and Fig.\,\ref{Phases}(d). The different coloured regions in Fig\,\ref{THE}(a) and\,\ref{THE}(b) denote distinct topological regions along the purple lines in Fig.\,\ref{Phases}(d) and Fig.\,\ref{Phases}(b) respectively. At the boundary of the different topological regions the band gap closes as shown in the inset of the figures. Generically, band closing occurs at the Dirac point, but sometime a semi-Dirac point is encountered. The type of semi-Dirac point at the boundary between green and purple topological regions in Fig.\,\ref{THE}(b) is also reported in the reference Ref.\,\cite{SemiDirac}.

Figs.\,\ref{THE}(c) and\,\ref{THE}(d) present the derivative of scaled thermal Hall conductivity  with respect to magnetic field $B$ and \gls{DMI} $D_{||,s}$ as a function of $B$ and the  
symmetric component of the in-plane \gls{DMI} $D_{||,s}$ respectively. At the boundary between two distinct topological phases, the derivative in the thermal Hall conductance has a logarithmic divergence. The origin of the divergence on the basis of Weyl-point also discussed in Ref.\,\cite{triangular}. It is observed from the figures that the logarithmic divergence is universal and independent of the type of band touching. We also have been shown analytically in the next subsections  that the nature of divergence is same for tilted Dirac point and Semi-Dirac point. Furthermore, it also can be seen from the figure that the peak height of divergence grows faster, if the band touching happens at the lower pair of bands,
due to the larger contribution to the thermal Hall conductivity from \gls{magnon}s in the lower bands. 
Finally, the sign of the divergence is positive (negative) if Chern number of lower band increases
(decreases) at the band topological transition.
In the next subsections we show that the nature of logarithmic divergence remain unchanged for different types of band-touching point.

\paragraph{Tilted Dirac point or Generalized Weyl point:}
 \label{TiltedDirac}
 The Hamiltonian corresponds to the tilted Dirac point or generalized Weyl point is given by,
\begin{equation}
\hat{H}=w^0_x k_x \mathbb{I}+w^0_y k_y \mathbb{I}+w_x k_x \sigma_x + w_y k_y \sigma_y + p \sigma_z +\epsilon\mathbb{I},
\label{Hamiltonian1}
\end{equation} 
where $k_x$, $k_y$ are the momentum with respect to the band touching point. $\sigma_i$ ($i=x,y,z$) are the Pauli's matrices. $p$ is a perturbation to open the gap. $\epsilon$ denotes the energy of band touching point. The Hamiltonian is general Hamiltonian for any linear dispersions.
The energies correspond to the Hamiltonian,
\begin{align}
E^{\pm}(\bk)=\epsilon+k_x w^0_x +k_y w^0_y \pm \sqrt{p^2+w^2_x k^2_x+w^2_y k^2_y}
=\epsilon_{0\bk} \pm \omega_{\bk},
\label{Energy1}
\end{align}  
where $\epsilon_{0\bk}=\epsilon_0+k_x w^0_x +k_y w^0_y $ and $\omega_{\bk}=\sqrt{p^2+w^2_x k^2_x+w^2_y k^2_y}$
 the corresponding eigenvectors are,
 \begin{equation}
 x^\pm=\frac{1}{N_{\pm}}\begin{pmatrix}
 \frac{p\pm \sqrt{p^2+w^2_x k^2_x+w^2_y k^2_y}}{w_x k_x+ i w_y k_y} &
 1
 \end{pmatrix}^T,
 \label{eigenvector1}
 \end{equation} 
 where $N_{\pm}$ are the normalization constant.  
 The expression of the Berry curvature of the lower band is given by,
 \begin{align}
 \Omega_1 &=\frac{1}{(E^+-E^-)^2}\text{Im}\left(\squeezeD{x_1}{\dXdY{\pazocal{H}}{k_x}}{x_2}\squeezeD{x_2}{\dXdY{\pazocal{H}}{k_y}}{x_1}\right),
 \label{BerryCurvature1}
 \end{align}
 where Im$(z)$ denotes imaginary part of $z$. 
 Using Eq.\,\ref{Hamiltonian1}, Eq.\,\ref{Energy1} and Eq.\,\ref{eigenvector1},
 \begin{equation}
 \squeezeD{x_1}{\dXdY{\pazocal{H}}{k_x}}{x_2}=\frac{2w_x}{\sqrt{N_1 N_2}} \left[ \frac{w_x k_x p-i w_y k_y \sqrt{p^2 w^2_x k^2_x+w^2_y k^2_y}}{w^2_x k^2_x+w^2_y k^2_y}\right]
 \label{auxilary1a}
 \end{equation}
Similarly,
 \begin{equation}
	 \squeezeD{x_2}{\dXdY{\pazocal{H}}{k_y}}{x_1}=\frac{2w_y}{\sqrt{N_1 N_2}} \left[ \frac{w_y k_y p-i w_x k_x \sqrt{p^2 w^2_x k^2_x+w^2_y k^2_y}}{w^2_x k^2_x+w^2_y k^2_y}\right] 
	 \label{auxilary1b}
 \end{equation} 
 Using Eq.\,\ref{auxilary1a} and Eq.\,\ref{auxilary1b} in Eq.\,\ref{BerryCurvature1}, the Berry curvature of the lower band,
 \begin{equation}
 \Omega_1=-\frac{2w_x w_y p}{(p^2+w^2_x k^2_x + w^2_y k^2_y)^{3/2}}
 \label{eq4.15}
 \end{equation}
 Similarly, for the upper band $\Omega_2=-\Omega_1$.
The Thermal Hall conductivity expression is given by,
\begin{equation}
\kappa_{xy}=\frac{k_B^2 T}{(2\pi)^2 \hbar} \sum_n \int_{BZ} c_2(\rho_{n\bk}) \Omega_n(\bk) d^2k, 
\label{ThermalHallEffect1}
\end{equation}
Using equations Eq.\,\ref{eq4.15} and $\Omega_2=-\Omega_1$, we get,
\begin{align}
\sum_n \Omega_n(\bk) c_2(\rho_{n\bk})&=\frac{2w_x w_y p}{(p^2+w^2_x k^2_x + w^2_y k^2_y)^{3/2}} \left[ c_2(\rho_{2\bk})-c_2(\rho_{1\bk})\right] \nonumber\\
&\propto \frac{2w_x w_y p}{(p^2+w^2_x k^2_x + w^2_y k^2_y)^{3/2}} \omega_\bk\quad \nonumber \\  \left[ where,c_2(\rho(\epsilon_{0\bk}+\right. & \left.\omega_{\bk}))-c_2(\rho(\epsilon_{0\bk}-\omega_{\bk}))\propto \omega_\bk \text{for small $p$ and $\bk$ } \right] \nonumber\\
&\propto \frac{p}{(p^2+w^2_x k^2_x + w^2_y k^2_y)}
\end{align}
 From Eq.\,\ref{ThermalHallEffect1}, integrating around the band touching points $k=\sqrt{w^2_x k^2_x + w^2_y k^2_y}<k_c$, we get,
 \begin{align}
 \dXdY{\kappa_{xy}}{p} &\propto \ddY{p}\int_{k<k_c}  \frac{p}{(p^2+w^2_x k^2_x + w^2_y k^2_y)} d^2k  \nonumber \\
 &\propto \int_{k<k_c} \frac{2\pi k dk}{(p^2+k^2)} - \int_{k<k_c} \frac{4p^2\pi k dk}{(p^2+k^2)^2}\nonumber \\
 &\propto \pi \ln\left(\frac{p^2+k^2_c}{p^2}\right) + 2p^2\pi  \left[\frac{1}{p^2+k^2_c}-\frac{1}{p^2}\right]\nonumber\\
 &\propto \ln(|p|) \quad \left[\text{Near phase transtion $k_c\gg p$}\right]
 \end{align}
 \paragraph{Semi-Dirac point:}
The dispersion of semi-Dirac point,
\begin{equation}
E^{\pm}(\bk)=\pm \sqrt{(w^2_x k^2_x)^2+w^2_yq^2_y}
\label{Energy2}
\end{equation}
  Out of many possibilities, two different possible Hamiltonians are,
 \begin{equation}
 \pazocal{H}_1=\begin{pmatrix}
 0 & w_x q_x  +i w_y q_y \\
 w_x q_x-i  w_y q_y  & 0
 \end{pmatrix}
 ,\,\,   
 \pazocal{H}_2=\begin{pmatrix}
w_x q_x & i w_y q_y\\
-i w_y q_y & -w_x q_x 
 \end{pmatrix},
 \label{Hamiltonian2}
 \end{equation} 
The Berry curvatures of the both Hamiltonian is given by,
\begin{equation}
\Omega^\pm_1=\pm \frac{4pq_x w_x w_y}{(p^2+w^4_x q^4_x +w^2_y q^2_y)^{3/2}}, \Omega^\pm_2=0,
\end{equation}  
 Thus, the second Hamiltonian in Eq.\,\ref{Hamiltonian2} does not produce any Berry curvature and so topological phase transition can not happen with this kind of Hamiltonian. It can be shown for the first Hamiltonian in equation Eq.\,\ref{Hamiltonian2},
\begin{equation}
 \dXdY{\kappa_{xy}}{p} \propto \ln(|p|)
 \end{equation}  
Particularly, in our study, we got semi-Dirac points with dispersion,
 \begin{equation}
 E^\pm(\bk)=w_0^2 k_y^2\pm \sqrt{w_x^2 k_x^2 + w_y^2 k_y^4}
 \end{equation}
Again particularly for phase transition point, near $D_{||,s}=0.4$ in Fig.\,\ref{THE}(d) is associated with Semi-Dirac dispersion with $w_0=w_y$, which is also encountered in reference Ref.\,\cite{SemiDirac}.

\begin{figure}[H]
	\centering
		\includegraphics[width=0.8\textwidth]{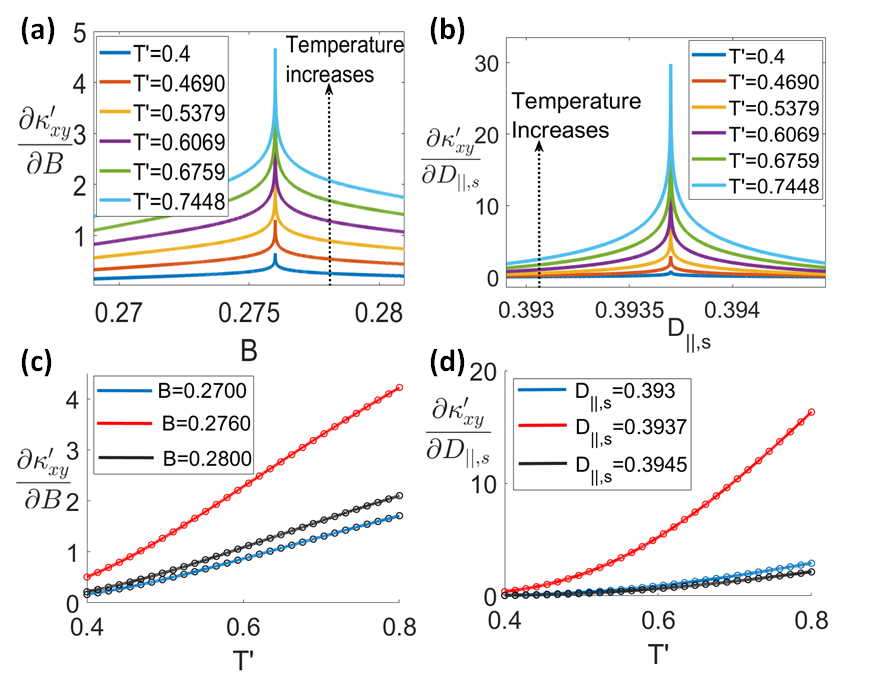}
	\caption{(a)-(b) The derivative of conductivity as a function of parameters $B$ and $D_{||,s}$ respectively, for different temperature. The method of plotting is described in the main text. (c)-(d) The derivative of conductance as a function of temperature is fitted using Eq.\,\ref{TemperatureDependenceEq}, at different $B$ and $D_{||,s}$ respectively. The circled points correspond to the calculated derivative of thermal Hall conductivity and the straight line represents the fitted curve. The fitting parameters $\left\lbrace A, \epsilon_0\right\rbrace$ for blue, red and black curves in (c) are $\left\lbrace 4.6886, 2.9243\right\rbrace$, $\left\lbrace 10.4323, 2.7539\right\rbrace$ and $\left\lbrace 5.5643, 2.8666\right\rbrace$. The fitting parameters $\left\lbrace A, \epsilon_0\right\rbrace$ for blue, red and black curves in (d) are $\left\lbrace 20.0830, 4.2135\right\rbrace$, $\left\lbrace 103.5474, 4.1025\right\rbrace$ and $\left\lbrace 15.4258, 4.2656\right\rbrace$.}
	\label{TemperatureDependenceFig}
\end{figure}

Here we show that that the logarithmic divergence follows simple analytical expression as a function of temperature.
The main contribution of derivative of thermal Hall conductance comes from the band touching point at energy $\epsilon_0$. Near this point at phase transition the Berry curvature is equal and opposite for the two bands$(\Omega^1_{xy}=-\Omega^2_{xy})$. Near the band touching point,

\begin{align}
\frac{\partial \kappa'_{xy}}{\partial p} &=\frac{T'}{2\pi^2}\frac{\partial}{\partial p} \left[ \Omega^1_{xy}(\bk) \left( c_2(\rho(\epsilon_0+\epsilon_{\bk}))-c_2(\rho(\epsilon_0-\epsilon_{\bk}))\right)\right]\nonumber\\
&=\frac{T'}{2\pi^2}\frac{\partial}{\partial p}\left[ \Omega^1_{xy}(\bk)\;\left( 2\left.\frac{dc_2}{d\rho}\right|_{\rho_0} \left.\frac{d\rho}{d\epsilon}\right|_{\epsilon_0}\right)\epsilon_{\bk}\right]\nonumber\\
&=A \exp(\frac{\epsilon_0}{T'}) \left[\ln\left(\frac{1+\rho_0}{\rho_0}\right)\right]^2\rho_0^2,\;
\label{TemperatureDependenceEq}
\end{align}
where, $\rho_0=1/(\exp(\frac{\epsilon_0}{T'})-1)$.
 $A$ is a constant independent of temperature and proportional to $\log(p-p_0)$, $p$ is a parameter of system(e.g. magnetic field etc.) and $p_0$ is the band topological transition point.
  Moreover, at a temperature, lower compared with the energy of band touching point, the equation\,\ref{TemperatureDependenceEq} transforms into,
\begin{equation}
\frac{\partial \kappa'_{xy}}{\partial p}=A\epsilon_0^2\exp(\frac{-\epsilon_0}{T}).
\end{equation}

To demonstrate the validity of Eq.\,\ref{TemperatureDependenceEq}, we have chosen the band topological transition points near $B=0.276$ and $D_{||,s}=0.3937$ of Fig.\,\ref{THE}(c) and Fig.\,\ref{THE}(d) respectively. The numerically calculation Berry-curvature around the transition point is computationally expensive and in-accurate. So, to correctly calculate the derivative of thermal Hall-conductivity, first the thermal Hall conductivity has been calculated near the transition point and fitted using the expression,
\begin{equation}
\kappa'_{xy}=m\ln(|p-p_0|)+m_0+m_1(p-p_0)+m_2(p-p_0)^2+m_3(p-p_0)^3,
\end{equation}
where $p_0$ is the band topological transition point.
 Then the derivative of the expression has been plotted and shown in Fig.\,\ref{TemperatureDependenceFig}(a) and Fig.\,\ref{TemperatureDependenceFig}(b). The divergent peak and nearby points increase with the temperature. In Fig.\,\ref{TemperatureDependenceFig}(c) and\,\ref{TemperatureDependenceFig}(d),  the derivative of conductivity is plotted and fitted as a function of temperature using the Eq.\,\ref{TemperatureDependenceEq}, considering $A$ and $\epsilon_0$ as fitting parameter. The band touches at $\epsilon_0=2.65$ for the band topological transition at $B=0.276$ in Fig.\,\ref{TemperatureDependenceFig}(a) and at $\epsilon_0=4.079$ for the band topological transition at $D_{||,s}=0.3937$ in Fig.\,\ref{TemperatureDependenceFig}(b). The values of fitting parameter $\epsilon_0$ described in Fig.\,\ref{TemperatureDependenceFig}(c) and Fig.\,\ref{TemperatureDependenceFig}(d) is quite near the band touching points.

\section{\label{sec4.4}Conclusion}
In conclusion we consider the flux state of \gls{ss} lattice and showed that in presence of in plane \gls{DMI} and magnetic field, the system gives rise to non-trivial topological \gls{magnon} bands. The canted flux state is a non-coplanar spin structure. This leads to a various topologically distinct \gls{magnon} band structure. Again, we observed the nature of first derivative of thermal Hall conductance is logarithmic divergent at the band topological transition, independent of the type of band touching. We have presented a simple temperature dependent parametric relation for the derivative of thermal Hall conductance, which might be useful to extract the energy of band touching during band topological transition for any generic spin Hamiltonian. We have also suggested an experimental realization of the model studied. In the present work, we have assumed a dilute gas of \gls{magnon}s without any interaction. At finite temperatures, as the density of thermally excited \gls{magnon}s increase, effects of interaction gain importance. Interaction between \gls{magnon}s further re-normalizes the bands and impart a finite life-time, which in turn can change the topological phase diagram obtained in this study. The study of the topological \gls{magnon} bands for this model in presence of interaction is planned for the future.

	
		
	\chapter{Weyl-triplons in \ce{SrCu2(BO3)2}}\label{chapter05}
	
		\epigraph{The results in this chapter is on the arXiv repository as D. Bhowmick and P. Sengupta. Weyl-triplons in \ce{SrCu2(BO3)2}. arXiv identifier: arXiv:2004.11551. (under review)}
	
	\section{\label{sec5.1}Introduction}
The successful detection of Weyl fermions in \ce{TaAs}\,\cite{TaAs1}, following 
theoretical prediction of the same\,\cite{TaAs_Theory1,TaAs_Theory2}, marks one of the latest 
milestones in the study of topological phases of matter, currently
the most active frontier in Condensed Matter Physics\,\cite{WSM1,WSM2,WSM3,WSM4,WSM5,WSM6,WSM7,TaAs2,WSM11}. Weyl fermions
are massless, linearly dispersing quasiparticles with finite chirality,
first proposed as solutions to massless Dirac equation in relativistic
particle physics\,\cite{Weyl}.  Pairs of Weyl fermions with opposite chirality may 
combine to form Dirac fermion. In condensed 
matter systems, non-relativistic analog of Weyl quasiparticles emerge at 
linear crossing of non-degenerate, topologically protected bands 
in three dimensional reciprocal space. Interest in these special band
crossings have increased since they act as sources of Berry flux and
impart topological character to the associated energy bands. 
Weyl nodes appear in pairs with opposite chirality and can be separated in
momentum space in systems with broken time reversal\,\cite{WSM4,WSM5,WSM6,WSM7} or inversion symmetry\,\cite{TaAs2,WSM11} or both\,\cite{WSM12}.

The appearance of Weyl points is governed by the geometry of the band structure
and symmetries of the Hamiltonian and lattice. As such, it is possible to
observe bosonic analogs of Weyl points. This has already been achieved in
artificially designed photonic\,\cite{photonics1,photonics2,photonics3,photonics4} and phononic crystals\,\cite{phonon1,phonon2}, and proposed for
magnons\,\cite{WeylPyrochlore4,WeylPyrochlore1,WM3,WM4,WM5,WM6}. 
 Weyl-points with toplogical charges $\pm 2$ are found in the phonon spectra\,\cite{DoublePhonon1,DoublePhonon2} and excitation spectra in phononic\,\cite{DoublePhononic1,DoublePhononic2} and photonic crystals\,\cite{DoublePhotonic1,DoublePhotonic2} which have no counterpart in high-energy physics. 
But no such unconventional Weyl-points have been reported in electronic or magnonic systems.
Our results reveal that magnetic excitations in the real quantum magnet \ce{SrCu2(BO3)2} is a promising platform to realise this unique doubly charged Weyl-points in magnetic excitations for the first time.

Quantum magnets are particularly promising since they have for long
been a versatile platform to realise complex quantum states of matter including
bosonic analogs of novel fermionic phases. The wide range of available
quantum magnets with different lattice geometries and the ability to tune
their properties readily by external magnetic field make them ideal testbed
for realising bosonic analogs of topological states of matter\,\cite{CoplanarIntro1,TopologicalMagnon3,ferroIntro,new2,TopologicalMagnon6,Triplon2D_1,StripeZigzag,Kane_Mele_Haldane,new4,new5,MagnonPolaron1,MagnonPolaron2,MagnonPolaron3,MagnonPolaron4,WeylPyrochlore4,WeylPyrochlore1,WM3,WM4,WM5,WM6}. However,
despite theoretical predictions, experimental observation of Weyl magnons
have remained elusive. In this work,
we present evidence for the existence of Weyl triplons in the geometrically
frustrated Shastry Sutherland compound, \ce{SrCu2(BO3)2}. In contrast to
previous studies that considered idealized Hamiltonians based on families
of quantum magnets\,\cite{WeylPyrochlore4,WeylPyrochlore1,WM3,WM5}, we focus on a realistic microscopic Hamiltonian of the
extensively studied geometrically frustrated quantum magnet, \ce{SrCu2(BO3)2}\,\cite{Romhanyi2,triplon2,triplon3}.
In this work, we have used experimentally determined Hamiltonian parameters\,\cite{ESR} 
for the microscopic model that have been demonstrated to reproduce faithfully 
the experimentally observed behavior of the material\,\cite{triplon2}.

\section{\label{sec5.2} Results}
\subsection{\label{sec5.2.1}Microscopic model.}

\begin{figure}[H]
\centering
\includegraphics[width=0.8\textwidth]{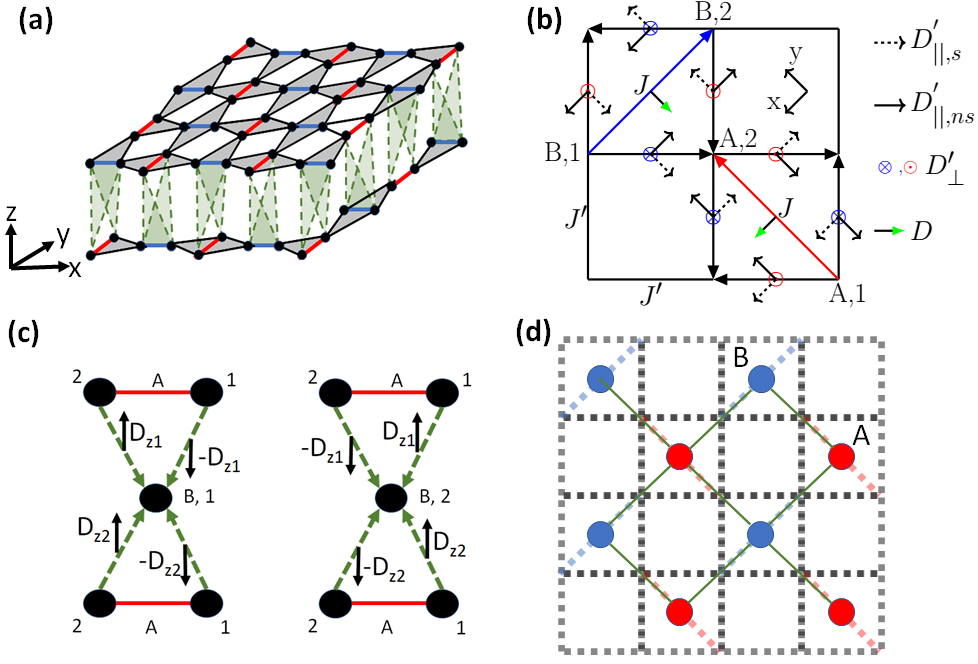} 
\caption{(a) The 3D-schematic lattice structure of compound \ce{SrCu2(BO3)2}. The red-bonds are dimer-A and blue-bonds are dimer-B. The green dotted-lines are the inter-layer bonds. (b) The intralayer Heisenberg and DM interactions. (c) The interlayer DM-interactions. (d) The effective square lattice structure after bond-operator transformation.}
\label{lattice}
\end{figure}

The figure Fig.\,\ref{lattice}(a) illustrates the three dimensional arrangements of \ce{Cu}-atoms of \ce{SrCu2(BO3)2} as a coupled layers of Shastry-Sutherland lattice. The Hamiltonian of the system is given by,
\begin{align}
\pazocal{H}=&J\sum_{\left\langle i,j \right\rangle,l} \bold{S}_{i,l}\cdot\bold{S}_{j,l} + J' \sum_{\left\langle\left\langle i,j \right\rangle\right\rangle} \bold{S}_{i,l}\cdot\bold{S}_{j,l}
  + \bold{D}\cdot\sum_{\left\langle i,j \right\rangle,l} \left(\bold{S}_{i,l}\times\bold{S}_{j,l}\right)+ \bold{D'}\cdot\sum_{\left\langle\left\langle i,j \right\rangle\right\rangle,l} \left(\bold{S}_{i,l}\times\bold{S}_{j,l}\right) \nonumber \\
  &-B\sum_{i,l}\bold{S_{i,l}^z}+J_z\sum_{\substack{i,j,\\ \left\langle l,l'\right\rangle}} \bold{S}_{i,l}\cdot\bold{S}_{j,l'}+\bold{D_z}\cdot\sum_{\substack{i,j ,\\ \left\langle l,l'\right\rangle}} \left(\bold{S}_{i,l}\times\bold{S}_{j,l'}\right),
  \label{eq1Ch5}
\end{align}
where, 
$\left\langle i,j \right\rangle$ and $\left\langle\left\langle i,j\right\rangle\right\rangle$ denote the summation over the sites belonging to intra-dimer and inter-dimer, bonds respectively in each layer and
$\langle l,l'\rangle$ denotes pairs of adjacent layers. 
The first four terms describe the intra-layer coupling terms and are depicted in 
Fig.\,\ref{lattice}(b), where $J$ and $J'$ are the intra-dimer and inter-dimer 
Heisenberg terms. $\bold{D}$ and  $\bold{D}'$ denote the intra-dimer and 
inter-dimer Dzyaloshinskii–Moriya (DM-interaction or DMI).  Finally, $J_z$ and $D_z$ are the inter-layer Heisenberg terms and DM-interactions along the green dotted bonds in Fig.\,\ref{lattice}(a). The fifth term is a Zeeman coupling of the spins 
with a magnetic field perpendicular to the Shastry-Sutherland layer. We
include DM-interactions that are symmetry allowed for \ce{SrCu2(BO3)2} at 
temperatures below $395$K\,\cite{structure1,structure2,Romhanyi2} in its low-symmetry
phase. In-plane components of the inter-layer DM-interaction is neglected,
even though it is allowed by the symmetry of the lattice, since it does not 
contribute to the low energy physics of the magnetic system. The 2D Hamiltonian describing the magnetic properties of each layer have been
extensively studied in the past and the

\noindent nature of triplon excitations above dimerized ground state and their topological
characters delineated using the bond operator formalism\,\cite{Romhanyi2,triplon2,triplon3}. We study the system with additional physically reliable interlayer Heisenberg and DM-interaction terms included in the Hamiltonian. The interlayer DM-interaction shown in Fig.\,\ref{lattice}(c), is taken in the z-direction and also symmetry allowed by the low temperature crystal-symmetry of \ce{SrCu2(BO3)2}. For simplicity, we assume $D_{z2}\approx D_{z1}=D_z$. The presence of interlayer DM-interaction drives a variety of topological phases in the system.

\subsection{\label{sec5.2.2}Triplon picture.}
The ground state is a product of singlet-dimer states on the red and blue-bonds(Fig.\,\ref{lattice}(a)) in a pure Shastry-Sutherland lattice. A singlet state is an entangled state of two spins such that the wave-function is defined by $\ket{s}=\left(\ket{\uparrow\downarrow}-\ket{\downarrow\uparrow}\right)/\sqrt{2}$. The lowest excitations above the ground states are direct product of triplets, which are $\ket{t_x}=i\left(\ket{\uparrow\uparrow}-\ket{\downarrow\downarrow}\right)/\sqrt{2}$, $\ket{t_y}=\left(\ket{\uparrow\uparrow}+\ket{\downarrow\downarrow}\right)/\sqrt{2}$, $\ket{t_z}=-i\left(\ket{\uparrow\downarrow}+\ket{\downarrow\uparrow}\right)/\sqrt{2}$. In presence of the small perturbative DM-interaction the  ground state is the product state of the rotated-singlets $\ket{\tilde{s}_A}$ and $\ket{\tilde{s}_B}$. The rotated singlets and rotated triplets are related to the original singlet and triplets, as written below,
\begin{equation}
\begin{pmatrix}
\ket{\tilde{s}_A} \\
\ket{\tilde{t}^x_A} \\
\ket{\tilde{t}^y_A} \\
\ket{\tilde{t}^z_A} \\
\end{pmatrix}=
\begin{pmatrix}
1 & -\alpha & 0 & 0\\
\alpha & 1 & 0 & 0\\
0 & 0 & 1 & 0\\
0 & 0 & 0 & 1
\end{pmatrix}
\begin{pmatrix}
\ket{s_A} \\
\ket{t^x_A} \\
\ket{t^y_A} \\
\ket{t^z_A} \\
\end{pmatrix}
,\,
\begin{pmatrix}
\ket{\tilde{s}_B} \\
\ket{\tilde{t}^x_B} \\
\ket{\tilde{t}^y_B} \\
\ket{\tilde{t}^z_B} \\
\end{pmatrix}=
\begin{pmatrix}
1 & 0 & \alpha & 0\\
0 & 1 & 0 & 0\\
-\alpha & 0 & 1 & 0\\
0 & 0 & 0 & 1
\end{pmatrix}
\begin{pmatrix}
\ket{s_B} \\
\ket{t^x_B} \\
\ket{t^y_B} \\
\ket{t^z_B} \\
\end{pmatrix},
\end{equation}
where, $\alpha=\frac{|D|}{2J}$. To study the low temperature excitation above the ground state, we used bond operator formalism\,\cite{Romhanyi1, Romhanyi2}, which is discussed in the section Sec.\,\ref{sec2.2.2}. In this formalism, we consider the product state of singlets as the vacuum-state and we treat triplets as the quasi-particle excitations named as triplon. The triplon quasi-particles hop around the effective square lattice consisting of two different sub-lattices as shown in Fig.\,\ref{lattice}(d). The low energy effective Hamiltonian in terms of triplons, is further transformed using unitary transformation, such that the two sub-lattices of the effective squared lattice become equivalent. The real-space triplon-Hamiltonian after neglecting the terms of order of $\alpha^2$ is given by,

\begin{align}
\pazocal{H}=&J\sum_{\bold{r}_i}\sum_{\mu=x,y} \tmudagn\tmun+ih_z\sum_{\bold{r}_i}\left[\txdagn\tyn-\tydagn\txn\right]
-\frac{iD'_\perp}{2} \sum_{\bold{r}_i}\sum_{\alpha=x,y} \left[\tydag{\alpha}\txn +\tydagn\tx{\alpha} - \text{h.c.}\right]\nonumber\\
+&\frac{i\tilde{D}'_{||,s}}{2}\sum_{\bold{r}_i}\left[+\tzdag{x}\tyn+\tydag{x}\tzn-\text{h.c.}
 -\tzdag{y}\txn-\txdag{y}\tzn - \text{h.c.}\right]\nonumber\\
-&iD_z\sum_{\bold{r}_i} \left[\tydag{z}\txn+\tydagn\tx{z}-\text{h.c.}\right].
\end{align}

The coupling terms $J_z$ and $D'_{||,ns}$ do not effect the energy of order of magnitude less than $\alpha^2$. Again the term $\tilde{D}'_{||,s}$ consists the interaction terms $D'_{||,s}$ and  $D$, such that $\tilde{D}'_{||,s}=D'_{||,s}-\frac{|D|J'}{2J}$.
The interplay between the DM-interactions $D_z$ and $D'_\perp$ generates different kinds of Weyl-triplons in the system. In the following sub-section, we discuss about the momentum space Hamiltonian and the toplogical Weyl-triplons in the system. 

\subsection{\label{sec5.2.3}Topological Weyl-triplons.}
The momentum space triplon-Hamiltonian is given as,
\begin{equation}
\pazocal{H}=\sum_\bk\sum_{\mu,\nu=x,y,z} \hat{\tilde{t}}^\dagger_{\mu,\bk} M_{\mu\nu}(\bk) \hat{\tilde{t}}_{\nu,\bk},
\end{equation}
where the matrix $M(\bk)$ is given by,
\begin{equation}
M(\bk)=\begin{pmatrix}
J & \begin{split} ig_z&h_z+2iD'_\perp\gamma_3\\&+2iD_z\gamma_4\end{split} & \tilde{D}'_{||}\gamma_2 \\[8pt]
\begin{split} -ig_z&h_z-2iD'_\perp\gamma_3\\&-2iD_z\gamma_4\end{split} & J & -\tilde{D}'_{||}\gamma_1 \\[8pt]
\tilde{D}'_{||}\gamma_2 & -\tilde{D}'_{||}\gamma_1 & J
\end{pmatrix},
\label{Matrix}
\end{equation}
where, $\gamma_1=\sin(k_x)$, $\gamma_2=\sin(k_y)$, $\gamma_3=\frac{1}{2}(\cos(k_x)+\cos(k_y))$, $\gamma_4=\cos(k_z)$.
The matrix $M(\bk)$ can be expressed in terms of Gell-Mann matrices,
\begin{equation}
    M(\bold{k})=J\mathbb{I}+ d_a \lambda_2 + d_b \lambda_4 + d_c \lambda_6,   
    \label{Eq10}
\end{equation}
where $d_a=-g_z h_z-2D'_\perp\gamma_3-2D_z\gamma_4$, $d_b=\tilde{D}'_{||}\gamma_2$, $d_c=-\tilde{D}'_{||}\gamma_1$ and $\lambda_2$, $\lambda_4$, $\lambda_6$ are the Gell-Mann matrices,
\begin{align}
    \lambda_2&=\begin{pmatrix}
    0 & -i & 0\\
    i & 0 & 0\\
    0 & 0 & 0
    \end{pmatrix},\quad
    \lambda_4=\begin{pmatrix}
    0 & 0 & 1\\
    0 & 0 & 0\\
    1 & 0 & 0
    \end{pmatrix}
    \lambda_6=\begin{pmatrix}
    0 & 0 & 0\\
    0 & 0 & 1\\
    0 & 1 & 0
    \end{pmatrix}
\end{align}
The eigen-values of the matrix $M(\bk)$ are $J$, $J+\frac{|d(\bk)|}{2}$ and $J-\frac{|d(\bk)|}{2}$, where $|d(\bk)|=\sqrt{d_a^2+d_b^2+d_c^2}$.
The Gell-Mann matrices span the Lie-algebra of $SU(3)$-group.
The $3\times 3$-matrix of the form in equation Eq.\,\ref{Eq10} contains only Gell-Mann matrices $\lambda_2$, $\lambda_4$, $\lambda_6$, which assures that the for a fixed $k_z$ if the bands are gapped then the Chern-number of the three-bands are of the following form $(-2C,0,2C)$, where $C$ is any integer\,\cite{Romhanyi2, MagnonPolaron3}.
In  this study $C=\pm 1$ and so the Chern number from lower to upper band is given by the set $(\pm 2,0,\mp 2)$.

 We study the model fixing the parameters $J=722$ GHz, $\left|\tilde{D}'_{||}\right|=20$ GHz, $D'_\perp=-21$ GHz and $g_z=2.28$\,\cite{Romhanyi2, ESR} and varying the parameters $h_z$ and $D_z$. At low energies, the system has three different triplon bands and the possible points at which the band-crossing happens are the high-symmetry points on the $k_x$-$k_y$ plane, which are $(\pi,0)$, $(0,\pi)$, $(0,0)$, $(\pi,\pi)$. The Weyl-points at the high symmetry points are triply degenerate, which has no equivalence in the high energy physics, because the quasi-paticle excitation triplons in this system do not follow the Poincare symmetry\,\cite{BeyondWeyl,BeyondWeylMagnon}.

\begin{figure}[H]
\includegraphics[width=\textwidth]{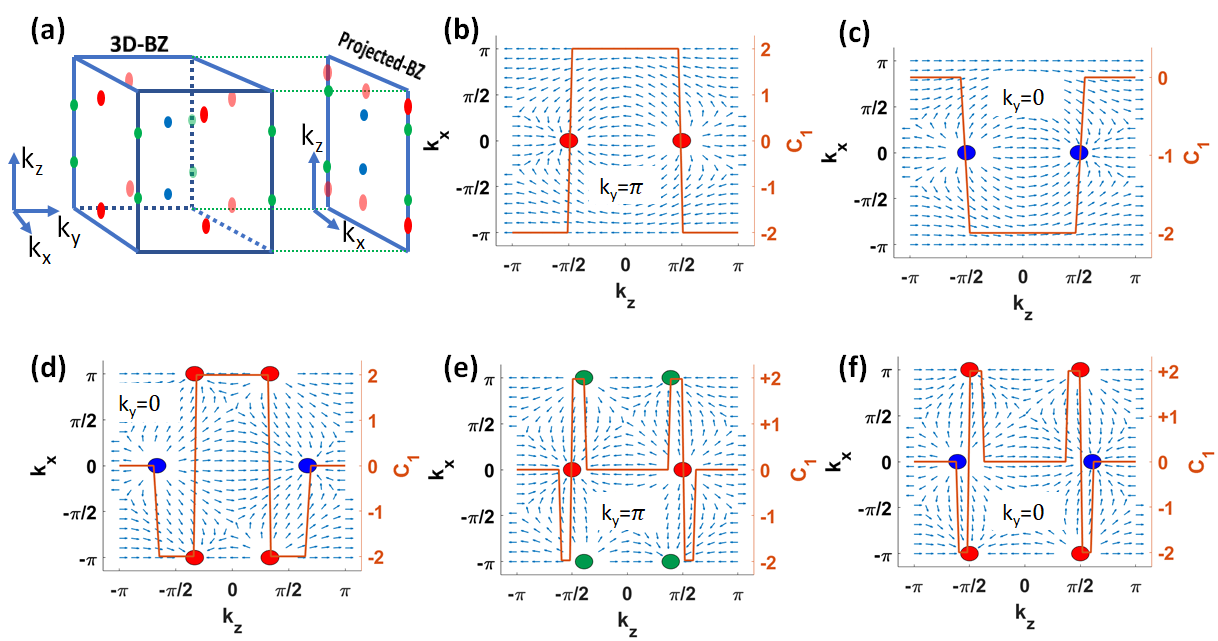} 
\caption{(a) A schematic picture for the presence of possible Weyl-points. The color coding is further described in the main text.The Weyl-points in BZ for parameters (b) $D_z=D'_\perp/2$, $h_z=0$,  (c) $D_z=D'_\perp/2$, $h_z=h_c$, (d) $D_z=D'_\perp$, $h_z=h_c/2$, (e) $D_z=3D'_\perp$, $h_z=0$, (f) $D_z=3D'_\perp$, $h_z=0$. Where, $h_c=\frac{2D'_\perp}{g_z}$. The blue arrows illustrate the direction of the Berry-curvature. The red curves show the change in the Chern number of the lowest band due to the Weyl-points.}
\label{Weyl}
\end{figure}

The schematic figure of different types of Weyl-points in the Brillouin-zone(BZ) is shown in Fig.\,\ref{Weyl}(a). The red-dotes illustrates the Weyl-points at positions $\left(0,\pi,k_{z1}\right)$ and $\left(\pi,0,k_{z1}\right)$, where $k_{z1}=\cos^{-1}\left(-\frac{h_zg_z}{2D_z}\right)$. The blue dotes denote the Weyl-points at position $(0,0,k_{z2})$, where $k_{z2}=\cos^{-1}\left(-\frac{h_zg_z+2D'_\perp}{2D_z}\right)$. Again, the green-points are Weyl-points at position $(\pi,\pi,k_{z3})$, where $k_{z3}=\cos^{-1}\left(\frac{2D'_\perp-h_zg_z}{2D_z}\right)$.

To verify the band crossings are topological Weyl-points, we plot the direction of Berry curvature and change in Chern-number within the first-BZ in Fig.\,\ref{Weyl}(b)-(f), for different parameter regions.
We note that the Chern number is defined strictly for a two dimensional band; in this study, the Chern number is defined for the lower band in two-dimensional $k_x-k_y$ planes at a fixed $k_z$-value in the 3D Brillouin zone and it is defined for $n$-th band as,
\begin{equation}
    C_n(k_z)=\frac{1}{2\pi}\int^\pi_{-\pi}\int^\pi_{-\pi} dk_x dk_y \Omega_n^z(\bk),
\end{equation}
where, $\Omega^z_n(\bk)$ is $z$-component of Berry-curvature of n-th band (n=1 denotes lowest band) at $\bk$-point in Brillouin-zone which is given by,

\begin{equation}
    \Omega^z_n(\bk) = i\sum_{m\neq n} 
    \frac{\squeezeD{m(\bk)}{\frac{\partial \pazocal{H}}{\partial k_x}}{n(\bk)} 
    \squeezeD{m(\bk)}{\frac{\partial \pazocal{H}}{\partial k_y}}{n(\bk)}-(k_x \leftrightarrow k_y)}{(E_n(\bk)-E_m(\bk))^2},
    \label{BerryCurvature}
\end{equation}

where $E_n(\bk)$ and $\ket{n(\bk)}$ denote the eigen-value and eigen-state of $n$-th band at $\bk$-point in Brillouin zone respectively.
Three-band tight binding models have previously been studied for two-dimensional systems and found to be have topologically gapped bands with Chern-numbers of three bands $(+c,0,-c)$ or $(+c,-2c,+c)$ with $c\in \mathbb{Z}$\,\cite{MagnonPolaron3,Romhanyi2,Extra1,Extra2,Extra3}.
In this study, the calculated Chern-numbers $C_n(k_z)$ of the three gapped bands at a fixed $k_z$-plane are found to be $(2c,0,-2c)$ with $c=\pm 1$ or $0$ which is similar to the two-dimensional counterpart of the model studied in the reference Ref.\,\cite{Romhanyi2}.
Weyl-points are band-topological transition points in three dimensional Brillouin zone resulting in change in Chern-numbers $C_n(k_z)$. 
It is found that the Chern number changes by $\pm 2$ for the Weyl-points present at $(0,0,\pm k_{z2})$ and $(\pi,\pi,\pm k_{z3})$ which indicates that the monopole charge associated with these Weyl-points are $\pm 2$. At momenta at $(0,\pi,\pm k_{z1})$ and $(\pi,0,\pm k_{z1})$, the Chern-number changes by $\pm 4$  due to the joint contributions from the Weyl-points , each of which carries a monopole charge of $\pm 2$.

\begin{figure}[H]
\centering
\includegraphics[width=\textwidth]{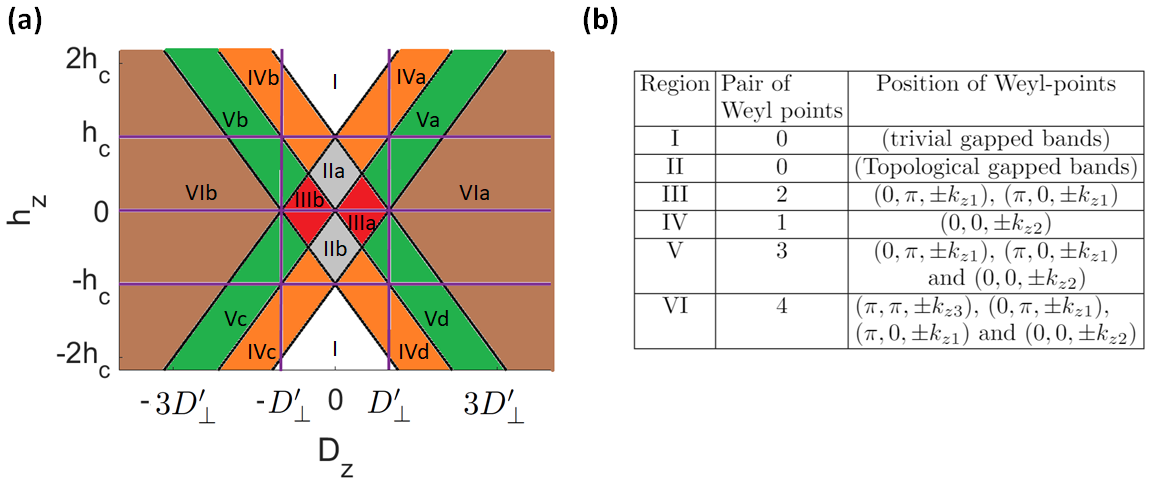} 
\caption{The color coded region of topological phase diagram is defined based on the number of Weyl-points and based on the position of the Weyl-points in the BZ. The subdivision of the regions a, b, c, d denotes the qualitative changes in the sign of monopole charge of Weyl-points. (b) The table shows the number and position of Weyl-points in different regions. The table is valid for $D'_\perp<0$. For the case $D'_\perp>0$, the Weyl-nodes at $(0,0,\pm k_{z2})$ are substituted by the Weyl-nodes at $(\pi,\pi,\pm k_{z3})$ and vice-versa.}
\label{Phase}
\end{figure}

Based on the number of Weyl-points and their positions in the $k_x$-$k_y$ plane, we categorize the $h_z$-$D_z$ parameter space in to several regions in Fig.\,\ref{Phase}(a).  Regions I and II contain no Weyl-points, but in region-II the triplon bands carry non-zero Chern numbers ($\pazocal{C}= -2, 0$ and $+2$ for bands with increasing energy) whereas in region-I, they are topologically trivial. For $D_z=0$, the transition in
band topology from region-I to region-II  occurs at critical magnetic field $h_c=\frac{2\left|D_z\right|}{g_z}$\,\cite{Romhanyi2}.
The remaining regions III, IV, V and VI feature multiple Weyl points (appearing in pairs), ranging from
one pair in region IV to four pairs in region VI. The locations of the Weyl points and their nature are listed in the table in figure Fig.\,\ref{Phase}(b).
The Weyl-points at different sub-regions (a), (b), (c), (d) are at the same position but, the monopole charges of some of the Weyl-points change by a sign.

\begin{figure}[H]
\includegraphics[width=\textwidth]{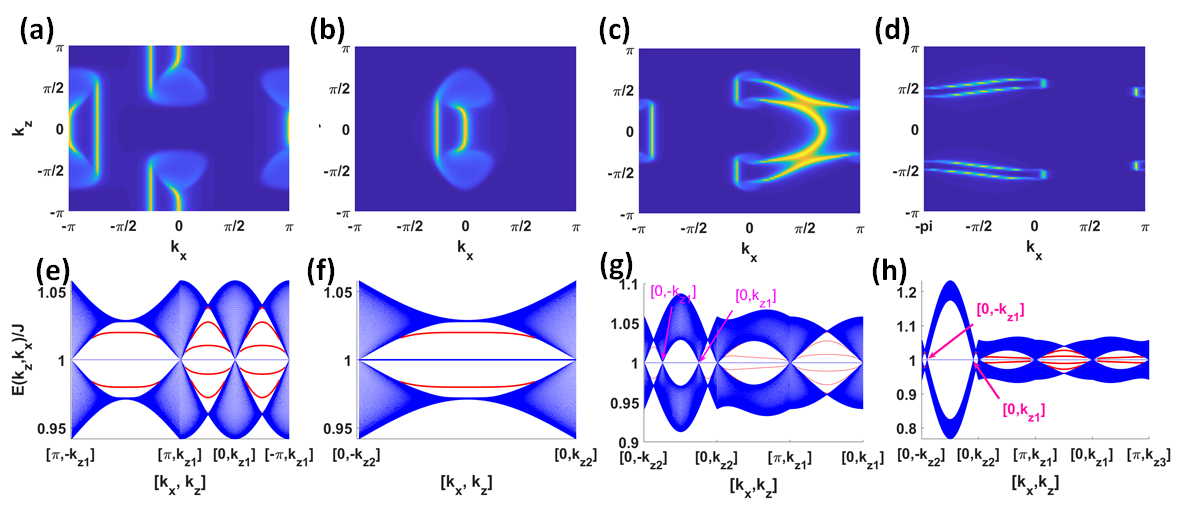} 
\caption{Triplon-arcs on the $x$-$z$ surface (a), (b), (c), (d)  for the parameters same as in (b), (c), (d), (e) in Fig.\,\ref{lattice}. The surface states on the $x$-$z$ surface for the parameters (e), (f), (g), (h) for the parameters same as in (a), (b), (c), (d) respectively.}
\label{Surface}
\end{figure}

\subsection{\label{sec5.2.4}Surface arcs and surface states.}
The topological nature of the system is revealed by the presence of the surface states in the material. In Fig.\,\ref{Surface}(a)-(d), we plot the surface spectral function of a slab geometry of a system extended along $x$-$z$ direction. Each of the projected bulk Weyl-points on the surface emits two triplon-arcs, which divulge that the monopole charge of a Weyl-point is $\pm 2$. The surface triplon-arcs of the system is quite different in the different regions of parameter space, because of the different position and different numbers of Weyl-points present in different sector in the parameter phase. The Fig.\,\ref{Surface}(a), (b), (c), (d) illustrates the surface triplon-arcs for the topological phase regime III, IV, V, VI respectively. For illustration, we describe the Fig.\,\ref{Surface}(a), which corresponds to the region-IIIa in phase diagram Fig.\,\ref{Phase}. There are two-pairs of Weyl-triplon in this region, at positions $(0,\pi,\pm k_{z1})$ and $(\pi,0,\pm k_{z1})$. So the projected Weyl-point on the $k_x$-$k_z$ surface exists at the positions $(\pi,\pm k_{z1})$ and $(0, \pm k_{z1})$. The pair of points along $k_z$ axis is connected by two surface triplon-arcs. The existence of surface triplon-arcs in the system can be detected using inelastic neutron scattering. The Fig.\,\ref{Surface}(e)-(h), describes that the surface states are chiral gap-less state present within the bulk gap in the system.

\subsection{\label{sec5.2.5}Thermal Hall effect for experimental detection.}  

Thermal Hall effect is the key experimental signature to detect topological excitations in a magnetic system. The characteristic features of Thermal Hall conductance of an Weyl-magnon is different from the usual gapped topological magnon bands, making it an ideal probe to detect Weyl points. We calculate the thermal Hall effect in different regimes with  Weyl points (phases III, IV, V or VI in Fig.\,\ref{Phase}), gapped
topological triplons (phases I, II in Fig.\,\ref{Phase}) and 
gapped topologically trivial triplon excitations to show that the 
thermal Hall conductivity exhibits distinct features identifying the 
different regimes. Since the Weyl-points in this system always occur in 
pairs aligned along the $z$-direction, a transverse current cannot 
be created along the $z$-axis. Similarly, a temperature gradient along
this direction cannot produce a transverse current along any other direction\,\cite{WeylPyrochlore4}. However, a transverse triplon current can be induced in $y$ (or $x$)-direction by applying a temperature gradient along the $y$ (or $x$)-direction. The thermal Hall conductance of the system in the $x-y$ plane is given by\,\cite{Katsura,THE6,THE5,BogoliubovBerryCurvature},
\begin{equation}
    \kappa_{xy}=\int^\pi_\pi \frac{dk_z}{2\pi}\kappa_{xy}^{2D}(k_z),
\end{equation}
where $\kappa_{xy}^{2D}(k_z)$ is the 2D-thermal Hall conductance contribution from the $k_x-k_y$-plane of fixed $k_z$-value in the Brillouin Zone, which is given by,
\begin{equation}
    \kappa_{xy}^{2D}(k_z)=-T\int \int \frac{dk_x dk_y}{(2\pi)^2} \sum_{n=1}^{N} c_2(f^B[E_n(\bold{k})]) \Omega^z_n(\bold{k}),
    \label{THEE}
\end{equation}
where, $\bold{k}=(k_x, k_y, k_z)$ and $c_2(x)=(1+x)(\ln\frac{1+x}{x})^2-(\ln x)^2-2\text{Li}_2(-x)$, with $\text{Li}_2(x)$ as bilogarithmic function.

\begin{figure}[H]
\centering
\includegraphics[width=0.7\textwidth]{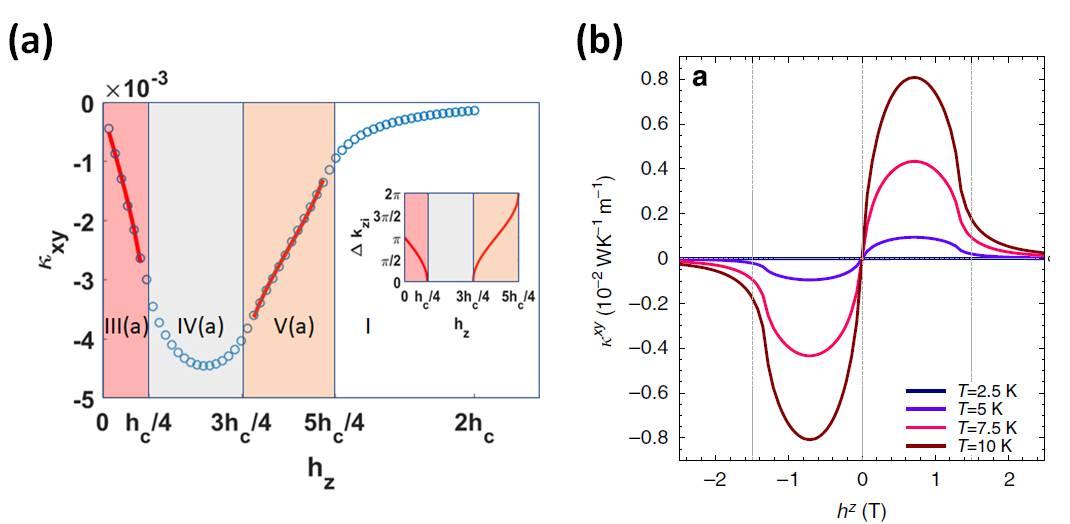} 
\caption{(a) The thermal Hall conductance as a function of the magnetic field is shown for $D_z=\frac{D'_\perp}{4}$. As magnetic field increases the system undergoes different phase regions as in Fig.\,\ref{Phase}. The dots are calculated Thermal Hall conductivity and the red-line is analytically fitted with the expressions $A \Delta k_{zi}(h_z) +B$ as discussed in the main text. The inset shows the distance between Weyl-nodes as a function of the magnetic field for phase regions III(a) and V(a) in Fig.\,\ref{Phase}. (b) The thermal Hall conductance as a function of magnetic field for non-zero inter-layer DMI is taken from the reference Ref.\,\cite{Romhanyi2}. The linear region near magnetic field $h_z=0$ is absent in these plots due to absence in Weyl-triplon regions. The signs in thermal Hall conductance for the plots (a) and (b) are opposite is due to different sign-conventions of the formula of thermal Hall conductance in Eq.\,\ref{THEE}.}
\label{THE2}
\end{figure}

While the nature and magnitude of interlayer DM-interaction ($D_z$) 
in \ce{SrCu2(BO3)2} has not been determined experimentally,
it is reasonable to expect finite $D_z$ as its presence is allowed by 
symmetry of the lattice. We assume a small, but finite, interlayer
DMI parallel to the layers (as allowed by the symmetry of the lattice),
such that the triplon bands of the system lies in region III(a) of Fig.\,\ref{Phase}. Assuming $D_z=\frac{D'_\perp}{4}$ the thermal Hall conductivity is plotted as a function of magnetic field in Fig.\,\ref{THE2}(a)
for $0 \le h_z \le 2h_c)$. The triplon bands undergo several topological phase transitions in this range of applied field -- $III(a) \rightarrow II(a) \rightarrow IV(a) \rightarrow I$ in Fig.\,\ref{Phase}. 
The triplon bands in region III(a) contains two pairs of Weyl points at $(0,\pi,\pm k_{z1})$ and $(\pi,0,\pm k_{z1})$, while there is one pair 
of Weyl points at $(0,0,\pm k_{z2})$ in the region V(a). The triplon
bands are fully gapped and topological in nature in region $II(a)$,
whereas they are gapped and topologically trivial in region $I$. It is noted that, although the Berry curvature at the Weyl point is ill-defined, the thermal Hall conductance $\kappa_{xy}^{2D}(k_z)$ is a continuous function of $k_z$, because the left and right hand limit $k_z\rightarrow k_{zw}+0^+ $ and  the left hand limit as $k_z \rightarrow k_{zw}+0^- $ are equal, where $k_{zw}$ denotes the position of the Weyl-point. 
The thermal Hall conductance exhibits a unique quasi-linear 
dependence as a function of magnetic field for a region with Weyl-points,
and quite different from the phase region without Weyl points.
Moreover the two-dimensional counterpart of the model without inter-layer interaction has no Weyl-points and the thermal Hall conductance for the two-dimensional model does not show any linear dependence as a function of magnetic-field as in figure Fig.\,\ref{THE2}(b).
This is because the thermal Hall conductance of magnetic excitations magnons
or triplons is proportional to the distance between the Weyl-nodes in momentum space\,\cite{WeylPyrochlore4}, analogous to that of (electrical) Hall conductance observed in Weyl semimetals\,\cite{HallConductance}, as a consequence the thermal Hall conductivity exhibits a quasi-linear dependence as a function of magnetic field.  In  regions III(a) and V(a) the distance between the pair of Weyl-nodes are $\Delta k_{z1}(h_z)=2\cos^{-1}(-\frac{h_z g_z}{2D_z})$  and $\Delta k_{z2}(h_z)=2\pi-2\cos^{-1}(-\frac{h_z g_z+2D'_\perp}{2D_z})$ respectively. The inset of the Fig.\,\ref{THE2}(a) shows the distance between the Weyl-nodes as a function of magnetic field has point of inflection at the points $h_z=0$ and $h_z=h_c$, which gives rise to the quasi-linear behaviour of the thermal Hall conductance. To show the quasi-linear dependence of the thermal Hall conductance in Weyl-triplon region as a function of magnetic field, the thermal conductivity is fitted with the analytical expression $A \Delta k_{zi}(h_z) +B$, where $A$ and $B$ are fitting parameters and $i=1$ or $2$. The quasi-linear characteristic in thermal Hall conductance in this system is solely due to the presence of the Weyl-nodes. On the
other hand, the field dependence of the thermal Hall conductance is
markedly different in region $IV(a)$ were the triplon bands are fully
gapped and carry finite Chern numbers, consistent with previous studies\,\cite{Romhanyi2,triplon3}. Thus the presence of quasi-linearity in the thermal Hall conductance as a function of magnetic field is a direct evidence of the presence of Weyl-nodes. Finally, in region $I$ with topologically trivial, gapped triplon bands, the thermall hall conductivity vanishes.

\section{\label{sec5.3} Conclusion and Discussion}
In conclusion we have demonstrated that \ce{SrCu2(BO3)2} is a possible host of Weyl-triplons. Our study shows that interlayer perpendicular DM-interaction (even if very small in magnitude) naturally give rise to the Weyl-triplons. Furthermore the nature of triplon bands at low temperature  depends neither on the interlayer Heisenberg interation (because of orthogonal Dimer arrangement) nor on the interlayer in-plane DM-interactions, which makes the appearance of Weyl nodes robust against small deviations from the idealized model. Finally, We have shown that the quasi-linear behaviour of thermal Hall conductance as a function of magnetic field is a possible experimental signature to detect the presence of Weyl nodes.

We also propose neutron scattering experiment as an alternative way to explore the Weyl-nodes in the triplon bands. 
The neutron scattering spectra is already studied in the reference Ref.\cite{triplon2}, which shows that triplon bound state contribution is important under their experimental conditions, which we have neglected in this study.
In presence of triplon bound state the three-fold degeneracy is absent at zero magnetic field.
But band-topological transitions still exist in the material as a function of magnetic field similar to that reported in Ref.\cite{Romhanyi2} and so the band touching happens as magnetic field is varied.
This band touching in 2D model is expected to translate into a Weyl-point in the 3D model in presence of interlayer Dzyaloshinskii–Moriya interaction (DMI).
Similar Weyl-triplon phases can also be investigated in the dimer phases of materials \ce{Rb2Cu3SnF12}\,\cite{OtherMaterial1,OtherMaterial2} and \ce{ZnCu3(OH)6Cl2}\,\cite{OtherMaterial3}.
	
		
	\chapter{Conclusions and future-work}\label{chapter06}
	\section{Conclusions}
	Throughout this thesis we have considered the two-dimensional Heisenberg model with \gls{DMI}, which in the \gls{LSWT}, \gls{sbmft} or bond-operator formalism transformed into another bosonic Hamiltonian representing a system of spin-excitations or spin quasi-particles.
	In this final chapter of the dissertation, we summarize the preceding chapters and present conclusions and outlook.
	In chapter Ch.\,\ref{chapter01}, we discuss about past studies and the open-questions, which is the starting point of the study.
	We describe the generic spin models and different methods to study the spin models in chapter Ch.\,\ref{chapter02}.
	The rest of the section is devoted in the discussion of the chapters Ch.\,\ref{chapter03}, Ch.\,\ref{chapter04} and Ch.\,\ref{chapter05}, which contains the main findings of our study.

	In chapter Ch.\,\ref{chapter03}, we investigate the presence of antichiral edge states in the honeycomb ferromagnets.
	We find that unequal \gls{DMI} for the two sublattices lead to an additional antichiral-\gls{DMI} and using \gls{spinon}-picture we determined the temperature range between which the antichiral edge states can be probed experimentally.
	Antichiral edge states in the system is found when antichiral-\gls{DMI} is dominant in compared with the chiral-\gls{DMI}.
	We show the way to engineer the materials \ce{CrSrTe3}, \ce{CrGeTe3}, \ce{AFe2(PO4)2} (A=Ba, Cs, K, La) to break the inversion symmetry at the center of the hexagon of honeycomb material which in turn induces the antichiral-\gls{DMI}. 
	Moreover the linear dispersion of antichiral edge states is only distorted in presence of other realistic interactions, but the presence of antichiral edge modes are not destroyed.  
	Moreover we show the possible experimental measurements to explore the existence of antichiral-edge-modes in the material.
	We investigated that the antichiral \gls{DMI} does not effect the spin Nernst effect qualitatively or quantitatively, thus spin-Nernst effect is not ideal measurement to detect this novel phenomenon.
	So we proposed some alternative techniques such as magnetic force microscopy, inelastic neutron scattering technique and inverse spin-Hall noise spectroscopy, which are the promising experimental technique to detect the anti-chiral edge states in the system.

	In chapter Ch.\,\ref{chapter04}, we studied the topological \gls{magnon}s in the non-coplanar magnetic canted-flux state in the frustrated Shastry-Sutherland lattice.
	We propose a model spin-Hamiltonian correspond to a distorted rare-earth tetraboride such that in-plane \gls{DMI} become symmetry allowed.
	In presence of perpendicular \gls{DMI} (greater than a critical value) and zero in-plane \gls{DMI} the ground state is described as a flux state and we find that the \gls{magnon} bands of this ground state is topologically trivial.
	But in presence of in-plane \gls{DMI} along with out of plane \gls{DMI} the magnetic ground state become a non-coplanar canted flux state.
	 In canted flux state, presence of out of plane magnetic field lifts the degeneracy of \gls{magnon} bands and the gapped out \gls{magnon} bands are topologically non-trivial due to non-zero Chern number of the bands.
	 We find many band-topological phase transitions in the parameter space due to non-coplanar spin configuration, in-plane \gls{DMI} and out of plane \gls{DMI}.
	 The thermal \gls{magnon} Hall effect and non-zero thermal Hall conductance are the signatures for topological \gls{magnon} bands in different band topological regions.
	 We further show that the derivative of thermal Hall conductance is logarithmic divergent at the band-topological transition point and independent of the type of the band touching point.
	 Moreover we derived a simple analytical expression for derivative of thermal Hall conductance as a function of temperature and propose that fitting of derivative of thermal Hall conductivity with the analytical expression near the band topological transition point gives the energy of the band touching point in experiments.

	 In chapter  Ch.\,\ref{chapter05} for the first time we show that Weyl-magnetic excitation exists for a dimerized ground states.
	   We propose a extended Shatry-Sutherland lattice with a interlayer coupling and show that \ce{SrCu2(BO3)2} is the ideal material for realization of Weyl-\gls{triplon}s.
	   We investigate the whole parameter space of the interlayer \gls{DMI} and  perpendicular magnetic field fixing the other parameter values same as of the material \ce{SrCu2(BO3)2}.
	   Any non-zero interlayer perpendicular \gls{DMI} ensures that the \gls{triplon} excitations must exhibit a Weyl-\gls{triplon} and the other realistic interlayer interaction like Heisenberg-interaction and in-plane \gls{DMI} do not alter the result.
	   It is found that there are many \gls{triplon} band topological phases and the phase are categorized by the numbers and positions of the Weyl-\gls{triplon}s present in the system.
	   We show that each Weyl-\gls{triplon} carries a topological charge $\pm 2$.
	   Again Fermi-arc like structure connecting pairs of Weyl-point is also present at the surface of the system denoting the topological Bulk-edge correspondence. 
	   Moreover we studied the thermal Hall conductivity as a function of magnetic field.
	   The system undergoes a several topological band transitions as the magnetic field is changed and specifically for the Weyl-\gls{triplon} region the thermal Hall conductance has a quasi-linear behaviour, which is a possible experimental signature for Weyl-\gls{triplon} region.

	   In conclusion we investigated various magnetic systems with different lattices to explore the topological magnetic excitations above the ground state.
	   Our result indicates emergence of antichiral edge states in honeycomb ferromagnets, topological magnetic excitations in the flux state of a distorted Shastry-Sutherland lattice and Weyl-\gls{triplon}s in the stacked Shastry-Sutherlnd material \ce{SrCu2(BO3)2}.
	   We study different spin models but as well as proposed material realizations and experimental techniques for detection.

	   \section{Future-Work}
	   The research in topological magnetic excitation is accelerating in last few years, denoting that there are much more to explore.
	There are many studies showing the magnonic analogue of phenomena which are generally observed in topological electronic system.
	For example Hofstadter butterfly\,\cite{Hofstadter}, Moir\'e magnons\,\cite{Moire} can also observed in the magnetic excitations of the ground state.
	However further study by varying the magnetic ground state and lattices is necessary to understand the in depth physics.
	  On the other hand Einstein-de Haas effect is theoretically proposed in the magnetic systems due to topological magnons\,\cite{Einstein-deHaas}.
	  However the effect is shown to be strong in the ferromagnetically ordered square-octagon lattice system which has no experimental counterpart and thus it is a challenging task to propose a experimentally possible magnetic material with a significant Einstein-de Haas effect.
	 Moreover, the studies of topological magnons are limited to the single-particle treatment of the magnon.
	  At high temperature the magnon-magnon interaction become important and the protection of the edge states against the magnon-magnon interaction is not well studied.
	  The topological protection of the edge states against interaction can be explored by using Landau-Lifsitz-Gilbert equation for ferromagnets\,\cite{MagnonTransport1}, whereas a non-equilibrium-Green function technique is suitable for any ground state\,\cite{MagnonTransport2}.

	  The topological magnonincs is a useful branch of magnon based spintronics.
	  But, to improve the applicability of the topological magnonics, it is important to understand how to increase the device efficiency and how to manipulate it externally.
	  \gls{DMI} is one of the key ingredient for the realization  of topological magnon bands.
	  However the magnitude of \gls{DMI} is quite low for some magnetic system.
	  It is shown that, when a circularly polarized incident perpendicularly on the honeycomb or Kagome lattice Heisenberg magnets, it induces synthetic \gls{DMI}\,\cite{PhotoInduced1,PhotoInduced2,PhotoInduced3,PhotoInduced4}.
	  Thus circularly polarized light is a useful tool for manipulate a band topological transition.
	  This method can be extended for other lattices for the future work.
	  Another practical limitation of topological magnonic system is the small signal strength from the edge modes and it is theoretically shown that the edge modes can be selectively populated by external light source\,\cite{TopologicalMagnonAmplification}.
	  This technique of amplification can also be applied in the system studied in this dissertation to understand the mechanism of amplification in each different magnetic system.
	   Magnon based interferometry also become popular\,\cite{Interferometry1,Interferometry2} and the use of topological systems in magnon interferometry is also needed to be explored.
	   Phonon driven topological magnons is also a interesting phenomena to understand in various systems\,\cite{MagnonPolaron1,MagnonPolaron2,MagnonPolaron3,MagnonPolaron4,PhononDrivenFloquette}.
	  Moreover recently the magnon based valleytronics is theoretically proposed\,\cite{Valleytronics}, however many prospects of the field need to be understood in details.

\phantomsection
\appendix
\newpage
\chapter{Appendix A : The properties of Schwinger boson bond operators}\label{appendixA}

There are two Schwinger boson bond operators that obeys $SU(2)$ symmetries of the isotropic Heisenberg Hamiltonian which are discussed in section\,\ref{SBMFT}.
The $SU(2)$-symmetric bond operators in terms of Schwinger boson operators are of the following form,
 \begin{equation}
	    	A_{ij}^\dagger=\frac{1}{2} \left(b_{i\uparrow}^\dagger b_{j\downarrow}^\dagger-b_{i\downarrow}^\dagger b_{j\uparrow}^\dagger\right), \quad
	       	B_{ij}^\dagger=\frac{1}{2} \left(b_{i\uparrow}^\dagger b_{j\uparrow}+b_{i\downarrow}^\dagger b_{j\downarrow}\right).
	    	\label{eqA1}
		\end{equation}
In this appendix we will discuss the properties of these bond operators.

\subsection*{$SU(2)$-symmetry of the bond operators}
The derivation is done for spin-half particles and it can be generalized for any spin.
A general spin-$\frac{1}{2}$ state can be rotated by an angle $\theta$ about $y$-axis using the following operator,
\begin{align}
	R_y&=e^{i\theta S_y}\nonumber\\
	   &=e^{i\theta \frac{\sigma_y}{2}}\nonumber\\
	   &=\cos\left(\frac{\theta}{2}\right)I-i\sin\left(\frac{\theta}{2}\right)\sigma_y,
\end{align}
where $I$ and $\sigma_y$ are identity operator and y-component of Pauli matrix respectively.

The Schwinger boson operators create the spin up or spin down state operating on a vacuum state $\left|0\right\rangle$ basis,
\begin{equation}
	c_{i\uparrow}^\dagger\left|0\right\rangle=\begin{pmatrix}
	1 \\
	 0
	\end{pmatrix},\quad
	c_{i\downarrow}^\dagger\left|0\right\rangle=\begin{pmatrix}
	0 \\
	 1
	\end{pmatrix}.
\end{equation}

So the transformation of operator $c_{i\uparrow}^\dagger$ by rotation of spin about $y$-axis,
\begin{align*}
	R_y c_{i\uparrow}^\dagger R_y^{-1} R_y \left|0\right\rangle
	&=R_y c_{i\uparrow}^\dagger \left|0\right\rangle \\
	&=\left[\cos\left(\frac{\theta}{2}\right)I-i\sin\left(\frac{\theta}{2}\right)\sigma_y\right] \begin{pmatrix}
	1 \\
	0
	\end{pmatrix}\\
	&=\cos\left(\frac{\theta}{2}\right)\begin{pmatrix}
	1\\
	0
	\end{pmatrix}+
	\sin\left(\frac{\theta}{2}\right)\begin{pmatrix}
	0\\
	1
	\end{pmatrix}\\
	&=\left[\cos\left(\frac{\theta}{2}\right) c_{i\uparrow}^\dagger +\sin\left(\frac{\theta}{2}\right)c_{i\downarrow}^\dagger\right]
\end{align*}
Similarly deriving for the $c_{i\downarrow}^\dagger$ operator we get the following sets of relations,
\begin{align}
	R_y c_{i\uparrow}^\dagger R_y^{-1}&=\left[\cos\left(\frac{\theta}{2}\right) c_{i\uparrow}^\dagger +\sin\left(\frac{\theta}{2}\right)c_{i\downarrow}^\dagger\right], \quad
	R_y c_{i\uparrow} R_y^{-1}=\left[\cos\left(\frac{\theta}{2}\right) c_{i\uparrow} +\sin\left(\frac{\theta}{2}\right)c_{i\downarrow}\right],\nonumber\\
	R_y c_{i\downarrow}^\dagger R_y^{-1}&=\left[\cos\left(\frac{\theta}{2}\right) c_{i\downarrow}^\dagger -\sin\left(\frac{\theta}{2}\right)c_{i\uparrow}^\dagger\right],\quad
	R_y c_{i\downarrow} R_y^{-1}=\left[\cos\left(\frac{\theta}{2}\right) c_{i\downarrow} -\sin\left(\frac{\theta}{2}\right)c_{i\uparrow}\right],
\label{A3}
\end{align}
Thus,
\begin{align*}
R_y A_{ij}^\dagger R_y^{-1}
&=\frac{1}{2} R_y \left( b_{i\uparrow}^\dagger b_{j\downarrow}^\dagger-b_{i\downarrow}^\dagger b_{j\uparrow}^\dagger\right) R_y^{-1}
\\
&=\frac{1}{2}\left[
\left(\cos\left(\frac{\theta}{2}\right) c_{i\uparrow}^\dagger +\sin\left(\frac{\theta}{2}\right)c_{i\downarrow}^\dagger\right)
\left(\cos\left(\frac{\theta}{2}\right) c_{j\downarrow}^\dagger -\sin\left(\frac{\theta}{2}\right)c_{j\uparrow}^\dagger\right)
\right.
\\
&\qquad\qquad \left.
-\left(\cos\left(\frac{\theta}{2}\right) c_{i\downarrow}^\dagger -\sin\left(\frac{\theta}{2}\right)c_{i\uparrow}^\dagger\right)
\left(\cos\left(\frac{\theta}{2}\right) c_{j\uparrow}^\dagger +\sin\left(\frac{\theta}{2}\right)c_{j\downarrow}^\dagger\right)
\right]
\\
&=A_{ij}^\dagger
\end{align*}
Using this relations\,\ref{A3} it is easy to prove that,
\begin{equation}
	R_y A_{ij}^\dagger R_y^{-1}=A_{ij}^\dagger, \quad
	R_y B_{ij}^\dagger R_y^{-1}=B_{ij}^\dagger.
\end{equation}
So the bond operators obeys the $SU(2)$-symmetry.

\subsection*{Bond operators as ferromagnetic or antiferromagnetic order parameter}

\subsubsection*{Some basic relations}
Before jumping into the derivation, here are some basic relations provided for the derivation.
The Schwinger boson transformation is given by,
\begin{equation}
		S^+_i=b_{i\uparrow}^\dagger b_{i_\downarrow},\quad
		S^-_i=b_{i\downarrow}^\dagger b_{i\uparrow},\quad
		S^z_i=\frac{1}{2}\left(b_{i\uparrow}^\dagger b_{i\uparrow}-b_{i\downarrow}^\dagger b_{i\downarrow}\right),
	\end{equation}
	where $b_{i\uparrow}^\dagger$ and $b_{i\downarrow}^\dagger$ are the two species of Schwinger boson at i-th site of the lattice.
Using the Schwinger boson transformation the following basic relations are derived,
\begin{align}
	&(S_i^{x})^2+(S_i^{y})^2+(S_i^{z})^2=S(S+1),\label{A7}\\
	&S_i^+ S_j^-=b_{i\uparrow}^\dagger b_{j\downarrow}^\dagger b_{i\downarrow} b_{j\uparrow},\label{A8}\\
	&S_i^- S_j^+=b_{i\downarrow}^\dagger b_{j\uparrow}^\dagger b_{i\uparrow} b_{j\downarrow},\label{A9}\\
	&S_i^z S_j^z=\frac{1}{4} \left(
	 b_{i\uparrow}^\dagger b_{i\uparrow} b_{j\uparrow}^\dagger b_{j\uparrow}
	 - b_{i\uparrow}^\dagger b_{i\uparrow} b_{j\downarrow}^\dagger b_{j\downarrow}
	 -b_{i\downarrow}^\dagger b_{i\downarrow} b_{j\uparrow}^\dagger b_{j\uparrow}
	 +b_{i\downarrow}^\dagger b_{i\downarrow} b_{j\downarrow}^\dagger b_{j\downarrow}
	  \right),
	  \label{A10}
	  \\
	  &S_i^z S_j^z=S^2-\frac{1}{2}\left(
	  b_{i\uparrow}^\dagger b_{j\downarrow}^\dagger b_{i\uparrow} b_{j\downarrow}
	  +b_{i\downarrow}^\dagger b_{j\uparrow}^\dagger b_{i\downarrow} b_{j\uparrow}
	  \right),
	  \label{A11}
	  \\
	  &S_i^z S_j^z=-S^2+\frac{1}{2}\left(b_{i\uparrow}^\dagger b_{i\uparrow} b_{j\uparrow}^\dagger b_{j\uparrow}
	  +b_{i\downarrow}^\dagger b_{i\downarrow} b_{j\downarrow}^\dagger b_{j\downarrow}
	  \right).
	  \label{A12}
\end{align}
The relations Eq.\,\ref{A10}, Eq.\,\ref{A11}, Eq.\,\ref{A12} are equivalent and they are connected by the equation denoting constraint over boson number,
\begin{equation}
		b_{i\uparrow}^\dagger b_{i_\uparrow}+b_{i\downarrow}^\dagger b_{i_\downarrow}=2S,
		\label{A13}
	\end{equation}
For example the equation Eq.\,\ref{A11} from Eq.\,\ref{A10} is derived in the following way,
\begin{align*}
	S_{i}^z S_j^z&=\frac{1}{4} \left[
	 b_{i\uparrow}^\dagger b_{i\uparrow} b_{j\uparrow}^\dagger b_{j\uparrow}
	 - b_{i\uparrow}^\dagger b_{i\uparrow} b_{j\downarrow}^\dagger b_{j\downarrow}
	 -b_{i\downarrow}^\dagger b_{i\downarrow} b_{j\uparrow}^\dagger b_{j\uparrow}
	 +b_{i\downarrow}^\dagger b_{i\downarrow} b_{j\downarrow}^\dagger b_{j\downarrow}
	  \right]
	  \\
	  &=\frac{1}{4} \left[
	 b_{i\uparrow}^\dagger b_{i\uparrow} \left(2S-b_{j\downarrow}^\dagger b_{j\downarrow}\right)
	 - b_{i\uparrow}^\dagger b_{i\uparrow} b_{j\downarrow}^\dagger b_{j\downarrow}
	 -b_{i\downarrow}^\dagger b_{i\downarrow} b_{j\uparrow}^\dagger b_{j\uparrow}
	 +b_{i\downarrow}^\dagger b_{i\downarrow} \left(2S-b_{j\uparrow}^\dagger b_{j\uparrow}\right)
	  \right]
	  \\
	  &=S^2-\frac{1}{2}\left(
	  b_{i\uparrow}^\dagger b_{j\downarrow}^\dagger b_{i\uparrow} b_{j\downarrow}
	  +b_{i\downarrow}^\dagger b_{j\uparrow}^\dagger b_{i\downarrow} b_{j\uparrow}
	  \right).
\end{align*}
In a similar way we can derive the equation Eq.\,\ref{A12} by transforming the second and third quartic terms of equation Eq.\,\ref{A10} using the constraint Eq.\,\ref{A13}.

\subsubsection*{Bond operators as ferromagnetic order parameter}
The $SU(2)$-symmetric Schwinger boson bond operators which is suitable for a ferromagnetic phase is given as,
\begin{equation}
	B_{ij}^\dagger=\frac{1}{2}\left( b_{i\uparrow}^\dagger b_{j\downarrow}+b_{i\downarrow}^\dagger b_{i\downarrow}\right).
\end{equation}
So,
\begin{align*}
	:B_{ij}^\dagger B_{ij}:&=
	:\frac{1}{2}(b_{i\uparrow}^\dagger b_{j\uparrow}
				+b_{i\downarrow}^\dagger b_{j\downarrow})
	\frac{1}{2}(b_{j\uparrow}^\dagger b_{i\uparrow}
				 +b_{j\downarrow}^\dagger b_{i\downarrow}):\\
				 &=
				 :\frac{1}{4}\left[
b_{i\uparrow}^\dagger b_{j\uparrow} b_{j\uparrow}^\dagger b_{i\uparrow}
+b_{i\uparrow}^\dagger b_{j\uparrow} b_{j\downarrow}^\dagger b_{i\downarrow}
+b_{i\downarrow}^\dagger b_{j\downarrow} b_{j\uparrow}^\dagger b_{i\uparrow}
+b_{i\downarrow}^\dagger b_{j\downarrow} b_{j\downarrow}^\dagger b_{i\downarrow}
				 \right]:
				 \\
				 &=
				 \frac{1}{4}\left[
b_{i\uparrow}^\dagger b_{j\uparrow}^\dagger b_{i\uparrow} b_{j\uparrow}+
b_{i\uparrow}^\dagger b_{j\downarrow}^\dagger b_{i\downarrow} b_{j\uparrow}+
b_{i\downarrow}^\dagger b_{j\uparrow}^\dagger b_{i\uparrow} b_{j\downarrow}+
b_{i\downarrow}^\dagger b_{j\downarrow}^\dagger b_{i\downarrow} b_{j\downarrow}
				 \right]
				 \\
				 &=\frac{1}{4}\left[
b_{i\uparrow}^\dagger b_{j\downarrow}^\dagger b_{i\downarrow} b_{j\uparrow}+
b_{i\downarrow}^\dagger b_{j\uparrow}^\dagger b_{i\uparrow} b_{j\downarrow}+
\left\lbrace
b_{i\uparrow}^\dagger b_{j\uparrow}^\dagger b_{i\uparrow} b_{j\uparrow}+
b_{i\downarrow}^\dagger b_{j\downarrow}^\dagger b_{i\downarrow} b_{j\downarrow}
\right\rbrace
\right]
\\
&=
\frac{1}{4}\left[
S_i^+ S_j^- +S_i^- S_j^+ +2S_i^z S_j^z +2S^2
\right] \qquad\left[\text{Using Eq.\,\ref{A8}, Eq.\,\ref{A9} and Eq.\,\ref{A12}}\right]
\\
&=
\frac{1}{4} \left[ 2\bold{S}_i\cdot\bold{S}_j+2S^2\right]
\\
&=
\frac{1}{4}
\left[
(S_i^x)^2+(S_i^y)^2+(S_i^z)^2-S(S+1)
+(S_j^x)^2+(S_j^y)^2+(S_j^z)^2-S(S+1)
\right.\\
&\left.
\qquad\qquad +2\bold{S}_i\cdot\bold{S}_j
+2S^2
\right]\qquad\left[\text{Using Eq.\,\ref{A7}}\right]
\\
&=
\frac{1}{4}\left[(\bold{S}_i+\bold{S}_j)^2-2S(S+1)+2S^2\right]
\\
&=
\frac{1}{4}(\bold{S}_i+\bold{S}_j)^2-\frac{S}{2}
\end{align*}
Thus the bond operator $B_{ij}^\dagger$ is related with the ferromagnet order parameter.

\subsubsection*{Bond operators as antiferromagnetic order parameter}
The $SU(2)$-symmetric Schwinger boson bond operators which is suitable for a antiferromagnetic phase is given as,
\begin{equation}
	A_{ij}^\dagger=\frac{1}{2} \left(b_{i\uparrow}^\dagger b_{j\downarrow}^\dagger-b_{i\downarrow}^\dagger b_{j\uparrow}^\dagger\right).
\end{equation}
So,
\begin{align*}
	:A_{ij}^\dagger A_{ij}:
	&= \frac{1}{4}
	(b_{i\uparrow}^\dagger b_{j\downarrow}^\dagger - b_{i\downarrow}^\dagger b_{j\uparrow}^\dagger) 
	(b_{i\uparrow} b_{j\downarrow} - b_{i\downarrow} b_{j\uparrow})
	\\
	&=
	\frac{1}{4}\left(
b_{i\uparrow}^\dagger b_{j\downarrow}^\dagger b_{i\uparrow} b_{j\downarrow}-
b_{i\uparrow}^\dagger b_{j\downarrow}^\dagger b_{i\downarrow} b_{j\uparrow}-
b_{i\downarrow}^\dagger b_{j\uparrow}^\dagger b_{i\uparrow} b_{j\downarrow}+
b_{i\downarrow}^\dagger b_{j\uparrow}^\dagger b_{i\downarrow} b_{j\uparrow}
	\right)
	\\
	&=\frac{1}{4}
\left[
\left\lbrace
b_{i\downarrow}^\dagger b_{j\uparrow}^\dagger b_{i\downarrow} b_{j\uparrow}+
b_{i\uparrow}^\dagger b_{j\downarrow}^\dagger b_{i\uparrow} b_{j\downarrow}\right\rbrace-
b_{i\uparrow}^\dagger b_{j\downarrow}^\dagger b_{i\downarrow} b_{j\uparrow}-
b_{i\downarrow}^\dagger b_{j\uparrow}^\dagger b_{i\uparrow} b_{j\downarrow}
\right]	
\\
&=
\frac{1}{4} \left(
2S^2-2S_i^z S_j^z-S_i^+ S_j^--S_i^- S_j^+
\right)\qquad\left[\text{Using Eq.\,\ref{A8}, Eq.\,\ref{A9} and Eq.\,\ref{A11}}\right]
\\
&=\frac{1}{4}\left[2S^2-2\bold{S}_i\cdot\bold{S}_j\right]
\\
&=\frac{1}{4}\left[
(S_i^x)^2+(S_i^y)^2+(S_i^z)^2-S(S+1)
+(S_j^x)^2+(S_j^y)^2+(S_j^z)^2-S(S+1)
\right.\\
&\left.
\qquad\qquad +2S^2-2\bold{S}_i\cdot\bold{S}_j\right] 
\qquad\left[\text{Using Eq.\,\ref{A7}}\right]
\\
&=\frac{1}{4}\left[2S^2-2S(S+1)+(\bold{S}_i-\bold{S}_j)^2\right]
\\
&=
\frac{1}{4}(\bold{S}_i-\bold{S}_j)^2-\frac{S}{2}
\end{align*}
Thus the bond operator $A_{ij}^\dagger$ is related with the antiferromagnet order parameter.

\subsection*{Different forms of isotropic Heisenberg Hamiltonian}

A isotropic Heisenberg Hamiltonian is given as,
\begin{equation}
	H=J\sum_{\left\langle i,j\right\rangle} \bold{S}_i\cdot\bold{S}_j.
	\label{A16}
\end{equation}

In the previous section we derived,
\begin{equation}
		:B_{ij}^\dagger B_{ij}:=\frac{1}{4}(\bold{S}_i+\bold{S}_j)^2-\frac{S}{2},\quad
		:A_{ij}^\dagger A_{ij}:=\frac{1}{4}(\bold{S}_i-\bold{S}_j)^2-\frac{S}{2}.
\end{equation}

From this we achieve,
\begin{equation}
	\bold{S}_i\cdot\bold{S}_j=S^2-2A_{ij}^\dagger A_{ij}
	=2:B_{ij}^\dagger B_{ij}:-S^2
	\label{A18}
\end{equation}

Using the relations Eq.\,\ref{A18} in Eq.\,\ref{A16}, we have different forms of Hamiltonian,
\begin{align}
	H_1&=J\sum_{\left\langle i,j\right\rangle} \left[2:B_{ij}^\dagger B_{ij}:-S^2\right]  \\
	&=J\sum_{\left\langle i,j\right\rangle} \left[S^2-2:A_{ij}^\dagger A_{ij}:\right] \\
	&=J\sum_{\left\langle i,j\right\rangle} \left[:B_{ij}^\dagger B_{ij}: -:A_{ij}^\dagger A_{ij}:\right].
	\end{align}

\chapter{Appendix B : Diagonalization of quadratic Hamiltonianin in k-space}\label{appendixB}

\subsection*{Tight binding Hamiltonian}
A tight binding Hamiltonian in k-space can be represented in the following form,
	\begin{align}
		H&=\sum_{\bold{k}} \Psi_\bold{k}^\dagger h(\bold{k}) \Psi_\bold{k},\nonumber\\
		&=\sum_{\bold{k}}\sum_{\alpha\beta} h_{\alpha\beta}(\bold{k}) c_{\bold{k},\alpha}^\dagger c_{\bold{k},\beta}.
		\label{B1}
	\end{align}
	where, $\Psi_{\bold{k}}^\dagger=\left(c_{\bold{k},1}^\dagger c_{\bold{k},2}^\dagger,\cdots ,c_{\bold{k},M}^\dagger \right)$ and $M$ is the number of species or degrees of freedom of a particle.
	$h(\bold{k})$ is $M\times M$ matrix.
	According to Bloch's theorem $\bold{k}$ is conserved quantity for a non-interacting system with translational invariance and so after digonalization the Hamiltonian can be written as,
	\begin{equation}
		H=\sum_{\bold{k},n} E_n(\bold{k}) \eta_{n\bold{k}}^\dagger \eta_{n\bold{k}},
		\label{B2}
	\end{equation}
	where $\eta_{n\bold{k}}^\dagger$ is the creation operator for the single particle state correspond to the energy $E_n(\bold{k})$. 
The operators after diagonalization can be written as linear superposition of the operators before diagonalization,
	\begin{equation}
		\eta_{n\bold{k}}^\dagger=\sum_\alpha u_{\alpha,n}(\bold{k}) c_{\bold{k},\alpha}^\dagger.
		\label{B3}
	\end{equation}
	Using equation Eq.\,\ref{B2}, we prove that,
	\begin{equation}
		\boxed{[H,\eta^\dagger_{n\bold{k}}]=E_n(\bold{k}) \eta_{n\bold{k}}^\dagger}.
		\label{B4}
	\end{equation}
	The L.H.S of the equation Eq.\,\ref{B4},
	\begin{align}
		\text{L.H.S(B4)}&=[H,\eta^\dagger_{n\bold{k}}]\nonumber\\
						&=\left[
\sum_{\bold{k}'}\sum_{\alpha'\beta'} h_{\alpha'\beta'}(\bold{k}') c_{\bold{k}',\alpha'}^\dagger c_{\bold{k}',\beta'},
						\sum_{\alpha} u_{\alpha,n}(\bold{k}) c_{\bold{k},\alpha}^\dagger
						\right]\nonumber\\
						&\qquad\qquad \left[\text{Using equations Eq.\,\ref{B1} and Eq.\,\ref{B3}}\right]\nonumber\\
						&=
						\sum_{\bold{k}}\sum_{\alpha'\alpha} h_{\alpha'\alpha}(\bold{k}) c_{\bold{k},\alpha'}^\dagger u_{\alpha,n}(\bold{k})\nonumber\\
						&\qquad \left[\text{Using bosonic commutation and fermionic anti-commutation relations}\right]\nonumber\\
						&=
\sum_{\bold{k}} \Psi_{\bold{k}}^\dagger h(\bold{k}) u_n(\bold{k})
\label{B5}
	\end{align}	
	From the R.H.S of the equation Eq.\,\ref{B4}, we get,
	\begin{align}
		\text{R.H.S(B4)}&=E_n(\bold{k}) \eta_{n\bold{k}}^\dagger\nonumber\\
		&=E_n(\bold{k}) \sum_\alpha u_{\alpha,n}(\bold{k}) c_{\bold{k},\alpha}^\dagger\nonumber\\
		&=E_n(\bold{k}) \Psi_{\bold{k}}^\dagger u_n(\bold{k}).
		\label{B6}
	\end{align}
	Equating equations Eq.\,\ref{B5} and  Eq.\,\ref{B6}, we get,
	\begin{equation}
		\boxed{h(\bold{k}) u_n(\bold{k})= E_n(\bold{k}) u_n(\bold{k})}
	\end{equation}
	Thus the diagonalization of a tight binding Hamiltonian Eq.\,\ref{B1} is simply a diagonalization problem of matrix $h(\bold{k})$.

\subsection*{Bogoliubov Hamiltonian}
In this section we show the procedure to diagonalize the Bogoliubov Hamiltonian for the bosons.
A general quadratic bosonic Hamiltonian in k-space is given by,
\begin{equation}
			H=\frac{1}{2}
			\begin{pmatrix}
				a_{\bold{k}} & a_{-\bold{k}}^\dagger 
			\end{pmatrix}
			\underbrace{
			\begin{pmatrix}
				\alpha(\bold{k}) & \gamma^\dagger(\bold{k})\\
				\gamma(\bold{k}) & \alpha^T(-\bold{k})
			\end{pmatrix}
			}_{M'(\bold{k})}
			\begin{pmatrix}
				a_{\bold{k}}^\dagger \\ a_{-\bold{k}}
			\end{pmatrix}.
		\end{equation}
		Matrix $M'$ is called the coefficient matrix.
		The explicit form of the Hamiltonian is given by,
		\begin{equation}
			H=\frac{1}{2}\sum_{ij} \sum_{\bold{k}} \left[
			\alpha_{ij}(\bold{k})a_{\bold{k}i} a_{\bold{k}j}^\dagger
			 +\alpha_{ij}(-\bold{k}) a_{-\bold{k}j}^\dagger a_{-\bold{k}i}
			 +\gamma_{ij}(\bold{k}) a_{-\bold{k}i}^\dagger a_{\bold{k}j}^\dagger
			 +\gamma^*_{ij}(\bold{k}) a_{\bold{k}j} a_{-\bold{k}i}
			  \right].
			  \label{B8}
		\end{equation}
		According to Bloch's theorem the diagonalized form of the Hamiltonian is given as,
		\begin{equation}
			H=\sum_{n,\bold{k}} E_n(\bold{k}) \eta_{\bold{k}n}^\dagger \eta_{\bold{k}n}.
			\label{B9}
		\end{equation}
		As discussed in the section Sec.\,\ref{sec2.4}, the operator $\eta_{\bold{k}n}$ can be expressed as linear combinations of the $a_{\bold{k}i}^\dagger$ operators,
		\begin{equation}
			\boxed{			
			\eta_{n\bold{k}}=\sum_{i} \left[
			u_{n,i}(\bold{k}) a_{\bold{k},i}^\dagger
			+v_{n,i}(\bold{k}) a_{-\bold{k},i}		
			\right]}
			.
			\label{B10}
		\end{equation}
		The equation eq.\,\ref{B4} is still valid in this case and rewriting the equation,
		\begin{equation}
		\boxed{[H,\eta^\dagger_{\bold{k}n}]=E_n(\bold{k}) \eta_{\bold{k}n}^\dagger}.
		\label{B11}
	\end{equation}
	The R.H.S. of the equation Eq.\,\ref{B11},
	\begin{align}
		\text{R.H.S(B11)}&=E_n(\bold{k})\eta_{\bold{k}n}^\dagger\nonumber\\
		&=E_n(\bold{k}) \sum_{\bold{k},i} \left[
			u_{n,i}(\bold{k}) a_{\bold{k},i}^\dagger
			+v_{n,i}(\bold{k}) a_{-\bold{k},i}		
			\right]
			\nonumber\\
		&=E_n(\bold{k}) \begin{pmatrix}
			 		a_{\bold{k}i}^\dagger & a_{-\bold{k}i}
				\end{pmatrix}
				\begin{pmatrix}
					u_n(\bold{k}) \\ v_n(\bold{k})
				\end{pmatrix}
				\label{B12}
	\end{align}
	The L.H.S. of the equation Eq.\,\ref{B11},
	\begin{align}
		\text{L.H.S(B11)}&=
		[H,\eta^\dagger_{\bold{k}n}]\nonumber\\
		=&\left[\frac{1}{2}\sum_{ij} \sum_{\bold{k}'} \left\lbrace
			\alpha_{ij}(\bold{k}')a_{\bold{k}'i} a_{\bold{k}'j}^\dagger
			 +\alpha_{ij}(-\bold{k}') a_{-\bold{k}'j}^\dagger a_{-\bold{k}'i}
			 +\gamma_{ij}(\bold{k}') a_{-\bold{k}'i}^\dagger a_{\bold{k}'j}^\dagger
			 +\gamma^*_{ij}(\bold{k}') a_{\bold{k}'j} a_{-\bold{k}'i}
			  \right\rbrace\right.
			  \nonumber\\ 
			  &\qquad\qquad\qquad\qquad 
			  \left.,\sum_{p}\left\lbrace
			u_{n,p}(\bold{k}) a_{\bold{k},p}^\dagger
			+v_{n,p}(\bold{k}) a_{-\bold{k},p}		
			\right\rbrace\right]\nonumber\\
			&\qquad\qquad
			\left[\text{Using Eq.\,\ref{B8} and Eq.\,\ref{B10}}\right]	
			\nonumber\\
			&=\sum_{i,j}
			\left[\left\lbrace
			\alpha_{ji}(\bold{k}) u_{nj} a_{\bold{k}i}^\dagger
			+\gamma^*_{ij}(\bold{k}) u_{nj}(\bold{k})a_{-\bold{k}i}
			\right\rbrace
			\right.
			\nonumber\\
			&\left.
			-\left\lbrace
			 \alpha_{ij}(-\bold{k}) v_{np}(\bold{k}) a_{-\bold{k}i}
			 +\gamma_{ji}(\bold{k}) v_{np}(\bold{k}) a_{\bold{k}i}^\dagger\right\rbrace
			 \right]
			 \nonumber\\
			 &[\text{Using the reletaion, }\quad \gamma(\bold{k})=\gamma(-\bold{k})^T]
			 \nonumber\\
			 &=\begin{pmatrix}
			 		a_{\bold{k}i}^\dagger & a_{-\bold{k}i}
				\end{pmatrix}			 
				\begin{pmatrix}
					\alpha^T(\bold{k}) & -\gamma^T(\bold{k})\\
					\gamma^*(\bold{k}) & -\alpha(-\bold{k})\\
				\end{pmatrix}					
				\begin{pmatrix}
					u_n(\bold{k}) \\ v_n(\bold{k})
				\end{pmatrix}
				\nonumber\\
				&=\begin{pmatrix}
			 		a_{\bold{k}i}^\dagger & a_{-\bold{k}i}
				\end{pmatrix}			 
				\begin{pmatrix}
					\alpha(\bold{k}) & \gamma^\dagger(\bold{k})\\
					-\gamma(\bold{k}) & -\alpha^T(-\bold{k})\\
				\end{pmatrix}^T					
				\begin{pmatrix}
					u_n(\bold{k}) \\ v_n(\bold{k})
				\end{pmatrix}
				\label{B13}
	\end{align}
	Equating Eq.\,\ref{B12} and Eq.\,\ref{B13},
	\begin{equation}
		\boxed{\begin{pmatrix}
					\alpha(\bold{k}) & \gamma^\dagger(\bold{k})\\
					-\gamma(\bold{k}) & -\alpha^T(-\bold{k})\\
				\end{pmatrix}^T					
				\begin{pmatrix}
					u_n(\bold{k}) \\ v_n(\bold{k})
				\end{pmatrix}
				=E_n(\bold{k})
				\begin{pmatrix}
					u_n(\bold{k}) \\ v_n(\bold{k})
				\end{pmatrix}
				}
				\label{B15}
	\end{equation}
	Thus the dynamic matrix in terms of coefficient matrix is $(\sigma_z M')^T$, where,
	\begin{equation}
		\sigma_z=\begin{pmatrix}
		\mathbb{I} & 0\\
		0 & -\mathbb{I}
		\end{pmatrix}.
	\end{equation}
	$\mathbb{I}$ is the identity matrix of size $N\times N$ if the coefficient matrix is of dimension $2N\times 2N$.
	Making complex conjugate of the both sides of equation Eq.\,\ref{B15}, we get,
	\begin{equation}
		\boxed{\begin{pmatrix}
					\alpha(-\bold{k}) & \gamma^\dagger(-\bold{k})\\
					-\gamma(-\bold{k}) & -\alpha^T(\bold{k})\\
				\end{pmatrix}^T					
				\begin{pmatrix}
					v^*_n(-\bold{k}) \\ u^*_n(-\bold{k})
				\end{pmatrix}
				=-E_n(-\bold{k})
				\begin{pmatrix}
					v^*_n(-\bold{k}) \\ u^*_n(-\bold{k})
				\end{pmatrix}
				}
				\label{B17}
	\end{equation}
	Thus if there is a positive eigen-value $E(\bold{k})$ present, then a negative eigen-value $-E(-\bold{k})$. 
	Moreover the eigenvectors of the two energies are also related with each other as in equations Eq.\,\ref{B15} and Eq.\,\ref{B17}.
	The negative eigenvalues have no physical significance as discussed in the section Sec.\,\ref{sec2.4}.
	However in numerical calculation presence of the equal negative eigenvalues verifies the correctness of the constructed Hamiltonian.
	The coefficient matrix $M'$ should be positive definite (A positive definite matrix is a Hermitian matrix with eigenvalues greater than zero), so that the dynamic matrix $(\sigma_zM')^T$ to be diagonalizable and that is the reason of the presence of negative and positive eigenvalues.

	Furthermore the normalization condition of the eigenvectors can be derived by using the bosonic commutation relation $\left[\eta_{n\bold{k}},\eta_{n'\bold{k}'}\right]=\delta_{nn'}\delta_{\bold{k}\bold{k}'}$ and equation Eq.\,\ref{B11}. The normalization condition is given by,
	\begin{align}
		\sum_i \left[|u_{n,i}(\bold{k})|^2-|v_{n,i}(\bold{k})|^2\right]&=1 \nonumber\\
		\text{Or,}\quad\begin{pmatrix}
			u_n^*(\bold{k}) & v_n^*(\bold{k})
		\end{pmatrix}
		\sigma_z
		\begin{pmatrix}
			u_n(\bold{k}) \\ v_n(\bold{k})
		\end{pmatrix}
		&=1
	\end{align}

	In a same manner using the anti-commutation relations of fermions, a similar Bogolubov-Valatin formalism can be developed and it can be shown that the coefficient matrix and dynamic matrix is same.

	\subsection*{Bogoliubov-Valatin transformation of Bosons as para-unitary transformation}
	
	In presence of pair-creation and annihilation operator in the Hamiltonian, the Hamiltonian can be represented as,
\begin{equation}
\pazocal{H}=\frac{1}{2} \sum_{\alpha, \beta, \bold{k}} \Psi^\dagger_{\alpha,\bold{k}} H_{\alpha,\beta}(\bold{k}) \Psi_{\alpha,\bold{k}},
\end{equation}
where, $\Psi_{\alpha,\bold{k}}=(\hat{a}_{\alpha\bold{k}},\hat{a}^\dagger_{\alpha ,-\bold{k}})^T$ is the Nambu-spinor. Here $\hat{a}_{\alpha\bold{k}}$ is the set of all annihilation operator.

The Hamiltonian is diagonalized by using para-unitary transformation as,
\begin{equation}
T^\dagger(\bold{k}) \pazocal{H}(\bold{k}) T(\bold{k})=\epsilon(\bold{k}),
\label{eqIa}
\end{equation}
where,
\begin{equation}
\epsilon(\bold{k})=\begin{pmatrix}
E_1(\bold{k}) & & & &\\
& ... & & & &\\
& & E_{N}(\bold{k}) & & &\\
& & &  E_1(-\bold{k}) & &\\
& & & & ...&\\
& & & & & E_{N}(-\bold{k})
\end{pmatrix}
\end{equation}
where $N$ is the number of sub-lattices multiplied by the number of types of bosons at each site. Moreover the para-unitary matrices follow the following rule,
\begin{equation}
T^\dagger(\bold{k}) \sigma_3 T(\bold{k})=T(\bold{k}) \sigma_3 T^\dagger(\bold{k})=\sigma_3,
\label{eqIIIa}
\end{equation}
using the above relation, we can re-write Eq.\,\ref{eqIa} as,
\begin{equation}
\sigma_3 \pazocal{H}(\bold{k}) T(\bold{k})=T(\bold{k})\sigma_3\epsilon(\bold{k})
\label{eqIIa}
\end{equation}
Taking the $(m,n)$-element of the matrix of the Eq.\,\ref{eqIIa}, we get,
\begin{align*}
\left(\sigma_3 \pazocal{H}_\bold{k} T_\bold{k}\right)_{mn}&=(T(\bold{k})\sigma_3\epsilon(\bold{k}))_{mn} \\
\text{or, }\sum_p (\sigma_3 \pazocal{H}_\bold{k})_{mp} (T_\bold{k})_{pn}&=\sum_p (T_\bold{k} \sigma_3)_{mp} (\epsilon_{\bold{k}})_{pn} \\
\text{or, } \sum_p (\sigma_3 \pazocal{H}_\bold{k})_{mp} (T_\bold{k})_{pn}&=\sum_p (T_\bold{k})_{mp} (\sigma_3)_{pp} (\epsilon_{\bold{k}})_{pn} \\,
\end{align*}
Define $\uuu{n}{m}=(T_{\bold{k}})_{mn}\label{Deq1a}$ and $\uuu{n}{m}^*=(T^\dagger_{\bold{k}})_{mn}\label{Deq2a}$,
\begin{align*}
\sum_p (\sigma_3 H_{\bk})_{mp} \uuu{n}{p}&=\sum_p \uuu{p}{m} (\sigma_3)_{pp} (\epsilon_{\bk})_{pn}\\
&=\sum_p \uuu{p}{m} (\sigma_3 \epsilon_{\bk})_{pn}\\
&=\sum_p \uuu{p}{m} (\sigma_3 \epsilon_{\bk})_{pn} \delta_{pn}\\
&=\uuu{n}{m}(\sigma_3 \epsilon_\bk)_{nn}\\
\text{or, } (\sigma_3 H_\bk)\ket{\uu{n}}&=(\sigma_3\epsilon_\bk) \ket{\uu{n}}
\end{align*}

So, we have,
\begin{equation}
\boxed{(\sigma_3 H_\bk)\ket{\uu{n}}=(\sigma_3\epsilon_\bk) \ket{\uu{n}}}
\label{Geq1a}
\end{equation}
The transpose of the equation gives,
\begin{equation}
\boxed{\bra{\uu{n}}\pazocal{H}_\bk\sigma_3=\bra{\uu{n}}(\sigma_3 \epsilon(\bk))}
\label{Geq2a}
\end{equation}

\subsubsection*{The completeness relation of the Bogoliubov-Nambu spinors}
From Eq.\,\ref{eqIIIa} we get the following ortho-normality relation,
\begin{equation}
\boxed{\bra{\uu{n}}\sigma_3\ket{\uu{m}}=\delta_{nm} (\sigma_3)_{nn}}
\end{equation} 

Thus, we can expand any state in the Hilbert space as,
\begin{equation}
\ket{\Phi}=\sum_n c_n \sigma_3 \ket{\uu{n}}
\label{eqIVa}
\end{equation}

Multiplying $\bra{\uu{m}}$ at the both sides,
\begin{align*}
\braket{\uu{m}}{\Phi}&=\sum_n c_n \bra{\uu{m}}\sigma_3\ket{\uu{n}}\\
&=\sum_n c_n \delta_{mn} (\sigma_3)_{nn}\\
&=c_m (\sigma_3)_{mm}\\
\therefore c_m&=(\sigma_3)_{mm} \braket{\uu{m}}{\Phi}
\end{align*}

Using the above equation in Eq.\,\ref{eqIVa}, we get,
\begin{align*}
\ket{\Phi}&=\sum_n c_n \sigma_3 \ket{\uu{n}}\\
&=\sum_n (\sigma_3)_{nn} \braket{\uu{n}}{\Phi} \sigma_3 \ket{\uu{n}}\\
&=\left[\sum_n (\sigma_3)_{nn} \sigma_3 \ket{\uu{n}}\bra{\uu{n}}\right]\ket{\Phi}
\end{align*}

So the completeness relation,
\begin{equation}
\boxed{\sum_n (\sigma_3)_{nn} \sigma_3 \ket{\uu{n}}\bra{\uu{n}}=\mathbb{I}}
\label{Geq3a}
\end{equation}

The equation also can be re-writen as,
\begin{equation}
\boxed{\sum_n (\sigma_3)_{nn} \ket{\uu{n}}\bra{\uu{n}}\sigma_3 =\mathbb{I}}
\label{Geq4a}
\end{equation}

\chapter{Appendix C : Berry-phase, Berry-curvature and Chern number}\label{appendixC}

\subsection*{Berry-phase in quantum system}
The Schrödinger's equation provides the time-evolution of the wave-function $\ket{\Psi}$ of a system with a Hamiltonian $H$ as,
\begin{equation}
	H(t) \ket{\Psi(t)}=i\hbar\frac{\partial }{\partial t} \ket{\Psi(t)}.
	\label{C1}
\end{equation}
The adiabatic process is a slow-process of evolving the Hamiltonian such that the eigenstates do not change and so an n-th eigen-state $\ket{\psi_n(t)}$ of the system is an instantaneous eigenstate of the system and follows the equation,
\begin{equation}
	H(\bold{R}(t))\ket{\psi_n(\bold{R}(t))}=E_n(\bold{R}(t)) \ket{\psi_n(\bold{R}(t))},
	\label{C2}
\end{equation}
where $\bold{R}(t)$ represents the set of time-dependent parameters of the Hamiltonian. 
The above equation is equivalent to the time-independent Schrödinger's equation for a time $t$.
Differentiating both sides of the equation Eq.\ref{C2} we get,
\begin{align}
	&\dot{H}(\bold{R}(t))\ket{\psi_n(\bold{R}(t))}
	+H(\bold{R}(t))\ket{\dot{\psi}_n(\bold{R}(t))}
	=\dot{E}_n(\bold{R}(t)) \ket{\psi_n(\bold{R}(t))}
	+E_n(\bold{R}(t)) \ket{\dot{\psi}_n(\bold{R}(t))}
	\nonumber\\
	&\text{Or,}\qquad\boxed{\braket{\psi_m}{\dot{\psi}_n}=\frac{\bra{\psi_m}\dot{H}\ket{\psi_n}}{\left|E_n-E_m\right|}.}
	\,\left[\text{squeezing $\bra{\psi_m}$ on both sides and $m\neq n$}\right]
\end{align}
According to adiabatic assumption the state $\ket{\psi_n}$ should not change to another state $\ket{\psi_m}$ and so $\braket{\psi_m}{\dot{\psi}_n}\rightarrow 0$ which gives,
\begin{equation}
	\boxed{\frac{\bra{\psi_m}\dot{H}\ket{\psi_n}}{\left|E_n-E_m\right|} 
	\rightarrow 0}
	\label{C4}
\end{equation}
The L.H.S. term is called characteristic time of transition from m-th eigenstate to n-th eigenstate and vice-versa.
For an adiabatic process the characteristic time of transition should be zero.

We assume the following ansatz is the solution of the time-dependent Schrödinger's equation Eq.\,\ref{C1},
\begin{equation}
	\ket{\psi_n(\bold{R}(t))}= 
	\sum_n c_n(\bold{R}(t)) \exp\left(-\frac{i}{\hbar}\int^t_0 E_n(\tau)d\tau\right) \ket{\psi_n(\bold{R}(t))}
	\label{C3}
\end{equation}
Substituting the ansatz into the equation Eq.\,\ref{C1} we get,
\begin{align}
   H(t)\sum_n c_n(\bold{R}(t)) &\exp\left(-\frac{i}{\hbar}\int^t_0 E_n(\tau)d\tau\right) \ket{\psi_n(\bold{R}(t))}
   \nonumber\\
   &=i\hbar\left[\sum_n \dot{c}_n(\bold{R}(t)) \exp\left(-\frac{i}{\hbar}\int^t_0 E_n(\tau)d\tau\right) \ket{\psi_n(\bold{R}(t))}
	\right.\nonumber\\   
   &+\sum_n c_n(\bold{R}(t)) \exp\left(-\frac{i}{\hbar}\int^t_0 E_n(\tau)d\tau\right) \ket{
   \dot{\psi}_n(\bold{R}(t))}
	\nonumber\\   
   &\left.-\frac{i}{\hbar}\sum_n E_n(t) c_n(\bold{R}(t)) \exp\left(-\frac{i}{\hbar}\int^t_0 E_n(\tau)d\tau\right) \ket{\psi_n(\bold{R}(t))}\right]
   \nonumber\\
	\text{OR, }\, 0=i\hbar\left[\sum_n \dot{c}_n(\bold{R}(t)) \right.&\exp\left(-\frac{i}{\hbar}\int^t_0 E_n(\tau)d\tau\right) \ket{\psi_n(\bold{R}(t))}
	\nonumber\\
   &\left.+\sum_n c_n(\bold{R}(t)) \exp\left(-\frac{i}{\hbar}\int^t_0 E_n(\tau)d\tau\right) \ket{
   \dot{\psi}_n(\bold{R}(t))}\right]
   \nonumber\\
   \left[\text{Using the Eq.\,\ref{C2}.2 }\right.& \left.\text{for the L.H.S., the L.H.S. and the last tem of R.H.S. cancels}\right]
   \nonumber\\
   \text{OR, }\,\dot{c}_n=-c_n\braket{\psi_n}{\dot{\psi}_n}
   &-\sum_{n\neq m} c_m \braket{\psi_n}{\dot{\psi}_m}\exp\left[-\frac{i}{\hbar} \int^t_0 \left(E_m(\tau)-E_n(\tau)\right)d\tau\right]
   \label{C6}
\end{align}
According to adiabatic assumption the last term of Eq.\,\ref{C6} drops out (see Eq.\,\ref{C3} and Eq.\,\ref{C4})and the solution of $c_n(t)$ is given by,
\begin{equation}
	\boxed{c_n(t)=c_n(0) e^{i\gamma_n(t)},}
\end{equation}
where,
\begin{equation}
	\boxed{\gamma_n(t)=i \int_0^t \bra{\psi_n\left(\bold{R}(\tau)\right)} 
	\frac{d}{d\tau} \ket{\psi_n(\bold{R}\left(\tau)\right)} d\tau
	}
\end{equation}
So for an adiabatic process the wavefunction of system gains only a phase factor $\gamma_n$, which is known as Berry-phase.
The berry-phase is independent of time, but depends on the path in the parameter space along which the system evolves, as shown below,
\begin{align}
	\gamma_n(t)&=i \int_0^t \bra{\psi_n\left(\bold{R}(\tau)\right)} 
	\frac{d}{d\tau} \ket{\psi_n(\bold{R}\left(\tau)\right)} d\tau
    \nonumber\\	
	&=i \int_0^t \bra{\psi_n\left(\bold{R}(\tau)\right)} 
	\nabla_{\bold{R}} \ket{\psi_n(\bold{R}\left(\tau)\right)} \frac{d\bold{R}}{d\tau} d\tau \quad \left[\text{Using chain rule of differentiation}\right]
	\nonumber\\
	 &\boxed{\therefore\gamma_n=i \int_\pazocal{C} \bra{\psi_n\left(\bold{R}\right)} 
	\nabla_{\bold{R}} \ket{\psi_n\left(\bold{R}\right)} d\bold{R},}
\end{align}
where $\pazocal{C}$ is the path in the parameter space along which the system is evolving.

\subsection*{Phase in Aharnov-Bohm effect}
In presence of magnetic vector potential $\pazocal{A}(\bold{r})$, the time dependent Schrödinger's equation is,
\begin{equation}
	\left[\frac{1}{2m}\left(i\hbar\nabla+e\pazocal{A}(\bold{r})\right)^2+V(\bold{r})\right]\Psi=i\hbar \frac{\partial \Psi}{\partial t},
	\label{C10}
\end{equation}
where $V(\bold{r})$ is a non-electric or non-magnetic potential.
Let us assume the solution of the equation in presence of magnetic vector potential,
\begin{equation}
	\boxed{\Psi(\bold{r},t)=e^{ig(\bold{r})} \Psi^\prime(\bold{r},t)},
\end{equation}
where $\Psi^\prime(\bold{r},t)$ is the wavefunction in absence of the magnetic vector potential and 
\begin{equation}
	g(\bold{r})=\frac{e}{\hbar}\int^\bold{r}_0 \pazocal{A}(\bold{r}') d\bold{r}'.
	\label{C12}
\end{equation}
It is important to note that when the particle moving in a space without a magnetic field $\bold{B}=\nabla\times \pazocal{A}=0$ i.e. $\pazocal{A}(\bold{r})$ is irrotational. 
So, $g(\bold{r})$ is independent of the path of the integration in Eq.\,\ref{C12} and only a function of position imposing that the gauge of $\pazocal{A}$ is fixed over the space.
 Using the Leibniz integral rule, the gradient of $g(\bold{r})$ is,
 \begin{equation}
 	\nabla g(\bold{r})=\frac{e}{\hbar} \pazocal{A}.
 	\label{C13}
 \end{equation}
 Taking gradient of both sides of equation Eq.\,\ref{C10} and using Eq,\,\ref{C13},
 \begin{equation}
	(-i\hbar\nabla-e\pazocal{A})\Psi=-i\hbar e^{ig(\bold{r})}\nabla\Psi'
 \end{equation}
 Again taking gradient of both sides,
 \begin{equation}
 	(-i\hbar\nabla-e\pazocal{A})^2\Psi=-\hbar^2 e^{ig(\bold{r})}\nabla^2\Psi'
 	\label{C15}
 \end{equation}
 Using Eq.\,\ref{C15} in Eq.\,\ref{C10} we get,
 \begin{equation}
 	\left[-\frac{\hbar^2}{2m}\nabla^2+V(\bold{r})\right]\Psi^\prime(\bold{r},t)=i\hbar \frac{\partial }{\partial t}\Psi^\prime(\bold{r},t).
 \end{equation}
 Thus we proved that $\Psi^\prime$ is the wave-function in absence of magnetic potential.
 So the relations Eq.\,\ref{C12} correctly relates the wavefunctions in absence and in presence of magnetic vector potential (note that the magnetic field is considered zero) respectively.

\subsection*{Berry-connection and Berry-curvature in k-space}
The derivations in this section are from the reference Ref.\cite{KaiSunLectureNote}.
Similar to the Fourier transformation the real-space wave function can be expressed in reciprocal space using Bloch-wave-functions as,
\begin{align}
	\Psi(\bold{r})&=\sum_n\int d\bold{k} \Psi(\bold{k}) \psi_{n,\bk}(\bold{r})\nonumber\\
	&=\sum_n\int d\bold{k} \Psi(\bold{k}) u_{n,\bold{k}}(\bold{r}) e^{i\bold{k}\cdot\bold{r}}.
\end{align}
Operating position-operator,
\begin{align*}
\hat{r}\Psi(\bold{r})&=\bold{r}\Psi(\bold{r})\\
                     &=\bold{r} \sum_n\int d\bold{k} \Psi(\bold{k}) u_{n,\bold{k}} e^{i\bold{k}\cdot\bold{r}}\\
                     &=\sum_n\int d\bold{k} \Psi(\bold{k}) u_{n,\bold{k}}(\bold{r}) \left(-i\nabla_{\bold{k}}e^{i\bold{k}\cdot\bold{r}}\right)\\
                     &=\sum_n\int d\bold{k} i\nabla_{\bold{k}}\left(\Psi(\bold{k}) u_{n,\bold{k}}(\bold{r})\right) e^{i\bold{k}\cdot\bold{r}}
                     \,\left[\text{Using integration by parts}\right]\\
                     &=\sum_n\int d\bold{k} \left(i\nabla_{\bold{k}}\Psi(\bold{k}) u_{n,\bold{k}}(\bold{r})+ \Psi(\bold{k}) i\nabla_{\bold{k}}u_{n,\bold{k}}(\bold{r})\right) e^{i\bold{k}\cdot\bold{r}}\\
                     &=\sum_n\int d\bold{k} \left[i\nabla_{\bold{k}}\Psi(\bold{k}) u_{n,\bold{k}}(\bold{r})+ \Psi(\bold{k}) \int d\bold{r}' \delta(\bold{r}-\bold{r}') i\nabla_{\bold{k}}u_{n,\bold{k}}(\bold{r}')\right] e^{i\bold{k}\cdot\bold{r}}\\
                     &=\sum_n\int d\bold{k} \left[i\nabla_{\bold{k}}\Psi(\bold{k}) u_{n,\bold{k}}(\bold{r})+ \Psi(\bold{k}) \int d\bold{r}' \delta(\bold{r}-\bold{r}') i\nabla_{\bold{k}}u_{n,\bold{k}}(\bold{r}')\right] e^{i\bold{k}\cdot\bold{r}}\\
                     &=\sum_n\int d\bold{k} \left[i\nabla_{\bold{k}}\Psi(\bold{k}) u_{n,\bold{k}}(\bold{r})+ \Psi(\bold{k}) \int d\bold{r}' \sum_m u_{m,\bold{k}}^*(\bold{r}') u_{m,\bold{k}}(\bold{r}) i\nabla_{\bold{k}}u_{n,\bold{k}}(\bold{r}')\right] e^{i\bold{k}\cdot\bold{r}}\\
                     &=\sum_n\int d\bold{k} \left[i\nabla_{\bold{k}}\Psi(\bold{k}) u_{n,\bold{k}}(\bold{r})+ \Psi(\bold{k}) \sum_m u_{m,\bold{k}}(\bold{r}) \left\lbrace \int d\bold{r}' u_{m,\bold{k}}^*(\bold{r}') i\nabla_{\bold{k}}u_{n,\bold{k}}(\bold{r}')\right\rbrace\right] e^{i\bold{k}\cdot\bold{r}}\\
                     &=\sum_n\int d\bold{k} \left[i\nabla_{\bold{k}}\Psi(\bold{k}) u_{n,\bold{k}}(\bold{r})+ \Psi(\bold{k}) u_{n,\bold{k}}(\bold{r}) \left\lbrace \int d\bold{r}' u_{n,\bold{k}}^*(\bold{r}') i\nabla_{\bold{k}}u_{n,\bold{k}}(\bold{r}')\right\rbrace\right] e^{i\bold{k}\cdot\bold{r}}\\
                    &\qquad\qquad \left[\int d\bold{r}' u_{m,\bold{k}}^*(\bold{r}') i\nabla_{\bold{k}}u_{n,\bold{k}}(\bold{r}')\rightarrow 0,\text{ for $m\neq n$, adiabatic assumption}\right]\\
                     &=\sum_n\int d\bold{k} \left[i\nabla_{\bold{k}}\Psi(\bold{k}) u_{n,\bold{k}}(\bold{r})+ \Psi(\bold{k})  u_{n,\bold{k}}(\bold{r}) \pazocal{A}_{n}(\bold{k})\right] e^{i\bold{k}\cdot\bold{r}}\\
                     &=\sum_n\int d\bold{k} \left[\left\lbrace i\nabla_{\bold{k}}+  \pazocal{A}_{n}(\bold{k})\right\rbrace\Psi(\bold{k})\right] \psi_{m,\bold{k}}.
\end{align*}
Thus the position operator in reciprocal space representation for a particle at $n$-th band in momentum space under adiabatic assumption is given by,
\begin{equation}
	\hat{r}=i\nabla_{\bold{k}}+  \pazocal{A}_{n}(\bold{k})\quad\text{and}\quad
	\pazocal{A}_n(\bold{k})=i\int d\bold{r}\, u_{n,\bold{k}}^*(\bold{r}) \nabla_{\bold{k}}u_{n,\bold{k}}(\bold{r})
	\label{C18}
\end{equation}
 The equation of motion of the position operator,
 \begin{align*}
 	\frac{d\hat{r}_x}{dt}&=\frac{i}{\hbar}\left[\hat{H},\hat{r}_x\right]\\
 	&=\frac{i}{\hbar}\left[\hat{H},i\partial_{k_x}+  \pazocal{A}^x_{n}(\bold{k})\right]\\
 	&=-\frac{1}{\hbar}\left[\hat{H},\partial_{k_x}\right]+  \frac{i}{\hbar}\left[\hat{H},\pazocal{A}^x_{n}(\bold{k})\right]\\
 	&=\frac{1}{\hbar}\frac{d\hat{H}(\hat{r},\bold{k})}{dk_x}+  \frac{i}{\hbar}\left[\hat{H},\pazocal{A}^x_{n}(\bold{k})\right]\\
 	&=\frac{1}{\hbar}\partial_{k_x} \hat{H}(\hat{r},\bold{k})
 	  +\frac{1}{\hbar}\sum_j \partial_{r_j} \hat{H} \frac{\partial r_j}{\partial k_x}
 	  +\frac{d\pazocal{A}^x_n}{dt}\\
 	&=\frac{1}{\hbar}\partial_{k_x} \hat{H}(\hat{r},\bold{k})
 	  +\frac{1}{\hbar}\sum_j \partial_{r_j} \hat{H} \frac{\partial \pazocal{A}^j_n}{\partial k_x}
 	  +\sum_j \frac{d\pazocal{A}^x_n}{\partial k_j} \frac{d k_j}{dt}\\
 	&=\frac{1}{\hbar}\partial_{k_x} \hat{H}(\hat{r},\bold{k})
 	  +\frac{1}{\hbar}\sum_j \frac{-i}{\hbar}\left[H,\hbar\hat{k}_j\right] \frac{\partial \pazocal{A}^j_n}{\partial k_x}
 	  +\sum_j \frac{d\pazocal{A}^x_n}{\partial k_j} \frac{d k_j}{dt}\\
 	  &=\frac{1}{\hbar}\partial_{k_x} \hat{H}(\hat{r},\bold{k})
 	  +\sum_j \left(-\frac{d k_j}{dt}\right) \frac{\partial \pazocal{A}^j_n}{\partial k_x}
 	  +\sum_j \frac{d\pazocal{A}^x_n}{\partial k_j} \frac{d k_j}{dt}\\
 	  &=\frac{1}{\hbar}\partial_{k_x} \hat{H}(\hat{r},\bold{k})
 	  +\sum_j \left(-\frac{d k_j}{dt}\right) \frac{\partial \pazocal{A}^j_n}{\partial k_x}
 	  +\sum_j \frac{d\pazocal{A}_n^x}{\partial k_j} \frac{d k_j}{dt}\\
 	  &=\frac{1}{\hbar}\partial_{k_x} \hat{H}(\hat{r},\bold{k})
 	  +\sum_j \frac{d k_j}{dt}\left(\frac{d\pazocal{A}_n^x}{\partial k_j} 
 	  - \frac{\partial \pazocal{A}_n^j}{\partial k_x}\right)\\
 	  &\approx \frac{1}{\hbar}\partial_{k_x} \epsilon_n(\bold{k})
 	  +\sum_j \frac{d k_j}{dt}\left(\frac{d\pazocal{A}_n^x}{\partial k_j} 
 	  - \frac{\partial \pazocal{A}_n^j}{\partial k_x}\right)\quad\left[\text{For low magnetic field $\hat{H}(\hat{r},\bold{k})\approx \epsilon(\bold{k})$}\right]\\
 \end{align*}
 The equation of motion become,
 \begin{equation}
 	\frac{d\bold{r}}{dt}=\frac{1}{\hbar}\nabla_{\bold{k}}\epsilon_n(\bold{k})+\frac{d\bold{k}}{dt}\cdot\left(\nabla_{\bold{k}}\times \pazocal{A}_n\right)
 	\label{C19}
 \end{equation}

 \begin{equation}
		\hat{p}=-i\nabla_{\bold{r}}-\frac{q}{c}\bold{A}(\bold{r}),\quad
		\frac{d\bold{r}}{dt}=-q\nabla_{\bold{r}}\phi+\frac{q}{c}\frac{d\bold{r}}{dt}\times \left(\nabla_{\bold{r}}\times \bold{A}(\bold{r})\right).
		\label{C20}
	\end{equation}
	Comparing equations Eq.\,\ref{C20} with the equations Eq.\,\ref{C18} and Eq.\,\ref{C19} we understand that the quantity $\pazocal{A}_n$ is equivalent to the magnetic vector potential $\bold{A}(\bold{r})$ and so $\pazocal{A}_n$ is the Berry-connection.

 \subsection*{Gauge-independent equations for Berry-curvature of 2D-materials}

 The Berry-curvature for 2D-lattices is,
\begin{align}
\boldsymbol{\Omega}_{m,\bold{k}}&=\gradK\times i\braket{\uu{m}}{\gradK\uu{m}} \nonumber \\
&=\left(\icap \ddkx+\jcap \ddky\right)\times i\braket{\uu{m}}{\left(\icap \ddkx+\jcap \ddky\right)\uu{m}}\nonumber \\
&=\kcap i \ddkx\braket{\uu{m}}{\dudky{m}}-\kcap i \ddky\braket{\uu{m}}{\dudkx{m}}\nonumber\\
&=\left[i\braket{\dudkx{m}}{\dudky{m}}-i\braket{\dudky{m}}{\dudkx{m}}\right]\kcap.\nonumber
\end{align}
 Because Berry curvature is single component for the case of 2D-materials so we can write down,
 \begin{equation}
 \Omega_{m,\bold{k}}=i\left[\braket{\dudkx{m}}{\dudky{m}}-\braket{\dudky{m}}{\dudkx{m}}\right].
\end{equation}   
This formula of Berry curvature is not a gauge independent as discussed in the section Sec.\,\ref{sec2.5.2}. 
Using the completeness relation $\mathbb{I}=\sum_n \ket{\uu{n}}\bra{\uu{n}}$, in the above expression we get,
\begin{equation}
\Omega_{m,\bold{k}}=i\left[\sum_{n\neq m} \braket{\dudkx{m}}{\uu{n}}\braket{\uu{n}}{\dudky{m}}-(k_x\leftrightarrow k_y)\right],
\label{eq1}
\end{equation}
where the summation over $n=m$ is neglected. Because,
when $n=m$, the expression at the R.H.S. become,
\begin{align*}
&\braket{\dudkx{n}}{\uu{n}}\braket{\uu{n}}{\dudky{n}}-\braket{\dudky{n}}{\uu{n}}\braket{\uu{n}}{\dudkx{n}} \\
=&\left[-\braket{\uu{n}}{\dudkx{n}} \right]\left[-\braket{\dudky{n}}{\uu{n}}\right]- \braket{\dudky{n}}{\uu{n}} \braket{\uu{n}}{\dudkx{n}}\\
=& 0,
\end{align*}
where the following equation at the second step is used,
\begin{align*}
\braket{\uu{n}}{\uu{n}}&=1 \\
\text{or, }\braket{\dudkx{n}}{\uu{n}}&+\braket{\uu{n}}{\dudkx{n}}=0\text{ [Taking derivative of both sides]}\\
\text{or, }\braket{\dudkx{n}}{\uu{n}}&=-\braket{\uu{n}}{\dudkx{n}}
\end{align*}
Again we have,
\begin{align*}
\pazocal{H}\ket{\uu{n}}&=E_n(\bold{k})\ket{\uu{n}} \\
\dXdY{\pazocal{H}}{k_x}\ket{\uu{n}}+\pazocal{H}\ket{\dudkx{\uu{n}}}&=\dXdY{E_n}{kx}\ket{\uu{n}}+E_n(\bold{k})\ket{\dudkx{\uu{n}}}\\
&\qquad\qquad \text{[Taking derivative of both sides w.r.t $k_x$]}\\
\bra{\uu{m}}\dXdY{\pazocal{H}}{k_x}\ket{\uu{n}}+\bra{\uu{m}}\pazocal{H}\ket{\dudkx{n}}&=\dXdY{E_n}{k_x}\braket{\uu{m}}{\uu{n}} + E_n(\bold{k}) \braket{\uu{m}}{\dudkx{n}}\\
&\qquad\qquad\text{[Multiplying $\bra{\uu{m}}$ on the both sides] }\\
\bra{\uu{m}}\dXdY{\pazocal{H}}{k_x}\ket{\uu{n}}+E_m(\bold{k})\braket{\uu{m}}{\dudkx{n}}&=E_n(\bold{k})\braket{\uu{m}}{\dudkx{n}} \\
\braket{\uu{m}}{\dudkx{n}}&=-\frac{1}{E_m(\bold{k})-E_n(\bold{k})} \bra{\uu{m}}\dXdY{\pazocal{H}}{k_x}\ket{\uu{n}}
\end{align*}
Similarly we get,
\begin{align}
\braket{\uu{m}}{\dudkx{n}}&=-\frac{1}{E_m(\bold{k})-E_n(\bold{k})} \bra{\uu{m}}\dXdY{\pazocal{H}}{k_x}\ket{\uu{n}}
\nonumber\\
\braket{\dudkx{n}}{\uu{m}}&=\frac{1}{E_n(\bold{k})-E_m(\bold{k})} \bra{\uu{n}}\dXdY{\pazocal{H}}{k_x}\ket{\uu{m}}
\label{eq3}
\end{align}
Using Eq.\,\ref{eq3} in Eq.\,\ref{eq1}, we get,
\begin{equation}
\boxed{\Omega_{m,\bold{k}}=i\sum_{n\neq m}\frac{\squeeze{\uu{m}}{\dXdY{\pazocal{H}}{k_x}}{\uu{n}}\squeeze{\uu{n}}{\dXdY{\pazocal{H}}{k_y}}{\uu{m}}-(k_x\leftrightarrow k_y)}{(E_m(\bold{k})-E_n(\bold{k}))^2}}
\end{equation}

 \subsection*{Gauge-independent equations for Berry-curvature for Bogoliubov-Valatin transformation}
 The Bogoliubov-Valatin transformation is discussed in the section Sec.\,\ref{sec2.4} and Appendix.\,\ref{appendixB}.
 The expression of the Berry curvature is given by\cite{BogoliubovBerryCurvature},
\begin{equation}
\Omega_{n\bk}=i\epsilon_{\mu\nu}\left[\sigma_3 \dXdY{T^\dagger_\bk}{k_\mu}\sigma_3\dXdY{T_\bk}{k_\nu}\right]_{nn} \qquad \left(n=1,2,3,...,2N\right)
\end{equation}
For 2D-lattice, it can be re-written as,
\begin{align*}
\Omega_{n\bk}&=i\left[\sigma_3 \dXdY{T^\dagger_\bk}{k_x}\sigma_3\dXdY{T_\bk}{k_y}-\sigma_3 \dXdY{T^\dagger_\bk}{k_y}\sigma_3\dXdY{T_\bk}{k_x}\right]_{nn} \\
&=i\sum_m \left(\sigma_3 \dXdY{T^\dagger_\bk}{k_x}\right)_{nm}\left(\sigma_3\dXdY{T_\bk}{k_y}\right)_{mn}-i\sum_m \left(\sigma_3 \dXdY{T^\dagger_\bk}{k_y}\right)_{nm}\left(\sigma_3\dXdY{T_\bk}{k_x}\right)_{mn}\\
&=i\sum_m (\sigma_3)_{nn} \dXdY{(T^\dagger_\bk)_{nm}}{k_x}(\sigma_3)_{mm}\dXdY{(T_\bk)_{mn}}{k_y}-i\sum_m (\sigma_3)_{nn} \dXdY{(T^\dagger_\bk)_{nm}}{k_y}(\sigma_3)_{mm}\dXdY{(T_\bk)_{mn}}{k_x}
\end{align*}
Using the definitions  $\uuu{n}{m}=(T_{\bold{k}})_{mn}$ and $\uuu{n}{m}^*=(T^\dagger_{\bold{k}})_{mn}$ as in the last subsection of Appendix.\,\ref{appendixB}, we get,
\begin{align*}
\Omega_{n,\bold{k}}=i\sum_{m} (\sigma_3)_{nn} \dXdY{\uuu{n}{m}^*}{k_x} (\sigma_3)_{mm} \dXdY{\uuu{n}{m}}{k_y}-i\sum_m (\sigma_3)_{nn} \dXdY{\uuu{n}{m}^*}{k_y}(\sigma_3)_{mm} \dXdY{\uuu{n}{m}}{k_x}
\end{align*}
\begin{equation}
\therefore \Omega_{n,\bold{k}}=i(\sigma_3)_{nn} \bra{\dudkx{n}}\sigma_3\ket{\dudky{n}}-i(\sigma_3)_{nn} \bra{\dudky{n}}\sigma_3\ket{\dudkx{n}} 
\end{equation}
Using the completeness-relation\,\ref{Geq4a},
\begin{align}
\Omega_{n,\bold{k}}&=i(\sigma_3)_{nn} \bra{\dudkx{n}}\sigma_3\mathbb{I}\ket{\dudky{n}}-i(\sigma_3)_{nn} \bra{\dudky{n}}\sigma_3\mathbb{I}\ket{\dudkx{n}}
\nonumber\\
\text{Or,}\,\Omega_{n,\bold{k}}&=\sum_{m\neq n} i (\sigma_3)_{nn}(\sigma_3)_{mm}\squeeze{\dudkx{n}}{\sigma_3}{\uu{m}}\squeeze{\uu{m}}{\sigma_3}{\dudky{n}}-(k_x\leftrightarrow k_y)
\label{EQ1}
\end{align}
The terms $n=m$ are zero and it can be shown using orthonormality relation Eq.\,\ref{eq2.71V2}.
Using Eq.\,\ref{Geq1},
\begin{align*}
&\quad(\sigma_3 H_\bk)\ket{\uu{n}}=(\sigma_3\epsilon_\bk) \ket{\uu{n}}\\
&\text{or, }\sigma_3 \dXdY{\pazocal{H}_\bk}{k_x}\ket{\uu{n}}+\sigma_3 \pazocal{H}_\bk \ket{\dudkx{n}}= \left(\sigma_3 \dXdY{\epsilon_\bk}{k_x}\right)_{nn} \ket{\uu{n}} +(\sigma_3 \epsilon_\bk)_{nn} \ket{\dudkx{n}}\\
&\qquad\qquad\qquad\qquad\qquad\qquad\text{ [Taking derivative of both sides w.r.t. $k_x$]}\\
&\text{or, }\squeeze{\uu{m}}{\dXdY{\pazocal{H}_\bk}{k_x}}{\uu{n}}+\squeeze{\uu{m}}{\pazocal{H}_\bk}{\dudkx{n}}=\left(\sigma_3 \dXdY{\epsilon_\bk}{k_x}\right)\squeeze{\uu{m}}{\sigma_3}{\uu{n}}\\
 &\qquad\qquad\qquad\qquad\qquad\qquad \qquad\qquad\qquad\qquad\qquad\qquad   +(\sigma_3\epsilon_\bk)_{nn} \squeeze{\uu{m}}{\sigma_3}{\dudkx{n}}\\
& \qquad\qquad\qquad\qquad\qquad\qquad \text{[Multiplying $\bra{\uu{m}}\sigma_3$ on both sides]}\\
&\text{or, } \squeeze{\uu{m}}{\dXdY{H_\bk}{k_x}}{\uu{n}}+(\sigma_3 \epsilon_\bk)_{mm} \squeeze{\uu{m}}{\sigma_3}{\dudkx{n}}=(\sigma_3\epsilon_\bk)_{nn} \squeeze{\uu{m}}{\sigma_3}{\dudkx{n}}
\end{align*}
\begin{equation}
\therefore\squeeze{\uu{m}}{\sigma_3}{\dudkx{n}}=-\frac{1}{(\sigma_3\epsilon_\bk)_{mm}-(\sigma_3\epsilon_\bk)_{nn}}\squeeze{\uu{m}}{\dXdY{\pazocal{H}_k}{k_x}}{\uu{n}
\label{EQ2}
}
\end{equation}
Similarly,
\begin{equation}
\squeeze{\dudkx{n}}{\sigma_3}{\uu{m}}=\frac{1}{(\sigma_3\epsilon_\bk)_{nn}-(\sigma_3\epsilon_\bk)_{mm}}\squeeze{\uu{n}}{\dXdY{\pazocal{H}_k}{k_x}}{\uu{m}}
\label{EQ3}
\end{equation}
  Using Eq.\,\ref{EQ3} and Eq.\,\ref{EQ2} in Eq.\,\ref{EQ1}, we get,
 \begin{equation}
 \boxed{
 \Omega_{n,\bold{k}}=\sum_{m\neq n} i (\sigma_3)_{nn}(\sigma_3)_{mm} \left[\frac{\squeeze{\uu{n}}{\dXdY{\pazocal{H}_\bk}{k_x}}{\uu{m}}\squeeze{\uu{m}}{\dXdY{\pazocal{H}_\bk}{k_y}}{\uu{n}}-(k_x\leftrightarrow k_y)}{((\sigma_3\epsilon_\bk)_{nn}-(\sigma_3\epsilon_\bk)_{mm})^2}\right]
 }
 \end{equation}

\chapter{Appendix D : Derivation of thermal Hall conductivity}\label{appendixD}

The out-line of the derivation of thermal Hall conductivity is provided in the section Sec.\,\ref{sec2.4.4}. In this appendix, the details of the derivation of some specific quantities are given.

\subsection*{Equation of motion of wave-packet}
The equation of motion of a wave-packet $\ket{W_0}=\int d^3\bk\, w(\bk,t)\ket{\psi_n(\bk)}$ as in equation Eq.\,\ref{wavepacket} is derived by using Lagrangian mechanics.
$w(\bk,t)=|w(\bk,t)| \exp(i\theta(\bk,t))$ is the envelope function such that the peak of the wave-packet in the real and momentum space are $\bold{r}_c$ and $\bold{k}_c$ respectively. Thus $|w(\bk,t)|^2$ can be approximated as $\delta(\bk-\bk_c)$ and so for an arbitrary function $f(\bk)$,
\begin{equation}
    \int d^3\bk |w(\bk,t)|^2 f(\bk)=f(\bk_c),
    \label{eqD1}
\end{equation}

The Lagrangian of the system is given by,
\begin{align}
    \pazocal{L}&=\squeezeD{W_0}{i\hbar \frac{\partial}{\partial t}-\pazocal{H}}{W_0}\nonumber\\
    &=\squeezeD{W_0}{\frac{\partial}{\partial t}}{W_0}-
    \squeezeD{W_0}{\pazocal{H}_{\text{lat}}+U(\bold{r})}{W_0}\qquad
    \left[\pazocal{H}_{\text{lat}}\,\text{is the Hamiltonian of the lattice system}\right]\nonumber\\
    &=\left[i\hbar \underbrace{\int d^3\bk |w(\bk,t)|\frac{\partial|w(\bk,t)|}{\partial t}}_{0}
    -\hbar\int d^3\bk |w(\bk,t)|^2 \frac{\partial \theta(\bk,t)}{\partial t}\right]-\left[\epsilon_n(\bk_c)+U(\bold{r}_c)\right]
    \nonumber\\
    &=\hbar\frac{\partial \theta(\bk_c,t)}{\partial t}-
    \left[\epsilon_n(\bk_c)+U(\bold{r}_c)\right]
    \nonumber\\
    &=-\hbar \left[-\dot{\bk}_c\cdot\boldsymbol{\nabla}_{\bk_c}\theta(\bk_c,t)+\frac{d\theta_c(\bk_c,t)}{dt}\right]-
    \left[\epsilon_n(\bk_c)+U(\bold{r}_c)\right]
    \nonumber\\
    &=-\hbar \left[\dot{\bk}_c\cdot\left(\bold{r}_c-\bold{A}_n(\bk_c)\right)+\frac{d\theta_c(\bk_c,t)}{dt}\right]-
    \left[\epsilon_n(\bk_c)+U(\bold{r}_c)\right]\qquad\left[\text{Using equation Eq.\,\ref{peak}}\right]
    \nonumber\\
    &=\hbar\bk_c\cdot\dot{\bold{r}}_c+\hbar\dot{\bk}_c\cdot\bold{A}_n(\bk_c)-
    \left[\epsilon_n(\bk_c)+U(\bold{r}_c)\right]+\frac{d\left(\theta_c(\bk_c,t)+\bold{r}_c\cdot\bk_c\right)}{dt}
    \nonumber\\
    &=\hbar\bk_c\cdot\dot{\bold{r}}_c+\hbar\dot{\bk}_c\cdot\bold{A}_n(\bk_c)-
    \left[\epsilon_n(\bk_c)+U(\bold{r}_c)\right]
    \qquad\left[\text{Omitting total time-derivative term due to gauge invariance}\right]
    \nonumber\\
\end{align}
One of the equation of motions,
\begin{align}
    &\frac{d}{dt}\left(\boldsymbol{\nabla}_{\dot{\bold{r}}_c}\pazocal{L}\right)=\boldsymbol{\nabla}_{\bold{r}_c}\pazocal{L}
    \nonumber\\
    &\therefore\hbar\dot{\bk}_c=-\boldsymbol{\nabla}_{\bold{r}_c}U(\bold{r}_c)
    \label{eqD3}
\end{align}
Another equation of motion,
\begin{align}
    &\frac{d}{dt}\left(\boldsymbol{\nabla}_{\dot{\bk}_c}\pazocal{L}\right)=\boldsymbol{\nabla}_{\bk_c}\pazocal{L}
    \nonumber\\
    &\therefore+\hbar \frac{d\bold{A}_n(\bk_c)}{dt}=\hbar \dot{\bold{r}}_c-\boldsymbol{\nabla}_{\bk_c}\epsilon_n(\bk_c)+\hbar \boldsymbol{\nabla}_{\bk_c}(\dot{\bk}_c\cdot \bold{A}_n(\bk_c))
    \nonumber\\
    &\text{or,}\quad\dot{\bold{r}}_c=\frac{1}{\hbar}\boldsymbol{\nabla}_{\bk_c}\epsilon_n(\bk_c)- \boldsymbol{\nabla}_{\bk_c}(\dot{\bk}_c\cdot \bold{A}_n(\bk_c))+\frac{d\bold{A}_n(\bk_c)}{dt}
    \nonumber\\
    &\text{or,}\quad\dot{\bold{r}}_c=\frac{1}{\hbar}\boldsymbol{\nabla}_{\bk_c}\epsilon_n(\bk_c)- \boldsymbol{\nabla}_{\bk_c}(\dot{\bk}_c\cdot \bold{A}_n(\bk_c))+\left(\dot{\bk}_c\cdot\boldsymbol{\nabla}_{\bk_c}\right)\bold{A}_n(\bk_c)
    \nonumber\\
    &\text{or,}\quad\dot{\bold{r}}_c=\frac{1}{\hbar}\boldsymbol{\nabla}_{\bk_c}\epsilon_n(\bk_c)-\dot{\bk}_c\cdot(\boldsymbol{\nabla}_{\bk_c}\times \bold{A}_n(\bk_c))
    \qquad\left[\text{Using vector calculus identity}\right]
    \nonumber\\
    &\text{or,}\quad\dot{\bold{r}}_c=\frac{1}{\hbar}\boldsymbol{\nabla}_{\bk_c}\epsilon_n(\bk_c)+\frac{1}{\hbar}\boldsymbol{\nabla}_{\bold{r}_c} U(\bold{r}_c)\times\Omega_n(\bk_c)
    \qquad\left[\text{Using equation Eq.\,\ref{eqD3}}\right]
    \label{eqMotion}
\end{align}

\subsection*{Thermal Orbital Magnetization due to rotational current along the edge}

\begin{figure}[H]
\centering
\includegraphics[width=0.5\textwidth]{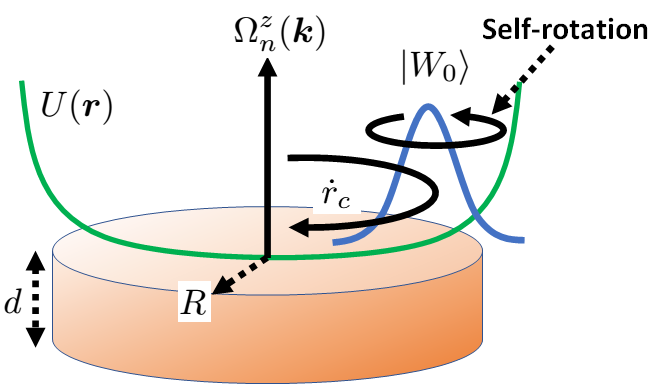} 
\caption{Schematic showing the rotational motion of the wave-packet. There are two contributions to rotational motion, one from self-rotation and another from rotation around the edges due to non-zero Berry-curvature.}
\label{WP}
\end{figure}

The origin of the magnetization is schematically shown in the figure Fig.\,\ref{WP}. The thermal orbital magnetization due to motion of wave-packets along the edge is given by,
\begin{align}
	     M_Q^\text{edge}&=
	     \frac{1}{(\pi\,R^2d)}\sum_n\int\int\int\int \frac{d^3\bk}{(2\pi)^3}
	     \underbrace{dz\,dr\, Rd\theta}_{\text{Volume element at edge}}
	     \underbrace{\rho(\epsilon_n(\bk)+U(\bold{r}))}_{\text{Bose-Einstein dist.}}
	     \underbrace{\frac{1}{2}\bold{R}\times\boldsymbol{J}_{Q}^n(\bk,\bold{r})}_{\substack{\text{thermal orbital magnetization} \\ \text{of a wave-packet at edge}}}
	     \label{eqD5}
	 \end{align}
	 where thermal orbital magnetization from a wave-packet at momentum $\bk$ and at position $\bold{r}$ and at $n$-th band is given by (using equation of motion Eq.\,\ref{eqMotion}),
	 \begin{align}
	     \frac{1}{2}\bold{R}\times\bold{J}^n_Q(\bk,\bold{r})&=\frac{1}{2}\bold{R}\times\left[\left(-\frac{1}{\hbar}\boldsymbol{\nabla}U(\bold{r})\times \Omega_n(\bk)\right)(\epsilon_n(\bk)+U(\bold{r})-\mu)\right]\nonumber\\
	     &=-\frac{1}{2}\frac{1}{\hbar}R\frac{\partial U(r)}{\partial r} \Omega_n(\bk)(\epsilon_n(\bk)+U(r)-\mu),
	     \label{eqD6}
	 \end{align}
	 where $\mu$ is the chemical potential. Using the fact that,
	 \begin{align}
	     U(r)&=0,\qquad\text{when,}\, r<R
	     \nonumber\\
	     &=\infty,\qquad\text{when,}\, r\rightarrow \infty
	 \end{align}
	 and using equations\,\ref{eqD5} and \ref{eqD6} and using $\frac{1}{V}\sum_{\bk}=\int\frac{d^3\bk}{(2\pi)^3}$ we get,
	 \begin{equation}
	     M_Q^{\text{edge}}=-\frac{1}{V\hbar}\sum_{n,\bk}\Omega_n(\bk)\left[\int^\infty_{\epsilon_n(\bk)}(\epsilon-\mu)\rho(\epsilon)d\epsilon\right]
	 \end{equation}

\chapter{Appendix D : Derivation for spin-Hall-noise-spectroscopy }\label{appendixE}

The out-line of the derivation of spin Hall noise spectrum is provided in the section Sec.\,\ref{sec2.5.5}. In this appendix, the details of the derivation of some specific equations are given.

\subsection*{spin-Current operator in metal}

  The interaction between metal and magnet is considered as isotropic Heisenberg interaction,
    \begin{equation}
    \hat{H}_c=-\eta a^3\sum_{p}\delta_{p_x,1}\hat{\bold{s}}(x=0,\bold{R}_p)\cdot\hat{\bold{S}}_p,
    \,\,\text{where,}\,\,
    \hat{\bold{s}}(\bold{R})=\frac{1}{2}\hat{\psi}_s^\dagger(\bold{R}) \boldsymbol{\tau}_{ss'} \hat{\psi}_s(\bold{R}).
    \end{equation}
    $\hat{\bold{s}}(\bold{R})$ and $\hat{\bold{S}}_p$ are the local spin-density operator in metal and spin-operator in magnet respectively.
    $\hat{\psi}_s(\bold{R})$ is the electronic field operator in metal and $\boldsymbol{\tau}_{ss'}$ is the vector representing Pauli-matrices.
    Moreover $\eta$ is the exchange constant and $a$ is the lattice scaling in the metal.

    Considering the spin-quantization axis in metal (axis for $\hat{s}^z$) aligned along the spin-polarization axis ($\boldsymbol{\sigma}=y$), the Hamiltonian can be explicitly written as,
    \begin{equation}
        \hat{H}_c=-\eta a^3\sum_{p}\delta_{p_x,1}
        \left(\hat{s}_p^y\hat{S}_p^x+
        \hat{s}_p^z\hat{S}_p^y+
        \hat{s}_p^x\hat{S}_p^z\right).
    \end{equation}
    The spin-current operator for $y$-component of spin is\,\cite{Transport},
    \begin{align}
        \hat{I}^y_s=\frac{d\hat{s}^z_q}{dt}
        &=i\left[\hat{H}_c,\hat{s}^z_q\right]
        \nonumber\\
        &=\frac{i\eta a^3}{2} \sum_{p} \delta_{p_x,1} \left[\hat{\psi}_\uparrow^\dagger\hat{\psi}_\downarrow (\hat{S}_p^z-i\hat{S}_p^x)
        -\hat{\psi}_\downarrow^\dagger\hat{\psi}_\uparrow (\hat{S}_p^z-i\hat{S}_p^x)\right]
        \nonumber\\
        &=\frac{i\eta a^3}{2} \hat{T}_y +\text{h.c.},
    \end{align}
    where, $\hat{T}_y=\sum_p \delta_{p_x,1} \hat{\psi}^\dagger_{\uparrow}(\bold{R}_p) \hat{\psi}_\downarrow(\bold{R}_p)
        (\hat{S}_p^z-i\hat{S}_p^x)$.

    \subsection*{Derivation for noise spectrum for spin current in metal}
    
    The noise spectrum for $y$-component of spin-current in metal is given as,
    \begin{align}
        \pazocal{I}_s^{y}(\Omega)&=\int dt
        \left\langle \hat{I}_s^{z}(0) \hat{I}_s^{z}(t)\right\rangle\exp(-i\Omega t)
        \nonumber\\
        &\approx\left(\frac{\eta a^3}{2}\right)^2\int dt
        \left[\underbrace{\left\langle \hat{T}_y^\dagger(0) \hat{T}_y(t)\right\rangle}_{T_1}+\underbrace{\left\langle \hat{T}_y(0) \hat{T}_y^\dagger(t)\right\rangle}_{T_2}\right]\exp(-i\Omega t).
        \nonumber\\
        &\left[\text{Taking the terms with lowest non-trivial order in $\eta$}\right]
        \label{eqE4}
    \end{align}
    The term $T_1$ can be explicitly given by,
    \begin{align}
        T_1&=\sum_{p,q}\delta_{p_x,1}\delta_{q_x,1}
        \left\langle \hat{\psi}_{\downarrow}^\dagger(\bold{R}_p)
        \hat{\psi}_{\uparrow}(\bold{R}_p)
        \hat{\psi}^\dagger_{\uparrow}(\bold{R}_q,t)
        \hat{\psi}_{\downarrow}(\bold{R}_q,t)
        \right\rangle
        \nonumber\\
        &\qquad\qquad\left[
        \left\langle
        \hat{S}_p^z\hat{S}_q^z(t)
        \right\rangle
        +\left\langle
        \hat{S}_p^x\hat{S}_q^x(t)
        \right\rangle
        -i\left\langle
        \hat{S}_p^z\hat{S}_q^x(t)
        \right\rangle
        +i\left\langle
        \hat{S}_p^x\hat{S}_q^z(t)
        \right\rangle
        \right]
        \nonumber\\
        &=\sum_{p,q}\delta_{p_x,1}\delta_{q_x,1}
        \left[
        \underbrace{
        \left\langle \hat{\psi}_{\downarrow}^\dagger(\bold{R}_p)
        \hat{\psi}_{\uparrow}(\bold{R}_p)
        \right\rangle
        \left\langle    
        \hat{\psi}^\dagger_{\uparrow}(\bold{R}_q,t)
        \hat{\psi}_{\downarrow}(\bold{R}_q,t)
        \right\rangle
        }_0
        +
        \left\langle \hat{\psi}_{\downarrow}^\dagger(\bold{R}_p)
        \hat{\psi}_{\downarrow}(\bold{R}_q,t)
        \right\rangle
        \left\langle    
        \hat{\psi}_{\uparrow}(\bold{R}_p)
        \hat{\psi}^\dagger_{\uparrow}(\bold{R}_q,t)
        \right\rangle
        \right]
        \nonumber\\
        &\qquad\qquad\left[
        \left\langle
        \hat{S}_p^z\hat{S}_q^z(t)
        \right\rangle
        +\left\langle
        \hat{S}_p^x\hat{S}_q^x(t)
        \right\rangle
        -i\left\langle
        \hat{S}_p^z\hat{S}_q^x(t)
        \right\rangle
        +i\left\langle
        \hat{S}_p^x\hat{S}_q^z(t)
        \right\rangle
        \right]
        \nonumber\\
        &\qquad\qquad\qquad\left[\text{Taking fully contracted terms using Wick's theorem\,\cite{book}}\right]
        \nonumber\\
        &=\sum_{p,q}\delta_{p_x,1}\delta_{q_x,1}
        iG^{<}(\bold{R}_{qp},t)
        (-i)G^{>}(\bold{R}_{pq},-t)
        \left[
        \left\langle
        \hat{S}_p^z\hat{S}_q^z(t)
        \right\rangle
        +\left\langle
        \hat{S}_p^x\hat{S}_q^x(t)
        \right\rangle
        -i\left\langle
        \hat{S}_p^z\hat{S}_q^x(t)
        \right\rangle
        +i\left\langle
        \hat{S}_p^x\hat{S}_q^z(t)
        \right\rangle
        \right]
        \nonumber\\
        &\left[\text{where } G^{>} \text{ and } G^{<} \text{ are advanced and retarded Green's function and } \bold{R}_{pq}=\bold{R}_p-\bold{R}_q\right]
        \nonumber\\
        &=4\sum_{p,q}\delta_{p_x,1}\delta_{q_x,1}
        \int\frac{d\nu}{2\pi} e^{i\nu t}
        \int\frac{d\omega}{2\pi} e^{i\omega t}
        \int\frac{d\rho}{2\pi} e^{i\rho t}
        \int \frac{d^3 \bold{k}}{(2\pi)^3} e^{-i\bold{k}_{\perp}\cdot \bold{R}_{pq}}
        \int \frac{d^3 \bold{k}'}{(2\pi)^3}
        e^{i\bold{k}^\prime_{\perp}\cdot \bold{R}_{qp}}
        \nonumber\\
        &
        G^{<}(\bold{k}',\omega)
        G^{>}(\bold{k},\rho)
        \left[
        i\pazocal{S}^{zz}_{pq}(\nu)
        +i\pazocal{S}^{xx}_{pq}(\nu)
        +\pazocal{S}^{zx}_{pq}(\nu)
        -\pazocal{S}^{xz}_{pq}(\nu)
        \right],
        \label{eqE5}
    \end{align}
    where $\bold{k}_\perp=(k_y,k_z)$ and $\pazocal{S}_{pq}^{\alpha\beta}$ is the \gls{dssf} defined in equation Eq.\,\ref{Eq:DSSF}. Moreover the Fourier transformations with periodic boundary condition along $y-z$ direction and open boundary condition along $x$-direction for Green's functions are defined as,
    \begin{equation*}
        G^{\alpha}(\bold{R}_{pq},t)
        =\int \frac{d^2\bold{k}_\perp}{(2\pi)^2}
        \frac{d k_x}{(\pi)^2}
        e^{i\bold{k}_\perp\cdot\bold{R}_{pq}} \cos(k_x R^x_{pq})
        \int \frac{d\omega}{2\pi}e^{i\omega t}
        G^{\alpha}(\bold{k},\omega).
    \end{equation*}
    In a similar way, the $T_2$ term in equation Eq.\,\ref{eqE4} can be derived and given by,
    \begin{align}
        T_2&=4\sum_{p,q}\delta_{p_x,1}\delta_{q_x,1}
        \int\frac{d\nu}{2\pi} e^{i\nu t}
        \int\frac{d\omega}{2\pi} e^{i\omega t}
        \int\frac{d\rho}{2\pi} e^{-i\rho t}
        \int \frac{d^3 \bold{k}}{(2\pi)^3} e^{-i\bold{k}_{\perp}\cdot \bold{R}_{pq}}
        \int \frac{d^3 \bold{k}'}{(2\pi)^3}
        e^{i\bold{k}^\prime_{\perp}\cdot \bold{R}_{qp}}
        \nonumber\\
        &
        G^{<}(\bold{k}',\omega)
        G^{>}(\bold{k},\rho)
        \left[
        +i\pazocal{S}^{zz}_{pq}(\nu)
        +i\pazocal{S}^{xx}_{pq}(\nu)
        -\pazocal{S}^{zx}_{pq}(\nu)
        +\pazocal{S}^{xz}_{pq}(\nu)
        \right].
        \label{eqE6}
    \end{align}
    Using equations Eq.\,\ref{eqE4}, Eq.\,\ref{eqE5}, Eq.\,\ref{eqE6}, we get,
    \begin{align}
        \pazocal{I}_s^y(\Omega)&=2i(\eta a^3)^2
        \int \frac{d\nu}{2\pi}
        \int \frac{d\omega}{2\pi}
        \int \frac{d^3 \bold{k}}{(2\pi)^3}
        \int \frac{d^3 \bold{k}'}{(2\pi)^3}
        \sum_{p,q} \delta_{p_x,1} \delta_{q_x,1}
        G^{>}(\bold{k},\omega+\nu-\Omega)
        G^{<}(\bold{k}',\omega)
        \nonumber\\
        &\qquad\qquad\qquad\qquad\qquad\qquad\qquad\qquad\qquad
        e^{-i(\bold{k}_\perp-\bold{k}'_\perp)\cdot (\bold{R}_p-\bold{R}_q)}
        \left[\pazocal{S}_{pq}^{xx}
        +\pazocal{S}_{pq}^{zz}\right].
        \label{EqE7}
    \end{align}
    The retarded and advanced Green's function are related to spectral function respectively as\,\cite{KaiSunLectureNote},
    \begin{align}
        G^{<}(\bold{k},\omega)&=A_\bold{k}(\omega)[1-f(\omega)]
        \nonumber\\
        G^{>}(\bold{k},\omega)&=A_\bold{k}(\omega)f(\omega),
        \label{EqE8}
    \end{align}
    where for non-interacting system the spectral function $A_k(\omega)=2\pi\delta(\omega-\epsilon_\bk/\hbar)$ and $f(\omega)$ is the Fermi-Dirac distribution. Using equation Eq.\,\ref{EqE8} in equation Eq.\,\ref{EqE7} we get,
    \begin{align}
        \pazocal{I}_s^y(\Omega)&=2i(\eta a^3)^2
        \int \frac{d\nu}{2\pi}
        \int \frac{d\omega}{2\pi}
        \int \frac{d^3 \bold{k}}{(2\pi)^3}
        \int \frac{d^3 \bold{k}'}{(2\pi)^3}
        \sum_{p,q} \delta_{p_x,1} \delta_{q_x,1}
        A_\bold{k}(\omega+\nu-\Omega)f(\omega+\nu-\Omega)
        A_{\bold{k}'}(\omega)f(\omega)
        \nonumber\\
        &\qquad\qquad\qquad\qquad\qquad\qquad\qquad\qquad\qquad
        e^{-i(\bold{k}_\perp-\bold{k}'_\perp)\cdot (\bold{R}_p-\bold{R}_q)}
        \left[\pazocal{S}_{pq}^{xx}
        +\pazocal{S}_{pq}^{zz}\right]
        \nonumber\\
        &\approx 2i(\eta a^3)^2
        \int \frac{d\nu}{2\pi}
        \int \frac{d\omega}{2\pi}
        \int \frac{d^3 \bold{k}}{(2\pi)^3}
        \int \frac{d^3 \bold{k}'}{(2\pi)^3}
        \sum_{p,q} \delta_{p_x,1} \delta_{q_x,1}
        A_\bold{k}(\mu)f(\omega+\nu-\Omega)
        A_{\bold{k}'}(\mu)f(\omega)
        \nonumber\\
        &\qquad\qquad\qquad\qquad\qquad\qquad\qquad\qquad\qquad
        e^{-i(\bold{k}_\perp-\bold{k}'_\perp)\cdot (\bold{R}_p-\bold{R}_q)}
        \left[\pazocal{S}_{pq}^{xx}
        +\pazocal{S}_{pq}^{zz}\right]
        \nonumber\\
        &\left[\text{Assuming electronic density of states do not vary near chemical potential }\mu.\text{thus,}\,A_\bold{k}(\omega)\approx A_{\bold{k}}(\mu)\right]
        \nonumber\\
        &= 2i(\eta a^3)^2
        \int \frac{d\nu}{(2\pi)^2}
        \int \frac{d^3 \bold{k}}{(2\pi)^3}
        \int \frac{d^3 \bold{k}'}{(2\pi)^3}
        \sum_{p,q} \delta_{p_x,1} \delta_{q_x,1}
        A_\bold{k}(\mu)
        A_{\bold{k}'}(\mu)
        \frac{\Omega-\nu}{1-e^{-\beta\hbar(\Omega-\nu)}}
        \nonumber\\
        &\qquad\qquad\qquad\qquad\qquad\qquad\qquad\qquad
        e^{-i(\bold{k}_\perp-\bold{k}'_\perp)\cdot (\bold{R}_p-\bold{R}_q)}
        \left[\pazocal{S}_{pq}^{xx}
        +\pazocal{S}_{pq}^{zz}\right]
        \qquad
        \left[\text{Integrating over variable }\omega\right]
        \nonumber\\
        &= 2i\left(\frac{\eta a^3 m k_F}{2\pi^2\hbar}\right)^2
        \int d\nu
        \sum_{p,q} \delta_{p_x,1} \delta_{q_x,1}
        \frac{\Omega-\nu}{1-e^{-\beta\hbar(\Omega-\nu)}}
        \left[\pazocal{S}_{pq}^{xx}
        +\pazocal{S}_{pq}^{zz}\right]
        \text{sinc}^2(k_F|\bold{R}_{pq}|)
        \nonumber\\
        &\left[\text{Integrating over }\bk\text{ and }\bk^\prime\text{,  using dispersion }\epsilon_\bk=\frac{\hbar k^2}{2m}. \,k_F= \text{Fermi-vector},\,m=\text{electron-mass}\right]
        \nonumber\\
        &\approx 2i\left(\frac{\eta a^3 m k_F}{2\pi^2\hbar}\right)^2
        \int d\nu
        \sum_{p} \delta_{p_x,1} 
        \frac{\Omega-\nu}{1-e^{-\beta\hbar(\Omega-\nu)}}
        \left[\pazocal{S}_{pp}^{xx}
        +\pazocal{S}_{pp}^{zz}\right]
        \\
        &\left[\because \text{for large $k_F,$ sinc$^2(k_F|\bold{R}_{pq}|)$ is non-zero when $p=q$} \right]
        \nonumber
    \end{align}

\renewcommand{\bibname}{References}
\bibliographystyle{apsrev4-1}

%

\addcontentsline{toc}{chapter}{References}

\end{document}